\documentclass[a4paper,11pt]{article} 
  
\usepackage[english]{babel}
\usepackage[a4paper]{geometry}
\usepackage{amsmath}
\usepackage{amsthm}
\usepackage{amssymb}
\usepackage{mathtools}
\usepackage{bbm}
\usepackage{color}
\usepackage{mathrsfs}
\usepackage{enumerate}
\usepackage{dsfont}
\usepackage[inline]{enumitem}

\usepackage{mathrsfs}

\usepackage[authormarkup=none,
deletedmarkup=xout
]{changes}

\definechangesauthor[name={Serena}, color={blue}]{Serena}

\definechangesauthor[name={Giulia}, color={purple}]{Giulia}

\usepackage[
           scr=esstix
           ]   
           {mathalpha}

\usepackage{hyperref}         

\usepackage{tikz}
\usetikzlibrary{fit}
\usetikzlibrary{shapes.geometric}
\usetikzlibrary{decorations.pathmorphing}

\tikzset{
point/.style={circle,fill=black,inner sep=1pt},
vertex/.style={circle,fill=black,inner sep=1.5pt},   
bvertex/.style={circle,fill=black,inner sep=2.8pt},
square/.style={regular polygon,regular polygon sides=4},
Bvertex/.style={circle,fill=black,inner sep=4pt}, 
specialEP/.style={rectangle,fill=white,draw,inner sep=3pt},  
whitevex/.style={circle,fill=white,draw, inner sep=2pt},
linelabel/.style={sloped,above,very near start, inner sep=1pt,execute at begin node=$\scriptstyle,execute at end node=$},
baseline=(current  bounding  box.center),doubled/.style={double distance= 1pt,line width=1.5pt},
th/.style={line width=0.5 pt, gray},  
med/.style={line width=1 pt}  
}

\xdefinecolor{myred}{rgb}{0.7,0,0.2} 
\xdefinecolor{mypurple}{rgb}{0.6,0.2,0.4} 
\xdefinecolor{myblue}{rgb}{0.259,0.2,0.6}   
\xdefinecolor{myorange}{rgb}{1,0.6,0.2}   
\definecolor{orange}{rgb}{1,0.5,0}
\xdefinecolor{purpleish}{cmyk}{0.75,0.75,0,0}

\def\bR{\mathbb{R}}

\def\bN{\mathbb{N}}
\def\NN{\mathbb{N}}
\def\bZ{\mathbb{Z}}

\def\cD{\mathcal{D}}

\def\cM{\mathcal{M}}

\def\cV{\mathcal{V}}

\def\cL{\mathcal{L}}

\def\cN{\mathcal{N}}
\def\cE{\mathcal{E}}
\def\cK{\mathcal{K}}
\def\cH{\mathcal{H}}
\def\cS{\mathcal{S}}

\def\eps{\varepsilon}
\def\ph{\varphi}

\def\wt{\widetilde}

\def\bbG{\mathbb{G}}

\def\indic{\hbox{\raise-2pt \hbox{\indbf 1}}}

\let\==\equiv

\let\io=\infty
\let\0=\noindent

\def\*{{\hfill\break\null\hfill\break}}

\def\ie{\hbox{\it i.e.\ }}

\def\red#1{\textcolor{red}{#1}} 
\def\blue#1{\textcolor{blue}{#1}}  

\def\tende#1{\,\vtop{\ialign{##\crcr\rightarrowfill\crcr
             \noalign{\kern-1pt\nointerlineskip}
             \hskip3.pt${\scriptstyle #1}$\hskip3.pt\crcr}}\,}
\def\otto{\,{\kern-1.truept\leftarrow\kern-5.truept\to\kern-1.truept}\,}

\def\Tr{{\rm Tr}}

\newtheorem{theorem}{Theorem}[section]  

\newtheorem{prop}[theorem]{Proposition}
\newtheorem{lemma}[theorem]{Lemma}

\numberwithin{equation}{section}

\def\tl#1{{\tilde{#1}}}
\def\be{\begin{equation}}
\def\ee{\end{equation}}

\newcommand{\hc}{\mbox{h.c.}}

\let\a=\alpha \let\b=\beta    \let\g=\gamma     \let\d=\delta     \let\e=\varepsilon
        \let\k=\kappa     \let\l=\lambda
                  \let\p=\pi        \let\r=\rho
\let\s=\sigma          \let\ph=\varphi   
   \let\o=\omega     
\let\G=\Gamma        \let\L=\Lambda    
             
\let\O=\Omega 

\newcommand{\A}{\beta}

\definecolor{lightblue}{rgb}{0, 0.33, 0.71}

\def\aa{\mathfrak{a}}
\def \sN{\mathscr N}
\def \sL{\mathscr L}
\def \Fock{\mathscr{F}}

\def \blue#1 {\textcolor{blue}{#1}}
\def \red#1 {\textcolor{red}{#1}}

\begin{document}

\title{A new upper bound on the specific free energy \\ of dilute Bose gases}

\author{Giulia Basti, Chiara Boccato, Serena Cenatiempo, Andreas Deuchert}

\date{\today}

\maketitle

\begin{abstract} 
We prove an upper bound for the free energy (per unit volume) of the dilute Bose gas in the thermodynamic limit, showing that the free energy at density $\varrho$ and inverse temperature $\beta$ differs from that of the non-interacting system by the correction term $4 \pi \aa (2 \varrho^2 - [\varrho-\varrho_{\mathrm{c}}(\beta)]_+^2)$. Here, $\aa$ denotes the scattering length of the interaction potential, $\varrho_{\mathrm{c}}(\beta)$ the critical density for Bose–Einstein condensation of the non-interacting gas, and $[\cdot]_+ = \max\{0,\cdot\}$. This result was previously established by Yin in \cite{Yin2010}. Our proof applies to a broader class of interaction potentials, yields a better rate, and we believe it has potential for further extensions.
\end{abstract}

\section{Introduction and main result}

We consider $\sN$  bosonic particles in the flat three-dimensional torus $\L_{\sL}$ with side length $\sL>0$. The interaction between the particles is described by a measurable, compactly supported function $V : [0,\infty) \to [0 , \infty]$, which satisfies $V(|\cdot|) \in L^3(\mathbb{R}^3)$. The Hamiltonian of the system is given by
\begin{equation}
	H_\sN = -\sum_{i=1}^\sN \Delta_i + \sum_{1 \leq i \leq j \leq \sN} V(d(x_i,x_j))\,.
	\label{eq:HL}
\end{equation}
Here $\Delta$ denotes the Laplacian on $\Lambda_\sL$ and $d(x,y)$ is the distance between two points $x,y \in \Lambda_\sL$. 
In the realization of $\L_\sL$ as the set $[0,\sL]^3$, $\Delta$ is the usual Laplacian with periodic boundary conditions and $d(x,y) = \min_{k \in \mathbb{Z}^3} | x- y - k\sL|$. Note also that $V(d(x,y)) = \sum_{k \in \mathbb{Z}^3} V(|x-y-k\sL|)$ holds provided the radius $R_0$ of the support of $V$ satisfies $R_0 < \sL/2$. The Hamiltonian acts on a dense domain in $L^2_{\mathrm{sym}}(\Lambda_\sL^\sN)$, the closed linear subspace of $L^2(\Lambda_\sL^\sN)$ consisting of those functions $\Psi(x_1,...,x_\sN)$ that are invariant under any permutation of the coordinates $x_1, ... , x_\sN \in \Lambda_\sL$. 

We are interested in investigating the equilibrium properties of the system at inverse temperature $\beta > 0$ and density $\varrho > 0$. In particular, we focus our attention on 
the {\it specific free energy} 
\begin{equation}\label{eq:freeenergy}
	f(\b, \varrho) = -\lim_{\substack{\sN,\, \sL \to \io\\ \sN/|\Lambda_\sL| = \varrho}} \frac{1}{\b |\Lambda_\sL|} \ln \left( \Tr[ \exp(-\b H_\sN )] \right)\,.
\end{equation}
The existence of the limit on the right-hand side, known as {\it thermodynamic limit}, and its independence of the boundary conditions imposed for the Laplacian in $H_\sN$ can be shown with standard techniques, see e.g. \cite{Robinson1971,Ruelle1999}.

In the case of a vanishing interaction potential, $V = 0$, the system is referred to as the {\it ideal Bose gas}, and its specific free energy $f_0(\b, \varrho)$ can be computed explicitly. It is given by:
\begin{equation} 
	f_0(\beta,\varrho) = \sup_{\mu \leq 0} \left\{ \frac{1}{\beta (2 \pi)^3} \int_{\mathbb{R}^3} \ln(1-\exp(-\beta(p^2-\mu))) \text{d}p + \mu \varrho \right\}\,.
	\label{eq:freeEnergyIdealGas}
\end{equation}
If the density $\varrho$ is larger than the critical density
\begin{equation}
	\varrho_{\mathrm{c}}(\beta) = \frac{1}{(2 \pi)^3} \int_{\mathbb{R}^3} \frac{1}{\exp(\beta p^2) - 1} \text{d}p
	\label{eq:criticalDensity}
\end{equation}
the supremum on the right-hand side of \eqref{eq:freeEnergyIdealGas} is attained at $\mu(\beta,\varrho) = 0$ and there is a macroscopic occupation of the zero momentum mode, \ie, Bose-Einstein condensation occurs. In contrast, if $\varrho < \varrho_{\mathrm{c}}(\beta)$ the chemical potential $\mu(\beta,\varrho)$ is strictly negative. Equivalently, assuming fixed density  $\varrho$ one sees from \eqref{eq:criticalDensity} that condensation takes place for inverse temperatures larger than
 the inverse critical temperature 		
		\begin{equation}
			\beta_{\mathrm{c}}(\varrho) = \frac{1}{4 \pi} \left( \frac{\varrho}{\zeta(3/2) } \right)^{-2/3}\,.
			\label{eq:criticalTemperatureIdealGas}
		\end{equation}
		 We also notice that the specific free energy in \eqref{eq:freeEnergyIdealGas} satisfies the scaling relation $f_0(\beta,\varrho) = \varrho^{5/3} f_0(\beta \varrho^{2/3},1)$. The only relevant parameter is therefore $\beta \varrho^{2/3}$.  

\smallskip

Let us now focus on the case $V \neq 0$, with $V$ satisfying the assumptions above \eqref{eq:HL}. 
Let $f : \mathbb{R}^3 \to \mathbb{R}$ be the solution to the scattering equation 
\begin{equation} \label{eq:scatteq}
	\Delta f(x) = \frac{V(|x|)}{2} f(x)
\end{equation}
subject to the boundary condition $\lim_{|x| \to \infty} f(x) = 1$ at infinity. For $|x|$ strictly larger than the radius $R_0$ of the support of $V$ we have $f(x) = 1 - \aa / |x|$, for a parameter $\mathfrak{a} > 0$ called the scattering length of $V$.  We will consider the dilute limit  $\varrho \aa^3$ small.

Our main result is the following upper bound for the specific free energy.

\begin{theorem}\label{thm:main}	
	Let $V : [0,\infty) \to [0,\infty]$ be a measurable, compactly supported function, which satisfies $V(|\cdot|) \in L^3(\mathbb{R}^3)$ and has scattering length $\aa$. Let  $f_0(\b, \varrho)$ and $\varrho_c(\b)$ denote the specific free energy and critical density of an ideal Bose gas, given in \eqref{eq:freeEnergyIdealGas} and \eqref{eq:criticalDensity}, respectively. Then, for any $c>0$, there exists $C>0$ such that the specific free energy of the interacting Bose gas in \eqref{eq:freeenergy} satisfies the bound
	\begin{equation}
		f(\b,\varrho) \leq f_0(\b, \varrho) + 4 \pi \aa (2 \varrho^2 - [\varrho - \varrho_{\mathrm{c}}(\beta) ]_+^2) + C \aa \varrho^{2} (\varrho\aa^3)^{1/7}
		\label{eq:mainResult}
	\end{equation} 
	for $\varrho \aa^3$ small enough and $\b \varrho^{2/3}\geq c$.
\end{theorem}
{\it Remarks.}
\begin{enumerate}[label=\roman*)]
\item A matching lower bound for the result in \eqref{eq:mainResult} was proved by Seiringer \cite{Sei-T} in 2008. His proof applies to repulsive interaction potentials that are measurable and integrable outside some ball. This, in particular, includes the case of a hard-core repulsion between the particles. An upper bound for nonnegative, bounded, piecewise continuous potentials with compact support was obtained by Yin \cite{Yin2010} in 2010. In this paper, we provide an alternative proof of the upper bound that applies to a broader class of interaction potentials and yields an explicit rate of convergence. Moreover, we believe that our approach, based on a conceptually simpler proof, has potential for further extensions.
\item For $\varrho \leq \varrho_{\mathrm{c}}(\beta)$ (or, equivalently, $\beta \leq \beta_{\mathrm{c}}(\varrho)$, \ie, in the non-condensed phase), the interaction energy in \eqref{eq:mainResult} is given to leading order by $8 \pi \aa \varrho^2$. This should be compared to its zero-temperature value, $4 \pi \aa \varrho^2$, see \cite{LiYng98}. The additional factor of two is an exchange effect resulting from the symmetrization of wave functions. It arises when particles occupy different one-particle wave functions, which is essentially always the case in the absence of condensation. However, particles within the condensate do not experience this effect, leading to the dependence of the interaction energy on $\varrho_{\mathrm{c}}(\beta)$.
\item At very large inverse temperatures, a more precise expansion of the free energy $f(\beta,\varrho)$  has recently been obtained in \cite{HHNST} (lower bound) and \cite{HHST} (upper bound). More precisely, for inverse temperatures satisfying $\beta \geq C \varrho^{-2/3} (\aa^3 \varrho)^{-1/3}$ for some constant $C > 0$ in the dilute limit, the authors establish an expansion of the free energy that resembles the Lee--Huang--Yang (LHY) formula \cite{LHY1957} (zero temperature case), previously proved in \cite{YauYin2009,BaCS2021} (upper bound) and \cite{FS,FS2} (lower bound). In \cite{HHNST,HHST} the authors find an additional temperature-dependent correction, given by the grand potential of an ideal gas of particles with a Bogoliubov dispersion relation. The analogue of the interaction energy in \eqref{eq:mainResult} is given in \cite{HHNST,HHST} by $4 \p \aa \varrho^2$ plus corrections on the LHY scale $\aa \varrho^2 (\aa^3 \varrho)^{1/2}$. This is because in their parameter regime $\varrho_{\mathrm{c}}(\beta)$ is very small when compared to $\varrho$. 
\end{enumerate}

In the following, we briefly sketch the main ideas of our proof of Theorem~\ref{thm:main}, which is obtained by exhibiting a suitable trial state, and comment on further connections to the literature. We apply a standard argument to show that it is sufficient to construct a trial state on a flat torus $\Lambda_L$ of fixed side length $L$ (independent of the thermodynamic limit) satisfying 
\begin{equation}
	L \gg \varrho^{-1/3} (\aa^3 \varrho)^{-1/3}	
	\label{eq:boxSize}
\end{equation}
in the dilute limit. The free energy of the trial state needs to satisfy an appropriate finite volume version of \eqref{eq:mainResult}. The condition on the size $L$ of the torus is related to ensuring that the remainder of a Riemann sum approximation of the integral in the definition of $f_0$ in \eqref{eq:freeEnergyIdealGas} is much smaller than the interaction energy in \eqref{eq:mainResult} (second term on the right-hand side). 

We first define an uncorrelated trial state with approximately $N=\rho L^3$ particles, 
and then add correlations. The construction takes place in the bosonic Fock space $\Fock$ over $L^2(\Lambda_L)$, which allows us to choose a state with a fluctuating particle number. This is justified by the equivalence of ensembles \cite{Ruelle1999}. Our trial state without correlations reads
	\begin{equation}
	\widetilde{\Gamma} = {\mathbb P}_{\xi_0} \otimes \Gamma_{\mathrm{G}} = \sum_{\alpha} \lambda_{\alpha} \,{\mathbb P}_{\psi_{\alpha}},
	\label{eq:proofSketch1}
\end{equation}
where the tensor product is taken with respect to the decomposition $\Fock \cong \Fock_0 \otimes \Fock_+$. Here, $\Fock_0$ and $\Fock_+$ denote the Fock spaces over the span of the constant function $\varphi_0= L^{-3/2}$ and its orthogonal complement, respectively. The vector $\xi_0$ describes a state with $N_0$ particles in $\varphi_0$, and ${\mathbb P}_{\xi_0}$ is the corresponding projection. By $\Gamma_{\mathrm{G}}$ we denote the grand canonical Gibbs state of the ideal gas, with the condensate removed by a partial trace. The right-hand side of \eqref{eq:proofSketch1} is an eigenfunction expansion of the state $\widetilde{\Gamma}$.

We use a Bijl--Dingle--Jastrow factor \cite{Bijl, Dingle, Jastrow1955} (or Jastrow factor for short) to add microscopic correlations to $\widetilde{\Gamma}$. This amounts to replacing each eigenfunction $\psi_{\alpha}$ in \eqref{eq:proofSketch1}, which can be chosen to have a definite particle number $N_{\alpha}$, by
\begin{equation}
	\frac{1}{Z_{\alpha}} \prod_{1 \leq i < j \leq N_{\alpha}} f(x_i-x_j) \psi_{\alpha}(x_1,...,x_{N_{\alpha}}).
	\label{eq:proofSketch2}
\end{equation}
Here, $Z_{\alpha}$ is a normalization constant and $f$ denotes the solution to a finite volume version of \eqref{eq:scatteq}. Trial states of the above form have been used in two and three space dimensions in \cite{DeuSeiYng2019, MaySei2020, DeuSei-20,CapDeu2024}. Earlier applications in the zero-temperature setting, where the state $\widetilde{\Gamma}$ is replaced by a suitable rank-one projection, can be found in \cite{Dyson1957,LiSeiYng2000}. For more recent applications in the same setting, we refer to \cite{BCOPS2023,FGJMO2024,BGOPS24}. In these references, computations with functions of the form \eqref{eq:proofSketch2} have been feasible either because the function $\psi_{\alpha}$ describes a pure condensate \cite{Dyson1957,LiSeiYng2000,BGOPS24} or because the volume of the torus is sufficiently small so that a perturbative expansion in powers of $w(x) = 1 - f(x)$ can be justified \cite{DeuSeiYng2019, MaySei2020,DeuSei-20,BCOPS2023,CapDeu2024,FGJMO2024}. In our setting, neither of these assumptions holds, due to the condition on $L$ in \eqref{eq:boxSize} and the form of $\widetilde{\Gamma}$ in \eqref{eq:proofSketch1}.

If one formally expands the expression in \eqref{eq:proofSketch2} in powers of $w$, one finds that it equals 
\begin{equation}
	\frac{1}{Z_{\alpha}} \bigg( 1 - \sum_{1 \leq i < j \leq N_{\alpha}} w(x_i-x_j) + \sum_{\substack{i<j, k<\ell \\ (i,j) \neq (k,\ell)}} w(x_i - x_j) w(x_k - x_\ell) - ... \bigg) \psi_{\alpha}(x_1,...,x_{N_{\alpha}}).
	\label{eq:proofSketch3}
\end{equation}
When written in second quantization, the second term in the bracket reads
\begin{equation}
	B = -\frac{1}{2 L^3} \sum_{p,u,v \in (2 \pi/L) \mathbb{Z}^3} \widehat{w}(p) a^*_{u+p} a^*_{v-p} a_u a_v.
	\label{eq:proofSketch4}
\end{equation}
Here, $a_p^*$ and $a_p$ denote the usual creation and annihilation operators of a particle in the plane wave state $\varphi_p(x) = L^{-3/2} e^{\mathrm{i}p \cdot x}$ and $\widehat{w}$ is the Fourier transform of $w$. In our case, only the first two terms in the bracket in \eqref{eq:proofSketch3} are relevant for an approximate computation of the energy. One could therefore replace \eqref{eq:proofSketch3} by
\begin{equation}
	\widetilde{Z}_{\alpha}^{-1} \exp\left( B \right) \psi_{\alpha}
	\label{eq:proofSketch5}
\end{equation}  
for some $\widetilde{Z}_{\alpha}$. The advantage of \eqref{eq:proofSketch5} is that it may be more amenable to computation than \eqref{eq:proofSketch3}. The main reason is that creation and annihilation operators in the second-quantized version $\mathcal{H}$ of our Hamiltonian in \eqref{eq:HL} act as derivatives on $B$\footnote{This follows from the canonical commutation relations and is made explicit by the representation of $a_p$, $a_p^*$ as $\frac{\partial}{\partial z_p}$, $z_p$, where $z_p$ denotes a complex variable, on the Segal--Bargmann space, see e.g. \cite[Chapter~1.6]{Foll1989}.}. Using this and the fact that the exponential function is preserved under differentiation, \eqref{eq:proofSketch5} may allow one to efficiently exploit cancellations between the numerator and the denominator in expressions of the form $\langle \exp(B) \psi_{\alpha}, \mathcal{H} \exp(B) \psi_{\alpha} \rangle/\widetilde{Z}_{\alpha}^2$. We note that this idea is also applicable in situations where perturbative expansions in powers of $w$ are not possible. However, putting it on a sound mathematical ground has proven challenging.

The first who made the above idea rigorous were Yau and Yin in \cite{YauYin2009}, where they obtained an upper bound for the ground state energy of the Bose gas with the correct LHY correction. Their approach was shortly thereafter extended by Yin \cite{Yin2010} to derive an upper bound for the free energy. While both papers are based on the above heuristics, the authors did not define their trial state in terms of the operator $\exp(B)$. Instead, they constructed a trial state by mimicking the action of $\exp(B)$ and applying suitable modifications, which constitute one of the main novelties of their work. 
These modifications include restrictions of the number and types of microscopic correlations that are incorporated in the trial state. The trial states in \cite{YauYin2009,Yin2010} are not quasi-free. That states in this class are insufficient to obtain the LHY correction follows from \cite{ErdSchlYau2008,NapReuSol2017}.

In the Gross--Pitaevskii (GP) limit, corresponding to the choice $L=(\rho \aa)^{-1/2}$,   where expansions in powers of $w$ are possible, the action of unitary operators resembling $\exp(B)$ (but where $B$ is cubic in creation and annihilation operators with non-zero momenta) was first  made rigorous in \cite{BocBreCeSchl2019,BBCS3}. This allowed the authors to compute the low-lying excitation spectrum of dilute Bose gases in this parameter regime. Using a second order truncation of $\exp(B)$, an upper bound for the free energy of the Bose gas in the GP limit for inverse temperature $\beta \sim \beta_{\mathrm{c}}$ that includes the influence of the Bogoliubov modes has been obtained in \cite{BocDeuSto2024}. See also \cite{CapDeu2024} for an alternative proof that applies to a broader class of interaction potentials including hard-core repulsions.

A novel approach to treating the cubic operator on larger boxes, allowing for an alternative proof of the result in \cite{YauYin2009}, was introduced in \cite{BaCS2021}; see also \cite{Basti2022} for a refined bound. To perform the computations, the authors multiplied their version of $B$
with a cut-off operator $\Theta$ that controls the number and type of excitations created by 
$\exp(B)$. This enabled them to provide a conceptually simpler proof, applicable to a broader class of interaction potentials, and to establish an explicit convergence rate. For a careful comparison of the trial states in \cite{YauYin2009,BaCS2021}, we refer to \cite[above Proposition 2.3]{BaCS2021}.

Extending this idea, the authors of \cite{HHST} used a unitary version of the exponential from \cite{BaCS2021} to prove an upper bound for $f(\beta,\varrho)$ at large inverse temperatures. In their approach the entropy, which is unitarily invariant, remains unaffected by the correlation structure. The computation of the energy is, however, more involved. This difficulty was overcome by working with a product of several exponentials, each of which adds correlations step by step. The fact that each exponential creates fewer excitations allowed them to apply Duhamel expansions to compute the energy approximately. In the recent preprint \cite{BOSAS}, the authors developed an alternative approach, defining a unitary version of $\exp(B)$ using a cut-off operator that locally controls the number of generated excitations in position space. This allowed them to obtain an upper bound for the ground state energy of the dilute Bose gas, including the third-order correction predicted in \cite{HuPi1959,Sa1959,Wu1959}. A related third-order expansion in the GP limit had previously been obtained in \cite{CarOlgSA2025}.

In this paper we follow the approach of \cite{BaCS2021,BocDeuSto2024} and use a trial state of the form   \begin{equation}
	\Gamma = \sum_{\alpha} \lambda_{\alpha} \mathbb P_{\Phi_{\alpha}}\,, \qquad  \Phi_{\alpha}= \frac{\widetilde{\mathcal{D}}_{\alpha} \psi_{\alpha}}{\Vert \widetilde{\mathcal{D}}_{\alpha} \psi_{\alpha} \Vert}
	\label{eq:trialStateIntro}
\end{equation}
with $\lambda_{\alpha}$, $\psi_{\alpha}$ in \eqref{eq:proofSketch1}. Here, 
\begin{equation}
	\widetilde{\mathcal{D}}_{\alpha} = \sum_{n=0}^{n_{\mathrm{c}}} \frac{\widetilde{D}_{\alpha}^n}{n!} \quad \text{ with } \quad \widetilde{D}_{\alpha} = -\frac{1}{2 L^3} \sum_{\substack{p \sim 1/\aa \\ u,v \in P_{\A} \cup \{ 0 \} }}  \widehat{w}(p) a^*_{u+p} a^*_{v-p} a_u a_v \Theta^{\alpha}_{p,u,v}.
	\label{eq:introCorrelationStructure}
	\end{equation}
The set $P_{\A}$ is chosen such that it contains the momenta of the majority of the particles described by $\Gamma_{\mathrm{G}}$. The parameter $n_{\mathrm{c}}$ is chosen sufficiently small to control the influence of the correlation structure on the entropy, but still large enough to approximate the full exponential. Our cut-off operator $\Theta^{\alpha}_{p,u,v}$ resembles the one in \cite{BaCS2021}, but is considerably more complex, reflecting the richer mathematical structure of our uncorrelated trial state. One of the restrictions imposed by $\Theta^{\alpha}_{p,u,v}$ is that only one particle per non-zero momentum mode can be annihilated by $\widetilde{D}_{\alpha}$. To ensure that we can apply $\widetilde{D}_{\alpha}$ up to $n_{\mathrm{c}}$ times to $\psi_{\alpha}$ and obtain an approximation of the full exponential, we introduce an additional cut-off operator acting on the uncorrelated state $\widetilde{\Gamma}$.  It guarantees that each eigenfunction $\psi_{\alpha}$ of $\widetilde{\Gamma}$ has at least $2 n_{\mathrm{c}}$ occupied momentum modes. This issue did not arise in \cite{BaCS2021} because the relevant cubic transformation contained only creation operators. 
Finding the correct cut-off operators is one of the main contributions of this article. Their definition (rescaled to the unit box) and a detailed explanation of how they enable the computation of the energy are given in Section~\ref{sec:correlation}. 

To conclude, let us comment on the main difference between the temperature regime considered in Theorem~\ref{thm:main} and the zero or low-temperature regimes studied in \cite{YauYin2009,BaCS2021,HHST,BOSAS}. In our parameter regime, where $\beta \sim \varrho^{-2/3}$, correlations among all particles (inside and outside the condensate) contribute to the interaction energy at the same order of magnitude. This explains the necessity of working with a transformation that is quartic in creation and annihilation operators, as well as the advantage of introducing all correlations through a single transformation, an idea previously employed in \cite{BocDeuSto2024}. In contrast, at low temperatures, the leading-order contribution to the energy arises from a quadratic (Bogoliubov) transformation, characterized by the choice $u=v=0$ in \eqref{eq:proofSketch4}, whose action is explicit. A subsequent cubic transformation (characterized by a single vanishing momentum) contributes only at a much lower order, and no quartic transformation is required. Consequently, in \cite{BaCS2021,HHST,BOSAS}, the authors first applied a quadratic transformation and then a cubic one. The absence of a clear hierarchy in the relevance of the different transformations is what makes our parameter regime significantly more challenging.

Our article is organized as follows. In Section~\ref{sec:setting}, we introduce our mathematical setting in more detail and show how Theorem~\ref{thm:main} follows from free energy bounds on a torus of side length one. In Section~\ref{sec:trialState}, we introduce our trial state and discuss some of its properties. In particular, we present our various cut-off operators and explain how they enable us to carry out computations. Section~\ref{sec:free-energy} constitutes the core of our paper, where we compute the free energy of our trial state. The proofs of several technical lemmas related to properties of our uncorrelated trial state and the function $\widehat{w}$ are deferred to Appendices~\ref{app:riemannSum}--\ref{app:Gamma0}.

\section{Setting}
\label{sec:setting}
We consider a system of bosons confined to the box $\Lambda_{\sL} = [0,\sL]^3$, whose one-particle Hilbert space is given by $L^2(\Lambda_{\sL})$. As will be explained below, it is convenient to consider a gas with a fluctuating particle number. The Hilbert space of the entire system is therefore given by the bosonic Fock space
\begin{equation*}
	\Fock(\L_\sL) 
	= \bigoplus_{n=0}^{\infty} L^2_{\mathrm{sym}}(\Lambda^n_\sL)\,,
	\label{eq:FockSpace}
\end{equation*}
where $L^2_{\mathrm{sym}}(\Lambda^n_\sL)$ for $n \geq 1$ is defined below \eqref{eq:HL} and  $L^2_{\mathrm{sym}}(\Lambda^0_\sL) = \mathbb{C}$. The creation operator $a^*(f)$ and the annihilation operator $a(f)$ of a particle described by the function $f \in L^2(\Lambda_{\sL})$ are defined by
\begin{align}
	(a^*(f) \psi)(x_1,...,x_{n+1}) &= \frac{1}{\sqrt{n+1}} \sum_{j=1}^n f(x_j) \psi(x_1,...,x_{j-1}, x_{j+1}, ..., x_n), \nonumber \\
	(a(f) \psi)(x_1,...,x_{n-1}) &= \sqrt{n} \int_{\Lambda} \overline{f(x)} \psi(x_1,...,x_{n-1},x) \mathrm{d} x,
	\label{eq:annihilationOperator}
\end{align}
and extended to $\mathscr{F}(\Lambda_{\sL})$ by linearity. As the notation suggests, the operator $a^*(f)$ is the adjoint of $a(f)$. The family of creation and annihilation operators satisfies the canonical commutation relations (CCR)
\begin{equation}
	[ a(f), a^*(g) ] = \langle f,g \rangle_{L^2(\Lambda_{\sL})} , \quad [a(f),a(g)] = 0 = [a^*(f),a^*(g)]
	\label{eq:CCR}
\end{equation} 
for all $f,g \in L^2(\Lambda_{\sL})$. If $\varphi_p(x) = \sL^{-3/2} e^{\mathrm{i}p \cdot x}$ with $p \in \Lambda_\sL^* = (2 \pi/\sL) \mathbb{Z}^3$ we write $a(\varphi_p) = a_p$ and $a^*(\varphi_p) = a^*_p$. In this case the first equation in \eqref{eq:CCR} reads $[a_p,a_q^*] = \delta_{p,q}$ with $p,q \in \Lambda_\sL^*$. 

\smallskip

The second quantization of the $\sN$-particle Hamiltonian $H_\sN$ in \eqref{eq:HL} reads
\begin{equation}
	\mathcal{H} = 0 \oplus \bigoplus_{n=1}^{\infty} H_{n} = \sum_{p \in \Lambda_\sL^*} p^2 a_p^* a_p + \frac{1}{2 \sL^3} \sum_{p,u,v \in \Lambda_\sL^*} \widehat{V}(p) a^*_{u+p} a^*_{v-p} a_u a_v
	\label{eq:secondQuantizedHamiltonian}
\end{equation}
where $\widehat{V}(p) = \int_{\Lambda_{\sL}} V(x) e^{- \mathrm{i} p \cdot x} \text{d} x$ denote the Fourier coefficients of $V$. The free energy in finite volume can be characterized via the Gibbs variational principle. Let 
\begin{equation}
	\mathcal{S}_{\sN,\sL} = \{\, \Gamma \in \mathcal{B}(\Fock(\L_\sL)) \ | \ \Gamma \geq 0,\, \Tr[ \Gamma] = 1,\, \mathcal{N}\, \Gamma = \sN\, \Gamma \,\}
	\label{eq:canonicalStates}
\end{equation}
be the set of all $\sN$-particle states on the bosonic Fock space $\Fock(\L_\sL)$. Here $\mathcal{B}(\Fock(\L_\sL))$ and  
\begin{equation*}
	\mathcal{N} = \bigoplus_{n=0}^{\infty} n = \sum_{p \in \Lambda_\sL^*} a_p^* a_p
	\label{eq:numberOperator}
\end{equation*}
denote the set of bounded operators and the number operator on $\Fock(\L_\sL)$, respectively. We also introduce the Gibbs free energy functional
\begin{equation*} 	\label{eq:GibbsFreeEnergyFunctional}
	\mathcal{F}(\Gamma) = \Tr[ \mathcal{H} \Gamma] - \frac 1 \beta  S(\Gamma) \quad
	\end{equation*}
with the von Neumann entropy
 \begin{equation*}
	S(\Gamma) = - \Tr[ \Gamma\,\ln(\Gamma)]\, .
\end{equation*}
Its infimum in the set $\mathcal{S}_{\sN,\sL} $ equals the free energy in $\Lambda_{\sL}$, that is,
\begin{equation}
	F(\beta,\sN,\sL) = \inf_{\Gamma \in \mathcal{S}_{\sN,\sL}} \mathcal{F}(\Gamma) = - \frac 1 \beta \ln\left(  \Tr [ P_{\sN,\sL} \exp(-\beta \mathcal{H} ) ] \right).
	\label{eq:finiteVolumeFreeEnergy}
\end{equation}
Here $P_{\sN,\sL}$ denotes the projection onto the $\sN$-particle sector of the bosonic Fock space $\Fock(\L_\cL)$. The right-hand side of this equation divided by $|\Lambda_\sL|$ equals the right-hand side of \eqref{eq:freeenergy} without the limit.

We find it mathematically more convenient to work with a grand canonical trial state instead of with a state in $\mathcal{S}_{\sN,\cL}$, and therefore define the set
\begin{equation}
	\mathcal{S}^{\mathrm{gc}}_{\sN, \sL} = \{ \Gamma \in \mathcal{B}(\Fock(\L_\sL)) \ | \ \Gamma \geq 0,\, \Tr[ \Gamma ] = 1,\, \Tr[ \mathcal{N} \Gamma ] = \sN \}
	\label{eq:grandCanonicalStates}
\end{equation}
of all states on $\Fock(\L_\sL)$ with an expected number of $\sN$ particles. It is possible to work with states in $\mathcal{S}^{\mathrm{gc}}_{\sN, \sL}$ because of the equivalence of ensembles, see e.g. \cite{Ruelle1999}. A precise statement including also the reduction to smaller boxes can be found in the next section 
in Proposition~\ref{prop:loc}.

\subsection{Localization to smaller boxes and change of boundary conditions}

An upper bound for $f(\b, \varrho)$ can be obtained by ``gluing'' together many copies of a trial state living on a torus whose side length is fixed in the thermodynamic limit but depends on the diluteness parameter $\varrho \aa^3$. This reduction to a smaller box and to a grand canonical setting is achieved through a standard argument (see, e.g. \cite{Robinson1971}). A precise statement is given in the following proposition. For a proof we refer to \cite[Prop.~2]{HHST}, which also applies in our setting.

\begin{prop}[Localization to finite size periodic boxes] \label{prop:loc}
Let $\cH$ be as defined in \eqref{eq:secondQuantizedHamiltonian} with $\sL$ substituted by $L$, and choose constants $c,R > 0$ such that $0 < R_0 < R < L$ with the radius $R_0$ of the support of $V$. Assume that $\G \in \mathcal{S}^{\mathrm{gc}}_{N, L}$ is a density matrix on the Fock space $\Fock(\L_L)$, satisfying periodic boundary conditions, and that its expected particle number is given by	
\be \label{eq:tilde-rho}
\Tr (\cN \G) =  \varrho \,(L + 2R + R_0)^3.
\ee
Then there exists a constant $C>0$ such that
\be \label{eq:f-tlrho-T}
f(\b, \varrho) \leq \frac{1}{(L + 2R + R_0)^3} \big[\Tr(\cH \G) -  \b^{-1} S(\G)\big] + \frac{C \varrho}{L R }\, 
\ee
holds for all $\b \geq c \varrho^{-2/3}$. 
\end{prop}

Thanks to Proposition~\ref{prop:loc}, the proof of Theorem~\ref{thm:main} reduces to establishing the existence of a trial state with the correct expected density $\tl \varrho:=  \varrho \,(1 + 2R/L + R_0/L)^3$ and the correct free energy per unit volume on boxes of size $L$. Here we have the following result. 
\begin{prop}[Trial state on a finite size periodic box] \label{prop:finitesizebox}
Let $\tl \varrho>0$, $\g>2/3$, and $c>0$ be given and set $L= \aa\, (\tl \varrho \aa^3)^{-\g}$, $N = \tilde{\varrho} L^3$. Then, for all $\beta > 0$ obeying the bound $ c \leq \b {\tl \varrho}^{2/3} \leq (\tl \varrho \aa^3)^{-1/3} $, there exists a density matrix $\G \in \cS_{N,L}^{\rm{gc}}$ on $\Fock(\L_L)$, satisfying periodic boundary conditions, such that
\begin{equation}\label{eq:mainUB}
\frac{1}{L^3} \big[\Tr(\cH \G) - \b^{-1} S(\G)\big]  \leq  f_0(\b, \tl \varrho) +  4 \pi \aa (2 \tl \varrho^2 - [\tl \varrho - \varrho_{\mathrm{c}}(\beta) ]_+^2) + {\cal E}
\end{equation}
holds with
\be \label{eq:cal-E}
 {\cal E} \leq C  \aa \tl \varrho^{2} \cdot (\tl \varrho \aa^3)^{\min \{\eps,\, 3(1-\g) - 2\eps-\d_1/3,\, 6(1-\g) - 6\eps,\,\g-2/3,\, \d_1/3 \}}
\ee
for some $C>0$ and all $\e \in (0, 1/3)$ and $\d_1>0$.
\end{prop}

Let us now conclude this section by showing how the proof of Theorem \ref{thm:main} follows from Propositions~\ref{prop:loc} and  \ref{prop:finitesizebox}.

\begin{proof}[Proof of Theorem \ref{thm:main}.] Let $\varrho > 0$ be given, denote by $R_0 >0$ the radius of the support of $V$, and define $R=\aa\, (L/\aa)^\alpha$ with some $\a \in (0,1)$. We apply Proposition~\ref{prop:finitesizebox} with the choice $\tl \varrho:=  \varrho \,(1 + 2R/L + R_0/L)^3$ so that $\Tr[\mathcal{N} \Gamma] = \tilde{\varrho} L^3$ implies \eqref{eq:tilde-rho}. 
	
Our choice of parameters guarantees the existence of a constant $C>0$ such that
\be \label{eq:rhotilde}
 \varrho \leq \tl \varrho \leq  \varrho  \,\big( 1 + C (\varrho \aa^3)^{\g(1-\a)}\big)\,.
\ee
With \eqref{eq:freeEnergyIdealGas} and $\b \geq c {\tl \varrho}^{\,-2/3}$, $c>0$, we check that there is a constant $\tl c>0$ such that $0\leq-\mu \leq \tl c \b^{-1}$ holds. Using this and \eqref{eq:freeEnergyIdealGas}, we show that $|f_0(\b, \tl \varrho)|\leq C (\b^{-5/2} + \tl \varrho\b^{-1})$. Hence, applications of Proposition~\ref{prop:loc} and \eqref{eq:rhotilde} give
\[ \begin{split}
f(\b, \varrho)   \leq\; & f_0(\b, \tl \varrho)  +  4 \pi \aa (2 \tl \varrho^2 - [\tl \varrho - \varrho_{\mathrm{c}}(\beta) ]_+^2)  \\
& + C \big[ \big(\varrho\b^{-1} +\aa \varrho^2 \big)  (\varrho \aa^3)^{\g(1-\alpha)}  + \aa  \varrho^2 ( \varrho \aa^3)^{\g(1+\alpha)-1} \big] + \cE\,
\end{split}
\]
with $\cE$ defined in \eqref{eq:cal-E}. 

Next, we notice that $\varrho \leq \tl \varrho$ implies $f_0(\b, \tl \varrho) \leq f_0(\b, \varrho)$ and that $|\tl \varrho - \varrho| \leq C \varrho (\varrho \aa^3)^{\g(1-\a)}$ follows from \eqref{eq:rhotilde}. With $ \b^{-1}\leq c\varrho^{2/3}$ we conclude that
\[ \begin{split}
f(\b, \varrho)  \leq \; &f_0(\b, \varrho) +  4 \pi \aa (2 \varrho^2 - [\varrho - \varrho_{\mathrm{c}}(\beta) ]_+^2)  + C \aa \varrho^2  (\varrho \aa^3)^{\min\{ \g(1-\a) -1/3, \g(1+\alpha)-1\} } + \cE\,.
\end{split}\]
Optimization over $\a$ leads to $\a=1/(3\g)$, which is consistent with the condition $\a \in (0,1)$ because $\g>2/3$. With the bound on $\cE$ we obtain
\[ \begin{split}
f(\b, \varrho)  \leq \; &f_0(\b, \varrho) +  4 \pi \aa (2 \varrho^2 - [\varrho - \varrho_{\mathrm{c}}(\beta) ]_+^2)  \\
& + C \aa \varrho^2  (\varrho \aa^3)^{\min\{ \g-2/3,\, \e,\, 3(1-\g)-2\e-\d_1/3,\, 6(1-\g)-6\e,\, \d_1/3 \} }.
\end{split}\]
Optimization over the parameters $\d_1 \in (0,1)$, $\e \in (0,1/3)$ and $\g>2/3$ leads to $\g=17/21$, $\e=1/7$, and $\d_1=3/7$. This proves the claim for all $\beta>0$ such that $ c \leq \b {\tl \varrho}^{2/3} \leq (\tl \varrho \aa^3)^{-1/3} $ holds.  

If $\b {\tl \varrho}^{2/3} \geq (\tl \varrho \aa^3)^{-1/3}$ the free energy of the ideal gas satisfies $|f_0(\beta,\varrho)| \leq C \beta^{-5/2} \leq C \varrho^2 \aa (\varrho \aa^3)^{1/2}$. We also have $4 \pi \aa (2 \varrho^2 - [\varrho - \varrho_{\mathrm{c}}(\beta) ]_+^2) \geq 4 \pi \aa \varrho^2$. When we use these bounds, $f(\beta,\varrho) \leq e(\varrho)$, where $e(\varrho)$ denotes the specific energy in the thermodynamic limit, and the upper bound $e(\varrho) \leq 4 \pi \aa \varrho^2(1+C (\varrho \aa^3)^{1/3})$ in \cite{Dyson1957}, we easily check that
\begin{equation*}
	f(\beta,\varrho) \leq f_0(\b, \varrho) +  4 \pi \aa (2 \varrho^2 - [\varrho - \varrho_{\mathrm{c}}(\beta) ]_+^2) (1+C (\varrho \aa^3)^{1/3})	
\end{equation*}
holds. That is, Theorem~\ref{thm:main} also holds in the parameter regime $ \b {\tl \varrho}^{2/3} \geq (\tl \varrho \aa^3)^{-1/3}$. 
\end{proof}

\subsection{Upper bound for free energy in finite volume}

In order to show Proposition \ref{prop:finitesizebox}, it is convenient to rescale variables $x_j \to x_j /L$, so that particles move in the unit torus $\L=[0,1]^3$. It follows that the Hamiltonian $\cH$ defined in \eqref{eq:secondQuantizedHamiltonian} with $\sL$ substituted by $L$ and acting on  $\Fock(\L_L)$ is unitarily equivalent to the operator $L^{-2} \cH_L$ on $\Fock(\L_1)$, where $\big( \cH_L \Psi\big)^{(n)}= \cH^{(n)}_L \Psi^{(n)}$ with
\[
\cH^{(n)}_L = \sum_{j=1}^n - \Delta_{x_j} + \sum_{1\leq i < j \leq n} L^2 V(L(x_i-x_j)).
\]
 Recalling the choice $L= \tilde{\varrho}^{-1/3} (\tilde{\varrho} \aa^3)^{\frac 1 3 -\g}$, with the density $\tilde{\varrho}$ defined below Proposition~\ref{prop:loc}, we have $N= (\tilde{\varrho} \aa^3)^{1-3\g}$. Hence, by setting $\k=(2\g-1)/(3\g-1)$, we find $L/ \aa = N^{1-\k}$, and we conclude that studying properties of $\cH$ is equivalent to considering the Hamiltonian  $\cH_N$ on $\Fock(\L_1)$, whose action is given by $\big( \cH_N \Psi\big)^{(n)}= \cH^{(n)}_N \Psi^{(n)}$ with 
\be \label{eq:HN}
\cH^{(n)}_N = \sum_{j=1}^n - \Delta_{x_j} + \sum_{1\leq i < j \leq n} N^{2(1-\kappa)} V(N^{1-\kappa}(x_i-x_j)).
\ee
Note that the requirement $\g>2/3$ in Proposition \ref{prop:finitesizebox} allows us to restrict attention to $\k \in [1/3;2/3)$. By a slight abuse of notation we kept the symbol $V$ for the interaction potential in \eqref{eq:HN} instead of using $\widetilde{V}(x) = \aa^{2} V(\aa x)$, whose scattering length equals one. We prefer to display $\aa$ explicitly in formulas, and therefore keep $V$ in \eqref{eq:HN}. Because the restriction of the original Hamiltonian to the $n$-particle sector of the Fock space is unitarily equivalent to $L^{-2} \mathcal{H}_L^{(n)}$, we also need to replace $\beta$ by $\beta_N = \beta/L^2 = \beta \varrho^{2/3} (\varrho \aa^3)^{-2/3 + 2 \gamma} = \beta \varrho^{2/3} N^{-2/3}$.

Let us denote by  
\begin{equation}
	F_0(\b, N) = \frac{1}{\beta} \sum_{p \in 2 \pi \mathbb{Z}^3} \ln( 1 - \exp(-\beta(p^2-\mu_0(\beta,N))) ) + \mu_0(\beta,N) N
	\label{eq:freeEnergyIdealGasTorus}
\end{equation}
the (grand canonical version of the) free energy of the ideal Bose gas on the unit torus. Here the chemical potential $\mu_0(\beta,N)$ is chosen such that the expected number of particles equals $N$, that is,
\begin{equation}
	N = \sum_{p \in 2 \pi \mathbb{Z}^3} \frac{1}{\exp(\beta(p^2 - \mu_0(\beta,N)))-1}
	\label{eq:chemicalPotentialIdealGasTorus}
\end{equation}
holds. The expected number of particles in the $p = 0$ mode (the condensate) will be denoted by 
\begin{equation}
	N_0^{\rm id}(\beta,N) = (\exp(-\beta \mu_0(\beta,N))-1)^{-1}.
	\label{eq:condensateNumberIdealGasTorus}
\end{equation}

\smallskip 

In the next Section we will construct a density matrix $\G_N$ belonging to the set $\cS^{\rm gc}_{N}:=\cS^{\rm gc}_{N,1}$ (see \eqref{eq:grandCanonicalStates}), which satisfies bounds compatible with those in Proposition~\ref{prop:finitesizebox} when we go back to the original variables. This result is captured in the following Proposition. 

\begin{prop}\label{prop:Hkappa} 
Let $V $ satisfy the assumption of Theorem \ref{thm:main}, define $\cH_N$ through \eqref{eq:HN}, and let $c,c'>0$ be given. For any $\beta_N$ with $c N^{-2/3} \leq \beta_N \leq c' N^{-\k} $ there exists a density matrix $\G_N \in \cS^{\rm gc}_{N}$ and a constant $C>0$ such that for all $\k \in (1/3; 2/3)$ and $N$ large enough the upper bound
\begin{equation}\label{eq:mainUB-kappa}
\begin{split}
\Tr(\cH_N \G_N) \; & -  \b_N^{-1} S(\G_N)   \leq  \ F_0(\b_N, N) +  4 \pi \aa N^{\k-1} (2 N^2 - (N_0^{\rm id}(\beta_N,N))^2) \\[0.2cm]
& + C N^{\k+1} \cdot \max\{ N^{-\eps}, N^{6\k-3 +2\e+\d_1/3}, N^{12\k-6 +6\e}, N^{-\d_1/3} \}
\end{split}
\end{equation}
holds with $F_0(\b, N)$ in \eqref{eq:freeEnergyIdealGasTorus}, $N_0^{\rm id}(\beta,N)$ in \eqref{eq:condensateNumberIdealGasTorus}, $\d_1>0$, and $\e>0$ such that $3\k-2+3\e <0$.
\end{prop}

To derive Proposition~\ref{prop:finitesizebox} from Proposition~\ref{prop:Hkappa}, we first replace the sum in the definition of the first term in $F_0(\beta_N,N)$ in \eqref{eq:freeEnergyIdealGasTorus} by an integral. An application of Lemma~\ref{lem:approximateMomentumSumsByIntegrals} in Appendix~\ref{app:riemannSum}, $-\mu_0(\beta_N,N) \beta_N = \ln(1+1/N_0^{\mathrm{id}})$, and $N_0^{\mathrm{id}}(\beta_N,N) \leq N$ show 
\begin{align}
	&\frac{1}{\beta_N} \sum_{p \in 2 \pi \mathbb{Z}^3} \ln(1-\exp(-\beta_N(p^2-\mu_0(\beta_N,N)))) \leq \frac{1}{\beta_N} \ln(1-\exp(\beta_N \mu_0(\beta_N,N))) \nonumber \\
	&\hspace{1cm}+ \frac{1}{(2 \pi)^3 \beta_N^{5/2}} \int_{|p| \geq 2 \pi \beta_N^{1/2}} \ln(1-\exp(-(p^2- \beta_N \mu_0(\beta_N,N)))) \left( 1 - \frac{3 \pi \beta_N^{1/2}}{|p|} \right) \mathrm{d}p \nonumber \\
	&\leq \frac{1}{(2 \pi)^3 \beta_N} \int_{\mathbb{R}^3} \ln(1-\exp(-\beta_N(p^2- \mu_0(\beta_N,N)))) \mathrm{d}p + C \left( \beta_N^{-2} + \beta_N^{-1} \ln(N) \right).
	\label{eq:proofPropsAndi1}
\end{align}
Next, we use the unitary equivalence of $\cH$ in \eqref{eq:secondQuantizedHamiltonian} and $L^{-2} \cH_N$ in \eqref{eq:HN} to write 
\begin{equation}
	 \Tr(\cH \G) - \b^{-1} S(\G) = L^{-2}   \big[\Tr(\cH_N \G_N) -  \b_N^{-1} S(\G_N)\big] 
	 \label{eq:proofPropsAndi1b}
\end{equation}
with the unitary image $\G$ of $\G_N$ and $\b/L^2=  \b_N$. With the definitions of $f_0$ in \eqref{eq:freeEnergyIdealGas}, $L$ and $N$ above \eqref{eq:HN}, we check that
\begin{align}
	&\frac{1}{(2 \pi)^3 \beta_N L^5} \int_{\mathbb{R}^3} \ln(1-\exp(-\beta_N(p^2- \mu_0(\beta_N,N)))) \mathrm{d}p - \mu_0(\beta_N,N) N/L^5 \nonumber \\
	&= \frac{1}{(2 \pi)^3 \beta} \int_{\mathbb{R}^3} \ln(1-\exp(-\beta(p^2- N^{-2/3} \tilde{\varrho}^{2/3} \mu_0(\beta_N,N)))) \mathrm{d}p - N^{-2/3} \tilde{\varrho}^{2/3} \mu_0(\beta_N,N) \tilde{\varrho} \nonumber \\
	&\leq f_0(\beta,\tilde{\varrho}). \label{eq:proofPropsAndi2}
\end{align}

Moreover, an application of Lemma~\ref{lem:boundRho0} shows
\begin{equation}
	\frac{N_0^{\mathrm{id}}(\beta_N,N)}{L^3} \geq \left[ \tilde{\varrho} - \varrho_{\mathrm{c}}(\beta) - \frac{C}{\beta L} | \ln(\beta^{1/2}/L) | \right]_+,
	\label{eq:proofPropsAndi3}
\end{equation}
and hence 
\begin{align}
	\frac{4 \pi N^{\k-1} ( 2 N^2 - (N_0^{\rm id}(\beta_N,N))^2)}{L^5} \leq& 4 \pi \aa \left( 2 \tilde{\varrho}^2 - [\tilde{\varrho} - \varrho_{\mathrm{c}}]_+^2 \right) \nonumber \\
	&+ C \aa \varrho^2 (\varrho \aa^3)^{\gamma-1/3} |\ln(\varrho \aa^3)|.
	\label{eq:proofPropsAndi4}
\end{align}
Note that we inserted $\aa=1$ on the left-hand side. This should be compared to the remark regarding $V$ and $\widetilde{V}$ below \eqref{eq:HN}. In combination, \eqref{eq:mainUB-kappa}--\eqref{eq:proofPropsAndi2}, \eqref{eq:proofPropsAndi4}, and $c \leq \beta \varrho^{2/3} \leq c' (\varrho \aa^3)$ show 
\begin{align}
	\frac{1}{L^3} \big(\Tr(\cH \G) - \b^{-1} S(\G)\big) \leq& f_0(\beta,\tilde{\varrho}) + 4 \pi \aa \left( 2 \tilde{\varrho}^2 - [\tilde{\varrho} - \varrho_{\mathrm{c}}]_+^2 \right)  \nonumber \\
	&+ C \varrho^{5/3} \left[ (\varrho \aa^3 )^{\gamma-1/3} + (\varrho \aa^3 )^{3\gamma-1} \left| \ln\left( \varrho \aa^3 \right) \right| \right] \label{eq:proofPropsAndi5} \\
	&+ C \aa \varrho^2 \left[  (\varrho \aa^3)^{\min\{ \eps,\, 3(1-\g) -2\e-\d_1/3,\, 6(1-\g)-6\e ,\, \delta_1/3  \}\,}  \right].
	\nonumber
\end{align}

\section{The trial state}
\label{sec:trialState}

In this Section we construct the density matrix satisfying Proposition \ref{prop:Hkappa}. To be able to distinguish between different parts of the system, as e.g. the condensate, the thermally excited particles, and the correlations induced by the interaction between the particles, we introduce a partition of the momentum space $\Lambda^* = 2 \pi \mathbb{Z}^3$. We recall that $\k \in (1/3, 2/3)$. We consider parameters $\d_1, \d_2, \eps>0$ that are assumed to satisfy $3\k-2+3\eps+\d_2<0$, and define
\begin{align}
	P_0 &\coloneqq \{ p \in \Lambda^* \ | \ p = 0 \}, \nonumber \\
	P_{\mathrm{G}} & \coloneqq \{ p \in \Lambda^* \ | \ 0 < |p| \leq \beta^{-1/2(1+\delta_2)} \}, \nonumber \\
	P_{\A} & \coloneqq \{ p \in \Lambda^* \ | \ \beta^{-1/2(1-\delta_1)} \leq |p| \leq \beta^{-1/2(1+\delta_2)} \}, \nonumber \\
	P_{\mathrm{H}} &\coloneqq \{ p \in \Lambda^* \ | \ N^{1 - \kappa - \eps} \leq |p| \}\,. \label{eq:momentumSets}
\end{align}
The labels $\rm G$ and $\rm H$ stand for ``Gibbs'' and ``high'', respectively. Indeed, the set $P_{\mathrm{G}}$ is large enough to contain the momenta of the majority of the thermally excited particles in the ideal Bose gas. This is also true for $P_{\A} \subset P_{\mathrm{G}}$. The set $P_{\mathrm{G}}$ will be used to describe the thermally excited particles in our uncorrelated trial state. In contrast, the particles with momenta in $P_0 \cup P_{\A}$ are those that will be affected by our correlation structure. 
 
Notice that the assumptions on $\d_1, \d_2, \eps$ guarantee $P_0 \cap P_{\mathrm{G}} = \emptyset = P_{\mathrm{G}} \cap P_{\mathrm{H}}$ provided $N$ is large enough. 
We will see that $\d_2$ can be chosen as small as we wish.

With the momentum sets in \eqref{eq:momentumSets} at hand, we decompose the one-particle Hilbert space  $\ell^2(\L^*)$ as  
\[
\ell^2(\L^*) = \ell^2(P_0) \oplus \ell^2(P_\mathrm{G}) \oplus  \ell^2(P_\mathrm{C})
\]  
with $P_\mathrm{C} = \L_* \setminus (P_0 \cup P_{\rm G})$. This decomposition induces a corresponding splitting of the Hilbert space $L^2(\Lambda)$ by unitary equivalence, and therefore of the Fock space $\Fock(\L)$. Let us now denote by $\Fock_{0}$, $\Fock_{\rm G} $ and $\Fock_{\rm C} $ the bosonic Fock spaces built on $\text{\rm Span}\{1\}$, $\text{\rm Span}\{e^{\mathrm{i} p \cdot x}, p \in P_{\rm G}\}$, and $\text{\rm Span}\{e^{\mathrm{i} p \cdot x}, p \in P_{\rm C}\}$, respectively. Then
\begin{equation}
	\Fock(\Lambda) \cong \Fock_{0} \otimes \Fock_{\mathrm{G}} 
	\otimes \Fock_{\mathrm{C}}
	\label{eq:decompositionFockSpace}
\end{equation}
where $\cong$ denotes unitary equivalence. We recall the definition of $a_p^*$ and $a_p$ below \eqref{eq:CCR}. By a slight abuse of notation we use the same symbol to denote the creation and annihilation operators acting on the spaces on the left- and on the right-hand side of \eqref{eq:decompositionFockSpace}. We are now ready to define our trial state.

\subsection{The uncorrelated state}

In this Section we describe our uncorrelated trial state and begin with the part describing the thermal cloud, which lives in $\mathscr{F}_{\mathrm{G}}$. We define $\mathcal{H}_{\mathrm{G}} = \sum_{p \in P_{\mathrm{G}}} p^2 a_p^* a_p $, $\mathcal{N}_{\mathrm{G}} = \sum_{p \in P_{\mathrm{G}}} a_p^* a_p$, and denote by 
\begin{equation}
	\wt \G_{\rm G} = \frac{\exp(-\beta_N( \mathcal{H}_{\mathrm{G}} - \mu_0(\beta_N,N) \mathcal{N}_{\mathrm{G}} ))}{\Tr_{\mathscr{F}_{\mathrm{G}}} \exp(-\beta_N( \mathcal{H}_{\mathrm{G}} - \mu_0(\beta_N,N) \mathcal{N}_{\mathrm{G}} ))}  
	\label{eq:GibbsStateIdealGas}
\end{equation} 
the Gibbs state of the ideal Bose gas when restricted to $\mathscr{F}_{\mathrm{G}}$ with a partial trace. Here $\mu_0(\beta_N,N)$  denotes the chemical potential of the ideal gas with $N$ particles at inverse temperature $\beta_N$. Let us also define the operator
\be \label{eq:opM}
\mathcal{M} = \sum_{p \in P_{\A} } \chi(a_p^* a_p > 0)\,,
\ee
counting the number of occupied momenta in $P_{\A}$. Here and below we use $\chi(t>0)$ to denote the indicator function of the set $\{t>0\}.$ Then, for some $M>0$ to be fixed later, we introduce the truncated Gibbs state on $\mathscr{F}_{\mathrm{G}}$ by
\begin{equation}
	\G_{\rm G} = \frac{\tl \G_{\rm G} \,\chi(\mathcal{M} \geq M )}{\Tr_{\mathscr{F}_{\mathrm{G}}}
	\big[ \tl \G_\mathrm{G}\, \chi(\mathcal{M} \geq M )\big]}\,.
	\label{eq:GibbsStateIdealGasWithCutoff}
\end{equation}
The meaning of the cut-off in \eqref{eq:GibbsStateIdealGasWithCutoff} is the following: the eigenfunctions of the Gibbs state $\tl \G_\mathrm{G}$ in \eqref{eq:GibbsStateIdealGas} can be chosen as symmetrized products of plane waves. The cut-off function $\chi(\mathcal{M} \geq M )$ restricts the eigenfunction expansion of $\G_\mathrm{G}$ to states, for which at least $M$ different momenta in $P_{\mathrm{G}}$ have an occupation number of at least one. The reason for this cut-off will be explained later in Section \ref{sec:correlation}, after we have introduced the correlation structure of our trial state. During our analysis we will assume that $M$ satisfies the bound
\begin{equation}
	M \leq \frac{1}{2} \sum_{p \in P_{\A}} \exp(-\beta_N(p^2 - \mu_0(\beta_N,N))),
	\label{eq:conditionOnM}
\end{equation}
which is motivated by the result in Lemma~\ref{lem:probabilisticLemma} in Appendix~\ref{app:Gamma0}. We highlight that, in contrast to $\tl \G_\mathrm{G}$, the state  $\G_\mathrm{G}$ is not quasi-free. 
The next lemma provides properties of the state $\G_\mathrm{G}$, which are proved in Appendix~\ref{app:Gamma0}.
\begin{lemma}\label{lm:cN}
We consider the limit $N \to \infty$, $1 \gg \beta_N \geq b \beta_{\mathrm{c}}(N)$ with some $b>0$ and $\beta_{\mathrm{c}}$ in \eqref{eq:criticalTemperatureIdealGas}, and assume that \eqref{eq:conditionOnM} holds. Let $\mathscr{F}_{\mathrm{G}}$ and $\G_{\mathrm{G}}$ be defined above \eqref{eq:decompositionFockSpace} and in \eqref{eq:GibbsStateIdealGasWithCutoff}, respectively. For any $p, q \in \mathbb{N}_0$ there is a constant $C_{p,q} > 0$ such that
	\begin{equation}
		\sum_{v \in P_{\A}} \Tr_{\mathscr{F}_{\mathrm{G}}}\big[ \mathcal{N}^p \left( a_v^* a_v \right)^q \G_\mathrm{G} \big]  \leq C_{p,q} \begin{cases}
			\beta_N^{-3(p+1)/2} \left( 1 + \beta_N^{-\delta_1(2q-3)/2} \right) & \text{ if } q > 0 \\
			\beta_N^{-3p/2} & \text{ if } q = 0
		\end{cases} 
		\label{eq:propParticleNumberPowers}
	\end{equation} 
holds. Moreover, we also have 
\begin{equation}
	\sum_{p \in P_{\mathrm{G}} \backslash P_{\A}}\Tr_{\mathscr{F}_{\mathrm{G}}}[a_p^*a_p \Gamma_{\mathrm{G}}] \leq C \beta^{-3/2 + \delta_1/2}
	\label{eq:numberOfUndressedParticles}
\end{equation}
and ($\mathcal{N}_{\A} = \sum_{p \in P_{\A}} a_p^* a_p$)
\begin{equation}
	\Tr_{\mathscr{F}_{\mathrm{G}}}[\cN_{\A}^2 \Gamma_{\mathrm{G}}]-(\Tr_{\mathscr{F}_{\mathrm{G}}}[\cN_{\A} \Gamma_\mathrm{G}])^2 \leq C\b_N^{-2}.
	\label{eq:propParticleNumberPowersKineticAndVariance}
\end{equation}

\end{lemma}

An application of the spectral theorem allows us to rewrite
\[
\G_{\rm G} = \sum_{\a} \l_\a {\mathbb P}_{g_{\alpha}}
\]
where ${\mathbb P}_{\ph}$ denotes the projection onto the subspace spanned by the function $\ph$, and we chose $g_\a$ as symmetrized products of plane wave states in the set $\{ e^{\mathrm{i} p \cdot x} \ | \ p \in P_{\mathrm{G}} \}$. 
Next, we add the condensate, by putting exactly $N_0$ particles in the zero momentum mode. 
Our trial state without microscopic correlations between the particles reads
\begin{equation} \label{eq:G0-def}
	\Gamma_0 = \sum_{\alpha}\lambda_{\alpha}\,   {\mathbb P}_{g^0_{\alpha}} 
\end{equation}
where  $g^0_\a:= (N_0!)^{-1/2}(a^*_0)^{N_0}\O_0 \otimes g_\alpha \otimes \O_{\rm C} $, and $\Omega_0$ and $\Omega_{\mathrm{C}}$ denote vacuum vectors of $\Fock_0$ and $\Fock_{\mathrm{C}}$, respectively. 

For any $p \in \L^*$ and  $\cN_p =a^*_p a_p$, we have $\mathcal{N}_p\, g^0_{\alpha} = n^{\alpha}_p g^0_{\alpha}$, where $n^\a_0= N_0$ for all $\alpha$, and $n^\a_p=0$ unless $p \in P_{\mathrm{G}} \cup \{0\}$. Notice that we treated the condensate differently from particles in $P_{\mathrm{G}}$. Indeed, the condensate cannot be described by the Gibbs state of the ideal gas because its self-interaction energy is twice as large as that of the occupation number state we have chosen. 
We choose $N_0=N_0(\beta,N)$ such that $\Tr [ \mathcal{N} \Gamma_0 ] = N$. As a direct consequence of $\Tr[a_0^* a_0 \Gamma_0] =N_0$ and \eqref{eq:numberOfUndressedParticles} we have
\be \begin{split} \label{eq:Nplus}
N-N_0 - C \beta_N^{-3/2 + \delta_1/2} \leq \sum_{\a}\l_\a \hskip -0.1cm \sum_{v \in P_{\A}} n_v^\a &\; \leq N-N_0.
\end{split}\ee
Note that $N_0$ is not necessarily an integer and should be replaced by e.g. $\lfloor N_0 \rfloor$. The resulting discrepancy in the expected particle number can be absorbed by a slight adjustment of the chemical potential $\mu_0$ in the definition of $\Gamma_{\mathrm{G}}$. For simplicity we will ignore this minor issue and leave the related details to the reader.

The following Lemma shows that the number of particles in the condensate and the free energy of $\Gamma_0$ are related to those of the Gibbs state of the ideal Bose gas. The proof is postponed to Appendix~\ref{app:Gamma0}. 
\begin{lemma} \label{lm:Gamma0} 
We have 
\begin{equation}
	 N_0(\beta_N,N) \geq N^{\rm id}_0(\beta_N,N) - 1
	\label{lem:N0UncorrelatedTrialState}
\end{equation} 
with $N_0^{\rm id}(\beta_N,N)$ in \eqref{eq:condensateNumberIdealGasTorus}. Moreover, in the limit $N \to \infty$, $N^{\s} \geq \beta_N \geq b \beta_{\mathrm{c}}(N)$ with $\beta_{\mathrm{c}}$ in \eqref{eq:criticalTemperatureIdealGas} and some $\s, b>0$, $\kappa > 1/3$ and for any choice of $\d_2>0$ in \eqref{eq:momentumSets}, the free energy of the state $\Gamma_0$ in \eqref{eq:G0-def} satisfies
\begin{align}
	\Tr (\cH_N \G_0) - \b^{-1}_N S(\G_0) \leq&  F_0(\b_N, N)  +\frac 12 N^{\k-1} \widehat V(0) \big(2 N^2 - [N_0^{\rm{id}}(\beta_N,N)]^2 \big)  \nonumber \\
	&+ C N^{1/3+\k} \label{lem:energyUncorrelatedTrialState}
\end{align} 
with $F_0(\b_N,N)$ in \eqref{eq:freeEnergyIdealGasTorus}. 
\end{lemma}
It is a trivial observation that the free energy of the state $\G_0$ is always larger than the right-hand side of \eqref{eq:mainUB}, being $8 \pi \aa < \widehat V(0)$. The correlation structure introduced in the next section will be responsible for the appearance of the scattering length.

\medskip

\subsection{The correlation structure} \label{sec:correlation}

In this Section we discuss the correlated trial state, which is built starting from $\G_0$ defined in \eqref{eq:G0-def}. For some $0 < \ell < 1/2$, we consider a finite volume version of the scattering equation in $\mathbb{R}^3$ introduced in \eqref{eq:scatteq}. Namely, we define $f_\ell$ as the lowest energy solution of the Neumann problem 
\be \label{eq:scatl}
	\Big[-\Delta+\frac12 V\Big]f_\ell=\lambda_\ell f_\ell
\ee
on the ball $|x|\leq N^{1-\k}\ell$, with the normalization $f_\ell(x)=1$ if $|x|=N^{1-\k}\ell$.
For $p\in \L^*$ we also define the coefficients
\be \label{eq:etap}
	\eta_p=-N(\widehat{1-f_{N,\ell}})(p),
\ee
where we introduced the rescaled solution $f_{N,\ell}(x)= f_{\ell}(N^{1-\k}x)$ of the scattering equation, and defined it on the whole torus $\L$ by setting $f_{N,\ell}(x)=1$ if $|x|>\ell$. Note that, by \eqref{eq:scatl}, $\eta_p$ satisfies
 \be  \label{eq:scatl-etap}
	p^2 \eta_p + \frac{N}{2} (\widehat{V}_N\ast \widehat f_{N,\ell})(p)  = N^{3-2\k}\l_\ell \big(\widehat{\chi}_\ell \ast \widehat{f}_{N,\ell} \big)(p),
\ee
where  $\widehat{V}_N(p)= N^{\k-1} \widehat V(p/N^{1-\k})$ and $\chi_\ell$ denotes the characteristic function of the ball of radius $\ell.$  The lemma below summarizes some of the properties of the solution of the eigenvalue problem in \eqref{eq:scatl}.

\begin{lemma} \label{lm:scatt-eq} Let $V \in L^3 (\bR^3)$ be non-negative, compactly supported and spherically symmetric. Fix $\ell > 0$ and let $f_\ell$ denote the solution of \eqref{eq:scatl}. Let $\eta_p$ be defined in \eqref{eq:etap}.  For $N\in\NN$ large enough the following properties hold true. 
	\begin{enumerate}
		\item [i)] We have 
		\begin{equation*}\label{eq:lambdaell} 
			\bigg| \lambda_\ell - \frac{3\aa }{N^{3-3\k}\ell^3} \bigg| \leq  \frac{1}{N^{3-3\k}\ell^3} \frac{C \aa^2}{\ell N^{1-\k}}\,.
		\end{equation*}
		\item[ii)] We have $0\leq f_\ell\leq1$. Moreover there exists a constant $C > 0$ such that  
		\begin{equation*} \label{eq:Vfa0} 
			\left|  \int  V(x) f_\ell (x) dx - 8\pi \aa   \right| \leq \frac{C \aa^2 }{\ell N^{1-\k}}.
		\end{equation*}
		\item[iii)] There exists a constant $C > 0$ such that, for all $p \in \bR^3$,  	
		\be \label{eq:etapbound} 
		|\eta_p|\leq \frac{CN^\kappa}{p^2}.
		\ee
		\item[iv)] Let $\Lambda_+^* = 2 \pi \mathbb{Z}^3 \backslash \{ 0 \}$. The following bound holds
		\be\label{BCS-bounds} 
\sup_{p \in \L^*}\sum_{r\in \L^*_+} \frac{\big|(\widehat V_N\ast\widehat f_{N,\ell})(r-p)\big|+ \big|\widehat V_N(r-p)\big|}{|r|^{2}}  \leq C. 
\ee
	\end{enumerate}        
\end{lemma}

In the following we will often consider $\eta_H$ defined as the restriction of $\eta$ to the set $P_{\mathrm{H}}$. With this notation the following holds.
\begin{lemma} \label{lm:eta} Let $\eta_H$ be the restriction of $\eta: \L^* \to \bR$ to the set $P_{\mathrm{H}}$  in \eqref{eq:momentumSets}. Then 
\be\label{eq:etaH}
\begin{split}
 \| \eta_H \|^2_2 &\leq C N^{3\k-1 +\eps}\,, \hskip 0.6cm  \| \check{\eta}_H \|^2_{H^1} \leq C  N^{1+\k}\,, \\
\| \eta_H \|_1 &\leq C N \,, \hskip 1.8cm  \| \eta_H \|_{\io} \leq C N^{3\k-2+2\eps}
\end{split}
\ee
where $\check{\eta}_H$ denotes the inverse Fourier transform of $\eta_{\mathrm{H}}$.
\end{lemma}
The proof of Lemmas \ref{lm:scatt-eq} and \ref{lm:eta} is deferred to Appendix~\ref{app:scatteringEquation}. \\

With $\eta_p$ defined in \eqref{eq:etap} and $n_{\mathrm{c}} \in \mathbb{N}$ a parameter that will be chosen to grow with $N$, we define 
\begin{equation} \label{eq:def_D}
	\mathcal{D}_{\alpha} = \sum_{n=0}^{n_{\mathrm{c}}} \frac{D_{\alpha}^n}{n!} \quad \text{ with } \quad D_{\alpha} = \frac{1}{2N} \sum_{\substack{r\in P_{\mathrm{H}},\\ (v,v')\in \bbG_r}} \eta_r a^*_{-r}a^*_{r+v+v'}a_va_{v'}\Theta^\a_{r,v,v'}
\end{equation}
where, for a given $r\in P_{\mathrm{H}}$, we introduce the set $\bbG_ r \coloneqq \bbG^{\rm q}_r \cup \bbG^{\rm c}_r \cup \{0,0\}$ with
\begin{equation}\begin{split}  \label{eq:P0G}
	\bbG^{\rm q}_r &\; \coloneqq\{(v,v') \in (P_{\mathrm{\A}})^2 \ | \ v\neq \pm v' \;\land\; r+v+v' \in P_{\mathrm{H}} \},  \\
				\bbG^{\rm c}_r&\; \coloneqq\{ (v, 0) \;|\; v  \in  P_{\A}\,,  r+v \in P_{\mathrm{H}} \}\cup \{ (0, v) \;|\; v  \in  P_{\A}\,,  r+v \in P_{\mathrm{H}} \}\,. \\
	\end{split}
\end{equation}
The momenta of the operators in the definition of $D_{\alpha}$ in \eqref{eq:def_D} differ slightly from those appearing in the related operator in \eqref{eq:introCorrelationStructure}. This difference is (almost) irrelevant for computations and allows an easier comparison with the zero temperature case. Finally, the operator $\Theta_{r,v,v'}^\a$ enforces suitable restrictions on the momenta of the excitations created by $\cD_\a$. More precisely, let $n^\a_p$ be defined below \eqref{eq:G0-def}, $\chi(t>0)$ be the indicator function of the set $\{t>0\}$ and introduce the notation $P_{\A}^0:= P_{\A} \cup \{0\}$. Then
\be \label{eq:definitionTheta}
\Theta^\a_{r,v,v'} \coloneqq  (\Theta_1)^\a_{r,v,v'}\, (\Theta_2)^\a_{r,v,v'}
\ee
with  
\[ 
	\begin{split}
		(\Theta_1)^\a_{r,v,v'}  \coloneqq \; &\prod_{\substack{s\in P_{\mathrm{H}}\\u\in P_{\A}^0}}\bigg[1-\chi(\cN_s>0)\chi(\cN_u<n^\a_u)\chi\Big(\sum_{\substack{w\in \{v,v'\}\\w\neq 0}}\cN_{-s+u+w}>0\Big)\bigg]\\
		& \times \prod_{\substack{u_1,u_2\in P_{\A}^0\\u_1\neq u_2}}\bigg[1-\chi(\cN_{u_1}<n^\a_{u_1})\chi(\cN_{u_2}<n^\a_{u_2})\\
		& \quad \qquad  \qquad\qquad\chi\Big(\sum_{p\in \{-r,r+v+v'\}}\cN_{-p+u_1+u_2}+\sum_{\substack{p\in \{-r,r+v+v'\}\\w\in \{v,v'\}}}\cN_{-p+u_1+w}>0\Big)\bigg]\,,
	\end{split}
\]
and
\[ \begin{split}
		(\Theta_2)^\a_{r,v,v'}  \coloneqq \; & \prod_{s\in P_{\mathrm{H}}}\bigg[1-\chi(\cN_s>0)(1-\d_{|v|+|v'|,0})\chi\Big(\cN_{-s+v+v'}>0\Big)\bigg] \\
		 & \times \prod_{w\in \{v,v'\}}\prod_{\substack{s\in P_{\mathrm{H}}\\u\in P_{\A}}}\bigg[1-\d_{w,0}\;\chi(\cN_0=N_0)\chi(\cN_s>0)\chi(\cN_u<n^\a_u)\chi(\cN_{-s+u}>0)\bigg]\\
		&\times \prod_{s\in P_{\mathrm{H}}}\bigg[1-\d_{|v|+|v'|,0}\;\chi(\cN_0\geq N_0-1)\chi(\cN_s+\cN_{-s}>2)\bigg] \,.
	\end{split}
\]

With $\eqref{eq:def_D}$ we define the state 
\begin{equation} \label{eq:GammaD}
 \Gamma  = \sum_{\alpha} \lambda_{\alpha}  {\mathbb P}_{\Phi_{\a}}\,,  \hskip 1cm \Phi_\a:= \frac{\mathcal{D}_{\alpha}g^0_{\alpha}}{\Vert \mathcal{D}_{\alpha} g^0_{\alpha} \Vert}\,.
\end{equation}
Let us briefly discuss the action of $\cD_\a$. Each of the operators $D_{\alpha}$ describes a scattering process among two excitations with high momenta $-r, r +v+v'$ and two particles with momenta in $P_{\A} \cup \{0\}$. 
Hence we find it convenient to describe with a single operator both, the processes with $(v,v')=(0,0)$, describing the creation of {\it pair excitations} from the condensate, and scattering processes involving the thermal cloud. In particular we will denote as {\it quadruplet} a process such that  $(v,v') \in \bbG^{\rm q}_r$, and as {\it triplet} one with $(v,v') \in \bbG^{\rm c}_r$. 

\smallskip

The cut-off operator $\Theta_{r,v,v'}^\a$ is defined so that the action of $(D_\a)^n$ on $ g^0_{\alpha}$ results in $n$ scattering processes with momenta $\{-r_\ell,r_\ell+v_\ell+v'_\ell,v_\ell,v'_\ell\}_{ \ell=1,\dots,n}$ such that, introducing the notations $w_\ell\in \{v_\ell,v'_\ell\}$ and $p_\ell\in \{r_\ell,-r_\ell+v_\ell+v'_\ell\}$, for all $i,j,h,k=1,\dots,n$ and $i \neq j$,  we have
\begin{equation}\label{eq:restrictions1}
	p_i\neq -p_j+w_h+w_k.
\end{equation}
Here we adopted the convention $w_h +w_k=v_h+v_h'$ for $h=k$. Moreover if $i=j$ and $h\neq i$
\begin{equation}\label{eq:restrictions2}
	v_i+v'_i\neq w_h+w_k
\end{equation}
with $w_h \neq 0$, and the same convention $w_h +w_k=v_h+v_h'$ if $h=k$ as above. 
 The restrictions in \eqref{eq:restrictions1} and \eqref{eq:restrictions2} will be crucial to simplify computations as discussed in detail in Section~\ref{sec:computationNorm} below.
 
 \medskip
{\it Remarks.}
\begin{enumerate}[label=\roman*)]
	\item Conditions similar to \eqref{eq:restrictions1} and \eqref{eq:restrictions2}, but considerably simpler, 
are present in the low temperature trial states introduced in \cite{BaCS2021}, and later extended in \cite{HHST}. 
In particular, the reader may compare $(\Theta_1)^\a_{r,v,0}$ with \cite[Eq.\,(2.15)]{BaCS2021}, keeping in mind that in the zero temperature case: a) the uncorrelated state is the complete condensate, hence $n_u^\a=0$ for all $u\neq 0$ and $\chi(\cN_u <n_u^\a)$ has to be replaced with the identity; b) low energy particles, being not present in the uncorrelated state, are always created by the operators implementing the correlation structure.

\item The introduction of $n_{\mathrm{c}}$ allows us to control the influence of the correlation structure on the entropy in a very efficient way, see Section~\ref{sec:entropy} below. We emphasize that introducing $n_{\mathrm{c}}$ does not introduce additional difficulties. This is because $g_\alpha^0$ contains only finitely many particles, thereby inducing an effective cut-off in the exponential sum -- an effect that must be accounted for even if $n_{\mathrm{c}} = +\infty$.

\item Let us comment on the role of the cut-off $\chi(\mathcal{M} > M)$ in the definition of $\Gamma_{\mathrm{G}}$ in \eqref{eq:GibbsStateIdealGasWithCutoff}. As explained below \eqref{eq:introCorrelationStructure}, it ensures that we can act with $D_{\alpha}$ on $g_{\alpha}^0$ sufficiently many times to approximate the full exponential. This is achieved by discarding those states $g_{\alpha}^0$ with fewer than $M$ occupied momentum modes $u \neq 0$. In Appendix~\ref{app:Gamma0}, we show that for a typical eigenfunction $g_\alpha^0$ of $\widetilde{\Gamma}_{\mathrm{G}}$ in \eqref{eq:GibbsStateIdealGas}, the number $M_\alpha$ of occupied momentum modes scales as $\beta_N^{-3/2} \sim N$ if $\beta_N \sim N^{-2/3}$. In contrast, to approximate the full exponential with $\mathcal{D}_{\alpha}$, we require $n_{\mathrm{c}} \gg \Vert \eta_H \Vert_2^2 \sim N^{3\kappa - 1 + \varepsilon}$. Thus, the assumption $3\k-2+\eps<0$ guarantees the possibility of choosing $M>n_\mathrm{c}\gg\|\eta_H\|_2^2$ for sufficiently large $N$.

\item In the following we will assume that $N_0 \geq 2 n_{\mathrm{c}}$, which allows us to apply $D_{\alpha}$ a sufficient number of times also to the condensate. If $N_0 < 2 n_{\mathrm{c}}$, we define $D_{\alpha}$ in \eqref{eq:def_D} with the set $\bbG_ r$ replaced by $\bbG^{\rm q}_r$. That is, the zero mode will not be dressed, and computations are simpler. This is justified because the contribution from $[\varrho - \varrho(\beta)]_+$ is a small perturbation in this case. The straightforward details of this analysis are left to the reader.  

\end{enumerate}


\subsection{Computation of the norm of $\mathcal{D}_{\alpha} g_\alpha^0$}
\label{sec:computationNorm}
The main feature of our trial state is that it allows for explicit computations of 
\begin{equation*} \label{eq:tr-cH-G}
\Tr[ \cH_N\, \Gamma ] = \sum_{\alpha} \lambda_{\alpha} \frac{\langle \mathcal{D}_{\alpha}  g^0_{\alpha} , \cH_N  \mathcal{D}_{\alpha}  g^0_{\alpha} \rangle}{\Vert \mathcal{D}_{\alpha}  g^0_{\alpha} \Vert^2}
\end{equation*}
based on a fine cancellation between the numerator and the denominator appearing on the right-hand side. Let us start discussing the denominator. For fixed 
$\a$:
\[       
	\|\cD_\a \,g^0_\a\|^2=\sum_{n=0}^{n_{\mathrm{c}}}\frac1{(n!)^2}\|(D_\alpha)^n g^0_\a\|^2\,.
\]
With \eqref{eq:def_D} and the shorthand notations $A_{r,v,v'}=a^*_{-r}a^*_{r+v+v'}a_va_{v'}$ 
we rewrite 
\begin{equation}\label{eq:norm}
    \begin{split}
        & \|\cD_\a \,g^0_\a\|^2 \\
        &= \! \sum_{n=0}^{n_{\mathrm{c}}}\frac{1}{(n!)^2(2N)^{2n}}\hspace{-0.3cm}
        \sum_{\substack{\scriptscriptstyle{r_1,\dots,r_n\in P_{\mathrm{H}}}\\\scriptscriptstyle\tilde r_1,\dots,\tilde r_n\in P_{\mathrm{H}}}}\sum_{\substack{\scriptscriptstyle{(v_1,v'_1)\in \bbG_{r_1}}\\\scriptscriptstyle{(\tilde v_1,\tilde v'_1)\in \bbG_{\tilde r_1}}}}\hspace{-0.2cm}\dots\hspace{-0.2cm} \sum_{\substack{\scriptscriptstyle{(v_n,v'_n)\in \bbG_{r_n}}\\\scriptscriptstyle{(\tilde v_n,\tilde v'_n)\in \bbG_{\tilde r_n}}}}\hspace{-0.4cm}\theta\big(\{r_j,v_j,v'_j\}_{j=1}^n\big)\theta\big(\{\tilde r_j,\tilde v_j,\tilde v'_j\}_{j=1}^n\big)
        \!\\
         & \hskip 2cm \times \big(\prod_{i=1}^n\eta_{r_i}\eta_{\tilde r_i}\big) \langle  \,g^0_\a, A^*_{\tilde{r}_n,\tilde{v}_n,\tilde{v}'_n}\dots A^*_{\tilde{r}_1,\tilde{v}_1,\tilde{v}'_1}A_{r_1,v_1,v'_1}\dots A_{r_n,v_n,v'_n} \,g^0_\a \rangle
    \end{split}
\end{equation}
with 
\be\label{eq:theta-small}
	\begin{split}
		\theta(\{r_j,v_j,v'_j\}_{j=1}^n)=\prod_{\substack{i,j,k,h\\i\neq j}}^n\,{\prod_{\substack{\scriptscriptstyle{\substack{p_i,p_j}}}}}\,
		{\prod_{\substack{w_h, w_k\\w_j}}}^{\hskip -0.15cm*} \Big(1- \d_{p_i, -p_j+w_h+w_k} \Big)  \Big(1- (1-\d_{w_j,0}) \d_{v_i+v_i', w_j+w_k}\Big)
	\end{split}
\ee
where in the above formula $p_\ell\in \{-r_\ell,r_\ell+v_\ell+v'_\ell\}$ and $w_\ell\in \{v_\ell,v'_\ell\}$ for $\ell=i,j,h,k$ and the $*$ in the product over $w_h$, $w_k$ and $w_j$ denotes the convention $w_\ell +w_k = v_k +v_k'$ for $\ell=k$ with $\ell=j,h$. Note that the function $\theta(\{r_j,v_j,v'_j\}_{j=1}^n)$ takes the restrictions that have been discussed in \eqref{eq:restrictions1} and \eqref{eq:restrictions2} into account.

It remains to evaluate the inner product in the last line of \eqref{eq:norm}. Since $g_{\alpha}^0$ has a fixed number of particles, each momentum associated with a creation operator must be paired with one from an annihilation operator. A priori, this results in approximately $(n!)^2$ terms. The operator $\Theta_\a$, which gives rise to the function $\theta$ in \eqref{eq:norm}, was introduced to tame this combinatorial growth. In particular, it enforces that the momenta appearing in the creation and annihilation operators within a given $A_{r,v,v'}$ must be matched with momenta from the corresponding creation and annihilation operators in the same $A^*_{\tl r,\tl v,\tl v'}$. This constraint reduces the combinatorial factor to $n!$.

\smallskip

\begin{figure}[t] 
\centering
\begin{tikzpicture}[scale =0.6,
    every path/.style = {},
 ]
  \begin{scope} 
\filldraw[black] (0,1.5) circle (3pt);
\node[] at (0, 1) {\scriptsize  $-r_1$};
\filldraw[black] (0.8,1.5) circle (3pt) node[below]{};
\filldraw[color=black, fill=white, thick] (1.6,1.4) rectangle (1.8,1.6);
\node[] at (1.7, 1) {\scriptsize  $v_1$};
\filldraw[color=black, fill=white, thick] (2.3,1.4) rectangle (2.5,1.6);
\node[] at (2.4, 1) {\scriptsize  $v'_1$};
\filldraw[black] (0,-1.5) circle (3pt);
\node[] at (0, -2) {\scriptsize  $-\tl{r}_1$};
\filldraw[black] (0.8,-1.5) circle (3pt) node[below]{};
\filldraw[color=black, fill=white, thick] (1.6,-1.4) rectangle (1.8,-1.6);
\node[] at (1.7, -2) {\scriptsize  $\tl{v}_1$};
\filldraw[color=black, fill=white, thick] (2.3,-1.4) rectangle (2.5,-1.6);
\node[] at (2.4, -2) {\scriptsize  $\tl{v}'_1$};
\filldraw[black] (4,1.5) circle (3pt);
\node[] at (4, 1) {\scriptsize  $-r_2$};
\filldraw[black] (4.8,1.5) circle (3pt) node[below]{};
\filldraw[color=black, fill=white, thick] (5.6,1.4) rectangle (5.8,1.6);
\node[] at (5.7, 1) {\scriptsize  $v_2$};
\filldraw[color=black, fill=white, thick] (6.3,1.4) rectangle (6.5,1.6);
\node[] at (6.4, 1) {\scriptsize  $v'_2$};
\filldraw[black] (4,-1.5) circle (3pt);
\node[] at (4, -2) {\scriptsize  $-\tl{r}_2$};
\filldraw[black] (4.8,-1.5) circle (3pt) node[below]{};
\filldraw[color=black, fill=white, thick] (5.6,-1.4) rectangle (5.8,-1.6);
\node[] at (5.7, -2) {\scriptsize  $\tl{v}_2$};
\filldraw[color=black, fill=white, thick] (6.3,-1.4) rectangle (6.5,-1.6);
\node[] at (6.4, -2) {\scriptsize  $\tl{v}'_2$};
\filldraw[black] (8,1.5) circle (3pt);
\node[] at (8, 1) {\scriptsize  $-r_3$};
\filldraw[black] (8.8,1.5) circle (3pt) node[below]{};
\filldraw[color=black, fill=white, thick] (9.6,1.4) rectangle (9.8,1.6);
\node[] at (9.7, 1) {\scriptsize  $v_3$};
\filldraw[color=black, fill=white, thick] (10.3,1.4) rectangle (10.5,1.6);
\node[] at (10.4, 1) {\scriptsize  $v'_3$};
\filldraw[black] (8,-1.5) circle (3pt);
\node[] at (8, -2) {\scriptsize  $-\tl{r}_3$};
\filldraw[black] (8.8,-1.5) circle (3pt) node[below]{};
\filldraw[color=black, fill=white, thick] (9.6,-1.4) rectangle (9.8,-1.6);
\node[] at (9.7, -2) {\scriptsize  $\tl{v}_3$};
\filldraw[color=black, fill=white, thick] (10.3,-1.4) rectangle (10.5,-1.6);
\node[] at (10.4, -2) {\scriptsize  $\tl{v}'_3$};
\node[below] at (12.2, 1.8) {\footnotesize  $\cdots$};
\node[below] at (12.2, -1.2) {\footnotesize  $\cdots$};
\filldraw[black] (14,1.5) circle (3pt);
\node[] at (14, 1) {\scriptsize  $-r_n$};
\filldraw[black] (14.8,1.5) circle (3pt) node[below]{};
\filldraw[color=black, fill=white, thick] (15.6,1.4) rectangle (15.8,1.6);
\node[] at (15.7, 1) {\scriptsize  $v_n$};
\filldraw[color=black, fill=white, thick] (16.3,1.4) rectangle (16.5,1.6);
\node[] at (16.4, 1) {\scriptsize  $v'_n$};
\filldraw[black] (14,-1.5) circle (3pt);
\node[] at (14, -2) {\scriptsize  $-\tl{r}_n$};
\filldraw[black] (14.8,-1.5) circle (3pt) node[below]{};
\filldraw[color=black, fill=white, thick] (15.6,-1.4) rectangle (15.8,-1.6);
\node[] at (15.7, -2) {\scriptsize  $\tl{v}_n$};
\filldraw[color=black, fill=white, thick] (16.3,-1.4) rectangle (16.5,-1.6);
\node[] at (16.4, -2) {\scriptsize  $\tl{v}'_n$};
\draw[color=orange, very thick, dashed, - ] (0,-1.2) -- (0,0.7);  
\draw[color=orange, very thick, dashed, - ] (0.9,1) -- (3.8,-1.2);
\draw[color=blue, very thick, dotted, - ] (7.9,0.7) -- (5,-1.2);  
\draw[color=blue, very thick, dotted, - ] (8.9,1) -- (14,-1.2);
\draw[color=green!80!blue, very thick,  -] (8,-1.2) -- (14,0.8);   
\draw[color=green!80!blue, very thick,  - ] (8.8,-1.2) -- (14.8,0.8);
\draw[fill=white, opacity=0.1] (-0.8,-2.4) rectangle (17.3,2);
\end{scope}
\end{tikzpicture}
\centering 
\begin{minipage}{13cm}
\caption{\small Schematic picture of possible contractions among the quartic operators $\{A_{r_i, v_i, v'_i} \}_{i=1,...,n} $ (upper set of points) and $\{A^*_{\tl r_i, \tl v_i, \tl v_i'} \}_{i=1,...,n} $ (lower set of points). Each dot (resp. square) represents a creation/annihilation operator with momenta in $P_{\mathrm{H}}$ (resp. in $P_{\A}^0$); each line connecting two dots represents a delta function between the momenta labeling the dots. The contractions depicted by the dashed and dotted lines above are forbidden due the restrictions encoded in \eqref{eq:restrictions1} and \eqref{eq:restrictions2}; differently, the contractions depicted by the solid lines are allowed and imply  $v_n+v'_n= \tl{v}_3 + \tl{v}'_3$.
} \label{Fig1}
\end{minipage}
\end{figure}
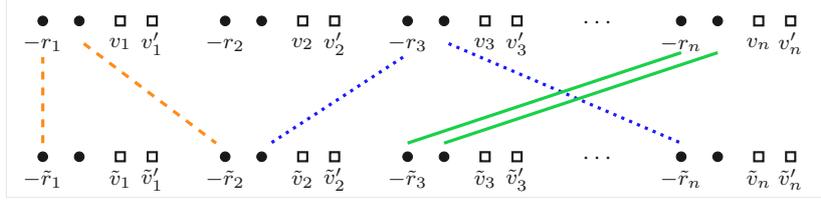

Let us see the computation in details. Recalling that the occupation number state $g^0_\a$ contains only momenta in $P_{\rm G} \cup\{0\}$, we conclude that each annihilation operator with high momentum appearing in $A^*_{\tilde{r}_n,\tilde{v}_n,\tilde{v}'_n}\dots A^*_{\tilde{r}_1,\tilde{v}_1,\tilde{v}'_1}$ has to be contracted with one among the creation operators with high momentum appearing in $A_{r_1,v_1,v'_1}\dots A_{r_n,v_n,v'_n}$. Since $g_\a^0$ has a fixed number of particles, for any creation operator $a^*_{\bar \o}$ with $\bar \o \in P_{\A}\cup \{0\}$ the right-hand side of \eqref{eq:norm2} is zero unless there  exists an annihilation operator with the same momentum.    Furthermore, due to the restrictions encoded in $\theta(\{r_j,v_j,v'_j\}_{i=1}^n)$ and $\theta(\{\tilde{r}_j,\tilde{v}_j,\tilde{v}_j'\}_{j=1}^n)$ we have that if $p_i=\tilde{p}_j$ then necessarily $-p_i+v_i+v_i'=-\tilde{p}_j+\tilde{v}_j+\tilde{v}'_j$ for any $p_i\in \{-r_i,r_i+v_i+v'_i\}$, $\tilde{p}_j\in \{-\tilde{r}_j,\tilde{r}_j+\tilde{v}_j+\tilde{v}_j'\}$ with $i,j=1,\dots,n$. Hence, this contraction implies $v_i+v'_i=\tilde v_j+\tilde v'_j$, see  Fig.~\ref{Fig1} for a schematic representation. Note that, due to the symmetry, we can always assume $i=j$ paying a factor $n!$. Thus, 
\begin{equation} \label{eq:norm2}
    \begin{split}
        \|\cD_\a  \,g^0_\a\|^2=\, &\sum_{n=0}^{n_{\mathrm{c}}}\frac{1}{n!(2N)^{2n}}
        \sum_{\substack{\scriptscriptstyle{r_1,\dots,r_n\in P_{\mathrm{H}}}\\\scriptscriptstyle\tilde r_1,\dots,\tilde r_n\in P_{\mathrm{H}}}}\sum_{\substack{\scriptscriptstyle{(v_1,v'_1)\in \bbG_{r_1}}\\\scriptscriptstyle{(\tilde v_1,\tilde v'_1)\in \bbG_{\tilde r_1}}}}\hspace{-0.2cm}\dots\hspace{-0.2cm} \sum_{\substack{\scriptscriptstyle{(v_n,v'_n)\in \bbG_{r_n}}\\\scriptscriptstyle{(\tilde v_n,\tilde v'_n)\in \bbG_{\tilde r_n}}}}\theta\big(\{r_j,v_j,v'_j\}_{j=1}^n\big)\\
         &\times \theta\big(\{\tilde r_j,\tilde v_j,\tilde v'_j\}_{j=1}^n\big) \prod_{i=1}^n\eta_{r_i}\eta_{\tilde r_j}  (\d_{\tilde{r}_i,r_i}+\d_{-\tilde{r}_i,r_i+v_i+v'_i}) \, \d_{v_i+v'_i,\tilde{v}_i+\tilde{v}_i'} \\
         & \times \langle g^0_\alpha, a^*_{\tilde v_n}a^*_{\tilde v'_n}\dots a^*_{\tilde v_1}a^*_{\tilde v'_1}a_{v_1}a_{v'_1}\dots a_{v_n}a_{v'_n} g^0_\a\rangle\,.
    \end{split}
\end{equation}

Let us focus now on the set of momenta $\{v_1, \ldots, v'_n\}$. We denote with $h$ and $k$ the number of low momenta pairs $(v_\ell, v'_\ell)$ coming from a quadruplet or a triplet respectively, in the sense of the definition above \eqref{eq:restrictions1}; then clearly $n-h-k$ is the number of pairs $(v_\ell, v'_\ell)=(0,0)$. 
We now claim that, due to the restrictions encoded in $\cD_\a$, the number of triplets (resp. quadruples) among the momenta $\{\tl v_1, \ldots, \tl v'_n\}$ has to be $h$ (resp. $k$) as well, and $\{v_\ell, v'_\ell\}=\{\tl v_\ell, \tl v'_\ell\}$ for any $\ell= 1, \ldots, n$. In fact:
\begin{enumerate}
\item If $v_\ell= v'_\ell=0$, due the presence of $\d_{v_\ell+v'_\ell,\tilde v_\ell+\tilde v'_\ell}$,  and recalling that by definition  $\tilde v'_\ell\neq -\tilde v_\ell$ if $\tl v_\ell \neq 0$, we conclude  $\tilde v_\ell=\tilde v'_\ell=0$. 
\item  If $v_\ell=0$ and $v'_\ell \neq 0$, we notice that $\d_{v'_\ell, \tl v_\ell  + \tl v'_\ell}=1$  only if $\{\tl v_\ell , \tl v'_\ell \} = \{0, v'_\ell \}$, due to the conditions encoded by the product $\theta(\{r_i,v_i,v'_i\}_{i=1}^n)\theta(\{\tilde r_i,\tilde v_i.\tilde v'_i\}_{i=1}^n)$. In fact, if by contradiction $\tl v_\ell =w_i \neq 0$ for $w_i \in \{v_i, v'_i\}$ and some $i \neq \ell $ and  $\tl v'_\ell =w_j$, we would get $v'_\ell  = w_i+w_j$ which is not allowed, see \eqref{eq:restrictions2}.  The case  $v'_\ell=0$ and $v_\ell \neq 0$ is analogous. 
\item Finally, if $v_\ell,v_\ell'\neq 0$, then  $\d_{v_\ell + v'_\ell, \tl v_\ell  + \tl v'_\ell}=1$  forces $\{v_\ell, v'_\ell\} =\{\tl v_\ell, \tl v'_\ell\}$. Any other possibility can be ruled out by \eqref{eq:restrictions2}.
\end{enumerate}

Moreover, exploiting the symmetry, we can assume the first $h$ indices in \eqref{eq:norm2} to be associated to quadruplets, and the following $k$ indices to triplets by putting the appropriate binomial factors. 
Based on the above considerations,  and using that $a^*_v a_v g^0_\a = n^\a_v g^0_\a$ for any $v \in P_{\A}$,  and 
 that for all $t \leq N_0$
\be \label{eq:a0m}
(a^*_0)^{t}(a_0)^{t} g^\a_{0}  = \frac{N_0!}{(N_0 - t)!} \, g^{\a}_0
\ee
we get (with $t= k + 2(n-k-h)$)
\[
	\begin{split}
		 &\|\cD_\a \,g^0_\a\|^2\\
		 &=\sum_{n=0}^{n_{\mathrm{c}}}\frac1{n!(2N)^{2n}}\sum_{h=0}^n\sum_{k=0}^{n-h}\binom{n}{h}\binom{n-h}{k} 2^{h+2k+(n-h-k)}  \frac{N_0!}{(N_0 - k-2(n-k-h))!}
		 \\
		 &\phantom{{}={}}\times\hspace{-0.3cm}\sum_{\substack{r_1\in P_{\mathrm{H}}\\(v_1,v_1')\in \bbG^{\rm q}_{r_1}}}\dots\hspace{-0.2cm}\sum_{\substack{r_h\in P_{\mathrm{H}}\\(v_h,v'_h)\in \bbG^{\rm q}_{r_h}}} \sum_{\substack{r_{h+1}\in P_{\mathrm{H}}\\(v_{h+1},0)\in \bbG^{\rm c}_{r_{h+1}}}}\dots\hspace{-0.2cm}\sum_{\substack{r_{h+k}\in P_{\mathrm{H}}\\(v_{h+k},0)\in \bbG^{\rm c}_{r_{h+k}}}}\sum_{r_{h+k+1},\dots,r_n\in P_{\mathrm{H}}}\hspace{-0.3cm}\theta(\{r_s,v_s,v'_s\}_{s=1}^n)\\
		 &\phantom{{}={}}\times\prod_{i=1}^h \eta_{r_i}(\eta_{r_i}+\eta_{r_i+v_i+v'_i})\,n^\a_{v_i}n^\a_{v'_i}\prod_{j=h+1}^{h+k}\eta_{r_j}(\eta_{r_j}+\eta_{r_j+v_j})\,n^\a_{v_j}\prod_{m=h+k+1}^n\eta_{r_m}^2\,.
	\end{split}
\]

 Note that in $\theta(\{r_s,v_s,v'_s\}_{s=1}^{n})$ we are adopting the convention $v_i'=0$ for any $i> h$ and $v_i=0$ for any $i>h+k.$ 
 Using the aforementioned properties that $(v,v')\in \bbG_{r}$ if and only if $(v,v')\in \bbG_{-r+v+v'}$ and the fact that $\theta(\{r_s,v_s,v_s'\}_{s=1}^{n})$ is invariant under the exchange $r_\ell\to -r_\ell+v_\ell+v_\ell'$ we can manipulate the sums over $r_1,\dots,r_{h+k}$ to rewrite:
\be   \label{eq:norm-key}
	 \begin{split}
         &\|\cD_\a \,g^0_\a\|^2\\
         &=\sum_{n= 0}^{n_{\mathrm{c}}}\frac{1}{N^{2n}}\sum_{h=0}^n\sum_{k=0}^{n-h}\frac{1}{h!k!(n-h-k)!}  \frac{N_0!}{(N_0 - k-2(n-k-h))!} \, 2^{-2h-k-(n-h-k)} \\
         &\phantom{{}={}}\times\hspace{-0.3cm}\sum_{\substack{\scriptscriptstyle{r_1\in P_{\mathrm{H}}}\\\scriptscriptstyle{(v_1,v_1')\in \bbG^{\rm q}_{r_1}}}}\dots\hspace{-0.2cm}\sum_{\substack{\scriptscriptstyle{r_h\in P_{\mathrm{H}}}\\\scriptscriptstyle{(v_h,v_h')\in \bbG^{\rm q}_{r_h}}}}\;\sum_{\substack{\scriptscriptstyle{r_{h+1}\in P_{\mathrm{H}}}\\\scriptscriptstyle{(v_{h+1},0)\in \bbG^{\rm c}_{r_{h+1}}}}}\hspace{-0.2cm}\dots\hspace{-0.2cm}\sum_{\substack{\scriptscriptstyle{r_{h+k}\in P_{\mathrm{H}}}\\\scriptscriptstyle{(v_{h+k},0)\in \bbG^{\rm c}_{r_{h+k}}}}}\;\sum_{\scriptscriptstyle{r_{h+k+1},\dots,r_n\in P_{\mathrm{H}}}}\hspace{-0.3cm}\theta(\{r_s,v_s,v'_s\}_{s=1}^n)\\
         &\phantom{{}={}}\times\prod_{i=1}^h(\eta_{r_i}+\eta_{r_i+v_i+v'_i})^2\,n^\a_{v_i}n^\a_{v'_i}\prod_{j=h+1}^{h+k}(\eta_{r_j}+\eta_{r_j+v_j})^2\,n^\a_{v_j}\prod_{m=h+k+1}^n\eta_{r_m}^2.
        \end{split}	
\ee
 We will use the expression \eqref{eq:norm-key} extensively in the following section to compute the energy of our trial state.  In particular, it leads to the lower bound 
 \be \label{eq:LB-norm}
 \|\cD_\a \,g^0_\a\|^2 \geq 1 
 \ee
 since the $n=0$ term on the right-hand side of \eqref{eq:norm-key} is equal to one, and all other terms are non negative.

\section{Free energy of the correlated state} \label{sec:free-energy}

In this Section we conclude the proof of Proposition \ref{prop:Hkappa} by showing that  there exist $\k \in (1/3; 2/3)$, $\e, \d_1, \d_2>0$ with $3\k-2+3\e +\d_2 <0$ and $n_{\mathrm{c}} \in \bN$ such that the  trial state $\G$ defined in \eqref{eq:GammaD} satisfies \eqref{eq:mainUB-kappa}. To this aim, we rewrite the Fock space Hamiltonian $\cH_N$, defined by \eqref{eq:HN}, in momentum space
\begin{equation} 	\label{eq:HN-Fock}
	\cH_N = \sum_{p \in \Lambda^*} p^2 a_p^* a_p + \frac{1}{2} \sum_{p,q,s \in \Lambda^*} \widehat{V}_N(s) a^*_{p+s} a^*_{q} a_{q+s} a_p := \cK + \cV_N
\end{equation}
where we introduced the notations $\widehat{V}_N(p) = N^{\k-1} \widehat V(p/N^{1-\k})$ and $\L^*:= \L^*_1= 2\pi \bZ^3$. 
Eq.\,\eqref{eq:mainUB-kappa} follows from the following two lemmas providing information on the energy and entropy of the correlated state respectively. 

\begin{lemma} \label{lm:corr} Let $\G$ be defined as in \eqref{eq:GammaD}, with $n_{\mathrm{c}} = N^{\d_2/3} \| \eta_H\|^2_2$ and $\eta_H$  as in Lemma \ref{lm:eta}.    Let $\cK$ and $\cV_N$ as in \eqref{eq:HN-Fock}.  
Then for all $\eps,\d_2>0$ so that $3\k-2 + 3 \eps+\d_2<0$ we have
\be \label{eq:corr-K}
\Tr (\cK \G)  \leq \Tr (\cK \G_0)+ \frac 1 {N^2} \sum_{r \in P_{\mathrm{H}}} r^2 \eta_r^2 \,\big(2N^2- N_0^2\big) +  C N^{\k+1-\eps}
\ee
and 
\be \label{eq:corr-VN}
\Tr (\cV_N \G)  \leq  \bigg[\frac {1} {2} \widehat V_N(0) + \frac 1 N \sum_{r \in P_{\mathrm{H}}} \widehat V_N(r) \eta_r +\frac{1}{2N^2} \sum_{r \in P_{\mathrm{H}}} \big(\widehat V_N\ast \eta\big)_r \eta_r \bigg]\big(2N^2- N_0^2\big) 
+  {\cal E}_N
\ee
with
\be  \label{eq:calEN}
{\cal E}_N \leq C N^{\k+1} \max \{N^{-\eps}, N^{6\k -3 +2\eps+\d_1/3}, N^{12\k- 6+6\eps}, N^{-\delta_1/3} \} 
\ee
for all $\k \in (1/3,2/3)$ and $N$ large enough.  
\end{lemma}

The following lemma establishes a lower bound for the entropy of our trial state. 

\medskip

\begin{lemma} \label{lm:entropy} 
Let $\G$ be defined as in \eqref{eq:GammaD}, with $n_{\rm c}$ such that $4 n_{\mathrm{c}} \leq | P_{\A} | $, and $P_{\A}$ defined as in \eqref{eq:momentumSets}. Then there exists a constant $C>0$ such that  
	\begin{equation}
		S(\Gamma) \geq -\sum_{\alpha} \lambda_{\alpha} \ln( \lambda_{\alpha} )  
		-C \left( 1 + n_{\mathrm{c}} (1 + \ln(|P_{\A}|))  \right)\,.
		\label{eq:entropyLowerBound}
		\end{equation}
\end{lemma}

In the following we show how Proposition \ref{prop:Hkappa} follows from the use of Lemma \ref{lm:corr} and \ref{lm:entropy}, together with Lemma \ref{lm:Gamma0}.

\medskip

\begin{proof}[Proof of Proposition \ref{prop:Hkappa}]  
We first notice that the choice $n_{\mathrm{c}}= N^{\d_2/3} \|\eta_H\|^2_2$ is compatible with the assumption in Lemma~\ref{lm:entropy}. Indeed, for any $\b_N$ satisfying $ N^{-\k}\geq \b_N \geq c N^{-2/3}$ and $N$ large enough we have 
$n_{\rm c} \ll |P_{\A}|\leq C \b_N^{-3/2(1+\d_2)}$,  
having set $3\k-2+3\eps +\d_2 <0$. 
Then, \eqref{eq:corr-K}--\eqref{eq:calEN}, together with \eqref{eq:scatl-etap} lead to the upper bound
\[ \begin{split}
\Tr(\cH_N \G) &\, \leq  \Tr(\cK \G_0) \\
&+ \frac 1 2 \bigg[ \widehat V_N(0) + \frac 1 {N} \sum_{r \in P_{\mathrm{H}}} \big(\widehat V_N(r) \eta_r +  N^{2-2\k}\l_\ell \big(\widehat{\chi}_\ell \ast \widehat{f}_{N,\ell} \big)(r)\eta_r\big) \bigg]\big(2N^2- N_0^2\big) +  {\cal E}'_N
\end{split}\]
with ${\cal E}'_N \leq C N^{\k+1}  \max\{N^{-\eps},N^{6\k-3+2\eps +\d_1/3}, N^{12\k-6+6\eps}, N^{-\delta_1/3} \}$.

Using that $N^{3-3\k}\l_\ell \leq C$ (Lemma~\ref{lm:scatt-eq}), and that $\|\widehat{\chi}_\ell \ast \widehat{f}_{N,\ell} \|_2 = \| \chi_\ell f_{N,\ell}\|_2 \leq C $, together with \eqref{eq:etaH}, we find
\be \label{eq:prop2.4-a}
N^{3-2\k}\l_\ell\sum_{r \in P_{\mathrm{H}}} \big(\widehat{\chi}_\ell \ast \widehat{f}_{N,\ell} \big)(r) \,\eta_r   \leq C N^\k \| \chi_\ell f_{N,\ell}\|_2 \| \eta_H \|_2 \leq C N^{\frac 5 2 \k - (1-\eps)/2}\,.
\ee
On the other hand, 
\be \label{eq:prop2.4-b}
N \sum_{\substack{r \in \L^*\setminus\{0\}\\
|r|\leq N^{1-\k -\eps}}} \widehat V_N(r) \eta_r  \leq  C N |\widehat V_N(0)|  \sum_{\substack{r \in \L^*\setminus\{0\}\\
|r|\leq N^{1-\k -\eps}}} \frac{N^\k}{|r|^2} \leq C N^{1+\k -\eps}
\ee
so that 
\be \label{eq:prop2.4-c}
	\begin{split}
		\frac 1 2 \bigg[ \widehat V_N(0) + \frac 1 {N} \sum_{r \in P_{\mathrm{H}}} \widehat V_N(r) \eta_r \bigg]&\big(2N^2- N_0^2\big)  \\
&\leq \frac12N^{\k-1}\int V(x)f_{\ell}(x)dx \big(2N^2- N_0^2\big)+CN^{1+\k -\eps}.
	\end{split}
\ee
Hence, with \eqref{eq:prop2.4-a}--\eqref{eq:prop2.4-c}, and part ii) in Lemma~\ref{lm:scatt-eq} we obtain
\[ 
\Tr(\cH_N \G) \, \leq  \Tr(\cK \G_0) + 4 \pi \aa \big(2 N^2- N_0^2\big) +{\cE''_N}
\]
with $\pm \cE''_N \leq C N^{\k+1}  \max\{ N^{-\eps},\, N^{6\k-3+2\eps+\d_1/3}, N^{12\k-6 +6\e}, N^{-\delta_1/3} \}$\,. 
It remains to discuss the contribution coming from the entropy of $\G$. With \eqref{eq:entropyLowerBound}, and choosing $n_{\mathrm{c}}= N^{\d_2/3} \|\eta_H\|_2^2 $, we obtain 
\[ 
\Tr(\cH_N \G) - T S(\G) \, \leq  \Tr(\cK \G_0) - T S(\G_0) + 4 \pi \aa N^{\k-1}\big(2 N^2- N_0^2\big) + \widetilde{\cE}_N
\]
with 
\[\pm \widetilde{\cE}_N \leq C N^{\k+1}  \max\{ N^{-\eps},\, N^{6\k-3+2\eps+\d_1/3}, N^{12\k-6 +6\e}, N^{-\delta_1/3} \}\,.
\] 
Using Lemma \ref{lm:Gamma0} to bound $\Tr(\cK \G_0) - T S(\G_0)$ from above, 
together with the lower bound $N_0\geq N_0^{\rm id}-1$ we conclude. 
\end{proof}
We are left with the proof of the Lemma \ref{lm:corr} and \ref{lm:entropy}. To compute the expectation of the kinetic and potential energy on the state $\G$, we will proceed similarly as in the computation leading to \eqref{eq:norm-key}, and 
 make repetitively use of Lemma \ref{lm:cN} 
 (with some additional care for non-positive terms, see Subsection \ref{sec:V_2a}). 
 To compute the entropy, we use the cutoffs $\Theta^\a$ and $n_{\rm c}$ appearing in $\cD_\a$ in \eqref{eq:def_D} to control the overlap (with respect to the Fock space inner product) between the vectors $\mathcal{D}_{\alpha} g_{\alpha}^0$ with different $\alpha$.

\subsection{Kinetic energy of the trial state.}

In this Section we prove Eq.\,\eqref{eq:corr-K}. We start noticing that for any $r\in P_{\mathrm{H}}$, and $v,v'\in P_{\A}^0=P_{\A} \cup \{0\}$  we have
\be \label{eq:cKcomm}
\cK \,a^*_{-r}a^*_{r+v+v'}a_va_{v'}=a^*_{-r}a^*_{r+v+v'}a_va_{v'}\big( \cK+(r^2+(r+v+v')^2-v^2-v
'^{\,2}\big)\,.
\ee
Moreover,  with the notations $\cK_H:= \sum_{p \in P_{\mathrm{H}}} p^2 a^*_p a_p$ and $\cK_G:= \sum_{p \in P_{\mathrm{G}}} p^2 a^*_p a_p,$ we have
\be \label{eq:cKidentities-1}
\sum_{n=0}^{n_{\mathrm{c}}}\frac1{(n!)^2}\langle D_\alpha^n \, g^0_\a,D_\alpha^n \,\cK_H g^0_\a\rangle=0 \\
\ee
and
\be \begin{split} \label{eq:cKidentities-2}
		\sum_{n=0}^{n_{\mathrm{c}}}\frac1{(n!)^2}\langle D_\alpha^n \, g^0_\a,D_\alpha^n \cK_G g^0_\a\rangle &\; =\sum_{w\in P_{\mathrm{G}}}w^2n^\a_w\,\| \cD_\a g^0_\a\|^2=\langle\cK_G\rangle_{g^0_\a}\| \cD_\a g^0_\a\|^2\,.
	\end{split}
\ee
Using \eqref{eq:cKcomm} - \eqref{eq:cKidentities-2}, and proceeding as in the computation of the norm we get
\[
	 \begin{split}
         & \langle \cD_\a g^0_\a,  \cK \cD_\a g^0_\a\rangle-\langle\cK_G\rangle_{g^0_\a}\, \| \cD_\a g^0_\a\|^2\\
         &=\sum_{n=1}^{n_{\mathrm{c}}}\frac1{(2N)^{2n}}\sum_{h=0}^{n}\sum_{k=0}^{n-h}\frac{1}{h!k!(n-h-k)!}\frac{N_0!}{(N_0-k-2(n-k-h))!}\,2^{h+ 2k +(n-h-k)}\\
         &\phantom{{}={}}\times\hspace{-0.2cm}\sum_{\substack{\scriptscriptstyle{r_1\in P_{\mathrm{H}}}\\\scriptscriptstyle{(v_1,v_1')\in \bbG^{\rm q}_{r_1}}}}\hspace{-0.2cm}\dots\hspace{-0.2cm}\sum_{\substack{\scriptscriptstyle{r_h\in P_{\mathrm{H}}}\\\scriptscriptstyle{(v_h,v_h')\in \bbG^{\rm q}_{r_h}}}}\;\sum_{\substack{\scriptscriptstyle{r_{h+1}\in P_{\mathrm{H}}}\\\scriptscriptstyle{(v_{h+1},0)\in \bbG^{\rm c}_{r_{h+1}}}}}\hspace{-0.2cm}\dots\hspace{-0.2cm}\sum_{\substack{\scriptscriptstyle{r_{h+k}\in P_{\mathrm{H}}}\\\scriptscriptstyle{(v_{h+k},0)\in \bbG^{\rm c}_{r_{h+k}}}}}\;\hskip -0.2cm\sum_{\scriptscriptstyle{r_{h+k+1},\dots,r_{n}\in P_{\mathrm{H}}}}\hspace{-0.4cm}\theta(\{r_s,v_s,v'_s\}_{s=1}^{n})\\
         &\phantom{{}={}}\times\prod_{i=1}^h\eta_{r_i}(\eta_{r_i}+\eta_{r_i+v_i+v'_i})n^\a_{v_i}n^\a_{v'_i}\prod_{j=h+1}^{h+k}\eta_{r_j}(\eta_{r_j}+\eta_{r_j+v_j})n^\a_{v_j}\prod_{m=h+k+1}^{n}\eta_{r_m}^2\\
            &\phantom{{}={}}\times 2\big[h(r_1^2+r_1 \cdot(v_1+v'_1) + v_1\cdot v'_1)+k(r_{h+1}^2+r_{h+1}\cdot v_{h+1})+(n-h-k)r_{n'+h+k}^2 \big].
        \end{split}	
\]
In the above formula we are denoting, similarly to what we did below \eqref{eq:norm2}, with $h,k,n-h-k$ the number of quadruplets, triplets and pairs respectively. Moreover, in $\theta(\{r_s,v_s,v'_s\}_{s=1}^{n})$ we are adopting again the convention  $v'_s=0$ for any $s>h$ and $v_s=0$ for any $\ell>h+k$. Splitting now the three terms in the square bracket in the last line of the above formula we rewrite
\be\label{eq:split-K}
	 \langle\cD_\a g^0_\a,  \cK \cD_\a g^0_\a\rangle-\langle\cK_G\rangle_{g^0_\a}\, \| \cD_\a g^0_\a\|^2=K_{\rm q}+K_{\rm c}+K_{\rm p}.
\ee
  The subindex  in the notation on the right-hand side refers to the fact that the three terms in the square brackets come from  a quartic, a triplet or a pair excitation respectively. Manipulating the sum over $r_1,\dots,r_{h+k}$ we get
\[
	\begin{split}
		K_{\rm q}&=\sum_{n=1}^{n_{\mathrm{c}}}\frac1{N^{2n}}\sum_{h=1}^{n}\sum_{k=0}^{n-h}\frac{1}{h!k!(n-h-k)!}\frac{N_0!}{(N_0-k-2(n-k-h))!}2^{-2h-k-(n-h-k)}\\
         &\phantom{{}={}}\times\hspace{-0.2cm}\sum_{\substack{\scriptscriptstyle{r_1\in P_{\mathrm{H}}}\\\scriptscriptstyle{(v_1,v_1')\in \bbG^{\rm q}_{r_1}}}}\hspace{-0.2cm}\dots\hspace{-0.2cm}\sum_{\substack{\scriptscriptstyle{r_h\in P_{\mathrm{H}}}\\\scriptscriptstyle{(v_h,v_h')\in \bbG^{\rm q}_{r_h}}}}\;\sum_{\substack{\scriptscriptstyle{r_{h+1}\in P_{\mathrm{H}}}\\\scriptscriptstyle{(v_{h+1},0)\in \bbG^{\rm c}_{r_{h+1}}}}}\hspace{-0.2cm}\dots\hspace{-0.2cm}\sum_{\substack{\scriptscriptstyle{r_{h+k}\in P_{\mathrm{H}}}\\\scriptscriptstyle{(v_{h+k},0)\in \bbG^{\rm c}_{r_{h+k}}}}}\;\hskip -0.2cm\sum_{\scriptscriptstyle{r_{h+k+1},\dots,r_{n}\in P_{\mathrm{H}}}}\hspace{-0.4cm}\theta(\{r_s,v_s,v'_s\}_{s=1}^{n})\\
         &\phantom{{}={}}\times\prod_{i=1}^h(\eta_{r_i}+\eta_{r_i+v_i+v'_i})^2n^\a_{v_i}n^\a_{v'_i}\prod_{j=h+1}^{h+k}(\eta_{r_j}+\eta_{r_j+v_j})^2n^\a_{v_j}\prod_{m=h+k+1}^{n}\eta_{r_m}^2\\
            &\phantom{{}={}}\times 2\big[h(r_1^2+r_1\cdot(v_1+v'_1)  + v_1\cdot v'_1)\big].
	\end{split}
\]
Then, using the fact that all terms in the sum are positive (note that $2\big(r_1^2+r_1\cdot(v_1+v'_1) + v_1\cdot v'_1\big)=r_1^2+(r_1+v_1+v'_1)^2-v_1^2-{v'_1}^2\geq 0$) for the sake of an upper bound we can ignore the restrictions in $\theta$ involving the indices $\{r_1,v_1,v'_1\}.$ This in turn allows us to move out the factor involving those indices which yields
\[
	\begin{split}
		K_{\rm q}&\leq 
          \frac1{N^2}\hspace{-0.2cm}\sum_{\substack{r\in P_{\mathrm{H}}\\(v,v')\in \bbG^{\rm q}_r}}\hspace{-0.2cm}\big(r^2\!+r\!\cdot\!(v+v') + v\cdot v'\big)\eta_r(\eta_r+\eta_{r+v+v'})\,n^\a_{v}n^\a_{v'}\\
         &\phantom{{}={}}\times \sum_{n=0}^{n_{\mathrm{c}}-1}\frac1{N^{2n}}\sum_{h=0}^{n}\sum_{k=0}^{n-h}\frac{1}{h!k!(n-h-k)!}\frac{N_0!}{(N_0-k-2(n-k-h))!}2^{-2h-k-(n-h-k)}\\
         &\phantom{{}={}}\times\hspace{-0.2cm}\sum_{\substack{\scriptscriptstyle{r_1\in P_{\mathrm{H}}}\\\scriptscriptstyle{(v_1,v_1')\in \bbG^{\rm q}_{r_1}}}}\hspace{-0.2cm}\dots\hspace{-0.2cm}\sum_{\substack{\scriptscriptstyle{r_h\in P_{\mathrm{H}}}\\\scriptscriptstyle{(v_h,v_h')\in \bbG^{\rm q}_{r_h}}}}\;\sum_{\substack{\scriptscriptstyle{r_{h+1}\in P_{\mathrm{H}}}\\\scriptscriptstyle{(v_{h+1},0)\in \bbG^{\rm c}_{r_{h+1}}}}}\hspace{-0.2cm}\dots\hspace{-0.2cm}\sum_{\substack{\scriptscriptstyle{r_{h+k}\in P_{\mathrm{H}}}\\\scriptscriptstyle{(v_{h+k},0)\in \bbG^{\rm c}_{r_{h+k}}}}}\;\hskip -0.2cm\sum_{\scriptscriptstyle{r_{h+k+1},\dots,r_{n}\in P_{\mathrm{H}}}}\hspace{-0.4cm}\theta(\{r_s,v_s,v'_s\}_{s=1}^{n})\\
         &\phantom{{}={}}\times\prod_{i=1}^h(\eta_{r_i}+\eta_{r_i+v_i+v'_i})^2n^\a_{v_i}n^\a_{v'_i}\prod_{j=h+1}^{h+k}(\eta_{r_j}+\eta_{r_j+v_j})^2n^\a_{v_j}\prod_{m=h+k+1}^{n}\eta_{r_m}^2
	\end{split}
\]
where we rescaled $n\to n-1,$ $h\to h-1$. Using again that all the terms are positive, we extend the sum over $n$ up to $n_{\mathrm{c}}$ in order to recover the expression \eqref{eq:norm-key}. Thus,
\be\label{eq:K_q}
	K_{\rm q}\leq  \frac1{N^2}\hspace{-0.2cm}\sum_{\substack{r\in P_{\mathrm{H}}\\(v,v')\in \bbG^{\rm q}_r}}\hspace{-0.2cm}\big(r^2\!+r\!\cdot\!(v+v')+v\cdot v'\big)\eta_r(\eta_r+\eta_{r+v+v'})\,n^\a_{v}n^\a_{v'}\|\cD_\a g^0_\a\|^2.
\ee
Proceeding analogously (rescaling $n\to n-1$, and $k\to k-1$)  we find

\[
	\begin{split}
		K_{\rm c}&\leq \frac{2N_{0}}{N^2}\hskip -0.2cm\sum_{\substack{r\in P_{\mathrm{H}}\\(v,0)\in \bbG^{\rm c}_r}}\hspace{-0.2cm}\big(r^2+r\cdot v\big)\eta_r(\eta_r+\eta_{r+v})\,n^\a_v\\
         &\phantom{{}={}}\times \sum_{n=0}^{n_{\mathrm{c}}-1}\frac1{N^{2n}}\sum_{h=0}^{n}\sum_{k=0}^{n-h}\frac{1}{h!k!(n-h-k)!}\frac{(N_0-1)!}{(N_0-1-k-2(n-k-h))!}2^{-2h-k-(n-h-k)}\\
         &\phantom{{}={}}\times\hspace{-0.2cm}\sum_{\substack{\scriptscriptstyle{r_1\in P_{\mathrm{H}}}\\\scriptscriptstyle{(v_1,v_1')\in \bbG^{\rm q}_{r_1}}}}\hspace{-0.2cm}\dots\hspace{-0.2cm}\sum_{\substack{\scriptscriptstyle{r_h\in P_{\mathrm{H}}}\\\scriptscriptstyle{(v_h,v_h')\in \bbG^{\rm q}_{r_h}}}}\;\sum_{\substack{\scriptscriptstyle{r_{h+1}\in P_{\mathrm{H}}}\\\scriptscriptstyle{(v_{h+1},0)\in \bbG^{\rm c}_{r_{h+1}}}}}\hspace{-0.2cm}\dots\hspace{-0.2cm}\sum_{\substack{\scriptscriptstyle{r_{h+k}\in P_{\mathrm{H}}}\\\scriptscriptstyle{(v_{h+k},0)\in \bbG^{\rm c}_{r_{h+k}}}}}\;\hskip -0.2cm\sum_{\scriptscriptstyle{r_{h+k+1},\dots,r_{n}\in P_{\mathrm{H}}}}\hspace{-0.4cm}\theta(\{r_s,v_s,v'_s\}_{s=1}^{n})\\
         &\phantom{{}={}}\times\prod_{i=1}^h(\eta_{r_i}+\eta_{r_i+v_i+v'_i})^2n^\a_{v_i}n^\a_{v'_i}\prod_{j=h+1}^{h+k}(\eta_{r_j}+\eta_{r_j+v_j})^2n^\a_{v_j}\prod_{m=h+k+1}^{n}\eta_{r_m}^2.
	\end{split}
\]
By positivity we add back the term corresponding to $n=n_{\mathrm{c}}$ and replace $\frac{(N_0-1)!}{(N_0-1-k-2(n-k-h))!}$ with $\frac{N_0!}{(N_0-k-2(n-k-h))!}$ arriving at
\be\label{eq:K_c}
	K_{\rm c}\leq  \frac{2N_{0}}{N^2}\hskip -0.2cm\sum_{\substack{r\in P_{\mathrm{H}}\\(v,0)\in \bbG^{\rm c}_r}}\hspace{-0.2cm}\big(r^2+r\cdot v\big)\eta_r(\eta_r+\eta_{r+v})\,n^\a_v \|\cD_\a g_\a^0\|^2.
\ee
Finally, rescaling $n\to n-1$, we get
\[
	\begin{split}
		K_{\rm p}&\leq \frac{N_0(N_0-1)}{N^2}\sum_{r\in P_{\mathrm{H}}}r^2\eta_r^2\\
		&\phantom{{}={}}\times \sum_{n=0}^{n_{\mathrm{c}}-1}\frac1{N^{2n}}\sum_{h=0}^{n}\sum_{k=0}^{n-h}\frac{1}{h!k!(n-h-k)!}\frac{(N_0-2)!}{(N_0-2-k-2(n-k-h))!}2^{-2h-k-(n-h-k)}\\
         &\phantom{{}={}}\times\hspace{-0.2cm}\sum_{\substack{\scriptscriptstyle{r_1\in P_{\mathrm{H}}}\\\scriptscriptstyle{(v_1,v_1')\in \bbG^{\rm q}_{r_1}}}}\hspace{-0.2cm}\dots\hspace{-0.2cm}\sum_{\substack{\scriptscriptstyle{r_h\in P_{\mathrm{H}}}\\\scriptscriptstyle{(v_h,v_h')\in \bbG^{\rm q}_{r_h}}}}\;\sum_{\substack{\scriptscriptstyle{r_{h+1}\in P_{\mathrm{H}}}\\\scriptscriptstyle{(v_{h+1},0)\in \bbG^{\rm c}_{r_{h+1}}}}}\hspace{-0.2cm}\dots\hspace{-0.2cm}\sum_{\substack{\scriptscriptstyle{r_{h+k}\in P_{\mathrm{H}}}\\\scriptscriptstyle{(v_{h+k},0)\in \bbG^{\rm c}_{r_{h+k}}}}}\;\hskip -0.2cm\sum_{\scriptscriptstyle{r_{h+k+1},\dots,r_{n}\in P_{\mathrm{H}}}}\hspace{-0.4cm}\theta(\{r_s,v_s,v'_s\}_{s=1}^{n})\\
         &\phantom{{}={}}\times\prod_{i=1}^h(\eta_{r_i}+\eta_{r_i+v_i+v'_i})^2n^\a_{v_i}n^\a_{v'_i}\prod_{j=h+1}^{h+k}(\eta_{r_j}+\eta_{r_j+v_j})^2n^\a_{v_j}\prod_{m=h+k+1}^{n}\eta_{r_m}^2\,.
	\end{split}•
\]
Adding the term corresponding to $n=n_{\mathrm{c}}$, bounding $N_0(N_0-1)\leq N_0^2$
and noticing  that $\frac{(N_0-2)!}{(N_0-2-k-2(n-k-h))!}\leq \frac{N_0!}{(N_0-k-2(n-k-h))!}$ we obtain
\be\label{eq:K_p}
	K_{\rm p}\leq  \frac{N_0^2}{N^2}\sum_{r\in P_{\mathrm{H}}}r^2\eta_r^2 \|\cD_\a g_\a^0\|^2.
\ee
With \eqref{eq:split-K}, \eqref{eq:K_q} - \eqref{eq:K_p}, we conclude
\be \label{eq:split-K-conclusion}
	\begin{split}
		& \frac{\langle \cD_\a g^0_\a \cK \, \cD_\a g^0_\a \rangle}{ \| \cD_\a g^0_\a\|^2}-\langle\cK_G\rangle_{g^0_\a}  \\
		& \leq\,\, \frac1{N^2}\hspace{-0.2cm}\sum_{\substack{r\in P_{\mathrm{H}}\\(v,v')\in \bbG^{\rm q}_r}}\hspace{-0.2cm}\big(r^2\!+r\!\cdot\!(v+v') +v\cdot v'\big)\eta_r(\eta_r+\eta_{r+v+v'})\,n^\a_{v}n^\a_{v'}\\
		&\hskip 3cm +\frac{2N_{0}}{N^2}\hskip -0.2cm\sum_{\substack{r\in P_{\mathrm{H}}\\(v,0)\in \bbG^{\rm c}_r}}\hspace{-0.2cm}\big(r^2+r\cdot v\big)\eta_r(\eta_r+\eta_{r+v})\,n^\a_v+\frac{N_0^2}{N^2}\sum_{r\in P_{\mathrm{H}}}r^2\eta_r^2\,.
	\end{split}
\ee
We now claim that for any $\eps,\d_2>0$ such that $3\k-2+3\eps+\d_2 <0$
\be \label{eq:eta-rv} 
\sup_{\substack{v,v'\in P_{\mathrm{G}}^{0}}}\sum_{\substack{r \in P_{\mathrm{H}}:\\r+v+v'\in P_{\mathrm{H}}}} |r|^2 |\eta_{r+v+v'}- \eta_{r}| |\eta_r|  
 \leq CN^{\k+1}N^{-\e}\,.
\ee
With Eqs. \eqref{eq:split-K-conclusion}--\eqref{eq:eta-rv} we get 
\be\label{eq:K_final}
	\begin{split}
		& \Tr(\cK\G)- \; \Tr(\cK\G_0)\\
		& \leq\frac1{N^2} \sum_{\substack{r\in P_{\mathrm{H}}}} r^2\eta_r^2 \bigg[ 2\sum_\a\l_\a \sum_{(v,v') \in \bbG^{\rm q}_r}n^\a_{v}n^\a_{v'} + 4N_0 \sum_\a\l_\a \sum_{(v,0) \in \bbG^{\rm c}_r} n^\a_{v} +N_0^2\bigg] +\cE_\cK
	\end{split}
\ee
where 
\[
	\begin{split}
		\cE_\cK \leq& \; \frac C {N^2}\hspace{-0.2cm}\sum_{\substack{r\in P_{\mathrm{H}}\\ (v,v') \in \bbG^{\rm q}_r}}\hspace{-0.2cm} |\eta_r|\big(|r|^2|\eta_r-\eta_{r+v+v'}|+ |v|(|r|+|v'|)|\eta_r+\eta_{r+v+v'}|\big)\sum_\a\l_\a     n^\a_{v}n^\a_{v'} \\
		&\;+ C \frac {N_0} {N^2}\hspace{-0.2cm}\sum_{\substack{r\in P_{\mathrm{H}}\\ (v,0) \in \bbG^{\rm c}_r}} |\eta_r|\big(|r|^2|\eta_{r}-\eta_{r+v}|+|r||v||\eta_r+\eta_{r+v}|\big)\sum_\a \l_\a   n^\a_{v} \\
		\leq & \; C N^{\k+1}N^{-\e}+N^{(1+\d_2)/3+\k}\sum_{r \in P_{\mathrm{H}}} \frac{|\eta_r|}{|r|} + N^{2(1+\d_2)/3}\|\eta_H\|^2_2 \leq  C N^{\k+1} N^{-\e}.
	\end{split}
\]
Here we used \eqref{eq:etaH},   \eqref{eq:Nplus}, Lemma \ref{lm:cN}, and the condition $3\k-2+3\eps+\d_2<0$ together with the bound
\be \label{eq:etar-over-r}
	\begin{split}
		N^{(1+\d_2)/3+\k}\sum_{r\in P_{\mathrm{H}}}\frac{|\eta_r|}{|r|}& \leq N^{(1+\d_2)/3+\k}\bigg[ 
		\sum_{\substack{r \in P_{\mathrm{H}}: \\ |r|\leq N^{1-\k}}}\frac{N^\k}{|r|^3}+\|\eta_H\|_{H^1}\Big(\sum_{|r|>N^{1-\k}}\frac1{|r|^4}\Big)^{1/2}\bigg]\\
		&\leq CN^{2\k+(1+\d_2)/3}\ln(N)+CN^{2\k+(1+\d_2)/3}\leq CN^{\k+1}N^{-\eps}.
	\end{split}
\ee
The proof of \eqref{eq:corr-K} follows adding the missing terms in the set $(v,v')\in \bbG^{\rm q}_r$ and $(v,0) \in \bbG^{\rm c}_r$ in the right-hand side of \eqref{eq:K_final} (note that all terms are positive) to reconstruct the sums over $v,v' \in P_{\mathrm{G}}$. Using \eqref{eq:Nplus} and noticing that, with Lemma \ref{lm:cN}, we have
\be \label{eq:Nplus2}
\Big| \sum_{\a}\l_\a \hskip -0.2cm\sum_{v,v' \in P_\b} n_v^\a n^\a_{v'} - (N-N_0)^2 \Big| \leq CN^{4/3}\,.
\ee
To conclude it remains to prove \eqref{eq:eta-rv}. Recalling the scattering equation \eqref{eq:scatl-etap} and introducing the notations $\widehat g_r= \frac N 2 \big(\widehat V_N \ast \widehat f_{\ell,N}\big)(r)$ and $\widehat{h}_r=N^{3-2\k} \l_\ell  \big(\widehat{\chi}_\ell \ast \widehat f_{\ell,N}\big)(r)$ we find
\be \label{eq:diff-eta}
	 |\eta_r - \eta_{r+v+v'}|\leq   (|\widehat g_{r+v+v'} - \widehat g_r| +|\widehat h_r- \widehat h_{r+v+v'}|)|r|^{-2} +CN^\k (|r+v+v'|^{-2} - |r|^{-2})
\ee
where in the second term we used $|\widehat g_{r+v+v'}|,|\widehat h_{r+v+v'}|\leq CN^\k.$ To bound the first term we notice that $\widehat g_r = N^\k \int e^{-i (r/N^{1-\k}) \cdot x} \, V(x) f_{\ell}(x) \mathrm{d}x$; hence
\be \label{eq:diff-eta-1}
 |\widehat g_{r+v+v'}- \widehat g_r|  \leq  C N^{2\k-1} |v+v'| \int |x| V(x) f_{\ell}(x) dx 
 \leq CN^{2\k-2/3+\d_2/3}\,.
\ee
In the second term instead we use that, for  any $r \in P_{\mathrm{H}}$, $v,v' \in P_\A^0$  and $N$ large enough we have $|r+v+v'|^2 \geq |r|^2/2$ yielding
\be \label{eq:diff-eta-2}
	\big(|r|^{-2}- |r+v+v'|^{-2}\big) \leq C |v+v'| |r|^{-3} \leq CN^{1/3+\d_2/3} |r|^{-3}\,.
\ee
With \eqref{eq:diff-eta}--\eqref{eq:diff-eta-2}, Lemma \ref{lm:eta} and \eqref{eq:prop2.4-a}, and using  $3\k-2 +3\e+\d_2<0,$ we conclude 
\[ \begin{split}
& \sum_{r\in P_{\mathrm{H}}}  |r|^2 |\eta_r| | \eta_r - \eta_{r+v+v'}|   \\
& \leq CN^{2\k-2/3+\d_2/3}\|\eta_H\|_1+C\|\widehat h\|_2\|\eta_H\|_2+CN^{\k+1/3+\d_2/3}\sum_{|r|\in P_{\mathrm{H}}}\frac{|\eta_r|}{|r|}
\leq CN^{\k+1}N^{-\eps}
\end{split}\]
where  in the last step we used \eqref{eq:etar-over-r}.

\medskip


\subsection{Interaction energy of the trial state}

The aim of this Section is to show \eqref{eq:corr-VN}. We note that due to the structure of $\G$ we have 
\[
\Tr(\cV_N \G) = \Tr(\cV_0 \G)+\Tr( (\cV_2 +\hc)\G)+\Tr(\cV_4 \G)
\]
with $\cV_j$ the interaction term involving $j$ high energy particles, namely
\be
\begin{split} \label{eq:spitV}
		\cV_0& \;=\frac1{2}\sum_{\substack{s\in \L^*,\,p,q\in P_\A^0:\\p+s,\,q+s\in P_\A^0}}\hskip -0.5cm\widehat{V}_N(s)a^*_{p+s}a^*_qa_{q+s}a_p\\
		\cV_2&\; =  \frac 12 \sum_{\substack{s\in \L^*,\,p \in P_\A^0,\, q \in P_{\mathrm{H}}: \\q+s\in P^0_\A\,,p+s\in P_{\mathrm{H}}}} \hskip -0.5cm\widehat{V}_N(s)a^*_{p+s}a^*_qa_{q+s}a_p  \\
		& \qquad + \frac 12 \sum_{\substack{s\in \L^*,\,p,q\in P_\A^0:\\q+s,p+s\in P_{\mathrm{H}}}} \hskip -0.5cm\big(\widehat{V}_N(s) +\widehat{V}_N(p-q) \big)a^*_{p+s}a^*_qa_{q+s}a_p:=\cV_2^{(a)}+\cV_2^{(b)}\\
		\cV_4&\; =\frac1{2}\sum_{\substack{s\in \L^*,\,p,q\in P_{\mathrm{H}}:\\p+s,\,q+s\in P_{\mathrm{H}}}} \hskip -0.5cm\widehat{V}_N(s)a^*_{p+s}a^*_qa_{q+s}a_p\,.
	\end{split}
	\ee
	In the following we analyse the contribution to the energy coming from each of $\cV_j$. 

\subsubsection{Contribution of $\cV_0$ on the correlated state}

We first analyse the contribution to the interaction energy coming from having all particles in $P_\A^0 = P_\A \cup \{0\}$. By definition of the state $\G$ (see Eq.\,\eqref{eq:GammaD}) we have
\[
	\Tr(\cV_0\G)=\sum_\a\l_\a\frac{\langle \cD_\a g^0_\a,\cV_0 \cD_\a g^0_\a\rangle}{\|\cD_\a g^0_\a\|^2}.
\]
To compute $\langle \cD_\a g^0_\a,\cV_0 \cD_\a g^0_\a\rangle$, we find it convenient to split $\cV_0$ in factors depending on the number of zero momenta appearing:  
\begin{equation}\label{eq:cV_Gsplit}
	\begin{split}
		\cV_0=\; &\frac{1}{2N^{1-\k}}\widehat{V}(0)a^*_0a^*_0a_0a_0+\frac1{N^{1-\k}}\sum_{s\in P_\A}\big(\widehat V(0)+\widehat V(s/N^{1-\k})\big)a^*_sa^*_0a_sa_0\\
		&+\frac1{2N^{1-\k}}\sum_{s\in P_\A}\widehat V(s/N^{1-\k})(a^*_{s}a^*_{-s}a_0a_0+\hc)\\
		&+\frac1{N^{1-\k}}\sum_{\substack{s,p\in P_\A:\\ p+s\in P_\A}}\widehat V(s/N^{1-\k})(a^*_{p+s} a^*_{-s}a_p a_0 + \hc)\\ 
		&+\frac1{2N^{1-\k}}\sum_{\substack{s\in \L^*\!,\,\,p,q\in P_\A:\\p+s,q+s\in P_\A}}\widehat{V}(s/N^{1-\k})  a^*_{p+s}a^*_qa_{q+s}a_p\\
		=&\; \cV_0^{(4)}+\cV_0^{(2,\mathrm{diag})}+\cV_0^{(2,\mathrm{off})}+\cV_0^{(1)}+\cV_0^{(0)}.
	\end{split}
\end{equation}
We start computing the expectation of $\cV_0^{(4)}$ on $\cD_\a g^0_\a.$ Using \eqref{eq:def_D} we write
\[
	\begin{split}
		&\langle \cD_\a g^0_\a,\cV^{(4)}_0 \cD_\a g^0_\a\rangle\\
		&=\sum_{n=0}^{n_{\mathrm{c}}}\frac{1}{(n!)^2(2N)^{2n}}\hspace{-0.2cm}  \sum_{\substack{\scriptscriptstyle{r_1,\dots,r_{n}\in P_{\mathrm{H}}}\\\scriptscriptstyle\tilde r_1,\dots,\tilde r_{n}\in P_{\mathrm{H}}}}\sum_{\substack{\scriptscriptstyle{(v_1,v'_1)\in \bbG_{r_1}}\\\scriptscriptstyle{(\tilde v_1,\tilde v'_1)\in \bbG_{\tilde r_1}}}}\hspace{-0.2cm}\dots\hspace{-0.2cm} \sum_{\substack{\scriptscriptstyle{(v_n,v'_n)\in \bbG_{r_n}}\\\scriptscriptstyle{(\tilde v_{n},\tilde v'_{n})\in \bbG_{\tilde r_n}}}} \theta\big(\{r_j,v_j,v'_j\}_{j=1}^{n}\big)\theta\big(\{\tilde r_j,\tilde v_j,\tilde v'_j\}_{j=1}^n\big)\\
         &\phantom{{}={}}\times\prod_{i=1}^n\eta_{r_i}\eta_{\tilde r_i} \langle g^0_\a, A^*_{\tilde r_n,\tilde v_n,\tilde v'_n}\dots A^*_{\tilde r_1,\tilde v_1,\tilde v'_1} \cV^{(4)}_0A_{r_1,v_1,v'_1}\dots A_{r_n,v_n,v'_n}g^0_\a\rangle
	\end{split}
\]
where we are again using the shorthand notation  $A_{r,v,v'}=a^*_{-r}a^*_{r+v+v'}a_va_{v'}$. 
We note that for all $t,m \in \bN$ such that $t+m \leq N_0$ we have (with \eqref{eq:a0m})
 \be \label{eq:a0m-t}
 (a^*_0)^t (a_0)^t  \,(a_0)^m  g_\a^0 = \frac{(N_0-m)!}{(N_0-m-t)!} \,(a_0)^m  g_\a^0 
 \ee
 being $\,(a_0)^m g_\a^0$ a state with $N_0-m$ particles in the condensate; hence
\[
	\begin{split}
		 \cV^{(4)}_0 &\, A_{r_1,v_1,v'_1}\dots A_{r_n,v_n,v'_n}g_\a^0\\
		&=\frac{\widehat V(0)}{2N^{1-\k}}\, \frac{(N_0-k-2(n-k-h))!}{(N_0-k-2(n-k-h)-2)!}\, A_{r_1,v_1,v'_1}\dots A_{r_n,v_n,v'_n}g^0_\a
	\end{split}
\]
with $k$ and $(n-h-k)$ counting the number of triplets and pairs excitations among the $n$ operators $A_{r_\ell,v_\ell,v'_\ell}$. Pairing then the momenta in $ A_{r_1,v_1,v'_1}\dots A_{r_n,v_n,v'_n}$ with those in $ A^*_{\tl r_1,\tl v_1,\tl v'_1}\dots A^*_{\tl r_n,\tl v_n,\tl v'_n}$ as in the computation leading to \eqref{eq:norm-key}, we obtain
\[\label{eq:cV_04_inter}
	 \begin{split}
         &\langle \cD_\a g^0_\a,\cV^{(4)}_0 \cD_\a g^0_\a\rangle\\
         &=\frac{N_0(N_0-1)}{2N^{1-\k}}\widehat V(0)\sum_{n= 0}^{n_{\mathrm{c}}}\frac{1}{N^{2n}}\sum_{h=0}^n\sum_{k=0}^{n-h}\frac1{h!k!(n-h-k)!}  \frac{2^{-2h-k-(n-h-k)}(N_0-2)!}{(N_0 - k-2(n-k-h)-2)!}  \\
         &\phantom{{}={}}\times\hspace{-0.3cm}\sum_{\substack{\scriptscriptstyle{r_1\in P_{\mathrm{H}}}\\\scriptscriptstyle{(v_1,v_1')\in \bbG^{\rm q}_{r_1}}}}\dots\hspace{-0.2cm}\sum_{\substack{\scriptscriptstyle{r_h\in P_{\mathrm{H}}}\\\scriptscriptstyle{(v_h,v_h')\in \bbG^{\rm q}_{r_h}}}}\;\sum_{\substack{\scriptscriptstyle{r_{h+1}\in P_{\mathrm{H}}}\\\scriptscriptstyle{(v_{h+1},0)\in \bbG^{\rm c}_{r_{h+1}}}}}\hspace{-0.2cm}\dots\hspace{-0.2cm}\sum_{\substack{\scriptscriptstyle{r_{h+k}\in P_{\mathrm{H}}}\\\scriptscriptstyle{(v_{h+k},0)\in \bbG^{\rm c}_{r_{h+k}}}}}\;\sum_{\scriptscriptstyle{r_{h+k+1},\dots,r_n\in P_{\mathrm{H}}}}\hspace{-0.3cm}\theta(\{r_s,v_s,v'_s\}_{s=1}^n)\\
         &\phantom{{}={}}\times\prod_{i=1}^h(\eta_{r_i}+\eta_{r_i+v_i+v'_i})^2\,n^\a_{v_i}n^\a_{v'_i}\prod_{j=h+1}^{h+k}(\eta_{r_j}+\eta_{r_j+v_j})^2\,n^\a_{v_j}\prod_{m=h+k+1}^n\eta_{r_m}^2.
        \end{split}	
\]
Since all terms are positive, we can bound $(N_0-2)!/(N_0-2-m!)$ by $N_0!/(N_0-m)!$, for all $m\leq N_0$. Hence, comparing with \eqref{eq:norm-key}, we conclude
\begin{equation}\label{eq:cV_04}
	\langle \cD_\a g^0_\a,\cV^{(4)}_0 \cD_\a \psi_\a\rangle\leq \frac{N_0^2}{2N^{1-\k}}\widehat V(0)\|\cD_\a g^0_\a\|^2.
\end{equation}
We focus now on $\cV_0^{(2,\mathrm{diag})}.$
By definition, 
\[
	\begin{split}
		&\langle \cD_\a g^0_\a,\cV^{(2,\mathrm{diag})}_0 \cD_\a g^0_\a\rangle\\
		&=\sum_{n=1}^{n_{\mathrm{c}}}\frac{1}{(n!)^2(2N)^{2n}}\hspace{-0.2cm}  \sum_{\substack{\scriptscriptstyle{r_1,\dots,r_{n}\in P_{\mathrm{H}}}\\\scriptscriptstyle\tilde r_1,\dots,\tilde r_{n}\in P_{\mathrm{H}}}}\sum_{\substack{\scriptscriptstyle{(v_1,v'_1)\in \bbG_{r_1}}\\\scriptscriptstyle{(\tilde v_1,\tilde v'_1)\in \bbG_{\tilde r_1}}}}\hspace{-0.2cm}\dots\hspace{-0.2cm} \sum_{\substack{\scriptscriptstyle{(v_n,v'_n)\in \bbG_{r_n}}\\\scriptscriptstyle{(\tilde v_{n},\tilde v'_{n})\in \bbG_{\tilde r_n}}}} \theta\big(\{r_j,v_j,v'_j\}_{j=1}^{n}\big)\theta\big(\{\tilde r_j,\tilde v_j,\tilde v'_j\}_{j=1}^n\big)\\
         &\phantom{{}={}}\times\prod_{i=1}^n\eta_{r_i}\eta_{\tilde r_i} \langle g^0_\a, A^*_{\tilde r_n,\tilde v_n,\tilde v'_n}\dots A^*_{\tilde r_1,\tilde v_1,\tilde v'_1} \cV^{(2,\mathrm{diag})}_0A_{r_1,v_1,v'_1}\dots A_{r_n,v_n,v'_n}g^0_\a\rangle.
	\end{split}
\]
Next, we use that (with Eq. \eqref{eq:a0m-t}) 
\[
	a^*_0a_0A_{r_1,v_1,v'_1}\dots A_{r_n,v_n,v'_n}g^0_\a=\big(N_0-k-2(n-k-h)\big)A_{r_1,v_1,v'_1}\dots A_{r_n,v_n,v'_n}g^0_\a
\]
where once again $k$ (respectively $n-k-h$) is the number of triplets (respectively pair excitations). We also rewrite, using canonical commutation relations,
\[
	\begin{split}
		&\sum_{s\in P_\A}\Big(\widehat V(0)+\widehat V(s/N^{1-\k})\Big)a^*_sa_sA_{r_1,v_1,v'_1}\dots A_{r_n,v_n,v'_n}g^0_\a\\
		&=\sum_{s\in P_\A}\Big(\widehat V(0)+\widehat V(s/N^{1-\k})\Big)n^\a_sA_{r_1,v_1,v'_1}\dots A_{r_n,v_n,v'_n}g^0_\a\\
		&\phantom{{}={}}-\sum_{i=1}^n\!\Big[\d_{v_i\neq0}\Big(\!\widehat V(0)\!+\!\widehat V(v_i/N^{1-\k})\Big)\!+\!\d_{v'_i\neq0}\Big(\!\widehat V(0)\!+\!\widehat V(v'_i/N^{1-\k})\!\Big)\!\Big]A_{r_1,v_1,v'_1}\dots A_{r_n,v_n,v'_n}g^0_\a.
	\end{split}
\]
Pairing the momenta appearing in $A_{r_1,v_1,v'_1}\dots A_{r_n,v_n,v'_n}$ and $A^*_{\tilde r_1,\tilde v_1, \tilde v'_1}\dots A_{\tilde r_n,\tilde v_n,\tilde v'_n}$ as  in Eq.\,\eqref{eq:norm-key}, 
we find
\be\label{eq:cV_02diag_inter}
	\begin{split}
		 \langle \cD_\a & g^0_\a,\cV^{(2,\mathrm{diag})}_0 \cD_\a g^0_\a\rangle\\ 
		 & =\frac{N_0}{N^{1-\k}}\sum_{s\in P_\A}\Big(\widehat{V}(0)+\widehat{V}(s/N^{1-\k})\Big)n^\a_s\\
		 &\times\sum_{n= 0}^{n_{\mathrm{c}}}\frac{1}{N^{2n}}\sum_{h=0}^n\sum_{k=0}^{n-h}\frac{2^{-2h-k-(n-h-k)}}{h!k!(n-h-k)!}  \frac{(N_0-1)!}{(N_0-1 - k-2(n-k-h))!} \,  \\
         &\times\hspace{-0.3cm}\sum_{\substack{\scriptscriptstyle{r_1\in P_{\mathrm{H}}}\\\scriptscriptstyle{(v_1,v_1')\in \bbG^{\rm q}_{r_1}}}}\dots\hspace{-0.2cm}\sum_{\substack{\scriptscriptstyle{r_h\in P_{\mathrm{H}}}\\\scriptscriptstyle{(v_h,v_h')\in \bbG^{\rm q}_{r_h}}}}\;\sum_{\substack{\scriptscriptstyle{r_{h+1}\in P_{\mathrm{H}}}\\\scriptscriptstyle{(v_{h+1},0)\in \bbG^{\rm c}_{r_{h+1}}}}}\hspace{-0.2cm}\dots\hspace{-0.2cm}\sum_{\substack{\scriptscriptstyle{r_{h+k}\in P_{\mathrm{H}}}\\\scriptscriptstyle{(v_{h+k},0)\in \bbG^{\rm c}_{r_{h+k}}}}}\;\sum_{\scriptscriptstyle{r_{h+k+1},\dots,r_n\in P_{\mathrm{H}}}}\hspace{-0.3cm}\theta(\{r_s,v_s,v'_s\}_{s=1}^n)\\
         &\times\prod_{i=1}^h(\eta_{r_i}+\eta_{r_i+v_i+v'_i})^2\,n^\a_{v_i}n^\a_{v'_i}\prod_{j=h+1}^{h+k}(\eta_{r_j}+\eta_{r_j+v_j})^2\,n^\a_{v_j}\prod_{m=h+k+1}^n\eta_{r_m}^2+\cE_{0,\a}^{(2,\mathrm{diag})}
	\end{split}
\ee
with
\be\label{eq:cE_02diag}
	\begin{split}
		\cE_{0,\a}^{(2,\mathrm{diag})}&=-\sum_{n= 0}^{n_{\mathrm{c}}}\frac{1}{N^{2n}}\sum_{h=0}^n\sum_{k=0}^{n-h}\frac{2^{-2h-k-(n-h-k)}}{h!k!(n-h-k)!}  \frac{(N_0-1)!}{(N_0-1 - k-2(n-k-h))!} \,  \\
         &\times\hspace{-0.3cm}\sum_{\substack{\scriptscriptstyle{r_1\in P_{\mathrm{H}}}\\\scriptscriptstyle{(v_1,v_1')\in \bbG^{\rm q}_{r_1}}}}\dots\hspace{-0.2cm}\sum_{\substack{\scriptscriptstyle{r_h\in P_{\mathrm{H}}}\\\scriptscriptstyle{(v_h,v_h')\in \bbG^{\rm q}_{r_h}}}}\;\sum_{\substack{\scriptscriptstyle{r_{h+1}\in P_{\mathrm{H}}}\\\scriptscriptstyle{(v_{h+1},0)\in \bbG^{\rm c}_{r_{h+1}}}}}\hspace{-0.2cm}\dots\hspace{-0.2cm}\sum_{\substack{\scriptscriptstyle{r_{h+k}\in P_{\mathrm{H}}}\\\scriptscriptstyle{(v_{h+k},0)\in \bbG^{\rm c}_{r_{h+k}}}}}\;\sum_{\scriptscriptstyle{r_{h+k+1},\dots,r_n\in P_{\mathrm{H}}}}\hspace{-0.3cm}\theta(\{r_s,v_s,v'_s\}_{s=1}^n)\\
         &\times\prod_{i=1}^h(\eta_{r_i}+\eta_{r_i+v_i+v'_i})^2\,n^\a_{v_i}n^\a_{v'_i}\prod_{j=h+1}^{h+k}(\eta_{r_j}+\eta_{r_j+v_j})^2\,n^\a_{v_j}\prod_{m=h+k+1}^n\eta_{r_m}^2\\
         &\times \frac{N_0}{N^{1-\k}}\big[h\big(2\widehat{V}(0)+\widehat{V}(v_{1}/N^{1-\k})+\widehat{V}(\tilde v_1/N^{1-\k})\big)+k\big(\widehat{V}(0)+\widehat{V}(v_{h+1}/N^{1-\k})\big)\big].
	\end{split}
\ee
The first term in \eqref{eq:cV_02diag_inter} gives a large  contribution. 
Since this term is not positive due to the presence of $\widehat{V}(s/N^{1-\k})$, we need to estimate the error coming from replacing $(N_0-1)!/ (N_0-1-k-2(n-h-k))!$ with $N_0! / (N_0-k-2(n-h-k))!$, a replacement which is needed to reconstruct the norm $\|\cD_\a g^0_\a\|^2$.
We obtain
\[
	\begin{split}
		&\Big|\langle \cD_\a g^0_\a,\cV^{(2,\mathrm{diag})}_0 \cD_\a g^0_\a\rangle-\cE_{0,\a}^{(2,\mathrm{diag})}-\frac{N_0}{N^{1-\k}}\sum_{s\in P_\A}\Big(\widehat{V}(0)+\widehat{V}(s/N^{1-\k})\Big)n^\a_s\|\cD_\a g^0_\a\|^2\Big|\\
		&\leq \frac{N_0}{N^{1-\k}}\sum_{s\in P_\A}\big|\widehat{V}(0)+\widehat{V}(s/N^{1-\k})\big|n^\a_s\\
		 &\quad \times\sum_{n= 0}^{n_{\mathrm{c}}}\frac{1}{N^{2n}}\sum_{h=0}^n\sum_{k=0}^{n-h}\frac{2^{-2h-k-(n-h-k)}}{h!k!(n-h-k)!}  \frac{(N_0-1)!}{(N_0 - k-2(n-k-h))!} \,  \\
         &\quad\times\hspace{-0.3cm}\sum_{\substack{\scriptscriptstyle{r_1\in P_{\mathrm{H}}}\\\scriptscriptstyle{(v_1,v_1')\in \bbG^{\rm q}_{r_1}}}}\dots\hspace{-0.2cm}\sum_{\substack{\scriptscriptstyle{r_h\in P_{\mathrm{H}}}\\\scriptscriptstyle{(v_h,v_h')\in \bbG^{\rm q}_{r_h}}}}\;\sum_{\substack{\scriptscriptstyle{r_{h+1}\in P_{\mathrm{H}}}\\\scriptscriptstyle{(v_{h+1},0)\in \bbG^{\rm c}_{r_{h+1}}}}}\hspace{-0.2cm}\dots\hspace{-0.2cm}\sum_{\substack{\scriptscriptstyle{r_{h+k}\in P_{\mathrm{H}}}\\\scriptscriptstyle{(v_{h+k},0)\in \bbG^{\rm c}_{r_{h+k}}}}}\;\sum_{\scriptscriptstyle{r_{h+k+1},\dots,r_n\in P_{\mathrm{H}}}}\hspace{-0.3cm}\theta(\{r_s,v_s,v'_s\}_{s=1}^n)\\
         &\quad\times\prod_{i=1}^h(\eta_{r_i}+\eta_{r_i+v_i+v'_i})^2\,n^\a_{v_i}n^\a_{v'_i}\prod_{j=h+1}^{h+k}(\eta_{r_j}+\eta_{r_j+v_j})^2\,n^\a_{v_j}\prod_{m=h+k+1}^n\hskip-0.2cm \eta_{r_m}^2\big[k+2(n-k-h)\big].
	\end{split}
\] 
We can now split the two terms in the square bracket in the last line. Then, we remove the restrictions in $\theta(\{r_s,v_s,s'_v\}_{s=1}^n)$ involving $r_{h+1},v_{h+1},v'_{h+1}$ in the first term (respectively $r_{h+k+1}$ in the second term) and move out the corresponding factor. Rescaling $n\to n-1,k\to k-1$ (respectively $n\to n-1$), bounding $\frac{(N_0-1)!}{(N_0-1-k-2(n-h-k))!}\leq \frac{N_0!}{(N_0-k-2(n-h-k))!}$ (respectively $\frac{(N_0-2)!}{(N_0-2-k-2(n-h-k))!}\leq \frac{N_0!}{(N_0-k-2(n-h-k))}$) and adding the term corresponding to $n=n_{\mathrm{c}}$ we conclude
\be\label{eq:cV_02diag_error1}
	\begin{split}
		&\Big|\langle \cD_\a g^0_\a,\cV^{(2,\mathrm{diag})}_0 \cD_\a g^0_\a\rangle-\cE_{0,\a}^{(2,\mathrm{diag})}-\frac{N_0}{N^{1-\k}}\sum_{s\in P_\A}\Big(\widehat{V}(0)+\widehat{V}(s/N^{1-\k})\Big)n^\a_s\|\cD_\a g^0_\a\|^2\Big|\\
		& \leq C\frac{N_0}{N^{3-\k}}\|V\|_1\|\eta_H\|_2^2N_\a(N_\a+N_0)\|\cD_\a g^0_\a\|^2.
	\end{split}
\ee
 To bound $\cE_{0,\a}^{(2,\mathrm{diag})}$ we proceed similarly taking the absolute value and splitting it according to the two terms in the bracket in the last line of \eqref{eq:cE_02diag}. We then remove the restrictions in $\theta$ involving $r_1,v_1,v'_1$ in the first term (respectively $r_{h+1},v_{h+1}$ in the second term) and move out the corresponding factor. Rescaling $n,h$ (respectively $n,k$) and bounding $ \frac{(N_0-1)!}{(N_0-1 - k-2(n-k-h))!} \leq  \frac{N_0!}{(N_0- k-2(n-k-h))!} $ (respectively $ \frac{(N_0-2)!}{(N_0- k-2-2(n-k-h))!} \leq  \frac{N_0!}{(N_0- k-2(n-k-h))!} $) and adding the term $n=n_{\mathrm{c}}$, we obtain
\be\label{eq:cE_02diag2}
	|\cE_{0,\a}^{(2,\mathrm{diag})}|\leq C\|V\|_1\frac{N_0}{N^{3-\k}}\|\eta_H\|_2^2N_\a(N_\a+N_0)\|\cD_\a g^0_\a\|^2\,.
\ee
The bounds \eqref{eq:cV_02diag_error1} and \eqref{eq:cE_02diag2} yield
\be\label{eq:cV_02diag_final}
	\begin{split}
		\Bigg|&\; \sum_\a \l_\a\frac{\langle \cD_\a g^0_\a,\cV^{(2,\mathrm{diag})}_0 \cD_\a g^0_\a\rangle}{\|\cD_\a g^0_\a\|^2}\,-\,\frac{N_0}{N^{1-\k}}\sum_\a \l_\a \sum_{s\in P_\A}\Big(\widehat{V}(0)+\widehat{V}(s/N^{1-\k})\Big)n^\a_s\Bigg|\\
		&\leq CN^{\k-2}\|\eta_H\|_2^2\sum_\a\l_\a N_\a(N_\a+N_0)\\
		& \leq CN^{4\k-3 +\eps}\big(\Tr(\cN^2\Gamma_G)+N_0\Tr(\cN\Gamma_G)\big)\leq CN^{4\k-1+\eps}
	\end{split}
\ee
where we used Eq.\,\eqref{eq:etaH}  and Lemma \ref{lm:cN}.

As for the expectation of $\cV_0^{(2,\mathrm{off})}$ we first contract the high momenta similarly to what we did in the computation of the norm to obtain 
\[
	\begin{split}
		   \langle \cD_\a g^0_\a,&\cV^{(2,\mathrm{off})}_0 \cD_\a g^0_\a\rangle=\sum_{n=0}^{n_{\mathrm{c}}}\frac{1}{n!(2N)^{2n}}\hspace{-0.3cm} \sum_{\substack{{r_1,\dots,r_{n}\in P_{\mathrm{H}}}\\\tilde r_1,\dots,\tilde r_{n}\in P_{\mathrm{H}}}}\sum_{\substack{{(v_1,v'_1)\in \bbG_{r_1}}\\{(\tilde v_1,\tilde v'_1)\in \bbG_{\tilde r_1}}}}\hspace{-0.2cm}\dots\hspace{-0.2cm} \sum_{\substack{{(v_n,v'_n)\in \bbG_{r_n}}\\{(\tilde v_{n},\tilde v'_{n})\in \bbG_{\tilde r_n}}}}\frac1{N^{1-\k}}\sum_{s\in P_\A}\widehat V(s/N^{1-\k})\\
		   &\times\prod_{i=1}^n\eta_{r_i}\eta_{\tilde r_i}  (\d_{\tilde{r}_i,r_i}+\d_{-\tilde{r}_i,r_i+v_i+v'_i})  \d_{v_i+v'_i,\tilde{v}_i+\tilde{v}_i'}  \theta\big(\{r_j,v_j,v'_j\}_{j=1}^{n}\big)\theta\big(\{\tilde r_j,\tilde v_j,\tilde v'_j\}_{j=1}^n\big)\\
		   &\times \langle g^0_\a,a^*_{\tilde{v}_{n}}a^*_{\tilde{v}'_{n}}\dots a^*_{\tilde{v}_{1}}a^*_{\tilde{v}'_{1}}a^*_sa^*_{-s}a_0a_0 a_{{v}_{1}}a_{{v}'_{1}}\dots a_{{v}_{n}}a_{{v}'_{n}}g^0_\a\rangle\,.
	\end{split}
\]
We now recall that $g_\a^0$ has a fixed number of particles in $P_\A\cup\{0\}$. 
Furthermore, due to the presence of $\prod_{i=1}^N\d_{v_i+v'_i,\tilde v_i+\tilde v'_i}$ we have $v_i=v'_i=0$ if and only if $\tilde v_i=\tilde v'_i=0$ ({due to \eqref{eq:P0G}}). 

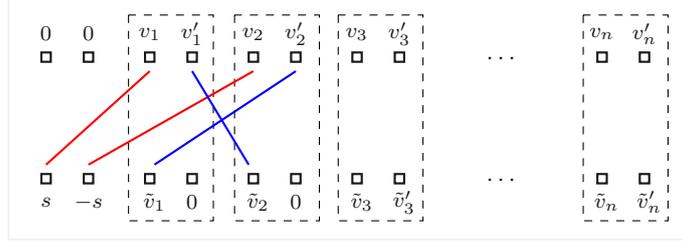
\begin{figure}[t] 
\centering
\begin{tikzpicture}[scale =0.62,
    every path/.style = {},
 ]
  \begin{scope} 
\filldraw[color=black, fill=white, thick] (1.2,1) rectangle (1.4,1.2);
\node[] at (1.3, 1.6) {\scriptsize  $0$};
\filldraw[color=black, fill=white, thick] (2.1,1) rectangle (2.3,1.2);
\node[] at (2.2, 1.6) {\scriptsize  $0$};
\filldraw[color=black, fill=white, thick] (1.2,-1.4) rectangle (1.4,-1.6);
\node[] at (1.3, -2) {\scriptsize  $s$};
\filldraw[color=black, fill=white, thick] (2.1,-1.4) rectangle (2.3,-1.6);
\node[] at (2.2, -2) {\scriptsize  $-s$};
\filldraw[color=black, fill=white, thick] (3.4,1) rectangle (3.6,1.2);
\node[] at (3.5, 1.6) {\scriptsize  $v_1$};
\filldraw[color=black, fill=white, thick] (4.3,1) rectangle (4.5,1.2);
\node[] at (4.4, 1.6) {\scriptsize  $v'_1$};
\filldraw[color=black, fill=white, thick] (3.4,-1.4) rectangle (3.6,-1.6);
\node[] at (3.6, -2) {\scriptsize  $\tl{v}_1$};
\filldraw[color=black, fill=white, thick] (4.3,-1.4) rectangle (4.5,-1.6);
\node[] at (4.4, -2) {\scriptsize  $0$};
\filldraw[color=black, fill=white, thick] (5.6,1) rectangle (5.8,1.2);
\node[] at (5.7, 1.6) {\scriptsize  $v_2$};
\filldraw[color=black, fill=white, thick] (6.5,1) rectangle (6.7,1.2);
\node[] at (6.6, 1.6) {\scriptsize  $v'_2$};
\filldraw[color=black, fill=white, thick] (5.6,-1.4) rectangle (5.8,-1.6);
\node[] at (5.8, -2) {\scriptsize  $\tl{v}_2$};
\filldraw[color=black, fill=white, thick] (6.5,-1.4) rectangle (6.7,-1.6);
\node[] at (6.6, -2) {\scriptsize  $0$};
\filldraw[color=black, fill=white, thick] (7.8,1) rectangle (8.0,1.2);
\node[] at (7.9, 1.6) {\scriptsize  $v_3$};
\filldraw[color=black, fill=white, thick] (8.7,1) rectangle (8.9,1.2);
\node[] at (8.8, 1.6) {\scriptsize  $v'_3$};
\filldraw[color=black, fill=white, thick] (7.8,-1.4) rectangle (8.0,-1.6);
\node[] at (8.0, -2) {\scriptsize  $\tl{v}_3$};
\filldraw[color=black, fill=white, thick] (8.7,-1.4) rectangle (8.9,-1.6);
\node[] at (8.9, -2) {\scriptsize  $\tl{v}'_3$};
\node[below] at (11, 1.4) {\footnotesize  $\cdots$};
\node[below] at (11, -1.2) {\footnotesize  $\cdots$};
\filldraw[color=black, fill=white, thick] (13.0,1) rectangle (13.2,1.2);
\node[] at (13.1, 1.6) {\scriptsize  $v_n$};
\filldraw[color=black, fill=white, thick] (13.9,1) rectangle (14.1,1.2);
\node[] at (14, 1.6) {\scriptsize  $v'_n$};
\filldraw[color=black, fill=white, thick] (13,-1.4) rectangle (13.2,-1.6);
\node[] at (13.2, -2) {\scriptsize  $\tl{v}_n$};
\filldraw[color=black, fill=white, thick] (13.9,-1.4) rectangle (14.1,-1.6);
\node[] at (14.1, -2) {\scriptsize  $\tl{v}'_n$};
\draw[fill=white, opacity=0.1] (0.5,-2.8) rectangle (15,2.4);
\draw[-, color=black,  dashed] (3.05, 2) rectangle (4.85, -2.4);
\draw[-, color=black, dashed] (5.3, 2) rectangle (7.1, -2.4);
\draw[-, color=black, dashed] (7.5, 2) rectangle (9.3, -2.4);
\draw[-, color=black, dashed] (12.7, 2) rectangle (14.5, -2.4);
\draw[color=red, thick] (1.3, -1.2)  -- (3.5, 0.8);
\draw[color=red, thick] (2.2, -1.2)  -- (5.7, 0.8);
\draw[color=blue, thick ] (3.6,-1.2)-- (6.6, 0.8);
\draw[color=blue, thick ] (5.6,-1.2)-- (4.4,0.8);
\end{scope}
\end{tikzpicture}
\centering 
\begin{minipage}{13cm}
\caption{\small Schematic picture of a non vanishing contraction among the operators on the right-hand side of \eqref{eq:cV0-start}. Upper (resp. lower) set of points represents annihilation (resp. creation) operators. Each line connecting two points represents a delta function between the moment labeling them.  Dashed boxes represent the conditions $v_\ell + v'_{\ell}=\tl v_\ell + \tl v'_{\ell}$ with $\ell=1, \ldots, n$ originating from the contraction of the momenta in $P_{\mathrm{H}}$.
} \label{Fig2}
\end{minipage}
\end{figure}

Hence, the expectation above is different from zero only if there exist  two indices $1\leq i,j\leq n$ such that $v_i,v'_i,v_j,v'_j\neq  0$ while each of the pairs $\{\tilde v_i,\tilde v'_i\}$ and $\{\tilde v_j,\tilde v'_j\}$ contains one zero momentum and one momentum in $P_\A$. Due to the restrictions encoded in  $\prod_{i=1}^N\d_{v_i+v'_i,\tilde v_i+\tilde v'_i}$ and $\theta\big(\{r_j,v_j,v'_j\}_{j=1}^{n-1}\big)\theta\big(\{\tilde r_j,\tilde v_j,\tilde v'_j\}_{j=1}^n\big)$, the only admissible contraction is such that $a^*_{s}a^*_{-s}$ are contracted with $a_{w_i}a_{w_j}$ where $w_\ell\in\{v_\ell,v'_\ell\},$ $\ell =i,j$ while $a_{-w_i+v_i+v'_i}$ 
 is contracted with the annihilation operator with non-zero momentum among $a^*_{\tilde v_j}$ and $a^*_{\tilde v'_j}$. Similarly, $a_{-w_j+v_j+v'_j}$ is contracted with the annihilation operator with non-zero momentum among $a^*_{\tilde v_i}$ and $a^*_{\tilde v'_i}.$  Furthermore, the sets $\{v_\ell,v'_\ell\}$ and $\{\tilde v_\ell,\tilde v'_\ell\}$ have to be equal for all $\ell=1,\dots,n$ such that $\ell\neq i,j$. By symmetry we can assume $i=1$ and $j=2$. A possible contration is depicted in Fig.\ref{Fig2}.

Based on the above considerations and denoting with $h$ (respectively $k$) the number of 	quartic (respectively triplets) among $A_{r_1,v_1,v'_1},\dots,A_{r_n,v_n,v'_n}$  we get  
\be \label{eq:cV0-start}
	\begin{split}
		&\langle \cD_\a g^0_\a\cV_{0}^{(2,\mathrm{off})}\cD_\a g^0_\a\rangle\\
		&=\sum_{n=2}^{n_{\mathrm{c}}}\frac1{N^{2n}}\sum_{h=2}^{n-2}\sum_{k=0}^{n-h-2}\frac{2^{-2(h-2)-k-(n-h-k)}}{(h-2)!k!(n-h-k)!}\frac{N_0!}{(N_0-2-k-2(n-k-h))!}\\
		&\phantom{{}={}}\times\hspace{-0.3cm}\sum_{\substack{r_1\in P_{\mathrm{H}}\\{(v_1,v'_1)\in \bbG^{\rm q}_{r_1}}\\{(\tilde v_1,0)\in \bbG^{\rm c}_{r_1}}}}\sum_{\substack{r_2\in P_{\mathrm{H}}\\{(v_2,v'_2)\in \bbG^{\rm q}_{r_2}}\\{(\tilde v_2,0)\in \bbG^{\rm c}_{r_2}}}}\sum_{\substack{r_3\in P_{\mathrm{H}}\\{(v_3,v_3')\in \bbG^{\rm q}_{r_3}}}}\hspace{-0.2cm}\dots\hspace{-0.2cm}\sum_{\substack{r_{h}\in P_{\mathrm{H}}\\(v_{h,}v_{h}')\in \bbG^{\rm q}_{r_{h}}}}\sum_{\substack{r_{h+1}\in P_{\mathrm{H}}\\(v_{h+1},0)\in \bbG^{\rm c}_{r_{h+1}}}}\hspace{-0.2cm}\dots\hspace{-0.2cm}\sum_{\substack{r_{h+k}\in P_{\mathrm{H}}\\(v_{h+k},0)\in \bbG^{\rm c}_{r_{h+k}}}}\sum_{r_{h+k+1},\dots,r_n\in P_{\mathrm{H}}}\\
         &\phantom{{}={}}\times \prod_{i=1}^{h}(\eta_{r_i}+\eta_{r_i+v_i+v'_i})^2n^\a_{v_i}n^\a_{v'_i}\prod_{j=h+1}^{h+k}(\eta_{r_j}+\eta_{r_j+v_j})^2n^\a_{v_j}\prod_{m=h+k+1}^n\hspace{-0.3cm}\eta_{r_m}^2 \;\d_{\tl v_1 ,v_1+v'_1}\\
         &\phantom{{}={}}\times\frac{1}{N^{1-\k}}\sum_{s\in P_\A}\widehat V(s/N^{1-\k})\d_{s,v_1}\d_{-s,v_2}\d_{\tilde v_1,v_2'}\d_{\tilde v_2,v'_1}\theta(\{r_s,v_s,v'_s\}_{s=1}^n)\theta(\{ r_s,\tilde v^\sharp_s,\tilde v'^\sharp_s\}_{s=1}^n)
	\end{split}
\ee
where we used \eqref{eq:a0m-t} with $t=2$ and $m=k+2(n-h-k)$. 
Above we introduced the convention $\tl v^\sharp_\ell=\tl v_\ell$  if $\ell =1,2$  and $\tl v^\sharp_\ell=v_\ell$ for $\ell\geq 3.$ Note that in this case we cannot get rid of $\theta(\{ r_s,\tilde v^\sharp_s,\tilde v'^\sharp_s\}_{s=1}^n)$ since it contains restrictions which are not included in $\theta(\{ r_s,\tilde v^\sharp_s,\tilde v'^\sharp_s\}_{s=1}^n)$. However, since \eqref{eq:cV0-start}
will turn out to be an error term we can estimate it in absolute value. Then, removing $\theta(\{\tl r_\ell,\tl v_\ell, \tl v'_\ell\}_{\ell=1}^n)$ and the dependence on $r_1,v_1,v'_1$ from $\theta(\{r_\ell,v_\ell,v'_\ell\}_{\ell=1}^n)$ we can move out the factor involving these indices. Scaling $n\to n-2,h\to h-2,$ bounding $\frac{(N_0-2)!}{(N_0-2-k-2(n-k-h))!}\leq \frac{N_0!}{(N_0-k-2(n-k-h))!}$  and adding the terms corresponding to $n=n_{\mathrm{c}}-1$ and $n=n_{\mathrm{c}},$ we obtain
\be\label{eq:cV_G2off}
	\begin{split}
		\sum_{\a}\l_\a \frac{|\langle \cD_\a g^0_\a\cV_{0}^{(2,\mathrm{off})}\cD_\a g^0_\a\rangle|}{\|\cD_\a g^0_\a\|^2}&\leq C\frac{N_0^2}{N^{5-\k}}\|V\|_1\|\eta_H\|_2^4\sum_{v\in P_\A}\sum_{\a}\l_\a n^\a_vn^\a_{-v}N_\a^2\\
		&\leq C 
	N^{7\k-5 +2\e}\sum_{v\in P_\A}\sum_\a \l_\a (n^\a_v)^2N_\a^2\\
	&\leq CN^{7\k-2+2\eps+\d_1/3}
	\end{split}
\ee
where in the last step we used Lemma \ref{lm:cN}.

We now compute the expectation of $\cV_0^{(1)}$. 
Contracting the high momenta we find 
\[
	\begin{split}
		   \langle \cD_\a g^0_\a,&\cV^{(1)}_0 \cD_\a g^0_\a\rangle=\sum_{n=0}^{n_{\mathrm{c}}}\frac{1}{n!(2N)^{2n}}\hspace{-0.3cm} \sum_{\substack{\scriptscriptstyle{r_1,\dots,r_{n}\in P_{\mathrm{H}}}\\\scriptscriptstyle\tilde r_1,\dots,\tilde r_{n}\in P_{\mathrm{H}}}}\sum_{\substack{\scriptscriptstyle{(v_1,v'_1)\in \bbG_{r_1}}\\\scriptscriptstyle{(\tilde v_1,\tilde v'_1)\in \bbG_{\tilde r_1}}}}\hspace{-0.2cm}\dots\hspace{-0.2cm} \sum_{\substack{\scriptscriptstyle{(v_n,v'_n)\in \bbG_{r_n}}\\\scriptscriptstyle{(\tilde v_{n},\tilde v'_{n})\in \bbG_{\tilde r_n}}}}\frac1{N^{1-\k}}\sum_{\substack{s,\,p\in P_\A:\\p+s\in P_\A }}\widehat V(s/N^{1-\k})\\
		   &\times\prod_{i=1}^n\eta_{r_i}\eta_{\tilde r_i}  (\d_{\tilde{r}_i,r_i}+\d_{-\tilde{r}_i,r_i+v_i+v'_i})  \d_{v_i+v'_i,\tilde{v}_i+\tilde{v}_i'}  \theta\big(\{r_j,v_j,v'_j\}_{j=1}^{n}\big)\theta\big(\{\tilde r_j,\tilde v_j,\tilde v'_j\}_{j=1}^n\big)\\
		   &\times \langle g^0_\a,a^*_{\tilde{v}_{n}}a^*_{\tilde{v}'_{n}}\dots a^*_{\tilde{v}_{1}}a^*_{\tilde{v}'_{1}}a^*_{p+s}a^*_{-s} a_pa_{0}a_{{v}_{1}}a_{{v}'_{1}}\dots a_{{v}_{n}}a_{{v}'_{n}}g^0_\a\rangle.
	\end{split}
\] 
Due to the restrictions enforced by $\theta\big(\{r_j,v_j,v'_j\}_{j=1}^{n-1}\big)\theta\big(\{\tilde r_j,\tilde v_j,\tilde v'_j\}_{j=1}^n\big)$ and taking into account the presence of $\prod_{i=1}^n\d_{v_i+v'_i,\tilde{v}_i+\tilde{v}_i'}$ we conclude that the in order to have a non-zero contribution there exists an index $i=1,\dots,n$ such that $v_i,v'_i\neq 0$ while $\{\tl v_i,\tl v'_i\}$ contains exactly one zero momentum ad one momentum in $P_\A$. Furthermore, the creation operators $a^*_{p+s}a^*_{-s}$ have to be contracted with $a_{ v_i}a_{ v'_i}$ and the annihilation operator $a_p$ has to be contracted with the operator with non-zero momentum between $a^*_{\tl v_i}$ and $a^*_{\tl v'_i}.$ Assuming by symmetry $i=1$ we obtain
\[
	\begin{split}
		&\langle \cD_\a g^0_\a,\cV^{(1)}_0 \cD_\a g^0_\a\rangle\\
		&=\sum_{n=1}^{n_{\mathrm{c}}}\frac1{(2N)^{2n}}\sum_{h=1}^n\sum_{k=0}^{n-h}\frac{2^{(h-1)+2k+(n-h-k)}}{(h-1)!k!(n-h-k)!}\frac{N_0!}{(N_0-1-k-2(n-h-k))!}\sum_{\substack{r_1\in P_{\mathrm{H}}\\(\tl v_1,\tl v'_1)\in \bbG^{\rm q}_{r_1}\\(\tl v_1,0)\in \bbG^{\rm c}_{r_1}}}\\
		&\times\hspace{-0.3cm}\sum_{\substack{r_2\in P_{\mathrm{H}}\\(v_2,v'_2)\in \bbG^{\rm q}_{r_2}}}\hspace{-0.2cm}\dots\sum_{\substack{r_{h}\in P_{\mathrm{G}}\\(v_h,v'_h)\in \bbG^{\rm q}_{r_h}}}\sum_{\substack{r_{h+1}\in P_{\mathrm{H}}\\(v_h,v'_h)\in \bbG^{\rm q}_{r_h}}}\sum_{\substack{r_{h+1}\in P_{\mathrm{H}}\\(v_{h+1},0)\in \bbG^{\rm c}_{r_{h+1}}}}\dots\hspace{-0.2cm} \sum_{\substack{r_{h+k}\in P_{\mathrm{H}}\\(v_{h+k},0)\in \bbG^{\rm c}_{r_{h+k}}}}\hspace{-0.1cm}\sum_{r_{h+k+1},\dots, r_n\in P_{\mathrm{H}}}\\
		&\times\prod_{i=1}^h\eta_{r_i}(\eta_{r_1}+\eta_{r_i+v_i+v'_i})n^\a_{v_i}n^\a_{v'_i}\prod_{j=h+1}^{h+k}\eta_{r_j}(\eta_{r_j}+\eta_{r_j+v_j})n^\a_{v_j}\prod_{m=h+k+1}^n\eta_{r_m}^2\\
		&\times\frac4{N^{1-\k}}\sum_{\substack{s, p\in P_\A:\\p+s\in P_\A}}\widehat V(s/N^{1-\k})\d_{p+s,v_1}\d_{-s,v'_1}\d_{p,\tl v_1}n^\a_{\tl v_1}  \theta\big(\{r_j,v_j,v'_j\}_{j=1}^{n}\big)
		\theta\big(\{ r_j,\tilde v^\sharp_j,\tilde v'^\sharp_j\}_{j=1}^n\big)
	\end{split}
\] 
where we used \eqref{eq:a0m-t} with $t=1$ and $m=k-1+2(n-h-k)$. Above we used the convention $\tl v^\sharp_1=\tl v_1$,  {$\tl v_1'^\sharp=\tl v_1'$ } and $\tl v^\sharp_\ell=v_\ell$, $\tl v'^\sharp_\ell=v'_\ell$ for $\ell>1$. 
Thus, thanks to Lemma \ref{lm:cN}, we can estimate
\be\label{eq:cV_G1}
\begin{split}
	\sum_\a\l_\a\frac{|\langle \cD_\a g^0_\a,\cV^{(1)}_0 \cD_\a g^0_\a\rangle|}{\|\cD_\a g^0_\a\|^2}& \leq C \frac {N_0}{N^{3-\k}}\|V\|_1\|\eta_H\|_2^2\sum_\a\l_\a\sum_{v,v'\in P_\A}n^\a_vn^\a_{v'}n^\a_{v+v'}\\
	&\leq CN^{4\k-1+\eps+\d_1/3}\,.
	\end{split}
\ee

To compute the expectation of $\cV^{(0)}_0$ we find it convenient to distinguish the terms where at least one of the conditions $p=q$ or $s=0$ are satisfied, from the term where none of these hold true. We correspondingly split: 
\begin{equation}\label{eq:cV_G0split}
	\begin{split}
		\cV^{(0)}_0
		=\; &\frac1{2N^{1-\k}}\sum_{\substack{\scriptscriptstyle{p,q\in P_\A:}\\\scriptscriptstyle{p\neq q}}}\Big(\widehat V(0)+\widehat V((p-q)/N^{1-\k})\Big)a^*_{p}a^*_qa_pa_{q}+\frac1{2N^{1-\k}}\sum_{\scriptscriptstyle{p\in P_\A}}\widehat V(0)a^*_{p}a^*_pa_pa_{p}\\
		&\; +\frac1{2N^{1-\k}}\sum_{\substack{\scriptscriptstyle{s\in \L^*_+,p,q\in P_\A:}\\\scriptscriptstyle{p\neq q,p+s,q+s\in P_\A}}}\widehat V(s/N^{1-\k})a^*_{p+s}a^*_qa_pa_{q+s}=\cV^{(0,a)}_0+\cV^{(0,b)}_0+\cV^{(0,c)}_0.
	\end{split}
\end{equation}
Let's start considering $\cV^{(0,a)}_0$. 

By definition  
\[
	\begin{split}
		&\langle \cD_\a g^0_\a,\cV^{(0,a)}_0 \cD_\a g^0_\a\rangle\\
		&=\sum_{n=0}^{n_{\mathrm{c}}}\frac{1}{(n!)^2(2N)^{2n}}\hspace{-0.2cm}  \sum_{\substack{\scriptscriptstyle{r_1,\dots,r_{n}\in P_{\mathrm{H}}}\\\scriptscriptstyle\tilde r_1,\dots,\tilde r_{n}\in P_{\mathrm{H}}}}\sum_{\substack{\scriptscriptstyle{(v_1,v'_1)\in \bbG_{r_1}}\\\scriptscriptstyle{(\tilde v_1,\tilde v'_1)\in \bbG_{\tilde r_1}}}}\hspace{-0.2cm}\dots\hspace{-0.2cm} \sum_{\substack{\scriptscriptstyle{(v_n,v'_n)\in \bbG_{r_n}}\\\scriptscriptstyle{(\tilde v_{n},\tilde v'_{n})\in \bbG_{\tilde r_n}}}} \theta\big(\{r_j,v_j,v'_j\}_{j=1}^{n}\big)\theta\big(\{\tilde r_j,\tilde v_j,\tilde v'_j\}_{j=1}^n\big)\\
         &\phantom{{}={}}\times\prod_{i=1}^n\eta_{r_i}\eta_{\tilde r_i} \langle g^0_\a, A^*_{\tilde r_n,\tilde v_n,\tilde v'_n}\dots A^*_{\tilde r_1,\tilde v_1,\tilde v'_1} \cV^{(0,a)}_0A_{r_1,v_1,v'_1}\dots A_{r_n,v_n,v'_n}g^0_\a\rangle.
	\end{split}•
\]
Moreover, for $p,q\in P_\A$, with $p\neq q,$ 
\[
	\begin{split}
		&a^*_{p}a^*_qa_pa_{q}A_{r_1,v_1,v'_1}\dots A_{r_n,v_n,v'_n}g^0_\a=n^\a_p n^\a_q A_{r_1,v_1,v'_1}\dots A_{r_n,v_n,v'_n}g^0_\a\\
		&-\sum_{i=1}^n\Big[n^\a_q\big(\d_{p, v_i}+\d_{p, v'_i}\big)+\big(\d_{q, v_i}+\d_{q, v'_i}\big)\big(n^\a_p-\sum_{j=1}^{n}(\d_{p,v_j}+\d_{p,v'_j})\big)\Big]A_{r_1,v_1,v'_1}\dots A_{r_n,v_n,v'_n}g^0_\a.		
	\end{split}•
\]
Hence,
\begin{equation}\label{eq:cV_00a}
		\langle \cD_\a g^0_\a,\cV^{(0,a)}_0 \cD_\a g^0_\a\rangle=\frac1{2N^{1-\k}}\sum_{\substack{p,q\in P_\A\\p\neq q}}\!\!\Big(\widehat V(0)+\widehat V((p-q)/N^{1-\k})\Big) n^\a_pn^\a_q\,\|\cD_\a g^0_\a\|^2+\cE^{(0,a)}_{0,\a}
\end{equation}
where
\begin{equation*}
	\begin{split}
		&\cE^{(0,a)}_{0,\a}=-\frac1{2N^{1-\k}}\sum_{\substack{p,q\in P_\A\\p\neq q}}\!\!\Big(\widehat V(0)+\widehat V((p-q)/N^{1-\k})\Big)\\
		&\times\sum_{n=0}^{n_{\mathrm{c}}}\frac1{(2N)^{2n}}\sum_{h=0}^n\sum_{k=0}^{n-h}\frac{2^{-2h-k-(n-h-k)}}{h!k!(n-h-k)!}\frac{N_0!}{(N_0-k-2(n-k-h))!}\\
		&\times\hspace{-0.3cm}\sum_{\substack{r_1\in P_{\mathrm{H}}\\(v_1,v_1')\in \bbG^{\rm q}_{r_1}}}\dots\hspace{-0.2cm}\sum_{\substack{r_h\in P_{\mathrm{H}}\\(v_h,v'_h)\in \bbG^{\rm q}_{r_h}}}\sum_{\substack{r_{h+1}\\(v_{h+1},0)\in \bbG^{\rm c}_{r_{h+1}}}}\dots\hspace{-0.2cm}\sum_{\substack{r_{h+k}\in P_{\mathrm{H}}\\(v_{h+k},0)\in \bbG^{\rm c}_{r_{h+k}}}}\hspace{-0.2cm}\sum_{r_{h+k+1}\dots,r_n\in P_{\mathrm{H}}}\hspace{-0.3cm}\theta(\{r_s,v_s,v'_s\}_{s=1}^n)\\
		&\times\prod_{i=1}^h (\eta_{r_i}+\eta_{r_i+v_i+v'_i})^2n^\a_{v_i}n^\a_{v'_i}\prod_{j=h+1}^{h+k}(\eta_{r_j}+\eta_{r_j+v_j})^2n^\a_{v_j}\prod_{m=h+k+1}^n\eta_{r_m}^2\\
		&\times\big[2h\d_{q,v_1}(2n^\a_p+\d_{p,v'_1}+4(h-1)\d_{p,v_2})+2k\d_{q,v_{h+1}}(n^\a_p+2h\d_{p,v_1}+(k-1)\d_{p,v_{h+2}})\big]\,.
	\end{split}
\end{equation*} 

Taking the absolute value and splitting the various terms appearing in the last line, rescaling appropriately the indices and adding back the missing terms to reconstruct \eqref{eq:norm-key}  one can check that for any $3\k-2+\eps<0$
\begin{equation}\label{eq:cE_G0a} 
	\begin{split}	
		\sum_\a\l_\a\frac{|\cE^{(0,a)}_{0,\a}|}{\|\cD_\a g^0_\a\|^2}
		\leq &\; CN^{\k-3}\|\eta_H\|_2^2\sum_\a\l_\a N_\a^2(N_0+N_\a+1)\\
		&+CN^{\k-5}\|\eta_H\|_2^4\sum_\a\l_\a N_\a^2(N_0^2+N_0N_\a+N_\a^2)
		\\ \leq &\;  CN^{4\k-1+\eps}
	\end{split}
\end{equation}
where in the last step we used Lemma \ref{lm:cN}. Similarly, we have

\[
	\begin{split}
		&\langle \cD_\a g^0_\a,\cV^{(0,b)}_0 \cD_\a g^0_\a\rangle\\
		&=\sum_{n=1}^{n_{\mathrm{c}}}\frac{1}{(n!)^2(2N)^{2n}}\hspace{-0.2cm}  \sum_{\substack{\scriptscriptstyle{r_1,\dots,r_{n}\in P_{\mathrm{H}}}\\\scriptscriptstyle\tilde r_1,\dots,\tilde r_{n}\in P_{\mathrm{H}}}}\sum_{\substack{\scriptscriptstyle{(v_1,v'_1)\in \bbG_{r_1}}\\\scriptscriptstyle{(\tilde v_1,\tilde v'_1)\in \bbG_{\tilde r_1}}}}\hspace{-0.2cm}\dots\hspace{-0.2cm} \sum_{\substack{\scriptscriptstyle{(v_n,v'_n)\in \bbG_{r_n}}\\\scriptscriptstyle{(\tilde v_{n},\tilde v'_{n})\in \bbG_{\tilde r_n}}}} \theta\big(\{r_j,v_j,v'_j\}_{j=1}^{n}\big)\theta\big(\{\tilde r_j,\tilde v_j,\tilde v'_j\}_{j=1}^n\big)\\
         &\phantom{{}={}}\times\prod_{i=1}^n\eta_{r_i}\eta_{\tilde r_i} \langle g^0_\a, A^*_{\tilde r_n,\tilde v_n,\tilde v'_n}\dots A^*_{\tilde r_1,\tilde v_1,\tilde v'_1} \cV^{(0,b)}_0A_{r_1,v_1,v'_1}\dots A_{r_n,v_n,v'_n}g^0_\a\rangle.
	\end{split}•.
\]
Using, 
\[
	\begin{split}
	 	&a^*_pa^*_pa_pa_pA_{r_1,v_1,v'_1}\dots A_{r_n,v_n,v'_n}g^0_\a=\!(n^\a_p-1)\Big(n^\a_p\!-\!2\sum_{i=1}^{ n }(\d_{p,v_i}+\d_{p,v'_i})\!\Big)A_{r_1,v_1,v'_1}\dots A_{r_n,v_n,v'_n}g^0_\a
	\end{split}•
\]
we rewrite
\[
	\begin{split}
		&\langle \cD_\a g^0_\a,\cV^{(0,b)}_0 \cD_\a g^0_\a\rangle=\sum_{n=1}^{n_{\mathrm{c}}}\frac1{N^{2n}}\sum_{h=0}^n\sum_{k=0}^{n-h}\frac{2^{-2h-k-(n-h-k)}}{h!k!(n-h-k)!}\frac{N_0!}{(N_0-k-2(n-k-h))!}\\
		&\times\hspace{-0.3cm}\sum_{\substack{r_1\in P_{\mathrm{H}}\\(v_1,v_1')\in \bbG^{\rm q}_{r_1}}}\dots\hspace{-0.2cm}\sum_{\substack{r_h\in P_{\mathrm{H}}\\(v_h,v'_h)\in \bbG^{\rm q}_{r_h}}}\sum_{\substack{r_{h+1}\in P_{\mathrm{H}}\\(v_{h+1},0)\in \bbG^{\rm c}_{r_{h+1}}}}\dots\hspace{-0.2cm}\sum_{\substack{r_{h+k}\in P_{\mathrm{H}}\\(v_{h+k},0)\in \bbG^{\rm c}_{r_{h+k}}}}\sum_{r_{h+k+1},\dots,r_n\in P_{\mathrm{H}}}\hspace{-0.3cm}\theta(\{r_s,v_s,v'_s\}_{s=1}^n)\\
		&\times\prod_{i=1}^h (\eta_{r_i}+\eta_{r_i+v_i+v'_i})^2 n^\a_{v_i}n^\a_{v'_i}\prod_{j=h+1}^{h+k}(\eta_{r_j}+\eta_{r_j+v_j})^2n^\a_{v_j}\prod_{m=h+k+1}^n\eta_{r_m}^2\\
		&\times\frac1{2N^{1-\k}}\sum_{p\in P_\A}\widehat V(0)\big[n^\a_p(n^\a_p-1)-4h\d_{p,v_1}(n^\a_{v_1}-1)-2k\d_{p,v_{h+1}}(n^\a_{v_{h+1}}-1)\big].
	\end{split}
\]
Hence, reasoning analogously to what we did before and using Lemma \ref{lm:cN} we get
\begin{equation}\label{eq:cV_G0b}
	\begin{split}
		\sum_\a\l_\a &\frac{|\langle \cD_\a g^0_\a,\cV^{(0,b)}_0 \cD_\a g^0_\a\rangle|}{\|\cD_\a g^0_\a\|^2} \\
		\leq &\; \frac{C}{N^{1-\k}}\|V\|_1\sum_\a\l_\a\sum_{v\in P_\A}n_\a(v)^2\\
		&\; +\frac{C}{N^{3-\k}}\|V\|_1\|\eta_H\|_2^2\sum_\a\l_\a\Big(\sum_{v\in P_\A}n_\a(v)^2+N_\a\Big)(N_0+N_\a)\\
		\leq &\; CN^{\k+\d_1/3}
	\end{split}
\end{equation}
where we used  $3\k-2+\eps<3\k-2+3\eps +\d_2<0$.

It remains to compute the expectation of $\cV^{(0,c)}_0.$ We first contract the high momenta to obtain
\[
	\begin{split}
		&\langle \cD_\a g^0_\a,\cV^{(0,c)}_0 \cD_\a g^0_\a\rangle\\
		&=\frac1{2N^{1-\k}}\sum_{\substack{\scriptscriptstyle{s\in \L^*_+,p,q\in P_\A:}\\\scriptscriptstyle{p\neq q,p+s,q+s\in P_\A}}}\widehat V(s/N^{1-\k})\sum_{n\geq0}\frac{1}{n!(2N)^{2n}}\hspace{-0.3cm} \sum_{\substack{\scriptscriptstyle{r_1,\dots,r_{n}\in P_{\mathrm{H}}}\\\scriptscriptstyle\tilde r_1,\dots,\tilde r_{n}\in P_{\mathrm{H}}}}\sum_{\substack{\scriptscriptstyle{(v_1,v'_1)\in \bbG_{r_1}}\\\scriptscriptstyle{(\tilde v_1,\tilde v'_1)\in \bbG_{\tilde r_1}}}}\hspace{-0.2cm}\dots\hspace{-0.2cm} \sum_{\substack{\scriptscriptstyle{(v_n,v'_n)\in \bbG_{r_n}}\\\scriptscriptstyle{(\tilde v_{n},\tilde v'_{n})\in \bbG_{\tilde r_n}}}}\\
		   &\phantom{{}={}}\times\prod_{i=1}^n\eta_{r_i}\eta_{\tilde r_i}  (\d_{\tilde{r}_i,r_i}+\d_{-\tilde{r}_i,r_i+v_i+v'_i})  \d_{v_i+v'_i,\tilde{v}_i+\tilde{v}_i'} \theta\big(\{r_j,v_j,v'_j\}_{j=1}^{n}\big)\theta\big(\{\tilde r_j,\tilde v_j,\tilde v'_j\}_{j=1}^n\big)\\
		   &\phantom{{}={}}\times  \langle g^0_\a,a^*_{\tilde{v}_{n}}a^*_{\tilde{v}'_{n}}\dots a^*_{\tilde{v}_{1}}a^*_{\tilde{v}'_{1}}a^*_{p+s}a^*_qa_pa_{q+s}a_{{v}_{1}}a_{{v}'_{1}}\dots a_{{v}_{n}}a_{{v}'_{n}}g^0_\a\rangle\,.
	\end{split}
\]
As in previous computations we note that each momentum appearing in an annihilation operator has to appear also in a creation operator. Since $p\neq q$ and $s\neq 0$ we can distinguish two cases: 
\begin{enumerate}
	\item there exists an index $1\leq i\leq n$ such that the set of momenta $\{p+s,q\}$ equals the set of momenta $\{v_i,v'_i\}$ and the set $\{q+s,p\}$ equals the set $\{\tilde v_i,\tilde v'_i\}$ (note that in particular $v_i,v'_i,\tilde v_i,\tilde v'_i\neq 0$). Note that this implies that the index $i$ corresponds to a quartic excitation. By symmetry we can assume $i=1$. All indices $2\leq \ell\leq n$ can be treated as in the computation of the norm.
	\item there exist two indices $1\leq i,j\leq n$, $i\neq j$ such that $p+s=w_i$ and $q=w_j$ with the usual notation $w_\ell\in \{v_\ell,v'_\ell\}, \ell=i,j$ (this implies in particular $w_i,w_j\neq 0$). Then, either $p=\tl w_i$ and $q+s=\tl w_j$, or $p=\tl w_j$ and $q+s=\tl w_i$, where $\tl w_\ell\in \{\tl v_\ell, \tl v'_\ell\}$, $\ell=i,j$. In both cases, since $p\neq q$ and $s\neq 0$ we have necessarily $v_i+v'_i-w_i=\tl v_j+\tl v'_j-\tl w_j$ and $v_j+v'_j-w_j=\tl v_i+\tl v'_i-\tl w_i$  (recall that after the high momenta pairing $v_\ell + v'_\ell = \tl v_\ell + \tl v'_\ell$, $\ell=1, \ldots, n$).
	 By symmetry we can assume $i=1, j=2.$ The operators with index $3\leq \ell\leq n$ can be treated as in the computation of the norm.
\end{enumerate}
 We split accordingly
\begin{equation}\label{eq:cV_G0c}
	\langle \cD_\a g^0_\a,\cV^{(0,c)}_0 \cD_\a g^0_\a\rangle=I^{(0,c)}_0+II^{(0,c)}_0\,.
\end{equation}
Here, $I^{(0,c)}_0$ corresponds to the first case described above, namely
\[
	\begin{split}
		I^{(0,c)}_0=&\sum_{n=1}^{n_{\mathrm{c}}}\frac1{N^{2n}}\sum_{h=1}^n\sum_{k=0}^{n-h}\frac{2^{-2h-k-(n-h-k)}}{h!k!(n-h-k)!}\frac{N_0!}{(N_0-k-2(n-k-h))!}\\
		&\times\sum_{\substack{r_1\in P_{\mathrm{H}}\\(v_1,v_1')\in \bbG^{\rm q}_{r_1}\\(\tilde v_1,\tilde v'_1)\in \bbG^{\rm q}_{r_1}}}\dots\hspace{-0.2cm}\sum_{\substack{r_h\in P_{\mathrm{H}}\\(v_h,v'_h)\in \bbG^{\rm q}_{r_h}}}\sum_{\substack{r_{h+1}\in P_{\mathrm{H}}\\(v_{h+1},0)\in \bbG^{\rm c}_{r_{h+1}}}}\hspace{-0.2cm}\dots\hspace{-0.2cm}\sum_{\substack{r_{h+k}\in P_{\mathrm{H}}\\(v_{h+k},0)\in \bbG^{\rm c}_{r_{h+k}}}}\sum_{r_{h+k+1},\dots,r_n\in P_{\mathrm{H}}}\hspace{-0.4cm}\theta(\{r_s,v_s,v'_s\}_{s=1}^n)\\
		&\times\theta(\{r_s,\tilde v^\sharp_s,\tilde v'^\sharp_s\}_{s=1}^n)\prod_{i=1}^h (\eta_{r_i}+\eta_{r_i+v_i+v'_i})^2n^\a_{v_i}n^\a_{v'_i}\prod_{j=h+1}^{h+k}(\eta_{r_j}+\eta_{r_j+v_j})^2n^\a_{v_j}\hspace{-0.2cm}\prod_{m=h+k+1}^n\hspace{-0.2cm}\eta_{r_m}^2 \\
		&\times\frac1{2N^{1-\k}}\sum_{\substack{s\in \L_+^*,p,q\in P_\A:\\p\neq q,p+s,q+s\in P_\A}}\hspace{-0.25cm}\widehat V(s/N^{1-\k})4h\d_{p+s,v_1}\d_{q,v'_1}\d_{q+s,\tilde v_1}\d_{v_1+v'_1,\tilde v_1+\tilde v'_1}\,n^\a_{\tilde v_1}\,n^\a_{\tilde v'_1}
	\end{split}
\]
where in $\theta(\{r_s,\tilde v^\sharp_s,\tilde v'^\sharp_s\}_{s=1}^n)$ we are adopting  the convention $\tilde v^\sharp_1=\tilde v_1$, $\tl v'^\sharp_1=\tilde v'_1$ and $\tilde v_s^\sharp=v_s$,  $v'^\sharp_s=v'_s$ for any $2\leq s\leq n.$ Similarly, $II^{(0,c)}_0$ corresponds to the second possible case and can be written explicitly as  
\[
	\begin{split}
		&II^{(0,c)}_0= \frac1{N^{1-\k}}\sum_{\substack{s\in \L^*_+,p,q\in P_\A:\\p\neq ,p+s,q+s\in P_\A}}\hspace{-0.3cm}\widehat V(s/N^{1-\k})\sum_{n=2}^{n_{\mathrm{c}}}\frac1{N^{2n}}\sum_{h=0}^{n-2}\sum_{k=0}^{n-2-h}\!\frac{2^{-2h-k-(n-2-h-k)}}{h!k!(n-2-h-k)!}\\
		&\times\frac{N_0!}{(N_0-k-(n-2-h-k))!}\hspace{-0.3cm}\sum_{\substack{r_1\in P_{\mathrm{H}}\\(v_1,v_1')\in \bbG_{r_1}\\(\tilde v_1,\tilde v'_1)\in \bbG_{r_1}}}\sum_{\substack{r_2\in P_{\mathrm{H}}\\(v_2,v_2')\in \bbG_{r_2}\\(\tilde v_2,\tilde v'_2)\in \bbG_{r_2}}}\sum_{\substack{r_3\in P_{\mathrm{H}}\\(v_3,v_3')\in \bbG^{\rm q}_{r_3}}}\hspace{-0.2cm}\dots\hspace{-0.3cm}\sum_{\substack{r_{h+2}\in P_{\mathrm{H}}\\(v_{h+2},v'_{h+2})\in \bbG^{\rm q}_{r_{h+2}}}}\\
		&\times \hspace{-0.3cm}\sum_{\substack{r_{h+3}\in P_{\mathrm{H}}\\(v_{h+3},0)\in \bbG^{\rm c}_{r_{h+3}}}}\hspace{-0.2cm}\dots\hspace{-0.2cm}\sum_{\substack{r_{h+k+2}\in P_{\mathrm{H}}\\(v_{h+k+2},0)\in \bbG^{\rm c}_{r_{h+k+2}}}}\hskip -0.2cm\sum_{r_{h+k+3},\dots,r_n\in P_{\mathrm{H}}}\hspace{-0.5cm}\theta(\{r_s,v_s,v'_s\}_{s=1}^n)\theta(\{r_s,\tilde v^\flat_s,\tilde v'^\flat_s\}_{s=1}^n)\\
		&\times\prod_{i=3}^{h} (\eta_{r_i}+\eta_{r_i+v_i+v'_i})^2n^\a_{v_i}n^\a_{v'_i}\prod_{j=h+3}^{h+k+2}\hskip -0.2cm(\eta_{r_j}+\eta_{r_j+v_j})^2n^\a_{v_j}\hspace{-0.2cm}\prod_{m=h+k+3}^n\hspace{-0.3cm}\eta_{r_m}^2 \eta_{r_1}\eta_{r_2}\eta_{\tl r_1}\eta_{\tl r_2}\sum_{p_1,p_2}\hskip -0.1cm \d_{\tl r_1,p_1} \hskip -0.05cm  \d_{\tl r_2,p_2}\\
		&\times \sum_{\substack{w_1,w_2\\\tl w_1,\tl w_2}}\d_{p+s,w_1}\d_{q,w_2}(\d_{p,\tl w_1}\d_{q+s,\tl w_2}+\d_{p,\tl w_2}\d_{q+s,\tl w_1})\d_{v_1+v'_1-w_1,\tl v_2+\tl v'_2-\tl w_2}n^\a_{p+s}n^\a_{q}n^\a_{q+s}n^\a_{p}\\
		&\times\Big\{n^\a_{v_1+v'_1-w_1}n^\a_{v_2+v'_2-w_2}\\
		&\hspace{1cm}+(N_0-k-2(n-2-k-h))\Big[(n^\a_{v_1+v'_1-w_1}\d_{v_2+v'_2-w_2,0}+\d_{v_1+v'_1-w_1,0}n^\a_{v_2+v'_2-w_2})\\
		&\hspace{1cm}+(N_0-k-2(n-2-h-k)-1)\d_{v_1+v'_1-w_1,0}\d_{v_2+v'_2-w_2,0}\Big]\Big\}
	\end{split}
\] 
where we used \eqref{eq:a0m-t} with $m=k+2(n-2-k-h)$ and $t=1,2$,  and in $\theta(\{r_s,\tilde v^\flat_s,\tilde v'^\flat_s\}_{s=1}^n)$ we are adopting  the convention $\tl v_\ell^\flat=\tl v_\ell$ and $\tl v'^{\flat}_\ell=\tl v'_\ell$ for $\ell=1,2$ while $\tl v_\ell^\flat=v_\ell$ and $\tl v'^{\flat}_\ell= v'_\ell$ for $\ell>2.$
Hence, using Lemma \ref{lm:cN}  
\begin{equation}\label{eq:I_G0c}
	\begin{split}
		\sum_\a\l_\a\frac{|I^{(0,c)}_0|}{\|\cD_\a g^0_\a\|^2}&\; \leq  \frac{C}{N^{3-\k}}\|\eta_H\|_2^2\|V\|_1\sum_\a\l_\a\sum_{v,v',\tl v\in P_{\mathrm{G}}}n^\a_{v}n^\a_{v'}n^\a_{\tl v}n^\a_{v+v'-\tl v} \\
			&\; \leq CN^{4\k-1+\eps+\d_1/3}
		\end{split}
\end{equation}
and
\begin{equation}\label{eq:II_G0c}
	\begin{split}
		\sum_\a\l_\a\frac{|II^{(0,c)}_0|}{\|\cD_\a g^0_\a\|^2}&\leq \frac{C}{N^{5-\k}}\|\eta_H\|_2^4\|V\|_1\sum_\a\l_\a\sum_{v,v',\tl v\in P_{\mathrm{G}}}n^\a_{v}n^\a_{v'}n^\a_{\tl v}n^\a_{v+v'-\tl v}(N_0+N_\a)^2\\
		&\leq CN^{7\k-2+2\eps+\d_1/3}\,.
	\end{split}
\end{equation}
Summarizing, \eqref{eq:cV_G0split}, \eqref{eq:cV_00a}, \eqref{eq:cE_G0a}, \eqref{eq:cV_G0b},\eqref{eq:cV_G0c},\eqref{eq:I_G0c} and \eqref{eq:II_G0c} yield
\begin{equation}\label{eq:cV_G0}
	\begin{split}
		\langle \cD_\a g^0_\a,\cV^{(0)}_0 \cD_\a g^0_\a\rangle=\frac1{2N^{1-\k}}\sum_{\substack{p,q\in P_\A\\p\neq q}}\!\!\Big(\widehat V\Big(\frac{p-q}{N^{1-\k}}\Big)+\widehat V(0)\Big)\,n^\a_{p}n^\a_{q}\,\|\cD_\a g^0_\a\|^2+\cE^{(0)}_{0,\a}
	\end{split}
\end{equation}
with
\begin{equation}\label{eq:cE_G0}
	\sum_\a\l_\a |\cE^{(0)}_{0,\a}| \|\cD_\a g^0_\a\|^{-2}
	\leq  C N^{7\k-2+2\eps+\d_1/3}
\end{equation}
for any $\k\in(1/3,2/3)$.
Therefore, from \eqref{eq:cV_Gsplit},\eqref{eq:cV_04},\eqref{eq:cV_02diag_final},\eqref{eq:cV_G2off}, \eqref{eq:cV_G1},\eqref{eq:cV_G0} and \eqref{eq:cE_G0}, we conclude
\[
	\begin{split}
	\langle \cD_\a g^0_\a,&\cV_0 \cD_\a g^0_\a\rangle=\frac1{2N^{1-\k}}\widehat{V}(0)\Big[N_0^2+2N_0\sum_{v\in P_\A}n^\a_v+\hspace{-0.3cm}\sum_{\substack{v,v'\in P_\A:\\v\neq v'}}n^\a_vn^\a_{v'}\Big]\|\cD_\a g^0_\a\|^2\\
			&+\frac1{2N^{1-\k}}\sum_{s\in P_\A}\widehat V(s/N^{1-\k})\Big[2N_0n^\a_{s}+\hspace{-0.2cm}\sum_{\substack{v\in P_\A:\\v+s\in P_\A}}\hspace{-0.2cm}n^\a_{v}n^\a_{v+s}\Big]\|\cD_\a g^0_\a\|^2+\cE_{0,\a}
	\end{split}
\]
with
\[
	\sum_\a\l_\a |\cE_{0,\a}| \|\cD_\a g^0_\a\|^{-2}\leq CN^{7\k-2+2\eps+\d_1/3 }\,.
\]
In the last term we use  $|\widehat{V}_N(s) - \widehat{V}_N(0)| \leq C N^{2(\k-1)} |s|$ and Lemma \ref{lm:cN} to bound
\[
\begin{split}
&\Big|\frac{N_0}{N^{1-\k}} \sum_\a \l_\a \sum_{s\in P_\A}\big(\widehat V(s/N^{1-\k})-\widehat V(0)\big)\Big[2N_0n^\a_{s}+\hspace{-0.2cm}\sum_{\substack{v\in P_\A:\\v+s\in P_\A}}\hspace{-0.2cm}n^\a_{v}n^\a_{v+s}\Big]\Big|
  \leq C N^{2\k-2/3} \,.
\end{split}\]
We conclude that for any $\k \in (1/3, 2/3)$ 
\[ \begin{split}
\sum_{\a}\l_\a \frac{\langle \cD_\a g^0_\a,\cV_0 \cD_\a g^0_\a\rangle}{\|\cD_\a g^0_\a\|^2} \leq &\;  \frac1{2}\widehat{V}_N(0) \sum_\a \l_\a \Big[N_0^2+4 N_0\sum_{v\in P_\A}n^\a_v+ 2\hspace{-0.3cm}\sum_{\substack{v,v'\in P_\A:\\v\neq v'}}n^\a_vn^\a_{v'}\Big]\\
&\; +  C N^{7\k-2+2\eps+\d_1/3 }.
\end{split}
\]
This leads to the first term on the right-hand side of \eqref{eq:corr-VN} using \eqref{eq:Nplus} and \eqref{eq:Nplus2},  and noticing that we can add the term $v=v'$ in the last sum on the right-hand side of the previous expression because it is positive.


\subsubsection{Contribution of $\cV_2^{(a)}$ on the correlated state}\label{sec:V_2a}

Since $\cV_{2}^{(a)}$ has the same form as $D_\a,$ namely it annihilates two momenta in $P_\A\cup\{0\}$ and creates two momenta in $P_{\mathrm{H}},$ one can easily check
\[
	\langle \cD_\a g^0_\a, \cV_{2}^{(a)}\, \cD_\a g^0_\a\rangle=\sum_{n= 1}^{n_{\mathrm{c}}}\frac1{n!(n-1)!} \langle D_\a^n g^0_\a,\cV_{2}^{(a)}\,D_\a^{n-1}g^0_\a\rangle\,.
\]
 Recalling \eqref{eq:def_D} we get
\[
	\begin{split}
		   & \langle 	\cD_\a g^0_\a, \cV^{(a)}_{2}\cD_\a g^0_\a\rangle\!\\
		   & =\frac1{N^{1-\k}}\hspace{-0.3cm}\sum_{\substack{s\in \L^*,\,p\in P_\A^0,q\in P_{\mathrm{H}}:\\q+s\in P_\A^0, p+s\in P_{\mathrm{H}}}}\hspace{-0.3cm}\widehat{V}(s/N^{1-\k})\sum_{n=1}^{n_{\mathrm{c}}}\frac{1}{n!(n-1)!(2N)^{2n-1}}\hspace{-0.3cm} \sum_{\substack{{r_1,\dots,r_{n-1}\in P_{\mathrm{H}}}\\ \tilde r_1,\dots,\tilde r_{n}\in P_{\mathrm{H}}}}\\
		  &\times\prod_{i=1}^{n-1}\eta_{r_i}\eta_{\tilde r_i}\eta_{\tilde r_{n}}\hspace{-0.3cm}\sum_{\substack{{(v_1,v'_1)\in \bbG_{r_1}}\\{(\tilde v_1,\tilde v'_1)\in \bbG_{\tilde r_1}}}}\hspace{-0.2cm}\dots\hspace{-0.2cm} \sum_{\substack{{(v_{n-1},v'_{n-1})\in \bbG_{r_{n-1}}}\\{(\tilde v_{n-1},\tilde v'_{n-1})\in \bbG_{\tilde r_{n-1}}}}}\sum_{{(\tl v_n,\tl v'_n)\in \bbG_{\tl r_n}}}\hspace{-0.3cm}\theta\big(\{r_j,v_j,v'_j\}_{j=1}^{n-1}\big)\theta\big(\{\tilde r_j,\tilde v_j,\tilde v'_j\}_{j=1}^n\big)\\
         &\times \langle g^0_\a, A^*_{\tilde{r}_{n}n,\tilde{v}_{n},\tilde{v}'_{n}}\dots A^*_{\tilde{r}_1,\tilde{v}_1,\tilde{v}'_1} a^*_{p+s}a^*_qa_{q+s}a_pA_{r_1,v_1,v'_1}\dots A_{r_{n-1},v_{n-1},v'_{n-1}}g^0_\a\rangle.
	\end{split}
\]
As in the previous computations we first focus on the high momenta. Due to the presence of $\theta\big(\{r_j,v_j,v'_j\}_{j=1}^{n-1}\big)\theta\big(\{\tilde r_j,\tilde v_j,\tilde v'_j\}_{j=1}^{n}\big)$ we have that the operators $a^*_{p+s},a^*_q$ are necessarily contracted with $a_{\tilde p_i},a_{-\tilde p_i+\tilde v_i+\tilde v'_i}$ for some $i=1,\dots,n$, where, as usual, $\tilde p_i\in\{-\tilde r_i,\tilde r_i+\tilde v_i+\tilde v'_i\} $. 
Indeed, assume by contradiction there exist $1\leq i,j\leq n,$ $i\neq j$ such that $a^*_{p+s}$ is contracted with $a_{\tilde p_i}$ and $a^*_{q}$ is contracted with $a_{\tilde p_j},$ then $a_{-\tilde p_j+\tilde v_j+\tilde v'_j}$ has to be contracted with $a^*_{p_{t}}$ for some $1\leq t\leq n-1$ with $p_{t}\in \{-r_{t},r_{t}+v_{t}+v'_{t}\}.$ One can check that all possible contractions of the momenta $\{-p_{t}+v_{t}+v'_{t},v_{t},v'_{t}\}$ would then result in a contradiction of \eqref{eq:restrictions1}--\eqref{eq:restrictions2}.
We can assume $i=n$ by symmetry. Pairing also the remaining high momenta as in the computation of the norm, we obtain 
\[
	\begin{split}
		  &\langle \cD_\a g^0_\a, \cV^{(a)}_{2}\cD_\a g^0_\a\rangle\\
		  &=\frac1{N^{1-\k}}\hspace{-0.3cm}\sum_{\substack{s\in \L^*,\,p\in P_\A^0, q\in P_{\mathrm{H}}:\\q+s\in P_\A^0, p+s\in P_{\mathrm{H}}}}\hspace{-0.3cm}\widehat{V}(s/N^{1-\k})\sum_{n=1}^{n_{\mathrm{c}}}\frac{1}{(n-1)!(2N)^{2n-1}}\hspace{-0.3cm}\sum_{\substack{\scriptscriptstyle{r_1,\dots,r_{n-1}\in P_{\mathrm{H}}}\\\scriptscriptstyle\tilde r_1,\dots,\tilde r_{n}\in P_{\mathrm{H}}}}\sum_{\substack{\scriptscriptstyle{(v_1,v'_1)\in \bbG_{r_1}}\\\scriptscriptstyle{(\tilde v_1,\tilde v'_1)\in \bbG_{\tilde r_1}}}}\hspace{-0.2cm}\dots\hspace{-0.2cm} \sum_{\scriptscriptstyle{(\tilde v_{n},\tilde v'_{n})\in \bbG_{\tilde r_n}}}\\
		  &\phantom{{}={}}\times \prod_{i=1}^{n-1}\eta_{r_i}\eta_{\tilde r_i}(\d_{\tilde r_i,r_i}+\d_{-\tilde r_i,r_i+v_i+v'_i})·\d_{v_i+v'_i,\tilde v_i+\tilde v'_i}\;\theta\big(\{r_j,v_j,v'_j\}_{j=1}^{n-1}\big)\theta\big(\{\tilde r_j,\tilde v_j,\tilde v'_j\}_{j=1}^n\big)\\
		&\phantom{{}={}}  \times\eta_{\tilde r_n}\sum_{\tilde p_n}\d_{\tilde p_n,p+s}\d_{-\tilde p_n+\tilde v_n+\tilde v'_n,q}\langle g^0_\a, a^*_{\tilde v_n}a^*_{\tilde v'_n}\dots a^*_{\tilde v_1}a^*_{\tilde v'_1}a_{q+s}a_pa_{v_1}a_{v'_1}\dots a_{v_{n-1}}a_{v'_{n-1}} g^0_\a\rangle
	\end{split}
\]
where in the sum in the last line $\tilde p_n\in \{-\tilde r_n,\tilde r_n+\tilde v_n+\tilde v'_n\}$. 
We now focus on low momenta. We note that, recalling \eqref{eq:restrictions1}--\eqref{eq:restrictions2}, the presence of $\theta\big(\{r_j,v_j,v'_j\}_{j=1}^{n-1}\big)\theta\big(\{\tilde r_j,\tilde v_j,\tilde v'_j\}_{j=1}^n\big)$ and $\prod_{i=1}^{n-1}\d_{v_i+v'_i,\tilde v_i+\tilde v'_i}\d_{\tl v_n+\tl v'_n,p+q+s}$ imply that the set of momenta $\{p,q+s\}$ coincides with the set of momenta $\{\tilde v_n,\tilde v_n'\}.$
Denoting with $h,k=0,\dots,n-1$ the number of quartic and triplets among the indices $1,\dots, n-1,$ taking into account the different contributions coming from the different number of zero momenta in the set $\{\tl v_n,\tl v'_n\},$ and rescaling $n\to n-1$ we get 
\be\label{eq:V2a-computation}
	\begin{split}
	  &\langle \cD_\a g^0_\a, \cV^{(a)}_{2} \cD_\a g^0_\a\rangle =\frac1{N^{2-\k}}\sum_{n=0}^{n_{\mathrm{c}}-1}\frac1{N^{2n}}\sum_{h=0}^{n}\sum_{k=0}^{n-h}\frac{N_0!\,2^{-2h-k-(n-h-k)}}{(N_0-k-2(n-h-k))!h!k!(n-h-k)!} \\
		  &\times\hspace{-0.2cm}\sum_{{r_{1},\dots,r_{n+1}\in P_{\mathrm{H}}}} \sum_{{(v_1,v_1')\in \bbG^{\rm q}_{r_1}}}\hspace{-0.2cm}\dots\hspace{-0.2cm}\sum_{{(v_h,v_h')\in \bbG^{\rm q}_{r_h}}}\sum_{{(v_{h+1},0)\in \bbG^{\rm c}_{r_{h+1}}}}\hspace{-0.2cm}\dots\hspace{-0.2cm}\sum_{{(v_{h+k},0)\in \bbG^{\rm c}_{r_{h+k}}}}\;\sum_{{(v_{n+1},v'_{n+1})\in \bbG_{r_n}}}\hspace{-0.3cm}\\
		  &\times\theta\big(\{r_j,v_j,v'_j\}_{j=1}^{n+1}\big)\prod_{i=1}^h(\eta_{r_i}+\eta_{r_i+v_i+v'_i})^2n^\a_{v_i}n^\a_{v'_i} \prod_{j=h+1}^{h+k}\hspace{-0.2cm}(\eta_{r_j}+\eta_{r_j+v_j})^2n^\a_{v_j} \prod_{m=h+k+1}^{n}\hspace{-0.3cm}\eta_{r_m}^2\\
		  &\times\Big[2\widehat V\Big(\frac{r_{n+1}+v_{n+1}}{N^{1-\k}}\Big)\eta_{r_{n+1}}n^\a_{v_{n+1}}n^\a_{v'_{n+1}}\\
		  &\phantom{\times\Big[}+2\frac{(N_0-k-2(n-h-k))!}{(N_0-k-2(n-h-k)-1)!}\widehat V\Big(\frac{r_{n+1}}{N^{1-\k}}\Big)(\eta_{r_{n+1}}+\eta_{r_{n+1}+v_{n+1}})n^\a_{v_{n+1}}\d_{v'_{n+1},0}\\
		  &\phantom{\times\Big[}+\frac{(N_0-k-2(n-h-k))!}{(N_0-k-2(n-h-k)-2)!}\widehat V\Big(\frac{r_{n+1}}{N^{1-\k}}\Big)\eta_{r_{n+1}}\d_{v_{n+1},0}\d_{v'_{n+1},0}\Big].
	\end{split}
\ee

In order to isolate the large term and recognize the norm of $\cD_\a g^0_\a$ needed to get the necessary cancellation  we would like to move out the sum over $r_{n+1},v_{n+1},v'_{n+1}$ but  to do that we have to get rid of the dependence on these indices in $\theta\big(\{r_j,v_j,v'_j\}_{j=1}^{n+1}\big)$  (note that differently from the kinetic term this term is not clearly positive). Therefore, we decompose 
\begin{equation}\label{eq:dec_theta}
	\theta(\{r_i,v_i,v'_i\}_{i=1}^{n+1})=\theta(\{r_i,v_i,v'_i\}_{1=1}^{n})[1+(\theta_{n+1}(\{r_i,v_i,v'_i\}_{i=1}^{n+1})-1)]
\end{equation}
where $\theta_{n+1}(\{r_i,v_i,v'_i\}_{i=1}^{n+1})$ encodes all the restrictions involving momenta with index $n+1,$ namely
\begin{equation}\label{eq:theta_n}
	\begin{split}
		&\theta_{n+1}(\{r_i,v_i,v'_i\}_{i=1}^{n+1})\\
		&=\prod_{\substack{j,h,k=1\\j\neq n+1}}^{n+1}\prod_{p_{n+1},p_j}{\prod_{\substack{w_h,w_k,\\w_j}}}^{\hskip -0.15cm*}(1-\d_{p_{n+1},-p_j+w_k+w_h})\Big(1-(1-\d_{w_j,0})\d_{v_{n+1}+v'_{n+1},w_j+w_k}\Big)\\
		&\phantom{{}={}}\times \prod_{\substack{k,j=1\\j\neq i}}^n\prod_{w_j,w_{n+1}}\Big(1-(1-\d_{w_{n+1},0})\d_{v_{k}+v'_{k},w_{n+1}+w_j}\Big)\Big(1-(1-\d_{w_{j},0})\d_{v_{k}+v'_{k},w_{n+1}+w_j}\Big)\,.
	\end{split}
\end{equation}
Here, similarly to \eqref{eq:theta-small}, $p_\ell\in \{-r_\ell,r_{\ell}+v_\ell+v'_\ell\}$ and $w_\ell\in \{v_\ell,v'_\ell\}$ for $\ell=j,k,h,n+1$ and the $*$ in the product over $w_h,w_k,w_j$ denotes the convention $w_\ell+w_k=v_k+v'_k$ when $\ell=k$ with $\ell=j,h$.
Correspondingly, we decompose
\be \label{eq:cV2a-split}
	\langle \cD_\a g^0_\a, \cV^{(a)}_{2}\cD_\a g^0_\a\rangle={V}^{(a)}_{2,\a}+\cE^{(a)}_{2,\a}
\ee
where ${V}^{(a)}_{2,\a}$ and $\cE^{(a)}_{2,\a}$ have the same expression as $\langle\cD_\a g^0_\a, \cV^{(a)}_{2}\cD_\a g^0_\a\rangle$ in \eqref{eq:V2a-computation} with \\$\theta(\{r_i,v_i,v'_i\}_{i=1}^{n+1})$ replaced by $\theta(\{r_i,v_i,v'_i\}_{1=1}^{n})$ and $\theta(\{r_i,v_i,v'_i\}_{1=1}^{n})(\theta_{n+1}(\{r_i,v_i,v'_i\}_{i=1}^{n+1})-1)$ respectively. From ${V}^{(a)}_{2,\a}$ we get a large contribution. Indeed we can split
\be\label{eq:I2a}
	{V}^{(a)}_{2,\a}=I_{\rm q}^{\a}+I_{\rm c}^{\a}+I_{\rm p}^{\a}
\ee
where $I_{\rm q}^{\a}, I_{\rm c}^{\a}, I_{\rm p}^{\a}$ correspond to the various terms appearing in the last square bracket in \eqref{eq:V2a-computation}, the subindices refer to the fact that they were obtained by contracting the observable with a quartic, a cubic and a pair excitation respectively. Moving out the term involving the indices $r_n,v_n,v'_n$ we have
\[
	\begin{split}
		I_{\rm q}^{\a}&=\frac2{N^{2-\k}}\sum_{\substack{r\in P_{\mathrm{H}}\\(v,v')\in \bbG^{\rm q}_r}}\widehat V\Big(\frac{r+v}{N^{1-\k}}\Big)\eta_r n^\a_vn^\a_{v'}\\
		  &\phantom{{}={}}\times\sum_{n=0}^{n_{\mathrm{c}}-1}\frac1{N^{2n}}\sum_{h=0}^{n}\sum_{k=0}^{n-h}\frac{N_0!2^{-2h-k-(n-h-k)}}{(N_0-k-2(n-h-k))!h!k!(n-h-k)!} \\
		  &\phantom{{}={}}\times\hspace{-0.2cm}\sum_{{r_{1},\dots,r_{n}\in P_{\mathrm{H}}}} \sum_{{(v_1,v_1')\in \bbG^{\rm q}_{r_1}}}\hspace{-0.2cm}\dots\hspace{-0.2cm}\sum_{{(v_h,v_h')\in \bbG^{\rm q}_{r_h}}}\sum_{{(v_{h+1},0)\in \bbG^{\rm c}_{r_{h+1}}}}\hspace{-0.2cm}\dots\hspace{-0.2cm}\sum_{{(v_{h+k},0)\in \bbG^{\rm c}_{r_{h+k}}}}\\
		  &\phantom{{}={}}\times\theta\big(\{r_j,v_j,v'_j\}_{j=1}^{n}\big)\prod_{i=1}^h(\eta_{r_i}+\eta_{r_i+v_i+v'_i})^2n^\a_{v_i}n^\a_{v'_i} \prod_{j=h+1}^{h+k}\hspace{-0.2cm}(\eta_{r_j}+\eta_{r_j+v_j})^2n^\a_{v_j} \prod_{m=h+k+1}^{n}\hspace{-0.3cm}\eta_{r_m}^2\,.
	\end{split}
\]
In order to reconstruct the norm of $\cD_\a g^0_\a$ we have to add and subtract the term with $n=n_{\mathrm{c}}.$ We obtain
\[
	\begin{split}
		&I_{\rm q}^{\a}-\frac2{N^{2-\k}}\sum_{\substack{r\in P_{\mathrm{H}}\\(v,v')\in \bbG^{\rm q}_r}}\widehat V\Big(\frac{r+v}{N^{1-\k}}\Big)\eta_r n^\a_vn^\a_{v'}	\|\cD_\a g^0_\a\|^2\\
		&=	\frac2{N^{2-\k}}\sum_{\substack{r\in P_{\mathrm{H}}\\(v,v')\in \bbG^{\rm q}_r}}\widehat V\Big(\frac{r+v}{N^{1-\k}}\Big)\eta_r n^\a_vn^\a_{v'}\\
		&\phantom{{}={}}\times\frac1{N^{2n_{\mathrm{c}}}}\sum_{h=0}^{n_{\mathrm{c}}}\sum_{k=0}^{n_{\mathrm{c}}-h}\frac{N_0!2^{-2h-k-(n_{\mathrm{c}}-h-k)}}{(N_0-k-2(n_{\mathrm{c}}-h-k))!h!k!(n_{\mathrm{c}}-h-k)!} \\
		  &\phantom{{}={}}\times\hspace{-0.2cm}\sum_{{r_{1},\dots,r_{n_{\mathrm{c}}}\in P_{\mathrm{H}}}} \sum_{{(v_1,v_1')\in \bbG^{\rm q}_{r_1}}}\hspace{-0.2cm}\dots\hspace{-0.2cm}\sum_{{(v_h,v_h')\in \bbG^{\rm q}_{r_h}}}\sum_{{(v_{h+1},0)\in \bbG^{\rm c}_{r_{h+1}}}}\hspace{-0.2cm}\dots\hspace{-0.2cm}\sum_{{(v_{h+k},0)\in \bbG^{\rm c}_{r_{h+k}}}}\\
		  &\phantom{{}={}}\times\theta\big(\{r_j,v_j,v'_j\}_{j=1}^{n_{\mathrm{c}}}\big)\prod_{i=1}^h(\eta_{r_i}+\eta_{r_i+v_i+v'_i})^2n^\a_{v_i}n^\a_{v'_i} \prod_{j=h+1}^{h+k}\hspace{-0.2cm}(\eta_{r_j}+\eta_{r_j+v_j})^2n^\a_{v_j} \prod_{m=h+k+1}^{n_{\mathrm{c}}}\hspace{-0.3cm}\eta_{r_m}^2.
	\end{split}
\]
Recalling that $|\eta_r|\leq CN^\k|r|^{-2}$ for any $r\in P_{\mathrm{H}}$ and using the bound \eqref{BCS-bounds} 
we find  
\[
	\begin{split}
		&\bigg|I_{\rm q}^{\a}-\frac2{N^{2-\k}}\sum_{\substack{r\in P_{\mathrm{H}}\\(v,v')\in \bbG^{\rm q}_r}}\widehat V\Big(\frac{r+v}{N^{1-\k}}\Big)\eta_r n^\a_vn^\a_{v'}	\|\cD_\a g^0_\a\|^2\bigg|\\
		&\leq	 CN^{-1+\k}N_\a^2\Big(\frac{N_0}{N}\Big)^{2n_{\mathrm{c}}}\frac1{n_{\mathrm{c}}!}\|\eta_H\|_2^{2n_{\mathrm{c}}}\sum_{h=0}^{n_{\mathrm{c}}}\sum_{k=0}^{n_{\mathrm{c}}-h}\binom{n_{\mathrm{c}}}{h}\binom{n_{\mathrm{c}}-h}{k}\Big(\frac{N_\a}{N_0}\Big)^k\Big(\frac{N_\a}{N_0}\Big)^{2h}\\
		&\leq CN^{-1+\k}N_\a^2\frac 1{n_{\mathrm{c}}!}\|\eta_H\|_2^{2n_{\mathrm{c}}}\frac1{N^{2n_{\mathrm{c}}}}\big(N_0^2+N_0N_\a+N_\a^2\big)^{n_{\mathrm{c}}}\|\cD_\a g^0_\a\|^2
	\end{split}•
\] 
where on the last line we used \eqref{eq:LB-norm}. 
Hence, by Lemma \ref{lm:cN}, we conclude
\be\label{eq:Iq-final}
	\sum_\a\frac{\l_\a}{\|\cD_\a g^0_\a\|^2}\bigg|I_{\rm q}^{\a}-\frac2{N^{2-\k}}\sum_{\substack{r\in P_{\mathrm{H}}\\(v,v')\in \bbG_r}}\widehat V\Big(\frac{r+v}{N^{1-\k}}\Big)\eta_r n^\a_vn^\a_{v'}	\|\cD_\a g^0_\a\|^2\bigg|\leq CN^{1+\k}\frac{\|\eta_H\|_2^{2n_{\mathrm{c}}}}{n_{\mathrm{c}}!}
\ee
which is much smaller than $N^{1+\k - \eps}$ choosing of $n_{\mathrm{c}} = N^{\d_2/3} \|\eta_H\|_2^{2}$. 

We now analyze $I_{\rm c}^{\a}$. We have
\[
	\begin{split}
		I_{\rm c}^{\a}&=\frac{2N_0}{N^{2-\k}}\sum_{\substack{r\in P_{\mathrm{H}}\\(v,0)\in \bbG^{\rm c}_r}}\widehat V\Big(\frac{r}{N^{1-\k}}\Big)(\eta_r+\eta_{r+v}) n^\a_v\\
		  &\phantom{{}={}}\times\sum_{n=0}^{n_{\mathrm{c}}-1}\frac1{N^{2n}}\sum_{h=0}^{n}\sum_{k=0}^{n-h}\frac{(N_0-1)!\,2^{-2h-k-(n-h-k)}}{(N_0-k-2(n-h-k)-1)!h!k!(n-h-k)!} \\
		  &\phantom{{}={}}\times\hspace{-0.2cm}\sum_{{r_{1},\dots,r_{n}\in P_{\mathrm{H}}}} \sum_{{(v_1,v_1')\in \bbG^{\rm q}_{r_1}}}\hspace{-0.2cm}\dots\hspace{-0.2cm}\sum_{{(v_h,v_h')\in \bbG^{\rm q}_{r_h}}}\sum_{{(v_{h+1},0)\in \bbG^{\rm c}_{r_{h+1}}}}\hspace{-0.2cm}\dots\hspace{-0.2cm}\sum_{{(v_{h+k},0)\in \bbG^{\rm c}_{r_{h+k}}}}\\
		  &\phantom{{}={}}\times\theta\big(\{r_j,v_j,v'_j\}_{j=1}^{n}\big)\prod_{i=1}^h(\eta_{r_i}+\eta_{r_i+v_i+v'_i})^2n^\a_{v_i}n^\a_{v'_i} \prod_{j=h+1}^{h+k}\hspace{-0.2cm}(\eta_{r_j}+\eta_{r_j+v_j})^2n^\a_{v_j} \prod_{m=h+k+1}^{n}\hspace{-0.3cm}\eta_{r_m}^2\,.
	\end{split}
\]
To reconstruct the norm of $\|\cD_\a g^0_\a\|^2$ 
{we rewrite $(N_0-k-2(n-h-k)-1)!=(N_0-k-2(n-h-k))! (N_0-k-2(n-h-k))^{-1} $. } This results in the error term  
\[
	\begin{split}
		\cE_{\rm c}^{\a}=&-\frac{2N_0}{N^{2-\k}}\sum_{\substack{r\in P_{\mathrm{H}}\\(v,0)\in \bbG^{\rm c}_r}}\widehat V\Big(\frac{r}{N^{1-\k}}\Big)(\eta_r+\eta_{r+v}) n^\a_v\\
		&\times\sum_{n=1}^{n_{\mathrm{c}}-1}\frac1{N^{2n}}\sum_{h=0}^{n}\sum_{k=0}^{n-h}\frac{(N_0-1)!2^{-2h-k-(n-h-k)}}{(N_0-k-2(n-h-k))!h!k!(n-h-k)!} (k+2(n-h-k))\\
		  &\times\hspace{-0.2cm}\sum_{{r_{1},\dots,r_{n}\in P_{\mathrm{H}}}} \sum_{{(v_1,v_1')\in \bbG^{\rm q}_{r_1}}}\hspace{-0.2cm}\dots\hspace{-0.2cm}\sum_{{(v_h,v_h')\in \bbG^{\rm q}_{r_h}}}\sum_{{(v_{h+1},0)\in \bbG^{\rm c}_{r_{h+1}}}}\hspace{-0.2cm}\dots\hspace{-0.2cm}\sum_{{(v_{h+k},0)\in \bbG^{\rm c}_{r_{h+k}}}}\\
		  &\times\theta\big(\{r_j,v_j,v'_j\}_{j=1}^{n}\big)\prod_{i=1}^h(\eta_{r_i}+\eta_{r_i+v_i+v'_i})^2n^\a_{v_i}n^\a_{v'_i} \prod_{j=h+1}^{h+k}\hspace{-0.2cm}(\eta_{r_j}+\eta_{r_j+v_j})^2n^\a_{v_j} \prod_{m=h+k+1}^{n}\hspace{-0.3cm}\eta_{r_m}^2\,.
	\end{split}
\]
We now take the absolute value which allows us to get an upper bound adding the term $n=n_{\mathrm{c}},$ then we consider separately the terms $k$ and $2(n-h-k)$ in the bracket in the first line. In the first term we rescale $k\to k-1, n\to n-1$ and bound {$\frac{(N_0-1)!}{(N_0-k-2(n-h-k)-1)!}\leq \frac{N_0!}{N_0-k-2(n-h-k))!}$}, in the second term we rescale $n\to n-1$ and bound {$\frac{(N_0-2)!}{(N_0-k-2(n-h-k)-)!}\leq \frac{N_0!}{N_0-k-2(n-h-k))!}$ }. 
Recalling \eqref{BCS-bounds} and using Lemma \ref{lm:cN} we bound
\be\label{eq:cE_ca}
	\begin{split}
		\sum_\a\l_\a\frac{|\cE_{\rm c}^{\a}|}{\|\cD_\a g^0_\a\|^2}\leq C N^{-3+\k}N_0\|\eta_H\|_2^2\sum_\a\l_\a N_\a(N_0+N_\a)\leq CN^{1+\k}N^{3\k-2+\eps}\,.
	\end{split}•
\ee
We then add the last term $n=n_{\mathrm{c}}$ to the large contribution of $I_{\rm c}^{\a}$ getting
\[
	\begin{split}
	 	\Big|I_{\rm c}^{\a}-\frac{2N_0}{N^{2-\k}}\sum_{\substack{r\in P_{\mathrm{H}}\\(v,0)\in \bbG^{\rm c}_r}}\widehat V\Big(\frac{r}{N^{1-\k}}\Big)&(\eta_r+\eta_{r+v}) n^\a_v\|\cD_\a g^0_\a \|^2-\cE_{\rm c}^{\a}\Big|\\
		&\leq C N^{-1+\k}N_0N_\a\|\eta_H\|_2^{2n_{\mathrm{c}}}\frac{1}{n_{\mathrm{c}}!N^{2n_{\mathrm{c}}}}(N_0^2+N_0N_\a+N_\a^2)^{n_{\mathrm{c}}},
	\end{split}•
\]
which together with \eqref{eq:cE_ca} yields
\be\label{eq:Ic-final}
\begin{split}
		& \sum_\a\frac{\l_\a}{\|\cD_\a g^0_\a\|^2}\Big|I_{\rm c}^{\a}-\frac{2N_0}{N^{2-\k}}\sum_{\substack{r\in P_{\mathrm{H}}\\(v,0)\in \bbG^{\rm c}_r}}\widehat V\Big(\frac{r}{N^{1-\k}}\Big)(\eta_r+\eta_{r+v}) n^\a_v\|\cD_\a g^0_\a \|^2\Big|\\
		& \leq CN^{1+\k}\bigg(\frac{\|\eta_H\|_2^{2n_{\mathrm{c}}}}{n_{\mathrm{c}}!}+N^{3\k-2 +\eps}\bigg).
\end{split}
\ee
We proceed similarly for $I_{\rm p}^{\a}.$ From
\[
	\begin{split}
		I_{\rm p}^{\a}&=\frac{N_0(N_0-1)}{N^{2-\k}}\sum_{r\in P_{\mathrm{H}}}\widehat V\Big(\frac{r}{N^{1-\k}}\Big)\eta_r\\
		  &\phantom{{}={}}\times\sum_{n=0}^{n_{\mathrm{c}}-1}\frac1{N^{2n}}\sum_{h=0}^{n}\sum_{k=0}^{n-h}\frac{(N_0-2)!\,2^{-2h-k-(n-h-k)}}{(N_0-k-2(n-h-k)-2)!h!k!(n-h-k)!} \\
		  &\phantom{{}={}}\times\hspace{-0.2cm}\sum_{{r_{1},\dots,r_{n}\in P_{\mathrm{H}}}} \sum_{{(v_1,v_1')\in \bbG^{\rm q}_{r_1}}}\hspace{-0.2cm}\dots\hspace{-0.2cm}\sum_{{(v_h,v_h')\in \bbG^{\rm q}_{r_h}}}\sum_{{(v_{h+1},0)\in \bbG^{\rm c}_{r_{h+1}}}}\hspace{-0.2cm}\dots\hspace{-0.2cm}\sum_{{(v_{h+k},0)\in \bbG^{\rm c}_{r_{h+k}}}}\\
		  &\phantom{{}={}}\times\theta\big(\{r_j,v_j,v'_j\}_{j=1}^{n}\big)\prod_{i=1}^h(\eta_{r_i}+\eta_{r_i+v_i+v'_i})^2n^\a_{v_i}n^\a_{v'_i} \prod_{j=h+1}^{h+k}\hspace{-0.2cm}(\eta_{r_j}+\eta_{r_j+v_j})^2n^\a_{v_j} \prod_{m=h+k+1}^{n}\hspace{-0.3cm}\eta_{r_m}^2
	\end{split}
\]
replacing $\frac{(N_0-2)!}{(N_0-k-2(n-h-k)-2)!}$ with $\frac{N_0!}{(N_0-k-2(n-h-k))!}$ we get the error  \newpage
\[
	\begin{split}
		\cE_{\rm p}^{\a}&=-\frac{N_0(N_0-1)}{N^{2-\k}}\sum_{r\in P_{\mathrm{H}}}\widehat V\Big(\frac{r}{N^{1-\k}}\Big)\eta_r\sum_{n=0}^{n_{\mathrm{c}}-1}\frac1{N^{2n}}\sum_{h=0}^{n}\sum_{k=0}^{n-h}\frac{2^{-2h-k-(n-h-k)}}{h!k!(n-h-k)!}\\
		&\phantom{{}={}}\times\Big[\frac{(N_0-1)!}{(N_0-k-2(n-h-k))!}+\frac{(N_0-2)!}{(N_0-k-2(n-k-h)-1)!} \Big](k+2(n-h-k))\\
		  &\phantom{{}={}}\times\hspace{-0.2cm}\sum_{{r_{1},\dots,r_{n}\in P_{\mathrm{H}}}} \sum_{{(v_1,v_1')\in \bbG^{\rm q}_{r_1}}}\hspace{-0.2cm}\dots\hspace{-0.2cm}\sum_{{(v_h,v_h')\in \bbG^{\rm q}_{r_h}}}\sum_{{(v_{h+1},0)\in \bbG^{\rm c}_{r_{h+1}}}}\hspace{-0.2cm}\dots\hspace{-0.2cm}\sum_{{(v_{h+k},0)\in \bbG^{\rm c}_{r_{h+k}}}}\\
		  &\phantom{{}={}}\times\theta\big(\{r_j,v_j,v'_j\}_{j=1}^{n}\big)\prod_{i=1}^h(\eta_{r_i}+\eta_{r_i+v_i+v'_i})^2n^\a_{v_i}n^\a_{v'_i} \prod_{j=h+1}^{h+k}\hspace{-0.2cm}(\eta_{r_j}+\eta_{r_j+v_j})^2n^\a_{v_j} \prod_{m=h+k+1}^{n}\hspace{-0.3cm}\eta_{r_m}^2.
	\end{split}•
\]
Appropriately rescaling the indices in the various terms analogously to what we did in the estimates of $\cE_{\rm c}^{\a}$, and using \eqref{BCS-bounds} (recall that $|\eta_r|\leq N^\k |r|^{-2}$) we obtain 
\[
	\begin{split}
		\sum_\a\l_\a\frac{\cE_{\rm p}^{\a}}{\|\cD_\a g^0_\a\|^2}\leq CN^{-3+\k}N_0^2\|\eta_H\|_2^2\sum_{\a}\l_\a(N_0+N_\a)\leq CN^{1+\k}N^{3\k-2+\eps}.
	\end{split}•
\]
Then, adding the term $n=n_{\mathrm{c}}$ in the large contribution in $I_{\rm p}^{\a}$ we find
\[
	\begin{split}
		\Big|I_{\rm p}^\a-\frac{N_0(N_0-1)}{N^{2-\k}}\sum_{r\in P_{\mathrm{H}}}\widehat V\Big(\frac{r}{N^{1-\k}}\Big)&\eta_r\|\cD_\a g^0_\a\|^2-\cE_{\rm p}^\a\Big|\\
		&\leq  CN^{-1+\k}N_0^2\|\eta_H\|_2^{2n_{\mathrm{c}}}(N_0^2+N_0N_\a+N_\a^2)^{n_{\mathrm{c}}}.
	\end{split}	
\]
We conclude
\be\label{eq:Ip-final}
	\sum_\a\frac{\l_\a}{\|\cD_\a g^0_\a\|^2}\Big|I_{\rm p}^\a-\frac{N_0(N_0-1)}{N^{2-\k}}\sum_{r\in P_{\mathrm{H}}}\widehat V\Big(\frac{r}{N^{1-\k}}\Big)\eta_r\|\cD_\a g^0_\a\|^2\Big|\leq CN^{1+\k}\Big(\frac{\|\eta_H\|_2^{2n_{\mathrm{c}}}}{n_{\mathrm{c}}!}+N^{3\k-2+\eps}\Big).
\ee
By  \eqref{eq:I2a}, \eqref{eq:Iq-final}, \eqref{eq:Ic-final} and \eqref{eq:Ip-final},  and the bound
\[
\Big| \frac{N_0}{N^{2-\k}}\sum_{r\in P_{\mathrm{H}}}\widehat V\Big(\frac{r}{N^{1-\k}}\Big)\eta_r \Big| \leq C N^\k
\]
following from \eqref{BCS-bounds} we obtain
\be \label{eq:V2a-final}
	\begin{split}
		\frac{V^{(a)}_{2,\a}}{\|\cD_\a g^0_\a\|^2}=\frac1{N^{2-\k}}\Big[&\;N_0^2
		\sum_{r\in P_{\mathrm{H}}}\eta_{r}\widehat V\Big(\frac{r}{N^{1-\k}}\Big)+2N_0\hskip -0.2cm\sum_{\substack{r\in P_{\mathrm{H}},\\(v,0)\in \bbG^{\rm c}_{r}}}\hskip -0.2cm\widehat V\Big(\frac{r}{N^{1-\k}}\Big)(\eta_{r}+\eta_{r+v})n^\a_v\\
		  &\hskip 1cm +\sum_{\substack{r\in P_{\mathrm{H}},\\(v,v')\in \bbG_r^{\rm q}}}2\widehat V\Big(\frac{r+v}{N^{1-\k}}\Big)\eta_{r}n^\a_{v}n^\a_{v'}\Big]\| \cD_\a g^0_\a\|^2+\frac{R_{2,\a}^{(a)}}{\|\cD_\a g^0_\a\|^2}
	\end{split}
\ee
where  
\[
	\begin{split}
		\sum_\a\l_\a \frac{R_{2,\a}^{(a)}}{\|\cD_\a g^0_\a\|^2}\leq CN^{1+\k}\Big(\frac{\|\eta_H\|_2^{2n_{\mathrm{c}}}}{n_{\mathrm{c}}!}+N^{3\k-2+\eps}\Big) \leq C N^{\k+1-\eps}
	\end{split}
\]
for all $\k \in (1/3; 2/3)$ such that $3\k-2+3\eps <0$. 
It remains to show that $\cE_{2,\a}^{(a)}$ defined in \eqref{eq:cV2a-split} gives a negligible contribution.
With
\begin{equation}\label{eq:theta-1}
	\begin{split}
		& |\theta_n(\{r_i,v_i,v'_i\}_{i=1}^{n+1})-1|\\
		&\leq \sum_{j=1}^{n}\sum_{p_j, p_{n+1}} \Big[ \d_{p_{n+1},p_j}+ \sum_{w_j, w_{n+1}}\d_{p_{n+1},-p_j+w_j+w_{n+1}})\Big]\\
		&+ \sum_{j=1}^{n}\sum_{w_{n+1}}(1-\d_{w_{n+1},0})\Big[ \sum_{w_j, w_{n+1}}\d_{w_{n+1},w_j}+\d_{v_j+v'_j,v_{n+1}+v'_{n+1}}\Big]\\
		&+\sum_{\substack{j,k=1\\j\neq k}}^{n}\sum_{w_j} \Big[ (1-\d_{w_{j},0})\big(\sum_{w_k}\d_{v_{n+1}+v_{n+1}',w_j+w_k} +\sum_{w_{n+1}}\d_{v_k+v'_k,w_{n+1}+w_j}\big)\\
		& \hskip 7.5cm +\sum_{w_{n+1}}(1-\d_{w_{n+1},0})\d_{v_k+v'_k,w_{n+1}+w_j}\Big]\\
		&+\sum_{\substack{j,k=1\\j\neq k}}^{n}\Big[\sum_{p_k,p_j}\d_{p_k,-p_j+v_{n+1}+v'_{n+1}}+\sum_{p_k,p_{n+1}}\d_{p_k,-p_{n+1}+v_j+v'_j}\\
		&
		\hskip 1.2cm + \sum_{\substack{p_j, p_{n+1}, \\ w_j, w_k}}\d_{p_{n+1},-p_j+w_j+w_k}+\hskip -0.1cm\sum_{\substack{p_j, p_k,\\ w_j, w_{n+1}}}\hskip -0.1cm\d_{p_k,-p_j+w_j+w_{n+1}}+\hskip -0.1cm\sum_{\substack{p_k, p_{n+1},\\ w_{n+1}}}\hskip -0.1cm\d_{p_k,-p_{n+1}+w_{n+1}+w_j}\Big]\\
		&+\sum_{\substack{j,k,h=1\\j\neq k\neq h\neq j}}^{n}\sum_{p_k, w_h} \Big[ \sum_{p_{n+1},w_j}\d_{p_{n+1},-p_k+w_j+w_h}+\sum_{p_j,w_{n+1}}\d_{p_k,-p_j+w_h+w_{n+1}}\Big]
	\end{split}
\end{equation}
we can bound $ |\cE^{(a)}_{2,\a}| \leq \sum_{j=1}^5 X^{\a}_j$ with $X^{\a}_j$ denoting the contribution arising from $j$-th term in \eqref{eq:theta-1}.\\ 
We can estimate
\[
	\begin{split}
	  &X_1^\a\leq \frac1{N^{2-\k}}\sum_{n=1}^{n_{\mathrm{c}}-1}\frac1{N^{2n}}\sum_{h=0}^{n}\sum_{k=0}^{n-h}\frac{N_0!\,2^{-2h-k-(n-h-k)}}{(N_0-k-2(n-h-k))!h!k!(n-h-k)!} \\
		  &\times\hspace{-0.2cm}\sum_{{r_{1},\dots,r_{n+1}\in P_{\mathrm{H}}}} \sum_{{(v_1,v_1')\in \bbG^{\rm q}_{r_1}}}\hspace{-0.2cm}\dots\hspace{-0.2cm}\sum_{{(v_h,v_h')\in \bbG^{\rm q}_{r_h}}}\sum_{{(v_{h+1},0)\in \bbG^{\rm c}_{r_{h+1}}}}\hspace{-0.2cm}\dots\hspace{-0.2cm}\sum_{{(v_{h+k},0)\in \bbG^{\rm c}_{r_{h+k}}}}\;\sum_{{(v_{n+1},v'_{n+1})\in \bbG_{r_n}}}\hspace{-0.3cm}\\
		  &\times\theta(\{r_i,v_i,v'_i\}_{i=1}^{n})\sum_{\ell=1}^{n}\bigg[\sum_{p_\ell,p_{n+1}}\d_{p_\ell,p_{n+1}}+\sum_{w_\ell,w_{n+1}}\d_{p_{n+1},-p_\ell+w_\ell+w_{n+1}}\bigg]\\
		  &\times\prod_{i=1}^h(\eta_{r_i}+\eta_{r_i+v_i+v'_i})^2n^\a_{v_i}n^\a_{v'_i} \prod_{j=h+1}^{h+k}\hspace{-0.2cm}(\eta_{r_j}+\eta_{r_j+v_j})^2n^\a_{v_j} \prod_{m=h+k+1}^{n}\hspace{-0.3cm}\eta_{r_m}^2\\
		  &\times\Big[2\Big|\widehat V\Big(\frac{r_{n+1}+v_{n+1}}{N^{1-\k}}\Big)\Big||\eta_{r_{n+1}}|n^\a_{v_{n+1}}n^\a_{v'_{n+1}}\\
		  &\phantom{\times\Big[}+2\frac{(N_0-k-2(n-h-k))!}{(N_0-k-2(n-h-k)-1)!}\Big|\widehat V\Big(\frac{r_{n+1}}{N^{1-\k}}\Big)\Big|(|\eta_{r_{n+1}}+\eta_{r_{n+1}+v_{n+1}}|)n^\a_{v_{n+1}}\d_{v'_{n+1},0}\\
		  &\phantom{\times\Big[}+\frac{(N_0-k-2(n-h-k))!}{(N_0-k-2(n-h-k)-2)!}\Big|\widehat V\Big(\frac{r_{n+1}}{N^{1-\k}}\Big)\Big||\eta_{r_{n+1}}|\d_{v_{n+1},0}\d_{v'_{n+1},0}\Big]\,.
	\end{split}
\]
Since $\theta\big(\{r_i,v_i,v'_i\}_{i=1}^{n}\big)$ can be bounded from above by $\displaystyle{\theta\big(\{r_i,v_i,v'_i\}_{\substack{i=1\\i\neq \ell}}^{n}\big)}$ for any $\ell\in\{1,\dots,n\}$ we can appropriately rescaling the indices and eventually bounding the factorials involving $N_0$ reconstructing $\|\cD_\a g_\a^0\|^2$. We end up with
\be \label{eq:X1}
	\begin{split}
			&\frac{X_1^{\a}}{\|\cD_\a g_\a^0\|^2}\\
			&\leq C N^{\k-4}\sum_{s\in P_{\mathrm{H}}}\sum_{(u,u')\in \bbG_s}(\eta_s+\eta_{s+u+u'})^2(n^\a_{u}n^\a_{u'}+N_0n^\a_{u}\d_{u',0}+N_0^2\d_{u,0}\d_{u',0})\\
			&\times\sum_{r\in P_{\mathrm{H}}}\sum_{{(v,v')\in \bbG_r}}\bigg[N_0^2\,\Big|\widehat V\Big(\frac{r}{N^{1-\k}}\Big)\Big||\eta_{r}|\d_{v,0}\d_{v',0}+N_0\Big|\widehat V\Big(\frac{r}{N^{1-\k}}\Big)\Big||\eta_{r}+\eta_{r+v}|n^\a_{v}\d_{v',0}\\
		  &\hskip 3cm +\Big|\widehat V\Big(\frac{r+v}{N^{1-\k}}\Big)\Big||\eta_{r}|n^\a_{v}n^\a_{v'}\bigg]\hspace{-0.1cm}\sum_{\substack{{p\in \{-r,r+v+v'\}}\\ {\tl p\in \{-s,s+u+u'\}}}}\sum_{\substack{{w\in \{v,v',0\}}\\{\tl w\in \{u,u',0\}}}}\hspace{-0.2cm}(\d_{p,\tl p}+\d_{p,-\tl p+w+\tl w})\\
		  &\leq CN^{\k-4}\|\eta_H\|_\infty^2\sup_{q\in \L^*}\sum_{r\in \L^*_+}\widehat{V}((q-r))/N^{1-\k})|\eta_r|(N_0+N_\a)^4\\
		  &\leq CN^{7\k-7+4\eps}(N_0+N_\a)^4
	\end{split}
\ee
where we used \eqref{eq:etaH} and \eqref{BCS-bounds}. 
Proceeding similarly we find 
\be \label{eq:X2}
	\begin{split}
		&\frac{X_2^{\a}}{\|\cD_\a g_\a^0\|^2}\leq CN^{\k-4}\sum_{s\in P_{\mathrm{H}}}\sum_{(u,u')\in \bbG_s}(\eta_s+\eta_{s+u+u'})^2(N_0n^\a_{u}\d_{u',0}+n^\a_{u}n^\a_{u'})\\
		&\times\sum_{r\in P_{\mathrm{H}}}\sum_{(v,v')\in \bbG_r}\hspace{-0.2cm}\Big[N_0\Big|\widehat V\Big(\frac{r}{N^{1-\k}}\Big)\Big||\eta_{r}+\eta_{r+v}|n^\a_{v}\d_{v',0}+\Big|\widehat V\Big(\frac{r+v}{N^{1-\k}}\Big)\Big||\eta_{r}|n^\a_{v}n^\a_{v'}\Big]\\
		&\times\Big[\d_{v+v',u+u'}+\sum_{\substack{w\in \{v,v'\}\\\tl w\in \{u,u'\}}}(1-\d_{w,0})\d_{w,\tl w}\Big]\\
		&\leq CN^{\k-4}\|\eta_H\|_2^2\sup_{q\in \L^*}\sum_{r\in \L^*_+}\widehat{V}((q-r))/N^{1-\k})|\eta_r|\\
		& \hskip 3cm \times \sum_{v,v',u'\in P_\A^0}n^\a_{v}(N_0\d_{v',0}+n^\a_{v'})(N_0\d_{u',0}+n^\a_{u'})(n^\a_{v}+n^\a_{v+v'-u'})\\
		&\leq CN^{4\k-4+\eps}\sum_{v\in P_{\b}^0}(n^\a_{v})^2(N_0+N_\a)^2 \,.
	\end{split}
\ee
Analogously, 
\be \label{eq:X3}
	\begin{split}
		&\frac{X_3^{\a}}{\|\cD_\a g_\a^0\|^2}\leq CN^{\k-6}\sum_{\substack{s_1\in P_{\mathrm{H}}\\(u_1,u_1')\in \bbG_{s_1}}}(\eta_{s_1}+\eta_{{s_1}+u_1+u_1'})^2(N_0^2\d_{u_1,0}\d_{u_1',0}+N_0n^\a_{u_1}+n^\a_{u_1}n^\a_{u_1'})\\
		&\times\hspace{-0.3cm} \sum_{\substack{s_2\in P_{\mathrm{H}}\\(u_2,u_2')\in \bbG_{s_2}}}(\eta_{s_2}+\eta_{s_2+u_2+u_2'})^2(N_0^2\d_{u_2,0}\d_{u_2',0}+N_0n^\a_{u_2}+n^\a_{u_2} n^\a_{u_2'}) \hspace{-0.3cm}\\
		&\times \sum_{\substack{r\in P_{\mathrm{H}}\\(v,v')\in \bbG_r}}\Big[N_0^2\, \Big|\widehat V\Big(\frac{r}{N^{1-\k}}\big)\Big|\,|\eta_r|\d_{v,0}\d_{v',0}+N_0\,\Big|\widehat V\Big(\frac{r}{N^{1-\k}}\Big)\Big|\,|\eta_{r}+\eta_{r+v}|\, n^\a_{v}\d_{v',0}\\[-0.3cm]
		&\hskip 9cm +\!\Big|\widehat V\Big(\frac{r+v}{N^{1-\k}}\Big)\!\Big||\eta_{r}|n^\a_{v}n^\a_{v'}\Big]\\
		&\times\hspace{-0.3cm}\sum_{\substack{w\in \{v,v'\}}}\sum_{\substack{ w_1\in \{u_1,u_1'\}\\ w_2\in\{u_2,u_2'\}}}\big[(1-\d_{ w_2,0})(\d_{v+v',w_1+ w_2}+\d_{u_1+u_1',w+ w_2})+(1-\d_{w,0})\d_{u_1+u_1',w+ w_2})\big]\\
		&\leq CN^{\k-6}\|\eta_H\|^4_2\sup_{q\in \L^*}\sum_{r\in \L^*_+}\widehat{V}((q-r))/N^{1-\k})|\eta_r|\hskip -0.2cm\sum_{v,v'\in P_\A^0}(N_0^2\d_{v,0}\d_{v',0}+N_0n^\a_v\d_{v',0}+n^\a_vn^\a_{v'})\\
		&\times\sum_{u,u'\in P_\A^0}(N_0^2\d_{u,0}\d_{u',0}+N_0n^\a_u\d_{u',0}+n^\a_un^\a_{u'})\Big(\sum_{\tl w\in \{u,u'\}}\hspace{-0.3cm}n^\a_{v+v'-w}+\hspace{-0.3cm}\sum_{w\in \{v,v'\}}\hspace{-0.3cm}n^\a_{u+u'-w}\Big)(N_0+N_\a)\\
		&\leq CN^{7\k-7+2\eps}(N_0+N_\a)^4\sum_{v\in P_\A}(n^\a_{v})^2
	\end{split}
\ee
where, as above, we used \eqref{eq:etaH} and \eqref{BCS-bounds}.

Along the same line of reasoning we find the bound 
\be \label{eq:X4}
	\begin{split}
		&\frac{X_4^{\a}}{\|\cD_\a g_\a^0\|^2}\leq CN^{\k-6}\sum_{\substack{s_1\in P_{\mathrm{H}}\\(u_1,u_1')\in \bbG_{s_1}}}(\eta_{s_1}+\eta_{s_1+u_1+u_1'})^2(N_0^2\d_{u_1,0}\d_{u_1',0}+N_0n^\a_{u_1}+n^\a_{u_1}n^\a_{u_1'})\\
		&\times \sum_{\substack{s_2\in P_{\mathrm{H}}\\(u_2,u_2')\in \bbG_{s_2}}}(\eta_{s_2}+\eta_{s_2+u_2+u_2'})^2(N_0^2\d_{u_2,0}\d_{u_2',0}+N_0n^\a_{u_2}+n^\a_{u_2}n^\a_{u_2'})\\
		&\times\sum_{\substack{r\in P_{\mathrm{H}}\\(v,v')\in \bbG_r}}\hspace{-0.2cm}\Big[N_0^2\Big|\widehat{V}\Big(\frac{r}{N^{1-\k}}\Big)\Big||\eta_{r}|\d_{v,0}\d_{v',0}+N_0\Big|\widehat V\Big(\frac{r}{N^{1-\k}}\Big)\Big||\eta_{r}+\eta_{r+v}|n^\a_{v}\d_{v',0}\\
		&+\Big|\widehat V\Big(\frac{r+v}{N^{1-\k}}\Big)\Big||\eta_{r_n}|n^\a_vn^\a_{v'}\Big]	\hspace{-0.4cm}
		\sum_{\substack{p\in \{-r,r+v+v'\}}}
		\sum_{\substack{p_1\in \{-s_1,s_1+u_1+u_1'\}\\ p_2\in \{-s_2,s_2+u_2+u_2'\}}}
		\Big(\d_{ p_1,-p_2+v+v'}+\d_{p,-p_2+u_1+u_1}\\
		&\hskip 2.4 cm +\sum_{\substack{w\in \{v,v'\}}} \sum_{\substack{ w_1\in \{u_1,u_1'\}\\ w_2\in \{u_2,u_2'\}}} \big(\d_{p=-p_1+ w_1+w_2}+\d_{ p_2=- p_1+ w_1+w}+\d_{p_2=-p+w+ w_1}\big)\Big)\\
		&\leq CN^{\k-6}\|\eta_H\|_\infty^2\|\eta_H\|^2_2\sup_{q\in \L^*}\sum_{r\in \L^*_+}\widehat{V}((q-r))/N^{1-\k})|\eta_r|(N_0+N_\a)^6\\
		&\leq CN^{10\k-10+5\eps}(N_0+N_\a)^6.
	\end{split}
\ee 
Finally,
\be \label{eq:X5}
	\begin{split}
		&\frac{X_5^{\a}}{\|\cD_\a g_\a^0\|^2}\leq CN^{\k-8}\sum_{\substack{s_1\in P_{\mathrm{H}}\\(u_1,u_1')\in \bbG_{s_1}}}(\eta_{s_1}+\eta_{s_1+u_1+u_1'})^2(N_0^2\d_{u_1,0}\d_{u_1',0}+N_0n^\a_{u_1}+n^\a_{u_1}n^\a_{u_1'})\\
		&\times \sum_{\substack{s_2\in P_{\mathrm{H}}\\(u_2,u_2')\in \bbG_{s_2}}}(\eta_{s_2}+\eta_{s_2+u_2+u_2'})^2(N_0^2\d_{u_2,0}\d_{u_2',0}+N_0n^\a_{u_2}+n^\a_{u_2}n^\a_{u_2'})\\
		&\times \sum_{\substack{s_3\in P_{\mathrm{H}}\\(u_3,u_3')\in \bbG_{s_3}}}(\eta_{s_3}+\eta_{s_3+u_3+u_3'})^2(N_0^2\d_{u_3,0}\d_{u_3',0}+N_0n^\a_{u_3}+n^\a_{u_3}n^\a_{u_3'})\\
		&\times\sum_{\substack{r\in P_{\mathrm{H}}\\(v,v')\in \bbG_r}}\Big[N_0^2\Big|\widehat{V}\Big(\frac{r}{N^{1-\k}}\Big)\Big||\eta_{r}|\d_{v,0}\d_{v',0}+N_0\Big|\widehat V\Big(\frac{r_n}{N^{1-\k}}\Big)\Big||\eta_{r_n}+\eta_{r_n+v_n}|n^\a_{v}\d_{v'_n,0}\\
		&+\Big|\widehat V\Big(\frac{r+v}{N^{1-\k}}\Big)\Big||\eta_{r}|n^\a_{v}n^\a_{v'}\Big]	\hspace{-0.4cm}\sum_{\substack{p\in \{-r,r+v+v'\}\\\tl p_1\in \{-s_1,s_1+u_1+u_1'\}\\\tl p_2\in \{-s_2,s_2+u_2+u_2'\}}}\sum_{\substack{w\in \{v,v'\}\\\tl w_1\in \{u_1,u_1\}\\\tl w_3\in \{u_3,u_3'\}}}\hspace{-0.3cm}\d_{\tl p_2,-p+\tl w_1+\tl w_3}+\d_{\tl p_2,-\tl p_1+w+\tl w_3}\\
		&\leq CN^{\k-8}\|\eta_H\|_\infty^2\|\eta_H\|^4_2\sup_{q\in \L^*}\sum_{r\in \L^*_+}\widehat{V}\Big(\frac{q-r}{N^{1-\k}}\Big)|\eta_r|(N_0+N_\a)^8\leq CN^{13\k-13+6\eps}(N_0+N_\a)^8\,.
	\end{split}
\ee
 With \eqref{eq:X1}, \eqref{eq:X2}, \eqref{eq:X3}, \eqref{eq:X4}, \eqref{eq:X5} and Lemma \ref{lm:cN} we conclude that for all $\k \in (1/3;2/3)$
\be \label{eq:cE-a-2}
 \sum_\a \l_\a  \frac{  \big|\cE^{(a)}_{2,\a}\big|}{\|\cD_\a g_\a^0\|^2} 
 \leq  C \max \{N^{7\k-2+2\e+\d_1/3}, N^{13\k-5+6\e}\}\,.
\ee

Eqs.\,\eqref{eq:cV2a-split}, \eqref{eq:V2a-final} and \eqref{eq:cE-a-2} lead to the bound
\be \label{eq:cV2a-s1}
	\begin{split}
& \sum_\a \l_\a  \frac{\langle \cD_\a g^0_\a,\cV_{2}^{(a)}\cD_\a g^0_\a\rangle}{\| \cD_\a g^0_\a\|^2}   \leq   \frac{1}{N^{2-\k}} \Big[ N_0^2
		\sum_{r\in P_{\mathrm{H}}}\widehat V\Big(\frac{r}{N^{1-\k}}\Big)\eta_{r}  \\
		&\hskip 1.5cm +2N_0\sum_{\substack{r\in P_{\mathrm{H}},\\(v,0)\in \bbG^{\rm c}_{r}}}\widehat V\Big(\frac{r}{N^{1-\k}}\Big)(\eta_{r}+\eta_{r+v})n^\a_v +\hskip -0.2cm \sum_{\substack{r\in P_{\mathrm{H}},\\(v,v')\in \bbG_r^{\rm q}}} \hskip -0.2cm 2\widehat V\Big(\frac{r}{N^{1-\k}}\Big)\eta_{r+v}n^\a_{v}n^\a_{v'}\Big]  \\
		  &\hskip 1.5cm   + C N^{1+\k}\max \{N^{-\eps}, N^{6\k-3+2\e+\d_1/3}, N^{12\k-6+6\e}\}\,.
	\end{split}
	\ee
We now claim that  
\be \label{eq:V-eta-rv} 
	\sup_{\substack{v,v'\in P_\A^{0}}} N^\k \hskip -0.2cm \sum_{\substack{r \in P_{\mathrm{H}}:\\r+v+v'\in P_{\mathrm{H}}}} \hskip -0.2cm \widehat V\Big(\frac{r}{N^{1-\k}}\Big)\, |\eta_{r+v+v'}- \eta_{r}|  \leq C N^{\k+1}N^{-\eps}
\ee
which can be proved similarly to \eqref{eq:eta-rv} as shown below. By \eqref{eq:diff-eta}-\eqref{eq:diff-eta-2} we find
\be\label{eq:diff-eta-3}
	\begin{split}
		N^\k \hskip -0.2cm \sum_{\substack{r \in P_{\mathrm{H}}:\\r+v+v'\in P_{\mathrm{H}}}}& \hskip -0.2cm \widehat V\Big(\frac{r}{N^{1-\k}}\Big)\, |\eta_{r+v+v'}- \eta_{r}| \leq CN^{3\k-2/3+\d_2/3} \sum_{r\in P_{\mathrm{H}}} \frac{\widehat V(r/N^{1-\k})}{|r|^2}\\
		&+CN^\k\|\widehat h\|_2\Big(\sum_{r\in P_{\mathrm{H}}}\frac1{|r|^4}\Big)^{1/2}+N^{2\k+1/3+\d_2/3}\sum_{r\in P_{\mathrm{H}}}\frac{\widehat V(r/N^{1-\k})}{|r|^3}
	\end{split}•
\ee
where we recall the notation $\widehat{h}_r=N^{3-2\k} \l_\ell  \big(\widehat \chi_\ell \ast \widehat f_{\ell,N}\big)(r)$. To bound the last term we notice that, using an integral approximation and rescaling variables, we have for any $q'>1$ 
\[ \begin{split}
	\sum_{r\in P_{\mathrm{H}}}\frac{\widehat V(r/N^{1-\k})}{|r|^3} &\; \leq C \sum_{\substack{r\in P_{\mathrm{H}}\\ |r|\leq N^{1-\k}}}\frac{1}{|r|^3} + C\int_{|\tl r|>1}\frac{\widehat V(\tl r)}{|\tl r|^3}d\tl r\\
& \leq C  \ln N + C \|\widehat V\|_q\Big(\int_{|\tl r|>1}\frac1{|r|^{3q'}}d\tl r\Big)^{1/q'}
	\leq C \big(\ln N + \|\widehat V\|_q \big)
\end{split}\]
where $q$ is the conjugate exponent of $q'.$ 
Using the Hausdorff-Young inequality together with the assumption $V\in L^{q'}(\bR^3)$\ for some $q'>1$ 
we conclude
\be\label{eq:HYineq}
	\sum_{r\in P_{\mathrm{H}}}\frac{\widehat V(r/N^{1-\k})}{|r|^3}\leq C \ln N \,. 
\ee
Inserting this bound in \eqref{eq:diff-eta-3} and recalling the condition $3k-2+3\e +\d_2<0$, \eqref{BCS-bounds} and  \eqref{eq:prop2.4-a} we obtain \eqref{eq:V-eta-rv}. 

With \eqref{eq:V-eta-rv} and \eqref{eq:Nplus2} we bound
\[  \begin{split}
 \Big| N^{\k-1} \sum_\a \l_\a \sum_{\substack{r\in P_{\mathrm{H}},\\(v,0)\in \bbG^{\rm c}_{r}}}\widehat V\Big(\frac{r}{N^{1-\k}}\Big)(\eta_{r}-\eta_{r+v}) n^\a_v\Big|   &\;\leq CN^{\k+1}N^{-\eps}\\
\Big| N^{\k-2} \sum_\a \l_\a  \sum_{\substack{r\in P_{\mathrm{H}},\\(v,v')\in \bbG_r^{\rm q}}} \widehat V\Big(\frac{r}{N^{1-\k}}\Big)(\eta_{r+v}-\eta_{r})n^\a_{v}n^\a_{v'} \Big|&\;\leq CN^{\k+1}N^{-\eps}
\end{split}\]
and hence obtain 
\be \label{eq:I-a-2}
	\begin{split}
	& \sum_\a \l_\a  \frac{\langle \cD_\a g^0_\a,\cV_{2}^{(a)}\cD_\a g^0_\a\rangle}{\| \cD_\a g^0_\a\|^2} \\
	&  \; \leq \frac 1 N \sum_{r\in P_{\mathrm{H}}}\widehat{V}_N(r)\eta_{r}  \Big[N_0^2 +  4N_0\sum_\a \l_\a \hskip -0.2cm\sum_{\substack{v \in P_\A:\\ r+v \in P_{\mathrm{H}}}} \hskip -0.2cm n^\a_v  +2 \sum_\a \l_\a \hskip -0.2cm \sum_{\substack{v,v' \in P_\A:\\ v\neq \pm v',\\ r+v+v'  \in P_{\mathrm{H}} \\}}\hskip -0.2cm n^\a_{v}n^\a_{v'}\Big] \\
	& \ + C \max N^{\k+1} \{N^{-\eps}, N^{6\k-3+2\e +\d_1/3}, N^{12\k-6+6\e}\}
	\end{split}
\ee
for all $\k$ such that $3\k-2 +3\eps <0$. To conclude we notice that if $r\in P_{\mathrm{H}}$ and $v,v'\in P_\A^0$ then $r+v+v'\notin P_{\mathrm{H}}$ implies $|r|\leq CN^{1-\k-\eps}(1+N^{\k-2/3+\d_2/3})$ which allow us to remove the corresponding restrictions in the sums in the second line of \eqref{eq:I-a-2} up to an error of the order $N^{\k+1-\eps}.$
Moreover, the bound
\[
\sum_\alpha\l_\alpha\sum_{v \in P_\A} n^\a_v (n^\a_v +n^\a_{-v}) \leq C N^{1+\d_1/3}
\]
coming from Lemma \ref{lm:cN} guarantees that we can remove the restrictions $v\neq\pm v'$ in the third term on the second line of \eqref{eq:I-a-2} up to an error which satisfies the bound \eqref{eq:calEN}. With \eqref{eq:Nplus} and \eqref{eq:Nplus2} we obtain the second term on the right-hand side of \eqref{eq:corr-VN}.\\


\subsubsection{Contribution of $\cV_2^{(b)}$ on the correlated state}

We analyse the contribution of the term $\cV_2^{(b)}$ on the third line of \eqref{eq:spitV}. We first notice that being $p \in P_\A^0$ and $p+s \in P_{\mathrm{H}}$, then $s \neq 0$. 
We find it convenient to rewrite
\be\label{eq:V2b-split}
	\begin{split}
		\cV^{(b)}_{2}&=\frac1{2N^{1-\k}}\hspace{-0.3cm}\sum_{\substack{s\in \L^*_+\!,\,p\in P^0_\A:\\p+s\in P_{\mathrm{H}}}}\hspace{-0.3cm}\Big(\widehat{V}(s/N^{1-\k})+\widehat V(0)\Big)a^*_{p+s}a_{p+s}a^*_{p}a_p\\
		&\phantom{{}={}}+\frac1{2N^{1-\k}}\hspace{-0.5cm}\sum_{\substack{s\in \L^*_+\!,\,p,q\in P_\A^0:\\p\neq q,q+s,p+s\in P_{\mathrm{H}}}}\hspace{-0.5cm}\Big(\widehat{V}(s/N^{1-\k})+\widehat V((p-q)/N^{1-\k})\Big)a_{q+s}a^*_qa^*_{p+s}a_p\\
		&\eqqcolon \cV_2^{(b),1}+\cV_2^{(b),2}.
	\end{split}
\ee
We first consider the contribution coming from $\cV_2^{(b),1}$. Recalling \eqref{eq:def_D} we get
\[
	\begin{split}
		& \langle 	\cD_\a g^0_\a, \hskip -0.05cm \cV^{(b),1}_{2}\cD_\a g^0_\a\rangle\! =\sum_{n=1}^{n_{\mathrm{c}}}\frac{1}{(n!)^2(2N)^{2n}}\hspace{-0.3cm} \sum_{\substack{{r_1,\dots,r_{n}\in P_{\mathrm{H}}}\\ \tilde r_1,\dots,\tilde r_{n}\in P_{\mathrm{H}}}}\sum_{\substack{{(v_1,v'_1)\in \bbG_{r_1}}\\{(\tilde v_1,\tilde v'_1)\in \bbG_{\tilde r_1}}}}\dots \sum_{\substack{{(v_{n},v'_{n})\in \bbG_{r_{n}}}\\{(\tilde v_{n},\tilde v'_{n})\in \bbG_{\tilde r_{n}}}}}\hspace{-0.4cm}\theta\big(\{r_j,v_j,v'_j\}_{j=1}^{n}\big)  \\[-0.35cm]
         &\quad \times \theta\big(\{\tilde r_j,\tilde v_j,\tilde v'_j\}_{j=1}^n\big) \prod_{i=1}^{n}\eta_{r_i}\eta_{\tilde r_i} \langle g^0_\a, A^*_{\tilde{r}_{n},\tilde{v}_{n},\tilde{v}'_{n}}\dots A^*_{\tilde{r}_1,\tilde{v}_1,\tilde{v}'_1} \cV^{(b),1}_2A_{r_1,v_1,v'_1}\dots A_{r_{n},v_{n},v'_{n}}g^0_\a\rangle.
	\end{split}
\]
Using canonical commutation relations we find, for any $p+s\in P_{\mathrm{H}}$,
\[
		a^*_{p+s}a_{p+s}A_{r_1,v_1,v'_1}\dots A_{r_n,v_n,v'_n}g^0_\a=\sum_{i=1}^n(\d_{p+s,-r_i}+\d_{p+s,r_i+v_i+v'_i})A_{r_1,v_1,v'_1}\dots A_{r_n,v_n,v'_n}g^0_\a
\]
and, for any $p\in P_\A^0$,
\[
	\begin{split}
		&a^*_pa_pA_{r_1,v_1,v'_1}\dots A_{r_n,v_n,v'_n}g^0_\a=\\
		&\d_{p,0}(N_0-k-2(n-h-k))A_{r_1,v_1,v'_1}\dots A_{r_n,v_n,v'_n}g^0_\a +(1-\d_{p,0})n^\a_pA_{r_1,v_1,v'_1}\dots A_{r_n,v_n,v'_n}g^0_\a \\
		&-(1-\d_{p,0})\sum_{i=1}^n(\d_{p,v_i}+\d_{p,v'_i})A_{r_1,v_1,v'_1}\dots A_{r_n,v_n,v'_n}g^0_\a
	\end{split}
\]
where on the first line we used \eqref{eq:a0m-t} with $t=1$ and $m=k+2(n-h-k)$ and we recall that  $k,n-h-k$ denote respectively the number of cubic and pair excitation among $A_{r_1,v_1,v'_1}\dots A_{r_n,v_n,v'_n}$. Thus, pairing the momenta appearing in $A^*_{\tilde{r}_{n},\tilde{v}_{n},\tilde{v}'_{n}}\dots A^*_{\tilde{r}_1,\tilde{v}_1,\tilde{v}'_1}$ with the momenta appearing in $A_{r_1,v_1,v'_1}\dots A_{r_{n},v_{n},v'_{n}}$ as in \eqref{eq:norm-key}, we obtain
\[
	\begin{split}
		 &\langle 	\cD_\a g^0_\a, \cV^{(b),1}_{2}\cD_\a g^0_\a\rangle=\frac1{2N^{1-\k}}\hspace{-0.3cm}\sum_{\substack{s\in \L^*_+,\, p\in P^0_\A:\\p+s\in P_{\mathrm{H}}}}\hspace{-0.3cm}\Big(\widehat V(s/N^{1-\k})+\widehat V(0)\Big)\sum_{n= 1}^{n_{\mathrm{c}}}\frac{1}{N^{2n}}\sum_{h=0}^n\sum_{k=0}^{n-h}\\ 
		&\times \frac{2^{-2h-k-(n-h-k)} }{h!k!(n-h-k)!}  \frac{N_0!}{(N_0 - k-2(n-k-h))!} \sum_{\substack{\scriptscriptstyle{r_1\in P_{\mathrm{H}}}\\\scriptscriptstyle{(v_1,v_1')\in \bbG^{\rm q}_{r_1}}}}\dots\hspace{-0.2cm}\sum_{\substack{\scriptscriptstyle{r_h\in P_{\mathrm{H}}}\\\scriptscriptstyle{(v_h,v_h')\in \bbG^{\rm q}_{r_h}}}}\; \\
         &\times\hspace{-0.3cm}\sum_{\substack{\scriptscriptstyle{r_{h+1}\in P_{\mathrm{H}}}\\\scriptscriptstyle{(v_{h+1},0)\in \bbG^{\rm c}_{r_{h+1}}}}}\hspace{-0.2cm}\dots\hspace{-0.2cm}\sum_{\substack{\scriptscriptstyle{r_{h+k}\in P_{\mathrm{H}}}\\\scriptscriptstyle{(v_{h+k},0)\in \bbG^{\rm c}_{r_{h+k}}}}}\sum_{\scriptscriptstyle{r_{h+k+1},\dots,r_n\in P_{\mathrm{H}}}}\hspace{-0.3cm}\theta(\{r_s,v_s,v'_s\}_{s=1}^n)\prod_{i=1}^h(\eta_{r_i}+\eta_{r_i+v_i+v'_i})^2\,n^\a_{v_i}n^\a_{v'_i}\\
        & \times\prod_{j=h+1}^{h+k}(\eta_{r_j}+\eta_{r_j+v_j})^2\,n^\a_{v_j}\prod_{m=h+k+1}^n\eta_{r_m}^2\Big\{\Big(\d_{p,0}(N_0+k+2(n-h-k))+(1-\d_{p,0})n^\a_p\Big)\\&
        \times\Big(h\sum_{p_1}\d_{p+s,p_1}+k\sum_{p_{h+1}}\d_{p+s,p_{h+1}}+2(n-h-k)\d_{p+s,r_n}\Big)\\
        & -k\d_{p,v_{h+1}}\Big(h\sum_{p_1}\d_{p+s,p_1}+2(n-h-k)\d_{p+s,r_n}\Big)
        \hskip -0.1cm -2h\d_{p,v_1}\Big(k\sum_{p_{h+1}}\d_{p+s,p_{h+1}}\hskip -0.1cm+2(n-h-k)\d_{p+s,r_n}\Big)\\
        &-2h\sum_{p_1}\d_{p+s,p_1}\Big(\d_{p,v_1}+(h-1)\d_{p,v_2}\Big)-k\sum_{p_{h+1}}\d_{p+s,p_{h+1}}\Big(\d_{p,v_{h+1}}-(k-1)\d_{p,v_{h+2}}\Big)\Big\}\,.
	\end{split}•
\]
Then we take the absolute value and rescale appropriately the indices in the various terms, bounding the factorial involving $N_0$ when necessary. Recalling the bounds on $\|\eta_H\|_2$ and $\|\eta_H\|_\infty$ in \eqref{eq:etaH}
\be\label{eq:V2b1}
	\begin{split}
		& \frac{|\langle 	\cD_\a g^0_\a, \cV^{(b),1}_{2}\cD_\a g^0_\a\rangle| }{\|\cD_\a g^0_\a\|^2}\\
		&\leq C  N^\k\|\eta_H\|_2^2
		  \Big[N^{-3}(N_0+N_\a)^3+N^{-5}\|\eta_H\|_2
		^2(N_0+N_\a)^4\Big]\\
		&\leq CN^{1+\k}\Big[N^{3\k-5+\eps}(N_0+N_\a)^3+N^{6\k-8+2\eps}(N_0+N_\a)^4\Big]\,.
	\end{split}
\ee

Next, we consider $\cV_2^{(b),2}.$ By definition of $\cD_\a$ we have
\[
	\begin{split}
		  & \langle 	\cD_\a g^0_\a, \cV^{(b),2}_{2}\cD_\a g^0_\a\rangle\\
		  & =\sum_{\substack{s\in \L^*,\,p,q\in P_\A^0:\\p\neq q,q+s,p+s\in P_{\mathrm{H}}}}\hspace{-0.3cm}\Big(\widehat{V}(s/N^{1-\k})+\widehat V((p-q)/N^{1-\k})\Big)\\
		  &\times\sum_{n=1}^{n_{\mathrm{c}}}\frac{1}{(n!)^2(2N)^{2n}}\hspace{-0.3cm} \sum_{\substack{{r_1,\dots,r_{n}\in P_{\mathrm{H}}}\\ \tilde r_1,\dots,\tilde r_{n}\in P_{\mathrm{H}}}}\prod_{i=1}^{n}\eta_{r_i}\eta_{\tilde r_i}\hspace{-0.3cm}\sum_{\substack{{(v_1,v'_1)\in \bbG_{r_1}}\\{(\tilde v_1,\tilde v'_1)\in \bbG_{\tilde r_1}}}}\hspace{-0.2cm}\dots\hspace{-0.2cm} \sum_{\substack{{(v_{n},v'_{n})\in \bbG_{r_{n}}}\\{(\tilde v_{n},\tilde v'_{n})\in \bbG_{\tilde r_{n}}}}}\hspace{-0.4cm}\theta\big(\{r_j,v_j,v'_j\}_{j=1}^{n}\big)\theta\big(\{\tilde r_j,\tilde v_j,\tilde v'_j\}_{j=1}^n\big)\\
         &\times \langle g^0_\a, A^*_{\tilde{r}_{n},\tilde{v}_{n},\tilde{v}'_{n}}\dots A^*_{\tilde{r}_1,\tilde{v}_1,\tilde{v}'_1}a_{q+s}a^*_qa^*_{p+s}a_pA_{r_1,v_1,v'_1}\dots A_{r_{n},v_{n},v'_{n}}g^0_\a\rangle.
	\end{split}
\]
We start pairing 
momenta in $P_{\mathrm{H}}$. Since $g^0_\a$ does not contain  any momentum 
in $P_{\mathrm{H}}$ and $p\neq q$ there exist $i,j=1,\dots,n$ such that $p+s=\tl p_i$ and $q+s=p_j$ with $\tl p_i\in \{-\tl r_i,\tl r_i+\tl v_i+\tl v'_i\},$ $p_j\in \{-r_j,r_j+v_j+v'_j\}.$  For simplicity we assume $i=j=1.$ We can then  distinguish two cases depending on the operators with whom $a_{-\tl p_1+\tl v_1+\tl v'_1}$ and $a^*_{-p_1+v_1+v'_1}$ are contracted. Namely, 
\begin{enumerate}
\item $a_{-\tl p_1+\tl v_1+\tl v'_1}$ and $a^*_{-p_1+v_1+v'_1}$ are contracted among themselves. In this case the restrictions implemented by $\theta\big(\{r_j,v_j,v'_j\}_{j=1}^{n}\big)\theta\big(\{\tilde r_j,\tilde v_j,\tilde v'_j\}_{j=1}^n\big)$ imply that the operators in $A_{r_2,v_2,v'_2}\dots A_{r_n,v_n,v'_n}$ are contracted with the operators in $A^*_{\tl r_2,\tl v_2,\tl v'_2}\dots A^*_{\tl r_n,\tl v_n,\tl v'_n}$ as in the computation of the norm while $a_p$ is contracted either with $a^*_{\tl v_1}$ or with $a^*_{\tl v'_1}$ and $a^*_q$ is contracted with one of the operators $a_{v_1},a_{v'_1}.$ The remaining creation operator between $a^*_{\tl v_1}, a^*_{\tl v'_1}$ is then necessarily contracted with the remaining annihilation operator between $a_{v_1}$ and $a_{v'_1}.$
\item there exists $k,h=2,\dots,n$ such that $a_{-\tl p_1+\tl v_1+\tl v'_1}$ is contracted with $a^*_{p_k}$ and $a^*_{-p_1+v_1+v'_1}$ is contracted with $a_{\tl P_{\mathrm{H}}}. $ For simplicity let $h=k=2.$ In this case necessarily: i) $a^*_{-p_2+v_2+v'_2}$ is contracted with $a_{-\tl p_2+\tl v_2 + \tl v'_2}$; ii) $a^*_q$ is contracted with $a_{v_2}$ or $a_{ v'_2}$; iii) 
$a_p$ is contracted either with $a^*_{\tl v_2}$ or with $a^*_{\tl v'_2}.$ Then, $a_{v_1}, a_{v'_1}$ and the remaining operator among $a_{v_2},a_{v'_2}$ can be contracted with $a^*_{\tl v_1},a^*_{\tl v'_1}$ and the remaining operator among $a^*_{\tl v_2}$ and $a^*_{\tl v'_2}$ in all possible ways.
\end{enumerate}
We split accordingly
\be\label{eq:V2b2-split}
	\langle 	\cD_\a g^0_\a, \cV^{(b),2}_{2}\cD_\a g^0_\a\rangle=I_2^{(b),2}+II_2^{(b),2}.
\ee
We have
\[
	\begin{split}
		&I_2^{(b),2}=\frac1{2N^{1-\k}}\sum_{\substack{s\in \L^*,\,p,q\in P_\A^0:\\p\neq q,q+s,p+s\in P_{\mathrm{H}}}}\hspace{-0.3cm}\Big(\widehat{V}(s/N^{1-\k})+\widehat V((p-q)/N^{1-\k})\Big)\\
		  &\times\hspace{-0.1cm}\sum_{n= 1}^{n_{\mathrm{c}}}\frac{1}{N^{2n}}\sum_{h=0}^{n-1}\sum_{k=0}^{n-1-h}\hspace{-0.2cm}\frac{1}{h!k!(n-1-h-k)!}  \frac{N_0!\,2^{-2h-k-(n-1-h-k)}}{(N_0 - k-2(n-1-k-h))!}\hspace{-0.3cm}\sum_{\substack{r_1\in P_{\mathrm{H}}\\(v_1,v'_1)\in \bbG_{r_1}}}\sum_{\substack{\tl r_1\in P_{\mathrm{H}}\\(\tl v_1,\tl v'_1)\in \bbG_{\tl r_1}}} \,  \\
         &\times\hspace{-0.3cm}\sum_{\substack{{r_2\in P_{\mathrm{H}}}\\{(v_2,v_2')\in \bbG^{\rm q}_{r_1}}}}\dots\hspace{-0.2cm}\sum_{\substack{{r_h\in P_{\mathrm{H}}}\\{(v_h,v_h')\in \bbG^{\rm q}_{r_h}}}}\;\sum_{\substack{{r_{h+1}\in P_{\mathrm{H}}}\\{(v_{h+1},0)\in \bbG^{\rm c}_{r_{h+1}}}}}\hspace{-0.2cm}\dots\hspace{-0.2cm}\sum_{\substack{{r_{h+k}\in P_{\mathrm{H}}}\\{(v_{h+k},0)\in \bbG^{\rm c}_{r_{h+k}}}}}\sum_{{r_{h+k+1},\dots,r_n\in P_{\mathrm{H}}}}\hspace{-0.3cm}\theta(\{r_s,v_s,v'_s\}_{s=1}^n)\\
         &\times\theta(\{\tl r^\sharp_s,\tl v^\sharp_s,\tl v'^\sharp_s\}_{s=1}^n)\prod_{i=2}^h(\eta_{r_i}+\eta_{r_i+v_i+v'_i})^2\,n^\a_{v_i}n^\a_{v'_i}\prod_{j=h+1}^{h+k}(\eta_{r_j}+\eta_{r_j+v_j})^2\,n^\a_{v_j}\prod_{m=h+k+1}^n\eta_{r_m}^2\\
         &\times\frac14\eta_{r_1}\eta_{\tl r_1}\sum_{p_1,\tl p_1}\d_{p+s,\tl p_1}\d_{q+s,p_1}\d_{-p_1+v_1+v'_1,-\tl p_1+\tl v_1+\tl v'_1}\\
         &\times\sum_{w_1,\tl w_1}\d_{p,\tl w_1}\d_{q,w_1}\bigg\{n^\a_pn^\a_qn^\a_{\tl v_1+\tl v'_1-p}+(N_0-k-2(n-1-h-k))\Big[n^\a_pn^\a_q\d_{\tl v_1+\tl v'_1-p,0}\\
        &+\big(n^\a_q\d_{p,0}+n^\a_p\d_{q,0}\big)\big(n^\a_{\tl v_1+\tl v'_1-p}+(N_0-k-2(n-1-h-k)-1)\d_{\tl v_1+\tl v'_1-p,0}\big)\Big]\bigg\}
    	\end{split}•
\]
where in the last two lines we used \eqref{eq:a0m-t} with $m= k+2(n-1-h-k)$ and $t=1,2$ respectively, and we are adopting the convention  $\tl r_1^\sharp=\tl r_1, \tl v_1^\sharp=\tl v_1,\tl v'^\sharp_1=\tl v'_1$ and $\tl r_\ell^\sharp=r_\ell, \tl v_\ell^\sharp=v_\ell, \tl v'^\sharp_\ell=v'_\ell$ for $\ell=2,\dots,n$. In the above formula the curly bracket takes into account the possible number of zero momenta among $p,q,\tl v_1+\tl v'_1-p.$ Note that at most one between $p$ and $q$ can be zero being $p\neq q.$ We rewrite
\be\label{eq:I2b-split}
	I_2^{(b),2}=A_0+A_1+A_2
\ee
where $A_i,$ $i=0,1,2$ contains the terms where there are exactly $i$ zero momenta in the set $\{p,q,\tl v_1+\tl v'_1-p\}.$ We have
\[
	\begin{split}
		&A_0=\frac1{2N^{1-\k}}\sum_{\substack{s\in \L^*,\,p,q\in P_\A^0:\\p\neq q,q+s,p+s\in P_{\mathrm{H}}}}\hspace{-0.3cm}\Big(\widehat{V}(s/N^{1-\k})+\widehat V((p-q)/N^{1-\k})\Big)\\
		 &\times\hspace{-0.1cm}\sum_{n= 1}^{n_{\mathrm{c}}}\frac{1}{N^{2n}}\sum_{h=0}^{n-1}\sum_{k=0}^{n-1-h}\hspace{-0.2cm}\frac{1}{h!k!(n-1-h-k)!}  \frac{N_0!\,2^{-2h-k-(n-1-h-k)}}{(N_0 - k-2(n-1-k-h))!}\hspace{-0.3cm}\sum_{\substack{r_1\in P_{\mathrm{H}}\\(v_1,v'_1)\in \bbG_{r_1}}}\sum_{\substack{\tl r_1\in P_{\mathrm{H}}\\(\tl v_1,\tl v'_1)\in \bbG_{\tl r_1}}} \,  \\
        &\times\hspace{-0.3cm}\sum_{\substack{{r_2\in P_{\mathrm{H}}}\\{(v_2,v_2')\in \bbG^{\rm q}_{r_1}}}}\dots\hspace{-0.2cm}\sum_{\substack{{r_h\in P_{\mathrm{H}}}\\{(v_h,v_h')\in \bbG^{\rm q}_{r_h}}}}\;\sum_{\substack{{r_{h+1}\in P_{\mathrm{H}}}\\{(v_{h+1},0)\in \bbG^{\rm c}_{r_{h+1}}}}}\hspace{-0.2cm}\dots\hspace{-0.2cm}\sum_{\substack{{r_{h+k}\in P_{\mathrm{H}}}\\{(v_{h+k},0)\in \bbG^{\rm c}_{r_{h+k}}}}}\sum_{{r_{h+k+1},\dots,r_n\in P_{\mathrm{H}}}}\hspace{-0.3cm}\theta(\{r_s,v_s,v'_s\}_{s=1}^n)\\
        &\times\theta(\{\tl r^\sharp_s,\tl v^\sharp_s,\tl v'^\sharp_s\}_{s=1}^n)\prod_{i=2}^h(\eta_{r_i}+\eta_{r_i+v_i+v'_i})^2\,n^\a_{v_i}n^\a_{v'_i}\prod_{j=h+1}^{h+k}(\eta_{r_j}+\eta_{r_j+v_j})^2\,n^\a_{v_j}\prod_{m=h+k+1}^n\eta_{r_m}^2\\
        &\times\frac14 \eta_{r_1}\eta_{\tl r_1}\sum_{p_1,\tl p_1}\d_{p+s,\tl p_1}\d_{q+s,p_1}\d_{-p_1+v_1+v'_1,-\tl p_1+\tl v_1+\tl v'_1}\sum_{w_1,\tl w_1}\d_{p,\tl w_1}\d_{q,w_1}n^\a_pn^\a_qn^\a_{\tl v_1+\tl v'_1-p}.
    	\end{split}•
\]
We take the absolute value, then we get an upper bound if we neglect $\theta(\{\tl r^\sharp_s,\tl v^\sharp_s,\tl v'^\sharp_s\}_{s=1}^n)$ and replace $\theta(\{r_s,v_s,v'_s\}_{s=1}^n)$ with $\theta(\{r_s,v_s,v'_s\}_{s=1}^{n-1}).$ Proceeding in this way the factor involving the index $1$ can be moved out of the sum. Rescaling then $n\to n-1,$ and recalling \eqref{eq:norm-key}, we conclude
\be\label{eq:A0}
	\begin{split}
 & \frac{|A_0|}{\|\cD_\a g^0_\a\|^2} \\
 	& \leq \frac C{N^{3-\k}}\sum_{\substack{s\in \L^*,\,p,q,v\in P_\A:\\q+s,p+s,-s+v\in P_{\mathrm{H}}}}\hspace{-0.3cm}\Big(|\widehat{V}(s/N^{1-\k})|+|\widehat V((p-q)/N^{1-\k})|\Big)\\
	&\quad \times(|\eta_{p+s}||\eta_{q+s}|+|\eta_{-s+v}||\eta_{q+s}|+|\eta_{p+s}||\eta_{-s+v}|+|\eta_{-s+v}|^2)n^\a_pn^\a_qn^\a_v\\[0.2cm]
	&\leq CN^{-3}\|\eta_H\|_\infty \sup_{q\in \L^*}\sum_{s\in \L^*_+} N^\k|\widehat V((q-s)/N^{1-\k})||\eta_s| N_\a^3\\
	& \leq CN^{4\k-4+2\eps} N_\a^3
	\end{split}
\ee
where in the last step we used \eqref{BCS-bounds} and \eqref{eq:etaH}. \\
Similarly, 
\[
	\begin{split}
		&A_1=\frac1{2N^{1-\k}}\sum_{\substack{s\in \L^*,\,p,q\in P_\A^0:\\p\neq q,q+s,p+s\in P_{\mathrm{H}}}}\hspace{-0.3cm}\Big(\widehat{V}(s/N^{1-\k})+\widehat V((p-q)/N^{1-\k})\Big)\\
		  &\times\hspace{-0.1cm}\sum_{n= 1}^{n_{\mathrm{c}}}\frac{1}{N^{2n}}\sum_{h=0}^{n-1}\sum_{k=0}^{n-1-h}\hspace{-0.2cm}\frac{1}{h!k!(n-1-h-k)!}  \frac{N_0!\,2^{-2h-k-(n-1-h-k)}}{(N_0 - k-2(n-1-k-h)-1)!}\hspace{-0.3cm}\sum_{\substack{r_1\in P_{\mathrm{H}}\\(v_1,v'_1)\in \bbG_{r_1}}}\sum_{\substack{\tl r_1\in P_{\mathrm{H}}\\(\tl v_1,\tl v'_1)\in \bbG_{\tl r_1}}} \,  \\
         &\times\hspace{-0.3cm}\sum_{\substack{{r_2\in P_{\mathrm{H}}}\\{(v_2,v_2')\in \bbG^{\rm q}_{r_1}}}}\dots\hspace{-0.2cm}\sum_{\substack{{r_h\in P_{\mathrm{H}}}\\{(v_h,v_h')\in \bbG^{\rm q}_{r_h}}}}\;\sum_{\substack{{r_{h+1}\in P_{\mathrm{H}}}\\{(v_{h+1},0)\in \bbG^{\rm c}_{r_{h+1}}}}}\hspace{-0.2cm}\dots\hspace{-0.2cm}\sum_{\substack{{r_{h+k}\in P_{\mathrm{H}}}\\{(v_{h+k},0)\in \bbG^{\rm c}_{r_{h+k}}}}}\sum_{{r_{h+k+1},\dots,r_n\in P_{\mathrm{H}}}}\hspace{-0.3cm}\theta(\{r_s,v_s,v'_s\}_{s=1}^n)\\
         &\times\theta(\{\tl r^\sharp_s,\tl v^\sharp_s,\tl v'^\sharp_s\}_{s=1}^n)\prod_{i=2}^h(\eta_{r_i}+\eta_{r_i+v_i+v'_i})^2\,n^\a_{v_i}n^\a_{v'_i}\prod_{j=h+1}^{h+k}(\eta_{r_j}+\eta_{r_j+v_j})^2\,n^\a_{v_j}\prod_{m=h+k+1}^n\eta_{r_m}^2\\
         &\times\frac14\eta_{r_1}\eta_{\tl r_1}\sum_{p_1,\tl p_1}\d_{p+s,\tl p_1}\d_{q+s,p_1}\d_{-p_1+v_1+v'_1,-\tl p_1+\tl v_1+\tl v'_1}\sum_{w_1,\tl w_1}\d_{p,\tl w_1}\d_{q,w_1}\\
         &\times\Big[n^\a_pn^\a_q\d_{\tl v_1+\tl v'_1-p,0}+\big(n^\a_q\d_{p,0}+n^\a_p\d_{q,0}\big)n^\a_{\tl v_1+\tl v'_1-p}\Big].
	\end{split}•
\]
To bound $|A_1|$ we neglect $\theta(\{\tl r^\sharp_s,\tl v^\sharp_s,\tl v'^\sharp_s\}_{s=1}^n)$, replace $\theta(\{r_s,v_s,v'_s\}_{s=1}^n)$ by $\theta(\{r_s,v_s,v'_s\}_{s=1}^{n-1})$, move out $N_0$ and the factor involving the index $1$. 
Noting that 
$\frac{(N_0-1)!}{(N_0-k-2(n-1-k-h)-1)!}\leq \frac{N_0!}{(N_0-k-2(n-1-k-h))!}$ and rescaling $n\to n-1$ we recover the norm $\|\cD_\a g^0_\a\|^2.$ Hence,
\be\label{eq:A1}
		\frac{|A_1|}{\|\cD_\a g^0_\a\|^2}\leq  CN_0 N^{-3}\|\eta_H\|_\infty \sup_{q\in \L^*}\sum_{s\in \L^*_+} N^\k|\widehat V((q-s)/N^{1-\k})||\eta_s| N_\a^2\leq CN^{4\k-4+2\eps}N_0N_\a^2.
\ee
Finally, we have
\[
	\begin{split}
		&A_2=\frac1{2N^{1-\k}}\sum_{\substack{s\in \L^*,\,p,q\in P_\A^0:\\p\neq q,q+s,p+s\in P_{\mathrm{H}}}}\hspace{-0.3cm}\Big(\widehat{V}(s/N^{1-\k})+\widehat V((p-q)/N^{1-\k})\Big)\\
		  &\times\hspace{-0.1cm}\sum_{n= 1}^{n_{\mathrm{c}}}\frac{1}{N^{2n}}\sum_{h=0}^{n-1}\sum_{k=0}^{n-1-h}\hspace{-0.2cm}\frac{1}{h!k!(n-1-h-k)!}  \frac{N_0!\,2^{-2h-k-(n-1-h-k)}}{(N_0 - k-2(n-1-k-h)-2)!}\hspace{-0.3cm}\sum_{\substack{r_1\in P_{\mathrm{H}}\\(v_1,v'_1)\in \bbG_{r_1}}}\sum_{\substack{\tl r_1\in P_{\mathrm{H}}\\(\tl v_1,\tl v'_1)\in \bbG_{\tl r_1}}} \,  \\
         &\times\hspace{-0.3cm}\sum_{\substack{{r_2\in P_{\mathrm{H}}}\\{(v_2,v_2')\in \bbG^{\rm q}_{r_1}}}}\dots\hspace{-0.2cm}\sum_{\substack{{r_h\in P_{\mathrm{H}}}\\{(v_h,v_h')\in \bbG^{\rm q}_{r_h}}}}\;\sum_{\substack{{r_{h+1}\in P_{\mathrm{H}}}\\{(v_{h+1},0)\in \bbG^{\rm c}_{r_{h+1}}}}}\hspace{-0.2cm}\dots\hspace{-0.2cm}\sum_{\substack{{r_{h+k}\in P_{\mathrm{H}}}\\{(v_{h+k},0)\in \bbG^{\rm c}_{r_{h+k}}}}}\sum_{{r_{h+k+1},\dots,r_n\in P_{\mathrm{H}}}}\hspace{-0.3cm}\theta(\{r_s,v_s,v'_s\}_{s=1}^n)\\
         &\times\theta(\{\tl r^\sharp_s,\tl v^\sharp_s,\tl v'^\sharp_s\}_{s=1}^n)\prod_{i=2}^h(\eta_{r_i}+\eta_{r_i+v_i+v'_i})^2\,n^\a_{v_i}n^\a_{v'_i}\prod_{j=h+1}^{h+k}(\eta_{r_j}+\eta_{r_j+v_j})^2\,n^\a_{v_j}\prod_{m=h+k+1}^n\eta_{r_m}^2\\
         &\times \frac14\eta_{r_1}\eta_{\tl r_1}\sum_{p_1,\tl p_1}\d_{p+s,\tl p_1}\d_{q+s,p_1}\d_{-p_1+v_1+v'_1,-\tl p_1+\tl v_1+\tl v'_1}\sum_{w_1,\tl w_1}\d_{p,\tl w_1}\d_{q,w_1}\big(n^\a_q\d_{p,0}+n^\a_p\d_{q,0}\big)\d_{\tl v_1+\tl v'_1-p,0}.
   \end{split}•
\]
Again, we take 
the absolute value,
neglect $\theta(\{\tl r^\sharp_s,\tl v^\sharp_s,\tl v'^\sharp_s\}_{s=1}^n)$ and replace $\theta(\{r_s,v_s,v'_s\}_{s=1}^n)$ with $\theta(\{r_s,v_s,v'_s\}_{s=1}^{n-1})$. Moving out the factor involving the index $1$ and a factor $N_0(N_0-1)$ and bounding $\frac{(N_0-2)!}{(N_0-k-2(n-1-k-h)-2)!}\leq \frac{N_0!}{(N_0-k-2(n-1-k-h))!}$ we recover $\cD_\a g^0_\a$ after rescaling $n\to n-1.$ Proceeding similarly to \eqref{eq:A0} we obtain
\be\label{eq:A2}
	\frac{|A_2|}{\|\cD_\a\|^2}\leq CN_0^2 N^{-3} \|\eta_H\|_\infty  \sup_{q\in \L^*}\sum_{s\in \L^*_+} N^\k|\widehat V((q-s)/N^{1-\k})||\eta_s| N_\a\leq CN^{4\k-4+2\eps}N_0^2N_\a.
\ee
Recalling \eqref{eq:I2b-split}, the bounds \eqref{eq:A0}, \eqref{eq:A1} and \eqref{eq:A2}, provide
\be\label{eq:I2b-final}
		\frac{|I_2^{(b),2}|}{\|\cD_\a g^0_\a\|^2}\leq CN^{4\k-4+2\eps}(N_0+N_\a)^3.
\ee
To conclude this section it remains to show that also $II_2^{(b),2}$ gives a negligible contribution. We have
\[
	\begin{split}
		&II_2^{(b),2}=\frac1{2N^{1-\k}}\sum_{\substack{s\in \L^*,\,p,q\in P_\A^0:\\p\neq q,q+s,p+s\in P_{\mathrm{H}}}}\hspace{-0.3cm}\Big(\widehat{V}(s/N^{1-\k})+\widehat V((p-q)/N^{1-\k})\Big)\sum_{n= 2}^{n_{\mathrm{c}}}\frac{1}{N^{2n}}\sum_{h=0}^{n-2}\sum_{k=0}^{n-2-h}\\
		  &\times\frac{1}{h!k!(n-2-h-k)!}  \frac{N_0!\,2^{-2h-k-(n-2-h-k)}}{(N_0 - k-2(n-2-k-h))!}\hspace{-0.3cm}\sum_{\substack{r_1,\tl r_1\in P_{\mathrm{H}}\\(v_1,v'_1)\in \bbG_{r_1}\\(\tl v_1,\tl v'_1)\in \bbG_{\tl r_1}}}\sum_{\substack{r_2,\tl r_2\in P_{\mathrm{H}}\\( v_2, v'_2)\in \bbG_{ r_2}\\(\tl v_2,\tl v'_2)\in \bbG_{\tl r_2}}} \,\sum_{\substack{{r_3\in P_{\mathrm{H}}}\\{(v_3,v_3')\in \bbG^{\rm q}_{r_1}}}}  \\
         &\times\dots\hspace{-0.2cm}\sum_{\substack{{r_h\in P_{\mathrm{H}}}\\{(v_h,v_h')\in \bbG^{\rm q}_{r_h}}}}\sum_{\substack{{r_{h+1}\in P_{\mathrm{H}}}\\{(v_{h+1},0)\in \bbG^{\rm c}_{r_{h+1}}}}}\hspace{-0.2cm}\dots\hspace{-0.2cm}\sum_{\substack{{r_{h+k}\in P_{\mathrm{H}}}\\{(v_{h+k},0)\in \bbG^{\rm c}_{r_{h+k}}}}}\sum_{{r_{h+k+1},\dots,r_n\in P_{\mathrm{H}}}}\hspace{-0.3cm}\theta(\{r_s,v_s,v'_s\}_{s=1}^n)\theta(\{\tl r^\flat_s,\tl v^\flat_s,\tl v'^\flat_s\}_{s=1}^n)\\
         &\times\prod_{i=2}^h(\eta_{r_i}+\eta_{r_i+v_i+v'_i})^2\,n^\a_{v_i}n^\a_{v'_i}\prod_{j=h+1}^{h+k}(\eta_{r_j}+\eta_{r_j+v_j})^2\,n^\a_{v_j}\prod_{m=h+k+1}^n\eta_{r_m}^2\frac18\eta_{r_1}\eta_{\tl r_1}\eta_{r_2}\eta_{\tl r_2}\\
         &\times\sum_{\substack{p_1,\tl p_1\\p_2,\tl p_2}}\d_{p+s,\tl p_1}\d_{q+s,p_1}\d_{-p_1+v_1+v'_1,\tl p_2}\d_{-\tl p_1+\tl v_1+\tl v'_1,p_2}\d_{-p_2+v_2+v'_2,-\tl p_2+\tl v_2+\tl v'_2}\sum_{\substack{w_1,\tl w_1\\w_2,\tl w_2}}\d_{p,\tl w_2}\d_{q,w_2}\d_{w_1,\tl w_1}\\
         &\times(\d_{v_1+v'_1-w_1,\tl v_1+\tl v'_1-\tl w_1}+\d_{v_1+v_1'-w_1,\tl v_2+\tl v'_2-p})\bigg\{n^\a_pn^\a_{v_2}n^\a_{v_1}n^\a_{v'_1}n^\a_{v_2'}+(N_0-k-2(n-h-k)+4)\\
         &\times\Big[2n^\a_pn^\a_{v_2}n^\a_{v_1}n^\a_{v'_2}\d_{v'_1,0}+n^\a_pn^\a_qn^\a_{v_1}n^\a_{v'_1}\d_{v_2+v'_2-q,0}+(n^\a_p\d_{q,0}+n^\a_q\d_{p,0})n^\a_{v_1}n^\a_{v'_1}n^\a_{v_2+v_2'-q}\\
         &+(N_0-k-2(n-h-k)+3)\Big(n^\a_pn^\a_{v_2}n^\a_{v'_2}\d_{v_1,0}\d_{v'_1,0}+2n^\a_pn^\a_qn^\a_{v_1}\d_{v'_1,0}\d_{v_2+v'_2-q,0}\\
         &+(n^\a_p\d_{q,0}+n^\a_q\d_{p,0})(2n^\a_{v_1}n^\a_{v_2+v'_2-q}\d_{v'_1,0}+n^\a_{v_1}n^\a_{v'_1}\d_{v_2+v'_2-q,0})+(N_0-k-2(n-h-k)+2)\\
         &\times(n^\a_pn^\a_q\d_{v_2+v'_2-q}\d_{v_1,0}\d_{v'_1,0}+(n^\a_p\d_{q,0}+n^\a_q\d_{p,0})(n^\a_{v_2+v'_2-q}\d_{v_1,0}\d_{v'_1,0}+2n^\a_{v_1}\d_{v'_1,0}\d_{v_2+v'_2-q,0})\Big)\Big]\bigg\}
    	\end{split}
\]
where $\tl r_\ell^\flat=\tl r_\ell, \tl v_\ell^\flat=\tl v_\ell,\tl v'^\flat_\ell=\tl v'_\ell$ for $\ell=1,2$ and $\tl r_\ell^\flat=r_\ell, \tl v_\ell^\flat=v_\ell, \tl v'^\flat_\ell=v'_\ell$ for $\ell=3,\dots,n.$ Again the curly brackets embrace the various terms arising depending on how many and which among the momenta $p,q,v_1,v'_1,v_2+v'_2-q$ are actually zero. Analogously to what we did for $II_2^{(b),2}$ we split
\be\label{eq:II2b-split}
	II_2^{(b),2}=\sum_{i=0}^4 B_i
\ee
where $B_i$ contains terms where exactly $i$ momenta in the set 
$\{p,q,v_1,v'_1,v_2+v'_2-q\}$ are zero. We have
\[
	\begin{split}
		&B_0=\frac1{2N^{1-\k}}\sum_{\substack{s\in \L^*,\,p,q\in P_\A^0:\\p\neq q,q+s,p+s\in P_{\mathrm{H}}}}\hspace{-0.3cm}\Big(\widehat{V}(s/N^{1-\k})+\widehat V((p-q)/N^{1-\k})\Big)\sum_{n= 2}^{n_{\mathrm{c}}}\frac{1}{N^{2n}}\sum_{h=0}^{n-2}\sum_{k=0}^{n-2-h}\\
		  &\times\frac{1}{h!k!(n-2-h-k)!}  \frac{N_0!\,2^{-2h-k-(n-2-h-k)}}{(N_0 - k-2(n-2-k-h))!}\hspace{-0.3cm}\sum_{\substack{r_1,\tl r_1\in P_{\mathrm{H}}\\(v_1,v'_1)\in \bbG_{r_1}\\(\tl v_1,\tl v'_1)\in \bbG_{\tl r_1}}}\sum_{\substack{r_2,\tl r_2\in P_{\mathrm{H}}\\( v_2, v'_2)\in \bbG_{ r_2}\\(\tl v_2,\tl v'_2)\in \bbG_{\tl r_2}}} \,\sum_{\substack{{r_3\in P_{\mathrm{H}}}\\{(v_3,v_3')\in \bbG^{\rm q}_{r_3}}}}  \\
         &\times\dots\hspace{-0.2cm}\sum_{\substack{{r_h\in P_{\mathrm{H}}}\\{(v_h,v_h')\in \bbG^{\rm q}_{r_h}}}}\sum_{\substack{{r_{h+1}\in P_{\mathrm{H}}}\\{(v_{h+1},0)\in \bbG^{\rm c}_{r_{h+1}}}}}\hspace{-0.2cm}\dots\hspace{-0.2cm}\sum_{\substack{{r_{h+k}\in P_{\mathrm{H}}}\\{(v_{h+k},0)\in \bbG^{\rm c}_{r_{h+k}}}}}\hspace{-0.2cm}\sum_{{r_{h+k+1},\dots,r_n\in P_{\mathrm{H}}}}\hspace{-0.3cm}\theta(\{r_s,v_s,v'_s\}_{s=1}^n)\theta(\{\tl r^\flat_s,\tl v^\flat_s,\tl v'^\flat_s\}_{s=1}^n)\\
         &\times\prod_{i=1}^h(\eta_{r_i}+\eta_{r_i+v_i+v'_i})^2\,n^\a_{v_i}n^\a_{v'_i}\prod_{j=h+1}^{h+k}(\eta_{r_j}+\eta_{r_j+v_j})^2\,n^\a_{v_j}\prod_{m=h+k+1}^n\eta_{r_m}^2\frac18\eta_{r_1}\eta_{\tl r_1}\eta_{r_2}\eta_{\tl r_2}\\
         &\times\sum_{\substack{p_1,\tl p_1\\p_2,\tl p_2}}\d_{p+s,\tl p_1}\d_{q+s,p_1}\d_{-p_1+v_1+v'_1,\tl p_2}\d_{-\tl p_1+\tl v_1+\tl v'_1,p_2}\d_{-p_2+v_2+v'_2,-\tl p_2+\tl v_2+\tl v'_2}\sum_{\substack{w_1,\tl w_1\\w_2,\tl w_2}}\d_{p,\tl w_2}\d_{q,w_2}\d_{w_1,\tl w_1}\\
         &\times(\d_{v_1+v'_1-w_1,\tl v_1+\tl v'_1-\tl w_1}+\d_{v_1+v_1'-w_1,\tl v_2+\tl v'_2-p})n^\a_pn^\a_{v_2}n^\a_{v_1}n^\a_{v'_1}n^\a_{v_2'}.
	\end{split}•
\]
We now proceed similarly as above \eqref{eq:A1}:
we take the absolute value, then we neglect $\theta(\{\tl r^\flat_s,\tl v^\flat_s,\tl v'^\flat_s\}_{s=1}^n)$ and replace $\theta(\{r_s,v_s,v'_s\}_{s=1}^n)$ with $\theta(\{r_s,v_s,v'_s\}_{s=1}^{n-2}).$ Finally we move out the factors involving the indices $1$ and $2$ and rescale $n\to n-2$ to recover $\|\cD_\a g^0_\a\|^2.$ We find
\be\label{eq:B0}
	\begin{split}
		&\frac{|B_0|}{\|\cD_\a g^0_\a\|^2}\\
		&\leq \frac C{N^{5-\k}}\hspace{-0.3cm}\sum_{\substack{s\in \L^*,\,p,q\in P_\A^0:\\q+s,p+s\in P_{\mathrm{H}}}}\hspace{-0.4cm}\Big(|\widehat{V}(s/N^{1-\k})|+|\widehat V((p-q)/N^{1-\k})|\Big)\hspace{-0.2cm}\sum_{\substack{r_1,\tl r_1\in P_{\mathrm{H}}\\(v_1,v'_1)\in \bbG_{r_1}\\(\tl v_1,\tl v'_1)\in \bbG_{\tl r_1}}}\sum_{\substack{r_2,\tl r_2\in P_{\mathrm{H}}\\( v_2, v'_2)\in \bbG_{ r_2}\\(\tl v_2,\tl v'_2)\in \bbG_{\tl r_2}}}\hspace{-0.4cm} |\eta_{r_1}||\eta_{\tl r_1}||\eta_{r_2}||\eta_{\tl r_2}|\\
		&\times\sum_{\substack{p_1,\tl p_1\\p_2,\tl p_2}}\d_{p+s,\tl p_1}\d_{q+s,p_1}\d_{-p_1+v_1+v'_1,\tl p_2}\d_{-\tl p_1+\tl v_1+\tl v'_1,p_2}\d_{-p_2+v_2+v'_2,-\tl p_2+\tl v_2+\tl v'_2}\sum_{\substack{w_1,\tl w_1\\w_2,\tl w_2}}\d_{p,\tl w_2}\d_{q,w_2}\d_{w_1,\tl w_1}\\
         &\times(\d_{v_1+v'_1-w_1,\tl v_1+\tl v'_1-\tl w_1}+\d_{v_1+v_1'-w_1,\tl v_2+\tl v'_2-p})n^\a_pn^\a_{v_2}n^\a_{v_1}n^\a_{v'_1}n^\a_{v_2'}\\
         &\leq CN^{-5}\|\eta_H\|^3_\infty  \sup_{q\in \L^*}\sum_{s\in \L^*_+} N^\k|\widehat V((q-s)/N^{1-\k})||\eta_s| N_\a^5\leq CN^{10\k-10+6\eps}N^5_\a
	\end{split}
\ee
where in the last step we used \eqref{eq:etaH} and \eqref{BCS-bounds}. \\

Similarly, we have
\[
	\begin{split}
			B_1
			& =\frac1{2N^{1-\k}}\sum_{\substack{s\in \L^*,\,p,q\in P_\A^0:\\p\neq q,q+s,p+s\in P_{\mathrm{H}}}}\hspace{-0.3cm}\Big(\widehat{V}(s/N^{1-\k})+\widehat V((p-q)/N^{1-\k})\Big)\\
		  &\times \sum_{n= 2}^{n_{\mathrm{c}}}\frac{1}{N^{2n}}\sum_{h=0}^{n-2}\sum_{k=0}^{n-2-h}\frac{1}{h!k!(n-2-h-k)!}  \frac{N_0!\,2^{-2h-k-(n-2-h-k)}}{(N_0 - k-2(n-2-k-h)-1)!}\hspace{-0.3cm}  \\
         &\times\sum_{\substack{r_1,\tl r_1\in P_{\mathrm{H}}\\(v_1,v'_1)\in \bbG_{r_1}\\(\tl v_1,\tl v'_1)\in \bbG_{\tl r_1}}}\sum_{\substack{r_2,\tl r_2\in P_{\mathrm{H}}\\( v_2, v'_2)\in \bbG_{ r_2}\\(\tl v_2,\tl v'_2)\in \bbG_{\tl r_2}}} \,\sum_{\substack{{r_3\in P_{\mathrm{H}}}\\{(v_3,v_3')\in \bbG^{\rm q}_{r_1}}}} \dots\hspace{-0.2cm}\sum_{\substack{{r_h\in P_{\mathrm{H}}}\\{(v_h,v_h')\in \bbG^{\rm q}_{r_h}}}}\sum_{\substack{{r_{h+1}\in P_{\mathrm{H}}}\\{(v_{h+1},0)\in \bbG^{\rm c}_{r_{h+1}}}}}\hspace{-0.2cm}\dots\hspace{-0.2cm}\sum_{\substack{{r_{h+k}\in P_{\mathrm{H}}}\\{(v_{h+k},0)\in \bbG^{\rm c}_{r_{h+k}}}}}\\
         & \times \sum_{{r_{h+k+1},\dots,r_n\in P_{\mathrm{H}}}}\theta(\{r_s,v_s,v'_s\}_{s=1}^n)\theta(\{\tl r^\flat_s,\tl v^\flat_s,\tl v'^\flat_s\}_{s=1}^n) \\
         &\times\prod_{i=1}^h(\eta_{r_i}+\eta_{r_i+v_i+v'_i})^2\,n^\a_{v_i}n^\a_{v'_i}\prod_{j=h+1}^{h+k}(\eta_{r_j}+\eta_{r_j+v_j})^2\,n^\a_{v_j}\prod_{m=h+k+1}^n\eta_{r_m}^2\frac18 \eta_{r_1}\eta_{\tl r_1}\eta_{r_2}\eta_{\tl r_2}\\
         &\times\sum_{\substack{p_1,\tl p_1\\p_2,\tl p_2}}\d_{p+s,\tl p_1}\d_{q+s,p_1}\d_{-p_1+v_1+v'_1,\tl p_2}\d_{-\tl p_1+\tl v_1+\tl v'_1,p_2}\d_{-p_2+v_2+v'_2,-\tl p_2+\tl v_2+\tl v'_2 }\\ & \times \sum_{\substack{w_1,\tl w_1\\w_2,\tl w_2}}\d_{p,\tl w_2}\d_{q,w_2}\d_{w_1,\tl w_1}(\d_{v_1+v'_1-w_1,\tl v_1+\tl v'_1-\tl w_1}+\d_{v_1+v_1'-w_1,\tl v_2+\tl v'_2-p})\\
         &\times\Big[2n^\a_pn^\a_{v_2}n^\a_{v_1}n^\a_{v'_2}\d_{v'_1,0}+n^\a_pn^\a_qn^\a_{v_1}n^\a_{v'_1}\d_{v_2+v'_2-q,0}+(n^\a_p\d_{q,0}+n^\a_q\d_{p,0})n^\a_{v_1}n^\a_{v'_1}n^\a_{v_2+v_2'-q}\Big].
	\end{split}•
\]
We can bound $|B_1|$ neglecting $\theta(\{\tl r^\flat_s,\tl v^\flat_s,\tl v'^\flat_s\}_{s=1}^n)$ and replacing $\theta(\{r_s,v_s,v'_s\}_{s=1}^n)$ with $\theta(\{r_s,v_s,v'_s\}_{s=1}^{n-2}).$ We then move out the factor involving the indices $1$ and $2$ together with a factor $N_0$ and bound $\frac{(N_0-1)!}{(N_0 - k-2(n-2-k-h)-1)!}\leq \frac{N_0!}{(N_0 - k-2(n-2-k-h)!}$. Rescaling $n\to n-2$ we finally recover $\|\cD_\a g^0_\a\|^2$. We get
\be\label{eq:B1}
	\begin{split}
		& \frac{|B_1|}{\|\cD_\a g^0_\a\|^2}\\
		&\leq CN_0N^{-5}\|\eta_H\|^3_\infty  \sup_{q\in \L^*}\sum_{s\in \L^*_+} N^\k|\widehat V((q-s)/N^{1-\k})||\eta_s| N_\a^4\\
		& \leq CN^{10\k-10+6\eps}N_0N^4_\a.
	\end{split}•
\ee 
The estimate of $B_2$ can be obtained analogously. By definition
\[
	\begin{split}
			&B_2=\frac1{2N^{1-\k}}\sum_{\substack{s\in \L^*,\,p,q\in P_\A^0:\\p\neq q,q+s,p+s\in P_{\mathrm{H}}}}\hspace{-0.3cm}\Big(\widehat{V}(s/N^{1-\k})+\widehat V((p-q)/N^{1-\k})\Big)\sum_{n= 2}^{n_{\mathrm{c}}}\frac{1}{N^{2n}}\sum_{h=0}^{n-2}\sum_{k=0}^{n-2-h}\\
		  &\times\frac{1}{h!k!(n-2-h-k)!}  \frac{N_0!\,2^{-2h-k-(n-2-h-k)}}{(N_0 - k-2(n-2-k-h)-2)!}\hspace{-0.3cm}\sum_{\substack{r_1,\tl r_1\in P_{\mathrm{H}}\\(v_1,v'_1)\in \bbG_{r_1}\\(\tl v_1,\tl v'_1)\in \bbG_{\tl r_1}}}\sum_{\substack{r_2,\tl r_2\in P_{\mathrm{H}}\\( v_2, v'_2)\in \bbG_{ r_2}\\(\tl v_2,\tl v'_2)\in \bbG_{\tl r_2}}} \,\sum_{\substack{{r_3\in P_{\mathrm{H}}}\\{(v_3,v_3')\in \bbG^{\rm q}_{r_1}}}}  \\
         &\times\dots\hspace{-0.2cm}\sum_{\substack{{r_h\in P_{\mathrm{H}}}\\{(v_h,v_h')\in \bbG^{\rm q}_{r_h}}}}\sum_{\substack{{r_{h+1}\in P_{\mathrm{H}}}\\{(v_{h+1},0)\in \bbG^{\rm c}_{r_{h+1}}}}}\hspace{-0.2cm}\dots\hspace{-0.2cm}\sum_{\substack{{r_{h+k}\in P_{\mathrm{H}}}\\{(v_{h+k},0)\in \bbG^{\rm c}_{r_{h+k}}}}}\hskip -0.3cm\sum_{{r_{h+k+1},\dots,r_n\in P_{\mathrm{H}}}}\hspace{-0.3cm}\theta(\{r_s,v_s,v'_s\}_{s=1}^n)\theta(\{\tl r^\flat_s,\tl v^\flat_s,\tl v'^\flat_s\}_{s=1}^n)\\
         &\times\prod_{i=1}^h(\eta_{r_i}+\eta_{r_i+v_i+v'_i})^2\,n^\a_{v_i}n^\a_{v'_i}\prod_{j=h+1}^{h+k}(\eta_{r_j}+\eta_{r_j+v_j})^2\,n^\a_{v_j}\prod_{m=h+k+1}^n\eta_{r_m}^2 \frac18 \eta_{r_1}\eta_{\tl r_1}\eta_{r_2}\eta_{\tl r_2}\\
         &\times\sum_{\substack{p_1,\tl p_1\\p_2,\tl p_2}}\d_{p+s,\tl p_1}\d_{q+s,p_1}\d_{-p_1+v_1+v'_1,\tl p_2}\d_{-\tl p_1+\tl v_1+\tl v'_1,p_2}\d_{-p_2+v_2+v'_2,-\tl p_2+\tl v_2+\tl v'_2}\sum_{\substack{w_1,\tl w_1\\w_2,\tl w_2}}\d_{p,\tl w_2}\d_{q,w_2}\d_{w_1,\tl w_1}\\
         &\times(\d_{v_1+v'_1-w_1,\tl v_1+\tl v'_1-\tl w_1}+\d_{v_1+v_1'-w_1,\tl v_2+\tl v'_2-p})\big(n^\a_pn^\a_{v_2}n^\a_{v'_2}\d_{v_1,0}\d_{v'_1,0}+2n^\a_pn^\a_qn^\a_{v_1}\d_{v'_1,0}\d_{v_2+v'_2-q,0}\\
         &+(n^\a_p\d_{q,0}+n^\a_q\d_{p,0})(2n^\a_{v_1}n^\a_{v_2+v'_2-q}\d_{v'_1,0}+n^\a_{v_1}n^\a_{v'_1}\d_{v_2+v'_2-q,0})\big).
	\end{split}•
\]
Once we take the absolute value we can estimate the functions $\theta$ exactly as we did for $B_0,B_1$ and move out the factors involving the indices 1 and 2 multiplied times $N_0(N_0-1)$. Bounding $ \frac{(N_0-2)!}{(N_0 - k-2(n-2-k-h)-2)!}\leq  \frac{N_0!}{(N_0 - k-2(n-2-k-h))!}$ and rescaling $n\to n-2$ we recognize $\|\cD_\a g^0_\a\|^2.$ Hence,
\be\label{eq:B2}
\begin{split}
	\frac{|B_2|}{\|\cD_\a g^0_\a\|^2}&\leq CN_0^2N^{-5}\|\eta_H\|^3_\infty  \sup_{q\in \L^*}\sum_{s\in \L^*_+} N^\k|\widehat V((q-s)/N^{1-\k})||\eta_s| N_\a^3\\
	& \leq CN^{10\k-10+6\eps}N_0^2N^3_\a.
	\end{split}
\ee
Proceeding along the same line, moving out $\frac{N_0!}{(N_0-3)!}$ and $\frac{N_0!}{(N_0-4)!}$ respectively and bounding appropriately the remaining factor involving $N_0$ we can bound $|B_3|$ and $|B_4|.$ We get
\be\label{eq:B3}
\begin{split}
	\frac{|B_3|}{\|\cD_\a g^0_\a\|^2}& \leq CN^3_0N^{-5}\|\eta_H\|^3_\infty  \sup_{q\in \L^*}\sum_{s\in \L^*_+} N^\k|\widehat V((q-s)/N^{1-\k})||\eta_s| N_\a^2\\
	& \leq CN^{10\k-10+6\eps}N_0^3N^2_\a
\end{split}
\ee
and
\be\label{eq:B4}
\begin{split}
	\frac{|B_4|}{\|\cD_\a g^0_\a\|^2}& \leq CN^4_0N^{-5}\|\eta_H\|^3_\infty  \sup_{q\in \L^*}\sum_{s\in \L^*_+} N^\k|\widehat V((q-s)/N^{1-\k})||\eta_s| N_\a\\
	&\leq CN^{10\k-10+6\eps}N_0^4N_\a	.
\end{split}
\ee
Thus, \eqref{eq:II2b-split}--\eqref{eq:B4}, yield
\be\label{eq:II2b-final}
	\frac{|II_2^{(b),2}|}{\|\cD_\a g^0_\a\|^2}\leq CN^{10\k-10+6\eps}(N_0+N_\a)^5.
\ee
Finally, putting together \eqref{eq:V2b-split},\eqref{eq:V2b1},\eqref{eq:V2b2-split},\eqref{eq:I2b-final} and \eqref{eq:II2b-final}, and using Lemma \ref{lm:cN}, we conclude that for $\eps>0$ such that $3\k-2+3\eps<0$
\[
	\sum_\a\l_\a\frac{\cV_2^{(b)}}{\|\cD_\a g^0_\a\|^2}\leq CN^{\k+1}N^{-\eps}.
\]

\subsubsection{Contribution of $\cV_4$ on the correlated state}
Let us now focus on the term $\cV_4$ on the right-hand side of \eqref{eq:spitV}. Writing the state 
$\cD_\a g_\a^0$ explicitly we have:
\[
	\begin{split}
		   \langle \cD_\a g_\a^0,\, \cV_4\cD_\a g_\a^0\rangle=\;&\frac1{2N^{1-\k}}\sum_{\substack{s\in \L^*,\,p,q\in P_{\mathrm{H}}:\\p+s,\,q+s\in P_{\mathrm{H}}}}\widehat{V}(s/N^{1-\k})\sum_{n=1}^{n_{\mathrm{c}}}\frac{1}{(n!)^2(2N)^{2n}} \sum_{\substack{\scriptscriptstyle{r_1,\dots,r_n\in P_{\mathrm{H}}}\\\scriptscriptstyle\tilde r_1,\dots,\tilde r_n\in P_{\mathrm{H}}}}\\
		   &\times\hspace{-0.3cm}\sum_{\substack{\scriptscriptstyle{(v_1,v'_1)\in \bbG^0_{r_1·}}\\\scriptscriptstyle{(\tilde v_1,\tilde v'_1)\in \bbG^0_{\tilde r_1}}}}\hspace{-0.2cm}\dots\hspace{-0.2cm} \sum_{\substack{\scriptscriptstyle{(v_n,v'_n)\in \bbG^0_{r_n}}\\\scriptscriptstyle{(\tilde v_n,\tilde v'_n)\in \bbG^0_{\tilde r_n}}}}\hspace{-0.2cm}\theta\big(\{r_j,v_j,v'_j\}_{j=1}^n\big)\theta\big(\{\tilde r_j,\tilde v_j,\tilde v'_j\}_{j=1}^n\big)\prod_{i=1}^n\eta_{r_i}\eta_{\tilde r_i}\\
         &\times \langle g^0_\a, A^*_{\tilde{r}_n,\tilde{v}_n,\tilde{v}'_n}\dots A^*_{\tilde{r}_1,\tilde{v}_1,\tilde{v}'_1} a^*_{p+s}a^*_qa_{q+s}a_pA_{r_1,v_1,v'_1}\dots A_{r_n,v_n,v'_n}g^0_\a\rangle
	\end{split}
\]
where we recall the notation $A_{r,v,v'}=a^*_{-r} a^*_{r+v+v'}a_v a_{v'}$. 
Since $g^0_\a$ does not contain any high momentum 
the creation operators $a^*_{p+s}a^*_{q}$ have to be contracted with two annihilation operators $a_{\tilde p_i}, a_{\tilde p_j}$ with $i,j\in\{1,\dots,n\}$ and $\tilde p_\ell\in \{-\tilde r_\ell,\tilde r_\ell+\tilde v_\ell+\tilde v'_\ell\},$  for $\ell=i,j$. Analogously, the annihilation operators $a_{q+s},a_p$ have to be contracted with  $a^*_{p_k}, a^*_{p_m}$ with $k,m\in\{1,\dots,n\}$ and $p_\ell\in \{-r_\ell,r_\ell+v_\ell+v'_\ell\}$ for $\ell=k,m$. 
Due to the presence of the restrictions encoded in $\theta(\{r_i,v_i,v'_i\}_{i=1}^n)\theta(\{\tilde r_j,\tilde v_j,\tilde v'_j\}_{j=1}^n)$ only two cases are possible: 
\begin{enumerate}
\item $i=j.$ Then, also $k=m.$ Indeed, assume by contradiction $k\neq m.$ Then, $a^*_{-p_k+v_k+ v'_k}$ has to be contracted with $a_{\tl p_{t_1}}$ for $t_1\in \{1,\dots,n\},\,t_1\neq i$ while $a_{-\tl p_{t_1}+\tl v_{t_1}+\tl v'_{t_1}}$ will be contracted with $a^*_{p_{t_2}}$ with $t_2\neq k$. This  immediately yields $-p_k+ v_k+ v'_k=-p_{t_2} +\tl v_{t_1}+\tl v'_{t_1}$.  On the other hand, the expectation above would be zero unless there exist  $ w_{t_3},w_{t_4}$ so that $\tl v_{t_1}=w_{t_3}$ and $\tl v'_{t_1}=w_{t_4}$, with $t_3,t_4\in \{1,\dots,n\}$ (recall the notation $\tilde w_{\ell}\in \{\tilde v_{\ell},\tilde v'_{\ell}\}$) . Hence the condition $-p_k+v_k+ v'_k=-  p_{t_2}+ w_{t_3}+w_{t_4}$ and  $k\neq  t_2$ would 
contradict the restrictions in \eqref{eq:restrictions1}. Hence, $k=m$ and by symmetry we can assume $i=k=n.$ Note that, the choice of $p$ and $q+s$ also fixes $q=-(p+s)+\tilde v_n+\tilde v'_n$  
and implies $v_n+v'_n=\tilde v_n+\tilde v'_n$.  
\item $i\neq j.$ Similarly as before one can show that this implies $k\neq m$ together with the fact that the creation operators $a^*_{-p_k+v_k+v'_k},a^*_{-p_m+v_m+v'_m}$ are contracted with the annihilation operators $a_{-\tilde p_i+\tilde v_i+\tilde v'_i}, a_{-\tilde p_j+\tilde v_j+\tilde v'_j}.$ By symmetry we can assume $i=k=n,j=m=n-1$.
\end{enumerate}

We thus split $\langle \cD_\a g_\a^0,\cV_4 \cD_\a g_\a^0\rangle$ into two contributions corresponding to the two cases listed above.
Pairing the remaining high momenta we write
\[
	\begin{split}
		\langle \cD_\a g_\a^0, \cV_4 \cD_\a g_\a^0\rangle=V_{4,\a}+\cE^{(1)}_{4,\a}
	\end{split}
\] 
where
\begin{equation}\label{eq:V4-a}
	\begin{split}
		V_{4,\a}
		=&\frac1{2N^{1-\k}}\sum_{\substack{s\in \L^*,\,p\in P_{\mathrm{H}}:\\p+s\in P_{\mathrm{H}}}}\widehat{V}(s/N^{1-\k})\sum_{n=1}^{n_{\mathrm{c}}}\frac{1}{(n-1)!(2N)^{2n}}\hspace{-0.3cm}
        \sum_{\substack{\scriptscriptstyle{r_1,\dots,r_n\in P_{\mathrm{H}}}\\\scriptscriptstyle\tilde r_1,\dots,\tilde r_n\in P_{\mathrm{H}}}}\sum_{\substack{\scriptscriptstyle{(v_1,v'_1)\in \bbG^0_{r_1}}\\\scriptscriptstyle{(\tilde v_1,\tilde v'_1)\in \bbG^0_{\tl r_1}}}}\hspace{-0.2cm}\dots\hspace{-0.2cm} \sum_{\substack{\scriptscriptstyle{(v_n,v'_n)\in \bbG^0_{r_n}}\\\scriptscriptstyle{(\tilde v_n,\tilde v'_n)\in \bbG^0_{\tl r_n}}}}\\
        &\times \eta_{r_n}\eta_{\tilde r_n}\sum_{p_n,\tilde p_n} \d_{p_n,p}\d_{\tilde p_n,p+s}\d_{v_n+v'_n,\tilde v_n+\tilde v'_n}
         \prod_{i=1}^{n-1}\eta_{r_i}\eta_{\tilde r_i}(\d_{\tilde r_i,r_i}+\d_{-\tilde r_i,r_i+v_i+v'_i})\d_{v_i+v'_i,\tilde v_i+\tilde v'_i}\\
        &\times\theta\big(\{r_j,v_j,v'_j\}_{j=1}^n\big)\theta\big(\{\tilde r_j,\tilde v_j,\tilde v'_j\}_{j=1}^n\big)\langle g^0_\a, a^*_{\tilde v_n}a^*_{\tilde v'_n}\dots a^*_{\tilde v_1}a^*_{\tilde v'_1}a_{v_1}a_{v'_1}\dots a_{v_n}a_{v'_n} g^0_\a \rangle \\[0.2cm]
	\end{split}
\end{equation}
and
\begin{equation}\label{eq:V2a}
	\begin{split}
		&\cE^{(1)}_{4,\a} \\ 
		&=\frac1{2N^{1-\k}}\sum_{\substack{s\in \L^*,\,p,q\in P_{\mathrm{H}}:\\p+s,q+s\in P_{\mathrm{H}}}}\widehat{V}(s/N^{1-\k})\sum_{n=2}^{n_{\mathrm{c}}}\frac{1}{(n-2)!(2N)^{2n}}\hspace{-0.3cm}
        \sum_{\substack{\scriptscriptstyle{r_1,\dots,r_n\in P_{\mathrm{H}}}\\\scriptscriptstyle\tilde r_1,\dots,\tilde r_n\in P_{\mathrm{H}}}}\sum_{\substack{\scriptscriptstyle{(v_1,v'_1)\in \bbG^0_{r_1}}\\\scriptscriptstyle{(\tilde v_1,\tilde v'_1)\in \bbG^0_{\tilde r_1}}}}\hspace{-0.2cm}\dots\hspace{-0.2cm} \sum_{\substack{\scriptscriptstyle{(v_n,v'_n)\in \bbG^0_{r_n}}\\\scriptscriptstyle{(\tilde v_n,\tilde v'_n)\in \bbG^0_{\tilde r_n}}}}\\
        &\times\prod_{i=1}^{n-2}\eta_{r_i}\eta_{\tilde r_i}(\d_{\tilde r_i,r_i}+\d_{-\tilde r_i,r_i+v_i+v'_i})\d_{v_i+v'_i,\tilde v_i+\tilde v'_i} \eta_{r_n}\eta_{\tilde r_n}\eta_{r_{n-1}}\eta_{\tilde r_{n-1}} \hskip -0.4cm\\
        & \times \sum_{\substack{p_n,\,\tilde p_n\\p_{n-1},\,\tilde p_{n-1}}}  \hskip -0.2cm \d_{p_n, p}\, \d_{\tilde p_n,p+s}\,\d_{p_{n-1},q+s}\,\d_{\tilde p_{n-1},q}\, \big(\d_{v_n+v'_n,\tilde v_n+\tilde v'_n-s}\d_{v_{n-1}+v'_{n-1},\tilde v_{n-1}+\tilde v'_{n-1}+s}\\[-0.5cm]
        & \hskip 5cm +\d_{-p+v_n+v'_n,-q+\tilde v_{n-1}+\tilde v'_{n-1}}\d_{-q+v_{n-1}+v'_{n-1},-p+\tilde v_n+\tilde v'_n}\big)\\[0.2cm]
        &\times\theta\big(\{r_j,v_j,v'_j\}_{j=1}^n\big)\theta\big(\{\tilde r_j,\tilde v_j,\tilde v'_j\}_{j=1}^n\big)\langle g^0_\a, a^*_{\tilde v_n}a^*_{\tilde v'_n}\dots a^*_{\tilde v_1}a^*_{\tilde v'_1}a_{v_1}a_{v'_1}\dots a_{v_n}a_{v'_n}g^0_\a\rangle\,.
	\end{split}
\end{equation}

The contribution to the energy coming from the term $\cE^{(1)}_{4,\a}$ is an error term. Indeed, to bound $|\cE^{(1)}_{4,\a}|$ one can proceed as in the computation of the norm, pairing the low momenta variables on the last line of \eqref{eq:V2a} and rescaling the variable $n\to n-2$. Since we are taking the term in absolute value we can remove the cutoff function involving the variables $\{r_j, v_j, v'_j\}_{j=n, n-1}$ and $\{\tl r_j, \tl v_j, \tl v'_j\}_{j=n, n-1}$ (bounding the part of the cutoff function involving the additional variables by one) and add the terms $n=n_{\mathrm{c}}-1, n_{\mathrm{c}}$ to reconstruct the norm. One obtains:
\be \begin{split} \label{eq:cE-1-4}
\sum_\a \l_\a \frac{|\cE^{(1)}_{4,\a}|}{\| \cD_\a g_\a^0\|^2} \leq \frac{C }{N^{5-\k}}  \| \eta_H\|^4_2 \,  \sum_\a \l_\a  \hskip -0.1cm \sum_{v \in P_\A^0} (n^\a_v)^2 (N_0+N_\a)^2  \leq C N^{\k+1}  N^{6\k -5 + 2 \eps +\d_1/3}
\end{split}\ee
where in the last line we used \eqref{eq:etaH}.

Let's focus now on the term $V_{4,\a}$, which is the one containing the large contribution. 
Pairing the low momenta as in the computation of the norm below  Eq.\,\eqref{eq:norm2} and denoting with $h,k=1,\dots,n-1$ respectively the number of quartic and triplet excitations, we get 
\be \label{eq:V4alpha-2}
	\begin{split}
		V_{4,\a} &= \frac1{2 N^{1-\k}}\hskip -0.3cm\sum_{\substack{s\in \L^*,\,p \in P_{\mathrm{H}}:\\p+s\in P_{\mathrm{H}}}}\hskip -0.3cm \widehat{V}(s/N^{1-\k})  \\
		& \times  \sum_{n=1}^{n_{\mathrm{c}}}\frac 1{N^{2n}} \sum_{h=0}^{n-1}\;\sum_{k=0}^{n-1-h} \frac{1}{h!k!(n-1-h-k)!} \frac{N_0!\, 2^{-2h -k-(n-1-h-k)}}{(N_0 - k-2(n-1-k-h))!} \\
      &\times  \hspace{-0.2cm}
      \sum_{\substack{\scriptscriptstyle{r_{1},\dots,r_{n}\in P_{\mathrm{H}}}\\ \scriptscriptstyle{\tl r_n \in P_{\mathrm{H}}}}}
      \sum_{\scriptscriptstyle{(v_1,v_1')\in \bbG^{\rm q}_{r_1}}}\hspace{-0.2cm}\dots\hspace{-0.2cm}\sum_{\scriptscriptstyle{(v_h,v_h')\in \bbG^{\rm q}_{r_h}}}\;\hspace{-0.2cm}\sum_{\scriptscriptstyle{(v_{h+1},0)\in \bbG^{\rm c}_{r_{h+1}}}}\hspace{-0.3cm}\dots\hspace{-0.3cm}\;\sum_{\scriptscriptstyle{(v_{h+k},0)\in \bbG^{\rm c}_{r_{h+k}}}}  \sum_{\substack{\scriptscriptstyle{(v_n,v'_n)\in \bbG^0_{r_n}\hskip -0.05cm \cap \bbG^0_{\tl r_n} }}} \hspace{-0.6cm}\theta\big(\{r_j,v_j,v'_j\}_{j=1}^n\big)
      \\  
		&\times\theta(\{\tl r^\sharp_s,v_s,v'_s\}_{s=1}^n)
		 \prod_{i=1}^{h}(\eta_{r_i}+\eta_{r_i+v_i+v'_i})^2n^\a_{v_i}n^\a_{v'_i}\prod_{j=h+1}^{h+k}(\eta_{r_j}+\eta_{r_j+v_j})^2n^\a_{v_j}\prod_{m=h+k+1}^{n-1} \hspace{-0.3cm}\eta_{r_m}^2\\
		 & \times \frac14\eta_{r_n}\eta_{\tilde r_n}   \sum_{p_n,\tilde p_n} \d_{p_n,p}\d_{\tilde p_n,p+s}\big[2n^\a_{v_n}n^\a_{v'_n}\\
		&+(N_0-k-2(n-1-k-h))(4n^\a_{v_n}\d_{v'_n,0}+(N_0-k-2(n-h-k)-1)\d_{v_n,0}\d_{v'_n,0})\big] \\[0.2cm]
	\end{split}
\ee
where in the last line we used \eqref{eq:a0m-t} with $m=k+2(n-1-k-h)$ and $t=1,2$ respectively, and 
where we are adopting  the convention $\tilde r^\sharp_n=\tilde r_n$,  and $\tilde r^\sharp_s=r_s$  for any $1\leq s\leq n-1$.
To reconstruct the norm of $\cD_\a g_\a^0$ we have to remove the restrictions involving momenta with index $n$. With the decomposition in \eqref{eq:dec_theta},\eqref{eq:theta_n} we write 
\be \begin{split} \label{eq:V4-thetasplit}
	&\; \theta(\{r_i,v_i,v'_i\}_{i=1}^n)\,\theta(\{\tl r^\sharp_s, v_s,v'_s\}_{s=1}^n)\\
	& \;=\theta(\{r_i,v_i,v'_i\}_{i=1}^{n-1})\Big[1+\big(\theta_n(\{r_i,v_i,v'_i\}_{i=1}^n)\,  \theta_n(\{\tl r^\sharp_s,v_s,v'_s\}_{s=1}^n) -1 \big)\Big]
\end{split}\ee
where we used that $\theta(\{r_j,v_j,v'_j\}_{j=1}^{n-1})=\theta(\{r^\sharp_s,v_s,v'_s\}_{s=1}^{n-1})$. With \eqref{eq:V4-thetasplit} we write 
$V_{4,\a} =I^\a + \cE_{4,\a}^{(2)}$, 
where $I^\a$ and $\cE_{4,\a}^{(2)}$ have the same expression than \eqref{eq:V4alpha-2} with the product of the two theta functions replaced by the first and second term on the right-hand side of \eqref{eq:V4-thetasplit} respectively. We further decompose $I^\a=I^\a_{\rm q} + I^\a_{\rm c} + I^\a_{\rm p}$ according to the contributions to $I^\a$ coming from quadruplets, triplets and pair excitations, and we proceed as below \eqref{eq:I2a}. Rescaling the variable $n \to n-1$, and adding and subtracting the term with $n=n_{\mathrm{c}}$ we obtain:  \\
\be \label{eq:Ia-q}
\begin{split}
& \frac{I^\a_{\rm q}}{\|\cD_\a g^0_\a\|^2} \\
 & = \frac1{2 N^{3-\k}}\hskip -0.2cm\sum_{\substack{s\in \L^*,\, p\in P_{\mathrm{H}}:\\p+s\in P_{\mathrm{H}}}} \hskip -0.2cm\widehat V\big(s/N^{1-\k}\big)  \hskip -0.3cm \sum_{\substack{(v,v')
\in \bbG^{\rm q}_{p}\cap \bbG^{\rm q}_{p+s}}} \hskip -0.3cm \eta_p  \big(\eta_{p+s}+\eta_{p+s+v+v'}\big)  n^\a_v n^\a_{v'} + \frac{\cE_{4,\a}^{(\rm q)}}{\|\cD_\a g^0_\a\|^2} 
\end{split}
\ee
where similarly to  \eqref{eq:Iq-final}, the term $\cE_{4,\a}^{(\rm q)}$ can be bounded by
\be \begin{split} \label{eq:Ia-qerr}
|\cE_{4,\a}^{(\rm q)}| 
 \leq& \frac C {N^{3-\k}} \sup_{w \in P_\A^0}\sum_{\substack{s\in \L^*,\,p \in P_{\mathrm{H}}:\\p+s\in P_{\mathrm{H}}}}\hskip -0.3cm |\widehat{V}(s/N^{1-\k})| |\eta_s||\eta_{p+s+w}| \sum_{v,v' \in P_\A} n^\a_{v_n}n^\a_{v'_n}\\
				& \times \Big(\frac{N_0}{N_\a}\Big)^{2n_{\mathrm{c}}} \frac{1}{n_{\mathrm{c}}!} \sum_{h=0}^{n_{\mathrm{c}}}\;\sum_{k=0}^{n_{\mathrm{c}}-h} \frac{n_{\mathrm{c}}!}{h!k!(n_{\mathrm{c}}-h-k)!} \Big(\frac{N_\a}{N_0}\Big)^{2h+k}   \|\eta_H\|_2^{2n_{\mathrm{c}}} \\
				\leq &C N^{\k-1} N_\a^2  \,\big(\frac{\|\eta_H\|^2_2}{n_{\mathrm{c}}}. \big)^{n_{\mathrm{c}}}
\end{split}\ee
Here, in the last line we used  the bound
\be \label{eq:5.15-BCS}
	\begin{split}
	& \sup_{w\in P_\A^0} \frac C{N^{2-\k}}\hskip -0.3cm\sum_{\substack{s\in \L^*,\,p \in P_{\mathrm{H}}:\\p+s\in P_{\mathrm{H}}}}\hskip -0.3cm |\widehat{V}(s/N^{1-\k})| |\eta_p||\eta_{p+s+w}| \\
	& \leq  C N^{2\k-2} \| \eta_H\|_1  \sup_{p \in \L^*}\sum_{\substack{r\in \L^*_+}}\frac{|\widehat{V}((r-p)/N^{1-\k})|}{ |r|^{-2}} \leq  C N^\k
	\end{split}
\ee
which follows from \eqref{BCS-bounds} and Lemma \ref{lm:eta}. The terms $I^\a_{\rm c}$ and $I^\a_{\rm p}$ can be discussed analogously, obtaining:
\be \label{eq:Ia-cp}
\begin{split}
 \frac{I^\a_{\rm c}}{\|\cD_\a g^0_\a\|^2} & = \frac1{2 N^{3-\k}}\hskip -0.2cm\sum_{\substack{s\in \L^*,\, p\in P_{\mathrm{H}}:\\p+s\in P_{\mathrm{H}}}} \hskip -0.2cm\widehat V\big(s/N^{1-\k}\big)  \hskip -0.3cm \sum_{\substack{v \in P_\A : \\ p+v,\, p+s+v \in P_{\mathrm{H}}
}} \hskip -0.3cm \eta_p  \big(\eta_{p+s}+\eta_{p+s+v}\big) 2N_0 n^\a_v  + \frac{\cE_{4,\a}^{(\rm c)}}{\|\cD_\a g^0_\a\|^2}  \\
 \frac{I^\a_{\rm p}}{\|\cD_\a g^0_\a\|^2} & = \frac1{2 N^{3-\k}}\hskip -0.2cm\sum_{\substack{s\in \L^*,\, p\in P_{\mathrm{H}}:\\p+s\in P_{\mathrm{H}}}} \hskip -0.2cm\widehat V\big(s/N^{1-\k}\big)  \eta_p  \eta_{p+s} N^2_0  + \frac{\cE_{4,\a}^{(\rm p)}}{\|\cD_\a g^0_\a\|^2}
\end{split}
\ee
with $\cE_{4,\a}^{(\rm c)}$ and $\cE_{4,\a}^{(\rm p)}$ so that
\be \label{eq:Ia-cp-err} 
|\cE_{4,\a}^{(\rm c)}|  + |\cE_{4,\a}^{(\rm p)}|  \leq C N^{\k-1} (N_\a N_0 + N_0^2) \,\big(\|\eta_H\|_2^2/n_{\mathrm{c}} \big)^{n_{\mathrm{c}}}\,.
\ee
With Eq. \eqref{eq:Ia-q}, \eqref{eq:Ia-qerr}, \eqref{eq:Ia-cp} and \eqref{eq:Ia-cp-err}, together with the bound \eqref{eq:LB-norm} we conclude
\be \begin{split} \label{eq:Ia-final}
\bigg|\sum_\a \l_\a& \frac{I^\a}{\| \cD_\a g_\a^0\|^2} - \frac1{2 N^{3-\k}}\hskip -0.3cm\sum_{\substack{s\in \L^*,\, p\in P_{\mathrm{H}}:\\p+s\in P_{\mathrm{H}}}} \hskip -0.3cm\widehat V\big(s/N^{1-\k}\big)   \Big[ \sum_{\substack{(v,v')
\in \bbG^{\rm q}_{p}\cap \bbG^{\rm q}_{p+s}}} \hskip -0.5cm \eta_p  \big(\eta_{p+s}+\eta_{p+s+v+v'}\big)  \sum_\alpha\l_\a n^\a_v n^\a_{v'} \\
& \hskip 1cm + \sum_{\substack{v \in P_\A : \\ p+v,\, p+s+v \in P_{\mathrm{H}}
}} \hskip -0.3cm \eta_p  \big(\eta_{p+s}+\eta_{p+s+v}\big) 2N_0 \sum_\a\l_\a n^\a_v +  \eta_p  \eta_{p+s} N^2_0 \Big]\bigg | \leq C  N^{\k+1-\eps}
\end{split}\ee
with the choice $n_{\mathrm{c}} = N^{\d_2/3}\|\eta_H\|^2_2$. \newpage

We now focus on the term arising by substituting in \eqref{eq:V4alpha-2} the second term on the right-hand side of \eqref{eq:V4-thetasplit}. We obtain:
\[
	\begin{split}
		\cE_{4,\a}^{(2)}&\; =   \frac1{2N^{1-\k}}\sum_{\substack{s\in \L^*,\,p\in P_{\mathrm{H}}:\\p+s\in P_{\mathrm{H}}}}\widehat{V}(s/N^{1-\k})\\
		& \times \sum_{n=2}^{n_{\mathrm{c}}}\frac{1}{N^{2n}}\sum_{h=0}^{n-1}\;\sum_{k=0}^{n-1-h} \frac{1}{h!k!(n-1-h-k)!} \frac{N_0!\, 2^{-2h -k-(n-1-h-k)}}{(N_0 - k-2(n-1-k-h))!}  \\
		&\times \sum_{\substack{\scriptscriptstyle{r_1,\dots,r_n\in P_{\mathrm{H}}}\\\scriptscriptstyle\tilde r_n\in P_{\mathrm{H}}}} \sum_{\substack{\scriptscriptstyle{(v_1,v'_1)\in \bbG^{\rm q}_{r_1}}}}\hspace{-0.2cm}\dots\hspace{-0.2cm} \sum_{\substack{\scriptscriptstyle{(v_h,v'_h)\in \bbG^{\rm q}_{r_h}}}}\sum_{\substack{\scriptscriptstyle{(v_{h+1},0)\in \bbG^{\rm c}_{r_{h+1}}}}} \hspace{-0.2cm}\dots\hspace{-0.2cm}\sum_{\substack{\scriptscriptstyle{(v_{h+k},0)\in \bbG^{\rm c}_{r_{h+k}}}}} \sum_{\substack{\scriptscriptstyle{(v_n,v'_n)\in \bbG^0_{r_n} \cap \bbG^0_{\tilde r_n}}}}\\
		& \times \theta\big(\{r_j,v_j,v'_j\}_{j=1}^{n-1}\big)\Big[\theta_n\big(\{r_i,v_i,v'_i\}_{i=1}^n\big))\theta_n\big(\{r^\sharp_i,v_i,v'_i\}_{i=1}^n\big)-1\Big]\\
		&\times \prod_{i=1}^{h}(\eta_{r_i}+\eta_{r_i+v_i+v'_i})^2n^\a_{v_i}n^\a_{v'_i}\prod_{j=h+1}^{h+k}\hspace{-0.1cm}(\eta_{r_j}+\eta_{r_j+v_j})^2n^\a_{v_j}\prod_{m=h+k+1}^{n-1}\hspace{-0.3cm}\eta_{r_m}^2\\
		&\times \frac14\eta_{r_n}\eta_{\tilde r_n}   \sum_{p_n,\tilde p_n} \d_{p_n,p}\d_{\tilde p_n,p+s}\big[2n^\a_{v_n}n^\a_{v'_n}\\
		&+(N_0-k-2(n-1-k-h))(4n^\a_{v_n}\d_{v'_n,0}+(N_0-k-2(n-h-k)-1)\d_{v_n,0}\d_{v'_n,0})\big]
	\end{split}
\]
where, in the argument of $\theta_n$ we used the notation $r^{\sharp}_i=r_i$ for $i=1,\dots,n-1$, $r^\sharp_n=\tl r_n$. 
To bound  $\cE_{4,\a}^{(2)}$ we notice that  for all $n\geq 2$
\[
\big |\theta_n(\{r_i,v_i,v'_i\}_{i=1}^n)\theta_n\big(\{r^\sharp_i,v_i,v'_i\}_{i=1}^n\big)-1\big | \leq \big|\theta_n(\{r_i,v_i,v'_i\}_{i=1}^n)-1 \big| + \big| \theta_n\big(\{r^\sharp_i,v_i,v'_i\}_{i=1}^n\big)-1\big|
\]
which can be bounded by the right-hand side of \eqref{eq:theta-1} with $n+1$ replaced by $n$ and $p_n$ substituted by the index $q_n \in \{-r_n,r_n+v_n+v'_n,-\tl r_n,\tl r_n+v_n+v'_n\}$. Similarly as above we can then bound $|\cE_{4,\a}^{(2)}| \leq \sum_{j=1}^5W_j^\a$
 with $W_j^\a$ denoting the contribution arising from the j-th term in \eqref{eq:theta-1} (after the rescaling $n+1\to n$ and the change $p_{n} \to q_n$). Proceeding as in the computation of \eqref{eq:X1}--\eqref{eq:X5} one concludes that
 \be \label{eq:cE-2-4}
 \sum_\a \l_\a  \frac{  \big|\cE^{(2)}_{4,\a}\big|}{\|\cD_\a g_\a^0\|^2} 
 \leq  C N^{\k+1} \max \{N^{-\eps}, N^{6\k-3+2\e+\d_1/3}, N^{12\k- 6+6\eps}\}
\ee
for all $\k \in (1/3, 2/3)$, and $\e \in (0,1)$ so that $3\k-2+3\e <0$. 
With Eqs.\,\eqref{eq:V4-a}, \eqref{eq:cE-1-4},\eqref{eq:Ia-final}, \eqref{eq:cE-2-4} we conclude
\be \begin{split} \label{eq:V4-final1}
\bigg|\sum_\a \l_\a \frac{\cV_4}{\| \cD_\a g_\a^0\|^2} - &\; \frac1{2 N^{3-\k}}\hskip -0.2cm\sum_{\substack{s\in \L^*,\, p\in P_{\mathrm{H}}:\\p+s\in P_{\mathrm{H}}}} \hskip -0.2cm\widehat V\big(s/N^{1-\k}\big)   \Big[ \hskip -0.2cm\sum_{\substack{(v,v') \in \\
\bbG^{\rm q}_{p}\cap \bbG^{\rm q}_{p+s}}} \hskip -0.3cm \eta_p  \big(\eta_{p+s}+\eta_{p+s+v+v'}\big)  \sum_\a\l_\a n^\a_v n^\a_{v'} \\
& \hskip 1.2cm + \sum_{\substack{v \in P_\A : \\ p+v,\, p+s+v \in P_{\mathrm{H}}
}} \hskip -0.3cm \eta_p  \big(\eta_{p+s}+\eta_{p+s+v}\big) 2N_0 \sum_\a\l_\a n^\a_v +  \eta_p  \eta_{p+s} N^2_0 \Big]\bigg | \\
&  \leq C N^{\k+1} \max \{N^{-\eps}, N^{6\k-3+2\eps +\d_1/3 }, N^{12\k- 6+6\eps}\}
\end{split}\ee
for the same parameters. 

In order to reconstruct the convolution we have to add in the sums on the right hat side of \eqref{eq:V4-final1} the corresponding terms  with $|p+s|<N^{1-\k-\eps}$ and/or $|p+s+v+v'|<N^{1-\k-\eps}$ with $v,v'\in P_\A^0.$ Since by \eqref{eq:etapbound} for any $p\in P_{\mathrm{H}}$
\[
	\sum_{\substack{s'\in\L^*:\\|p+s|<N^{1-\k-\eps}+CN^{1/3+\d_2/3}}} \widehat V(s/N^{1-\k})|\eta_{p+s}|\leq CN^\k \sum_{\substack{s'\in \L^*:\\|s'|< CN^{1-\k-\eps}}}\frac1{|s'|^2}\leq CN^{1-\eps}
\]
we obtain,  using Lemma \ref{lm:cN} and $\|\eta_H\|_1\leq CN$ in Lemma \ref{lm:eta}, 
\[
	 \begin{split} 
		& \bigg|\sum_\a \l_\a \frac{\cV_4}{\| \cD_\a g_\a^0\|^2} - \; \frac1{2 N^{2}}\hskip -0.2cm\sum_{\substack{p\in P_{\mathrm{H}}\\(v,v')
\in \bbG^{\rm q}_{p}}} \hskip -0.2cm   \Big((\widehat V_N\ast\eta)(p)+(\widehat V_N\ast\eta)(p+v+v')\Big) \eta_p \sum_\a\l_\a n^\a_v n^\a_{v'} \\
		& + \frac1{ N^{2}}\hskip -0.2cm\sum_{\substack{ p\in P_{\mathrm{H}}\\(v,0)\in \bbG^{\rm c}_p}} \hskip -0.2cm    \Big((\widehat V_N\ast\eta)(p)+(\widehat V_N\ast\eta)(p+v)\Big)\eta_p N_0 \sum_\a\l_\a n^\a_v +\frac1{N^2}\sum_{p\in P_{\mathrm{H}}}( \widehat V_N\ast \eta)(p)\eta_p N^2_0\bigg | \\
		& \leq C N^{\k+1} \max \{N^{-\eps}, N^{6\k-3+2\eps +\d_1/3 }, N^{12\k- 6+6\eps}\}\,.
	\end{split}
\]
Next, we notice that for any $v,v'\in P_\A^0$
\be\label{eq:V*eta-rv}
	\sup_{v,v' \in P_\A^0}\sum_{p\in P_{\mathrm{H}}}(\widehat V_N\ast \eta)(p)|\eta_{p+v+v'}-\eta_p|\leq CN^{\k+1}N^{-\eps}.
\ee
This can be proved similarly to \eqref{eq:V-eta-rv} taking into account that by \eqref{BCS-bounds} $|(\widehat V_N\ast \eta)(p)|\leq CN^\k$ and 
\[
	\sum_{p\in P_{\mathrm{H}}} \frac{(\widehat V_N\ast \eta)(p)}{|p|^3} \leq C N^\k \ln N
\]
whose proof is analogous to the one of \eqref{eq:HYineq} (using $\|\widehat V_N\ast \eta\|_{q'}\leq N^\k\|V\|_{q'}$). 
By \eqref{eq:V*eta-rv} and Lemma \ref{lm:cN} we find
\be\label{eq:V4-final}
	\begin{split}
			&\bigg|\sum_\a \l_\a \frac{\cV_4}{\| \cD_\a g_\a^0\|^2} \\
			&- \frac1{ N^{2}}\sum_{p\in P_{\mathrm{H}}}   (\widehat V_N\ast\eta)(p) \eta_p \Big[N_0^2+4N_0\sum_\a\l_\a \sum_{\substack{v\in P_\A:\\r+v\in P_{\mathrm{H}}}}n^\a_v+2\sum_\a\l_\a \sum_{\substack{v,v'\in P_\A:\\v\neq \pm v'\\r+v+v'\in P_{\mathrm{H}}}} n^\a_v n^\a_{v'} \Big]\bigg | \\
		& \leq C N^{\k+1} \max \{N^{-\eps}, N^{6\k-3+2\eps +\d_1/3 }, N^{12\k- 6+6\eps}\}\,.
	\end{split}
\ee
To conclude we notice that $|r|\leq CN^{1-\k-\eps}(1+N^{\k-2/3+\d_2/3})$ for any $v,v'\in P_\A^0$ such that $r+v+v'\notin P_{\mathrm{H}}$ which allows us to remove the corresponding restrictions in the sums in the second line of \eqref{eq:V4-final} up to an error of the order $N^{\k+1-\eps}.$
Moreover, by Lemma \ref{lm:cN} we can remove the restrictions $v\neq\pm v'$ in the third term on the second line of \eqref{eq:V4-final} up to an error which satisfies the bound \eqref{eq:calEN}. With \eqref{eq:Nplus} and \eqref{eq:Nplus2} we obtain the third term on the right-hand side of \eqref{eq:corr-VN}.

\subsection{Entropy of the trial state}
\label{sec:entropy}

Let $\cD_\a$ be defined in Eq.~\eqref{eq:def_D}. Recall the spectral decomposition of the uncorrelated state $\G_0$ in \eqref{eq:G0-def} and the definition of the correlated state $\G$ in \eqref{eq:GammaD}.
 To estimate the influence of the correlation structure introduced by $\cD_\a$ on the entropy of $\G_0$, we use the following lemma, which appeared for the first time in \cite[Lemma~2]{Seiringer2006}. 

\begin{lemma}
	\label{lem:entropy}
	Let $\Gamma$ be a density matrix on some Hilbert space with eigenvalues $\{ \lambda_{\alpha} \}_{\alpha \in \mathbb{N}}$, let $\{ P_{\alpha} \}_{\alpha \in \mathbb{N}}$ be a family of one-dimensional orthogonal projection (for which $P_{\alpha_1} P_{\alpha_2} = \delta_{\alpha_1,\alpha_2} P_{\alpha_1}$ need not necessarily be true), and define $\widehat{\Gamma} = \sum_{\alpha} \lambda_{\alpha} P_{\alpha}$. Then we have
	\begin{equation*}
		S(\widehat{\Gamma}) \geq S(\Gamma) - \ln \mathrm{Tr}\left( \sum_{\alpha} P_{\alpha} \widehat{\Gamma} \right).
		\label{eq:LemEntropyCorr}
	\end{equation*}
\end{lemma}

With this Lemma we are now ready to show the lower bound for the entropy of $\G$ appearing in Lemma~\ref{lm:entropy}. 

\begin{proof}[Proof of Lemma~\ref{lm:entropy}]
An application of Lemma~\ref{lem:entropy} leads to 
\begin{equation}
	S(\Gamma) \geq -\sum_{\alpha} \lambda_{\alpha} \ln( \lambda_{\alpha} ) - \ln \left( \sum_{\alpha, \alpha'} \lambda_{\alpha} \frac{| \langle \mathcal{D}_{\alpha} g^0_{\alpha}\,, \mathcal{D}_{\alpha'} g^0_{\alpha'} \rangle |^2}{\Vert \mathcal{D}_{\alpha}g^0_{\alpha} \Vert^2 \Vert \mathcal{D}_{\alpha'} g^0_{\alpha'} \Vert^2}   \right). 
	\label{eq:entropy1}
\end{equation}
Since the vectors $g^0_{\alpha}$ contain no particles with momenta in $P_{\mathrm{H}}$ we have
\begin{equation}
	\langle \mathcal{D}_{\alpha} g^0_{\alpha}, \mathcal{D}_{\alpha'}g^0_{\alpha'} \rangle = \sum_{n=0}^{n_{\mathrm{c}}} \frac{1}{(n!)^2} \langle D_{\alpha}^n g^0_{\alpha}, D_{\alpha'}^n g^0_{\alpha'} \rangle.
	\label{eq:entropy2}
\end{equation}
We introduce for $\alpha$ and $n \in \mathbb{N}$ the set 
\begin{equation*}
	\mathcal{R}_{\alpha}^{n} = \{ \alpha'  \ : \ \langle D_{\alpha}^n g^0_{\alpha}, D_{\alpha'}^n g^0_{\alpha'} \rangle \neq 0 \}
\end{equation*}
and note that $ \mathcal{R}_{\alpha}^{n} \subseteq \mathcal{R}_{\alpha}^{n_{\mathrm{c}}} $ holds for all $n \leq n_{\mathrm{c}}$. Accordingly, \eqref{eq:entropy2} remains true if we multiply the right-hand side by the characteristic function $\chi(\alpha' \in \mathcal{R}_{\alpha}^{n_{\mathrm{c}}})$ of the set $\mathcal{R}_{\alpha}^{n_{\mathrm{c}}}$. The term inside the logarithm on the right-hand side of \eqref{eq:entropy1} therefore equals
\begin{equation*}
	\sum_{\alpha, \alpha'} \lambda_{\alpha} \chi(\alpha' \in \mathcal{R}_{\alpha}^{n_{\mathrm{c}}}) \frac{| \langle \mathcal{D}_{\alpha} g_{\alpha}^0, \mathcal{D}_{\alpha'} g_{\alpha'}^0 \rangle |^2}{\Vert \mathcal{D}_{\alpha} g_{\alpha}^0 \Vert^2 \Vert \mathcal{D}_{\alpha'} g_{\alpha'}^0 \Vert^2} \leq \sum_{\alpha, \alpha' } \lambda_{\alpha} \chi(\alpha' \in \mathcal{R}_{\alpha}^{n_{\mathrm{c}}})\,.
	\label{eq:entropy3}
\end{equation*}
To find a bound for the sum on the right-hand side, we have a closer look at the set $\mathcal{R}_{\alpha}^{n_{\mathrm{c}}}$.

Our goal is to find an upper bound for its cardinality. If $\langle D_{\alpha}^n g_{\alpha}^0, D_{\alpha'}^n g_{\alpha'}^0 \rangle \neq 0$ then the number of particles created by $D_{\alpha}^n$ and by $D_{\alpha'}^n$ in each mode with momentum in $P_{\mathrm{H}}$ need to be equal. Moreover, both operators annihilate $2^n$ particles with momenta in $P_{\A} \cup \{ 0 \}$, and afterwards, the resulting vectors need to agree. If we take the cut-off $\Theta^\a_{r,v,v'}$ in \eqref{eq:def_D} into account, we can additionally assume that only one particle is annihilated in each momentum mode $p \in P_{\A}$. This restriction does not apply to the condensate. However, both, $D_{\alpha}^n$ and $D_{\alpha'}^n$, need to annihilate the same number of particles in the $p=0$ mode. Therefore, if we count all possible ways, in which $k$ fermionic particles can be distributed over $| P_{\A} |$ energy levels (fermions because of the cut-off $\Theta^\a$), sum the result over $k$ from $0$ up to $2 n_{\mathrm{c}}$ (because each quartic operator can annihilate at most two particles with momentum in $P_{\A}$), and then take the square of the result (the operators $D^n_{\alpha}$ and $D^n_{\alpha'}$ alter both, $g_{\alpha}$ and $g_{\alpha'}$), we obtain an upper bound for $|\mathcal{R}_{\alpha}^{n_{\mathrm{c}}}|$.

Since there are $m!/(k!(m-k)!)$ ways to distribute $k$ fermionic particles over $m$ energy levels and we have $m = | P_{\A} |$, this implies the upper bound
\begin{align}
	&| \mathcal{R}_{\alpha}^{n_{\mathrm{c}}} | \leq \left( \sum_{k=0}^{2 n_{\mathrm{c}}} \frac{ | P_{\A} | ! }{k! ( | P_{\A} | - k )! } \right)^2 \leq \left( (2n_{\mathrm{c}}+1) \frac{ | P_{\A} | ! }{(2 n_{\mathrm{c}})! ( | P_{\A} | - 2 n_{\mathrm{c}} )! } \right)^2 \nonumber \\
	&\hspace{3cm} \leq C (2 n_{\mathrm{c}}+1)^2 \left( \frac{| P_{\A} |}{| P_{\A} | - 2 n_{\mathrm{c}}} \right)^{2(| P_{\A} | - 2 n_{\mathrm{c}} )} \left( \frac{|P_{\A}|}{2n_{\mathrm{c}}} \right)^{4 n_{\mathrm{c}}}.
	\label{eq:entropy4}	
\end{align}
To obtain the first bound we used $\binom{n}{k} \geq \binom{n}{k-1}$ for any $0 \leq k \leq \lceil n/2 \rceil$, and we used Stirling's approximation to obtain the second. Note that the bound in \eqref{eq:entropy4} is uniform in $\alpha$. In combination, the above considerations show that the second term on the right-hand side of \eqref{eq:entropy1} satisfies
\begin{equation}
	- \ln \left( \sum_{\alpha, \alpha'} \lambda_{\alpha} \frac{| \langle \mathcal{D}_{\alpha} g_{\alpha}^0, \mathcal{D}_{\alpha'} g_{\alpha'}^0 \rangle |^2}{\Vert \mathcal{D}_{\alpha} g_{\alpha}^0 \Vert^2 \Vert \mathcal{D}_{\alpha'} g_{\alpha'}^0 \Vert^2}   \right) \geq -C \big( 1 + n_{\mathrm{c}}(1 + \ln(|P_{\A}|))   \big).
	\label{eq:entropy5}
\end{equation}
In combination, \eqref{eq:entropy1} and \eqref{eq:entropy5} prove \eqref{eq:entropyLowerBound}.
\end{proof}

\medskip

\textbf{Acknowledgments.} A.D. gratefully acknowledges funding from the Swiss National Science Foundation (SNSF) through the Ambizione grant PZ00P2 185851.  G.B. and S.C. gratefully acknowledge financial support of the European Research Council through the ERC StG MaTCh, grant agreement n.\ 101117299.  G.B. gratefully acknowledges funding from the Italian Ministry of University and Research (MUR) and Next Generation EU through the Prin project 2022 - 2022CHELC7. C.B. gratefully acknowledges funding from the Italian Ministry of University and Research (MUR) and Next Generation EU through the PRIN 2022 project PRIN202223CBOCC\_01,
project code 2022AKRC5P.
C.B. also acknowledges support of Grant PID2024-156184NB-I00 funded by MICIU/AEI/10.13039/501100011033 and cofunded by the European Union.
G.B., C.B. and S.C. also warmly acknowledge the GNFM (Gruppo Nazionale per la Fisica Matematica) - INDAM.

\medskip

\appendix 

\begin{center}
	\huge \textsc{--- Appendix ---}
\end{center}	

\section{Approximating momentum sums by integrals}
\label{app:riemannSum}
In this section we prove the following lemma, which allows us to replace momentum sums by integrals. 

\begin{lemma}
	\label{lem:approximateMomentumSumsByIntegrals}
	Let $f : (0,\infty) \to \mathbb{R}$ be a nonnegative and monotone decreasing function and choose some $L > 0$, $\kappa > 0$. Then we have
	\begin{align}
		&\left( \frac{L}{2 \pi} \right)^3 \int_{|p| \geq \kappa} f(|p|) \left( 1 - \frac{3 \pi}{L |p|} \right) \mathrm{d} p \leq \sum_{p \in  \frac{2 \pi}{L} \mathbb{Z}^3 } f(|p|) \chi(|p| \geq \kappa) \nonumber \\
		&\hspace{3cm}\leq \left( \frac{L}{2 \pi} \right)^3 \int_{|p| \geq [\kappa - \sqrt{3} (2\pi/ L) ]_+} f(|p|) \left( 1 + \frac{3 \pi}{L |p|} + \frac{6 \pi}{L^2 p^2} \right) \mathrm{d} p.
		\label{eq:approximateSumsByIntegrals}
	\end{align}
\end{lemma}
\begin{proof}
	The proof of the upper bound in \eqref{eq:approximateSumsByIntegrals} can be found in \cite[Lemma~3.3]{DeuSei-20}, and we therefore only need to prove the lower bound. 
	
	Let us define the set $A = \{ p \in \mathbb{R}^3 \ | \ -2\pi/L \leq p_j \leq 0 \text{ for at least one } j \in \{1,2,3 \} \}$. We consider a decomposition of $\mathbb{R}^3$ into cubes with side length $2 \pi /L$. The corners of the cubes coincide with the points of the lattice $(2\pi/L) \mathbb{Z}^3$. Since $f$ is monotone decreasing, the smallest value of $f(|p|)$ in such a cube is obtained at the corner that is farthest away from the origin. Interpreting the sum in \eqref{eq:approximateSumsByIntegrals} as a Riemann sum, this allws us to show that
	\begin{equation}
		\sum_{p \in  (2 \pi/L) \mathbb{Z}^3 } f(|p|) 
		\chi(|p| \geq \kappa) \geq \left( \frac{L}{2 \pi} \right)^3 \int_{\mathbb{R}^3 \backslash A} f(|p|) \chi(|p| \geq \kappa) \mathrm{d} p.
		\label{eq:approximateSumsByIntegrals2}
	\end{equation}
	Using radial symmetry and monotonicity, we check that
	\begin{align}
		\int_{A } f(|p|) \chi(|p| \geq \kappa) \mathrm{d} p &\leq 3 \int_{ -2\pi/L \leq p_1 \leq 0} f(|p|) \chi(|p| \geq \kappa) \mathrm{d} p \nonumber \\
		&\leq \frac{6 \pi}{L} \int_{\mathbb{R}^2} f(|p|) \chi(|p| \geq \kappa) \mathrm{d} p. 
		\label{eq:approximateSumsByIntegrals3}
	\end{align} 
	We also have the identity
	\begin{equation}
		\left( \frac{L}{2\pi} \right)^2 \int_{\mathbb{R}^2}  f(|p|) \chi(|p| \geq \kappa) \mathrm{d} p = \left( \frac{L}{2\pi} \right)^3 \int_{\mathbb{R}^3}  f(|p|) \chi(|p| \geq \kappa) \frac{\pi}{L |p|} \mathrm{d} p.
		\label{eq:approximateSumsByIntegrals4}
	\end{equation}
	In combination, \eqref{eq:approximateSumsByIntegrals2}--\eqref{eq:approximateSumsByIntegrals4} prove the lower bound in \eqref{eq:approximateSumsByIntegrals}.
\end{proof}

\section{Properties of the ideal gas}
In this section we provide an estimate relating the expected number of particles in the condensate of the ideal gas in finite volume to the infinite volume expression $[\varrho - \varrho_{\mathrm{c}}]_+$. The precise statement is captured in the following lemma.

\begin{lemma}
	\label{lem:boundRho0}
	Let $N, L, \beta > 0$ and assume that $\mu_0(\beta,N,L) < 0$ is chosen such that
	\begin{equation}
		N = \sum_{p \in \frac{2 \pi}{L} \mathbb{Z}^3} \frac{1}{\exp(\beta(p^2 - \mu_0(\beta,N,L)))-1}
		\label{eq:appCondensateDensity1}
	\end{equation}
	holds. We denote the term in the sum on the right-hand side with $p=0$ times $L^{-3}$ by $\varrho_0(\beta,N,L)$. In the limit $N, L \to \infty$ with $N/L^3 = \varrho$ and $ c \varrho^{-2/3} \leq \beta \ll L^2 $ for some $c>0$, we have
	\begin{equation}
		\varrho_0(\beta,N,L) \geq \left[ \varrho - \varrho_{\mathrm{c}}(\beta) - \frac{C}{\beta L} | \ln(\beta^{1/2}/L) | \right]_+
		\label{eq:appCondensateDensity2}
	\end{equation}
	with $\varrho_{\mathrm{c}}$ in \eqref{eq:criticalDensity}. 
\end{lemma}
\begin{proof}
	We start our investigation by noting that an application of Lemma~\ref{lem:approximateMomentumSumsByIntegrals} shows
	\begin{align}
		&\left| \frac{1}{L^3} \sum_{p \in \frac{2 \pi}{L} \mathbb{Z}^3 \backslash \{ 0 \}} \frac{1}{\exp(\beta p^2)-1} - \left( \frac{1}{2 \pi} \right)^3 \int_{\mathbb{R}^3} \frac{1}{\exp(\beta p^2)-1} \mathrm{d} p \right| \nonumber \\ 
		&\hspace{3cm}\leq  C \left[ \frac{1} {\beta L}  + \frac{1}{\beta^{3/2}}  \int_{|p| \geq \frac{c \beta^{1/2} }{L}} \frac{1}{\exp(p^2)-1} \left( \frac{\beta^{1/2}}{L |p|} + \frac{\beta}{L^2 p^2} \right) \mathrm{d}p \right] \nonumber \\
		&\hspace{3cm}\leq \frac{C}{\beta L} | \ln(\beta^{1/2}/L) |. \label{eq:appCondensateDensity4}
	\end{align}
	Using \eqref{eq:appCondensateDensity1}, $\mu_0 < 0$, the monotonicity of the Bose distribution, and \eqref{eq:appCondensateDensity4}, we estimate
	\begin{align}
		\varrho_0(\beta,N,L) &\geq \varrho - \frac{1}{L^3} \sum_{p \in \frac{2 \pi}{L} \mathbb{Z}^3 \backslash \{ 0 \}} \frac{1}{\exp(\beta p^2)-1} \nonumber \\
		&\geq \varrho - \left( \frac{1}{2 \pi} \right)^3 \int_{\mathbb{R}^3} \frac{1}{\exp(\beta p^2)-1} \mathrm{d} p - \frac{C}{\beta L} | \ln(\beta^{1/2}/L) | \nonumber \\
		&= \varrho - \varrho_{\mathrm{c}}(\beta) - \frac{C}{\beta L} | \ln(\beta^{1/2}/L) |.
	\end{align}
	In combination with $\varrho_0 \geq 0$, this proves the claim.

\end{proof}
\section{Properties of the uncorrelated states} \label{app:Gamma0}
The aim of this appendix is to provide proofs for Lemmas~\ref{lm:cN} and \ref{lm:Gamma0}.
\subsection{Proof of Lemma \ref{lm:cN}}
In this section, we prove the bounds stated in Lemma~\ref{lm:cN}, which allow us to estimate expectations of number operators in the state $\G_\mathrm{G}$ defined in \eqref{eq:GibbsStateIdealGasWithCutoff}. This is not a straightforward task because $\Gamma_{\mathrm{G}}$ is not quasi-free. To overcome this difficulty, we exploit the following lemma, which allows us to quantify the influence of the operator $\chi(\mathcal{M} \geq M)$ in the definition of the state $\G_\mathrm{G}$ in \eqref{eq:GibbsStateIdealGasWithCutoff}. For the sake of simplicity of our notation, we write $\beta$ instead of $\beta_N$.

\begin{lemma}\label{lem:probabilisticLemma}
We recall the definition of $\cM$ in \eqref{eq:opM} and assume that the parameter $M>0$ satisfies \eqref{eq:conditionOnM}. Then we have
	\begin{equation}
		\Tr[ \chi(\mathcal{M} < M ) \,\wt{\G}_\mathrm{G} ] \leq \exp\left( - \frac{\left( \sum_{p \in P_{\A}} \exp(-\beta(p^2 - \mu_0(\beta,N))) - M \right)^2}{2 \sum_{p \in P_{\A}} \exp(-\beta(p^2 - \mu_0(\beta,N)))} \right).
		\label{eq:probabilisticLemma}
	\end{equation}
\end{lemma}
\begin{proof}
	An occupation number configuration is of the form $\mathbf{n}= ( n_p )_{p \in P_{\mathrm{G}}}$ with $n_p \in \mathbb{N}_0$ for all $p$. We also use the notation $|\textbf{n} \rangle \in \mathscr{F}_{\mathrm{G}}$ for the related occupation number vector. By
	\begin{equation}
		\textbf{P}(\mathbf{n}) = \Tr_{\mathscr{F}_{\mathrm{G}}}[ | \mathbf{n} \rangle \langle \mathbf{n} | \ \wt{\G}_\mathrm{G} ]
		\label{eq:probabilisticLemma1}
	\end{equation}
	we denote the probability that the occupation number configuration $\mathbf{n}$ occurs in the grand canonical ensemble described by the Gibbs state $\wt{\G}_\mathrm{G}$ in \eqref{eq:GibbsStateIdealGas}. The probability that the set $\mathcal{A}$ of occupation number configurations occurs is denoted by $\textbf{P}(\mathcal{A}) = \sum_{\bf{n} \in \mathcal{A}} \textbf{P}(\bf{n})$. 
	
	For $p \in P_{\mathrm{G}}$ we define the random variable $N_p(\textbf{n}) = n_p$, i.e., $N_p$ singles out the occupation number of the momentum $p$. We note that $\{ N_p \}_{p \in P_{\mathrm{G}} }$ is an independent family of geometric random variables with laws 
	\begin{equation}
		\textbf{P}(N_p = k) = \exp(-\beta(p^2-\mu_0)k)(1-\exp(-\beta(p^2-\mu_0))), \quad k \in \mathbb{N}_0,
		\label{eq:probabilisticLemma2}
	\end{equation}
	which is a direct consequence of the identity
	\begin{equation}
		\wt{\G}_\mathrm{G} = \prod_{p \in P_{\mathrm{G}}} \exp(-\beta(p^2-\mu_0) a_p^* a_p)(1-\exp(-\beta(p^2-\mu_0))).
	\end{equation}
	Note that we used the shortcut $\textbf{P}(N_p = k)$ to denote $\textbf{P}(\{{\bf n}\,|\, N_p({\bf n})=k\})$.
	
	This allows us to write 
	\begin{equation}
		\Tr[ \chi(\mathcal{M} < M ) \widetilde{\G}_G 
		 ]  = \textbf{P}( Y < M ) \quad \text{ with } \quad Y = \sum_{p \in P_{\A}} \chi(N_p \neq 0).
		\label{eq:probabilisticLemma2b}
	\end{equation}
	By definition, $\{ Z_p = \chi(N_p \neq 0) \}_{p \in P_{\mathrm{G}}}$ is an independent family of Bernoulli random variables with laws
	\begin{align}
		\mathbf{P}(Z_p = 0) &= 1 - \exp(-\beta(p^2 - \mu_0)), \nonumber \\
		\mathbf{P}(Z_p = 1) &= \exp(-\beta(p^2-\mu_0)). 
		\label{eq:probabilisticLemma3}
	\end{align}
	Using $Y=\sum_{p\in P_{\A}}Z_p$ and \eqref{eq:probabilisticLemma3}, we compute the expectation of $Y$ and find
	\begin{equation}
		\mathbf{E}(Y) = \sum_{p \in P_{\A}} \mathbf{E}(Z_p) = \sum_{p \in P_{\A}} \exp(-\beta(p^2 - \mu_0)).
		\label{eq:probabilisticLemma4}
	\end{equation}
	
	Our goal is to prove a tail bound for $Y$. For given $t > 0$ we estimate
	\begin{align}
		\mathbf{P}( Y < M) &\leq \mathbf{E} [ \exp(t( M - Y )) ] = \exp(t M) \prod_{p \in P_{\A}} \mathbf{E}[\exp(-t Z_p)] \nonumber \\
		&= \exp(t M ) \prod_{p \in P_{\A}} \left[ 1 + \exp(-\beta(p^2-\mu_0))(e^{-t}-1) \right] \nonumber  \\
		&= \exp\bigg(tM + \sum_{p \in P_{\A}} \left\{ \ln\left[ 1 + \exp(-\beta(p^2 - \mu_0))(e^{-t} - 1) \right] \right\} \bigg). \label{eq:probabilisticLemma5}
	\end{align}  
	To obtain the first and the second equality, we used the independence of the family $\{ Z_p \}_{p \in P_{\mathrm{G}}}$ and \eqref{eq:probabilisticLemma3}, respectively. Using $\ln(1+x) \leq x$ for $x > -1$ and $\exp(-t)-1 \leq t^2/2 - t$ for $t \geq 0$, we see that the term in the sum in the last line of \eqref{eq:probabilisticLemma5} is bounded from above by
	\begin{equation}
		\exp(-\beta(p^2-\mu_0))(e^{-t} - 1) \leq \exp(-\beta(p^2-\mu_0)) (t^2/2 - t).
		\label{eq:probabilisticLemma6}
	\end{equation}
	Putting \eqref{eq:probabilisticLemma4}--\eqref{eq:probabilisticLemma6} together, we find
	\begin{equation}
		\mathbf{P}( Y < M) \leq \inf_{t>0} \exp(t M + \mathbf{E}(Y)(t^2/2-t)).
		\label{eq:probabilisticLemma7}
	\end{equation}
	When we minimize the right-hand side of \eqref{eq:probabilisticLemma7} under the assumption $M < \mathbf{E}(Y)$ over $t>0$, we find that the minimum is attained at $t = (\mathbf{E}(Y)-M)/\mathbf{E}(Y)$, which yields the bound $\mathbf{P}( Y < M) \leq \exp(-(\mathbf{E}(Y)-M)^2/(2 \mathbf{E}(Y)))$. In combination, this bound, \eqref{eq:probabilisticLemma2b}, and \eqref{eq:probabilisticLemma4} prove the claim.
\end{proof}

\medskip

With the help of Lemma \ref{lem:probabilisticLemma}, we are well prepared to give the proof of Lemma~\ref{lm:cN}.

\begin{proof}[Proof of Lemma~\ref{lm:cN}]
	We start with the proof of \eqref{eq:propParticleNumberPowers}. In the first step we reduce the problem to the same problem with $\G_\mathrm{G}$ replaced by $\wt{\G}_\mathrm{G}$ in \eqref{eq:GibbsStateIdealGas}. We claim that $0 \leq -\beta \mu_0(\beta,N) \leq C$ for some $C>0$. If $N_0^{\mathrm{id}} \geq C' > 0$ this follows from the identity $\mu_0 = \beta^{-1}\ln(1+(N_0^{\mathrm{id}})^{-1})$. In contrast, if $N_0^{\mathrm{id}} \leq C'$ we use the bound $\beta \geq c \beta_{\mathrm{c}}$ with some $c > 0$ to see that 
	\begin{equation}
		0 < c \leq \beta^{3/2} N \simeq \left( \frac{1}{2\pi} \right)^3 \int_{\mathbb{R}^3} \frac{1}{\exp((p^2-\beta \mu_0)) - 1} \text{d} p
	\end{equation}
	(Here $\simeq$ denotes leading order asymptotics.). The right-hand side goes to zero for $-\beta \mu_0 \to \infty$. This causes a contradiction and allows us to conclude the claim. With our bound for $\mu_0$ it is not difficult to see that $\sum_{p \in P_{\A}} \exp(-\beta(p^2-\mu_0)) \sim \beta^{-3/2}$ ($\sim$ denotes asymptotic behavior without the correct constant).
	
	The above asymptotics of the momentum sum and an application of Lemma~\ref{lem:probabilisticLemma} allow us to obtain the bound 
	\begin{align}
		&\Tr_{\mathscr{F}_{\mathrm{G}}}[\chi(\mathcal{M} \geq M)  \exp(-\beta( \mathcal{H}_{\mathrm{G}} - \mu_0(\beta,N) \mathcal{N}_{\mathrm{G}} ))]  \nonumber \\
		&\hspace{4cm}\geq \Tr_{\mathscr{F}_{\mathrm{G}}}[ \exp(-\beta( \mathcal{H}_{\mathrm{G}} - \mu_0(\beta,N) \mathcal{N}_{\mathrm{G}} ))](1-e^{-c \beta^{-3/2}}) 
		\label{eq:propParticleNumberPowers1a}
	\end{align}
	for the partition function. Moreover, applications of \eqref{eq:propParticleNumberPowers1a} and the Cauchy--Schwarz inequality show
	\begin{align}
		&\sum_{v \in P_{\A}} \Tr_{\mathscr{F}_{\mathrm{G}}}\left[ \mathcal{N}^p \left( a_v^* a_v \right)^q \G_\mathrm{G} \right] \leq (1 + C e^{-c \beta^{-3/2} }) \sum_{v \in P_{\A}} \Tr_{\mathscr{F}_{\mathrm{G}}}\left[ \mathcal{N}^p \left( a_v^* a_v \right)^q \wt{\G}_\mathrm{G} \right] \label{eq:propParticleNumberPowers1} \\
		&\hspace{0.5cm}\leq (1 + C e^{-c \beta^{-3/2} }) \left( \Tr_{\mathscr{F}_{\mathrm{G}}}[ \mathcal{N}^{2p}\, \wt{\G}_\mathrm{G} ] \right)^{1/2} \left(  \sum_{v,w \in P_{\A}} \Tr_{\mathscr{F}_{\mathrm{G}}}[ (a^*_v a_v)^q (a^*_w a_w)^q \,\wt{\G}_\mathrm{G} ] \right)^{1/2}. \nonumber
	\end{align}
	A short computation yields $\Tr_{\mathscr{F}_{\mathrm{G}}}[\mathcal{N} \,\wt{\G}_\mathrm{G}] \leq C \beta^{-3/2}$. We use this and	\cite[Lemma~2.9]{CapDeu2024} to show that for any $p > 0$ there exists a constant $C_p$ such that 
	\begin{equation}
		\Tr_{\mathscr{F}_{\mathrm{G}}}[ \mathcal{N}^{2 p} \,\wt{\G}_\mathrm{G}] \leq C_p \beta^{-3 p }.
		\label{eq:propParticleNumberPowers2}
	\end{equation}
	In the proof of \cite[Lemma~2.9]{CapDeu2024} the dispersion relation $p \mapsto p^2 - \mu_0$ is replaced by a different dispersion relation for certain low momenta. It, however, applies also in our case. Eqs.~\eqref{eq:propParticleNumberPowers1} and \eqref{eq:propParticleNumberPowers2} prove \eqref{eq:propParticleNumberPowers} in the case $q=0$. From now on we therefore assume $q>0$.
	
	To obtain a bound for the third factor on the right-hand side of \eqref{eq:propParticleNumberPowers1}, we first estimate (for the notation see the proof of Lemma~\ref{lem:probabilisticLemma}) 
	\begin{align}
		\sum_{v,w \in P_{\A}} \Tr_{\mathscr{F}_{\mathrm{G}}}[ (a^*_v a_v)^q (a^*_w a_w)^q \,\wt{\G}_\mathrm{G} ] =& \sum_{v,w \in P_{\A}} \mathbf{E}( N_v^q N_w^q ) \nonumber \\
		\leq& \left( \sum_{v\in P_{\A}} \mathbf{E}( N_v^q ) \right)^2 + \sum_{v \in P_{\A}} \mathbf{E}( N_v^{2q} ). 
		\label{eq:propParticleNumberPowers3}
	\end{align}
	Note that we used the independence of the family $\{ N_p \}_{p \in P_{\mathrm{G}}}$ of random variables in \eqref{eq:probabilisticLemma2}. 
	An application of Lemma~\ref{lem:approximateMomentumSumsByIntegrals} shows
	\begin{align}
		&\sum_{v\in P_{\A}} \mathbf{E}( N_v^q ) = \sum_{n=1}^{\infty} n^q  \sum_{p \in P_{\A}} (1-\exp(-\beta(v^2-\mu_0))) \exp(-\beta(v^2-\mu_0) n) \nonumber  \\
		&\hspace{0.4cm}\leq C\left( \frac{1}{2 \pi \beta^{1/2}} \right)^{3} \sum_{n=1}^{\infty} n^q \int_{|v| \geq c \beta^{\delta_1/2}} (1-\exp(-v^2)) \exp(-v^2 n) \mathrm{d}v \label{eq:propParticleNumberPowers3b} \\
		&\hspace{0.4cm}\leq C \beta^{-3/2} \left[ \sum_{n=1}^{\infty} n^q \exp(-n/2) \int_{1}^{\infty} \exp(-v^2/2) v^2 \mathrm{d} v + \sum_{n=1}^{\infty} n^q \exp(-\beta^{\delta_1}n) \int_{\beta^{\delta_1/2}}^1 v^4 \mathrm{d}v \right]. \nonumber
	\end{align}
	The first term in the bracket on the right-hand side is bounded by a constant. When we bound the sum in the second term by an integral, we find that it is smaller than a constant times
	\begin{equation}
		\beta^{5\delta_1/2} \int_0^{\infty} x^q \exp(-\beta^{\delta_1}x) \mathrm{d}x \leq C \beta^{-\delta_1(2q-3)/2}.
		\label{eq:propParticleNumberPowers3c}
	\end{equation}
	In combination, \eqref{eq:propParticleNumberPowers3b} and \eqref{eq:propParticleNumberPowers3c} show
	\begin{equation}
		\sum_{v \in P_{\A}} \mathbf{E}( N_v^q ) \leq C_q \beta^{-3/2} \left( 1 + \beta^{-\delta_{1} (2q-3)/2} \right)
		\label{eq:propParticleNumberPowers8}
	\end{equation}
	for $q>0$. 
	
	In the last step we put \eqref{eq:propParticleNumberPowers1}--\eqref{eq:propParticleNumberPowers3}, and \eqref{eq:propParticleNumberPowers8} together, use $\delta_1 < 1$  and find \eqref{eq:propParticleNumberPowers} in the case $q>0$. This proves \eqref{eq:propParticleNumberPowers} and it remains to consider \eqref{eq:numberOfUndressedParticles} and \eqref{eq:propParticleNumberPowersKineticAndVariance}.
	
	We start the proof of \eqref{eq:propParticleNumberPowersKineticAndVariance} by noting that 
	\begin{align}
		\Tr_{\mathscr{F}_{\mathrm{G}}}\left[ \mathcal{N}_{\A} \G_\mathrm{G} \right] &\geq \Tr_{\mathscr{F}_{\mathrm{G}}}[ \mathcal{N}_{\A} \widetilde{\G}_\mathrm{G} ] - \frac{\Tr_{\mathscr{F}_{\mathrm{G}}}\left[ \mathcal{N}_{\A} \exp(-\beta( \mathcal{H}_{\mathrm{G}} - \mu_0(\beta,N) \mathcal{N}_{\mathrm{G}} )) \chi(\mathcal{M} < M) \right]}{\Tr_{\mathscr{F}_{\mathrm{G}}}[ \exp(-\beta( \mathcal{H}_{\mathrm{G}} - \mu_0(\beta,N) \mathcal{N}_{\mathrm{G}} )) ]} \nonumber \\
		&\geq \Tr_{\mathscr{F}_{\mathrm{G}}}[ \mathcal{N}_{\A} \widetilde{\G}_\mathrm{G} ] - \sqrt{ \Tr_{\mathscr{F}_{\mathrm{G}}}[ \mathcal{N}_{\A}^2 \widetilde{\G}_\mathrm{G} ] \ \Tr_{\mathscr{F}_{\mathrm{G}}}[ \chi(\mathcal{M} < M) \widetilde{\G}_\mathrm{G} ] } \nonumber \\
		&\geq \Tr_{\mathscr{F}_{\mathrm{G}}}[ \mathcal{N}_{\A} \widetilde{\G}_\mathrm{G} ] - C e^{-c \beta^{-3/2}}
		\label{eq:propParticleNumberPowers8b}
	\end{align}
	holds. To obtain the second bound we applied the Cauchy--Schwarz inequality, and the third bound follows from applications of Lemma~\ref{lem:probabilisticLemma} and \eqref{eq:propParticleNumberPowers}. Another application of Lemma~\ref{lem:probabilisticLemma}, \eqref{eq:propParticleNumberPowers} and \eqref{eq:propParticleNumberPowers8b} shows
	\begin{equation}
		\Tr_{\mathscr{F}_{\mathrm{G}}}[\cN_{\A}^2 \Gamma_{\mathrm{G}}]-(\Tr_{\mathscr{F}_{\mathrm{G}}}[\cN_{\A} \Gamma_\mathrm{G}])^2 \leq \Tr_{\mathscr{F}_{\mathrm{G}}}[\cN_{\A}^2 \widetilde{\Gamma}_{\mathrm{G}}]-(\Tr_{\mathscr{F}_{\mathrm{G}}}[\cN_{\A} \widetilde{\Gamma}_\mathrm{G}])^2 + C e^{-c \beta^{-3/2}}.
		\label{eq:propParticleNumberPowers8c}
	\end{equation}
	The variance of $\mathcal{N}_{\A}$ in the state $\widetilde{\Gamma}_{\mathrm{G}}$ is given by $(-1/\beta)$ times the second derivative of the grand potential (when the sums in all formulas are restricted to momenta in $P_{\A}$) with respect to $\mu_0$, in formulas,
	\begin{equation}
		\Tr_{\mathscr{F}_{\mathrm{G}}}[\cN_{\A}^2 \widetilde{\Gamma}_{\mathrm{G}}]-(\Tr_{\mathscr{F}_{\mathrm{G}}}[\cN_{\A} \widetilde{\Gamma}_\mathrm{G}])^2 = -\frac{1}{\beta^2} \frac{\partial^2}{\partial \mu_0^2} \sum_{p\in P_{\A}} \ln\left( 1-\exp(-\beta(p^2-\mu_0)) \right).
	\end{equation}
	Using this and $\sinh(x) \geq x$ for $x \geq 0$, we check that 
	\begin{equation}
		\Tr_{\mathscr{F}_{\mathrm{G}}}[\cN_{\A}^2 \widetilde{\Gamma}_{\mathrm{G}}]-(\Tr_{\mathscr{F}_{\mathrm{G}}}[\cN_{\A} \widetilde{\Gamma}_\mathrm{G}])^2 = \sum_{p \in P_{\A}} \frac{1}{4 \sinh^2\left( \frac{\beta(p^2-\mu_0)}{2} \right)} \leq \frac{1}{\beta^2} \sum_{p \neq 0} \frac{1}{|p|^4}.
		\label{eq:propParticleNumberPowers8d}
	\end{equation}
	In combination, \eqref{eq:propParticleNumberPowers8c} and \eqref{eq:propParticleNumberPowers8d} prove \eqref{eq:propParticleNumberPowersKineticAndVariance}. 
	
	To prove \eqref{eq:numberOfUndressedParticles} we apply Lemma~\ref{lem:probabilisticLemma} to estimate
	\begin{equation}
		\sum_{p \in P_{\mathrm{G}} \backslash P_{\A}}\Tr_{\mathscr{F}_{\mathrm{G}}}[a_p^*a_p \Gamma_{\mathrm{G}}] \leq \sum_{p \in P_{\mathrm{G}} \backslash P_{\A}}\Tr_{\mathscr{F}_{\mathrm{G}}}[a_p^*a_p \widetilde{\Gamma}_{\mathrm{G}}](1 + C \exp(-c\beta^{-3/2})). 
	\end{equation}
	When combined with the bound
	\begin{align}
		\sum_{\substack{p \in 2 \pi \mathbb{Z}^3: \\ 0 < |p| < \beta^{-1/2(1-\delta_1)}} } \frac{1}{\exp(\beta(p^2-\mu_0(\beta,N)))-1} &\leq \frac{1}{\beta} \sum_{\substack{p \in 2 \pi \mathbb{Z}^3: \\ 0 < |p| < \beta^{-1/2(1-\delta_1)}} } \frac{1}{p^2} \nonumber \\
		&\leq C \beta^{-3/2+\delta_1/2},
	\end{align}
	which follows from $(\exp(x)-1)^{-1} \leq x^{-1}$ for $x > 0$ and $\mu_0 < 0$, this proves \eqref{eq:numberOfUndressedParticles}. 
\end{proof}

\subsection{Proof of Lemma \ref{lm:Gamma0}}
In this Section we give the proof of Lemma~\ref{lm:Gamma0}. As a preparation, we first prove the following lemma, which provides us with a lower bound for the entropy of $\Gamma_{\mathrm{G}}$ in terms of that of $\widetilde{\Gamma}_{\mathrm{G}}$ in \eqref{eq:GibbsStateIdealGas}. 

\begin{lemma}
	\label{lem:removeCutoffEntropy}	 
	We consider the limit $N \to \infty$, $1 \gg \beta \geq b \beta_{\mathrm{c}}(N)$ with some $b>0$ and $\beta_{\mathrm{c}}$ in \eqref{eq:criticalTemperatureIdealGas}, and assume that \eqref{eq:conditionOnM} holds. Then there are two constants $C,c > 0$ such that the entropies of the states $\Gamma_{\mathrm{G}}$ and $\widetilde{\Gamma}_{\mathrm{G}}$ are related by the inequality
	\begin{equation} 
		S( \Gamma_{\mathrm{G}} ) \geq S(\widetilde{\Gamma}_{\mathrm{G}}) - C \exp( - c \beta^{-3/2} ).
		\label{eq:probabilisticEntropy1}
	\end{equation}
\end{lemma}
\begin{proof}
	Let us denote the right-hand side of \eqref{eq:probabilisticLemma} by $R$, which satisfies $R \leq c < 1$ with some constant $c$ by assumption. We also introduce the notation $Z(M) = \Tr[ \chi(\mathcal{M} \geq M) \widetilde{\Gamma}_{\mathrm{G}}]$ and note that $0 \leq 1 - Z(M) \leq R$. The entropy of $\Gamma_{\mathrm{G}}$ can be written as
	\begin{align}
		S( \Gamma_{\mathrm{G}} ) &= - \Tr[ \chi(\mathcal{M} \geq M) \widetilde{\Gamma}_{\mathrm{G}} \ln( \widetilde{\Gamma}_{\mathrm{G}}/Z(M) ) ]/Z(M) \label{eq:probabilisticEntropy2} \\
		&= - \Tr[ \widetilde{\Gamma}_{\mathrm{G}} \ln( \widetilde{\Gamma}_{\mathrm{G}} ) ]/Z(M) + \Tr[ \chi(\mathcal{M} < M) \widetilde{\Gamma}_{\mathrm{G}} \ln( \widetilde{\Gamma}_{\mathrm{G}} ) ]/Z(M) + \ln(Z(M)). \nonumber
	\end{align}
	The first term on the right-hand side is bounded from below by $S(\widetilde{\Gamma}_{\mathrm{G}})$ because $0 < Z(M) \leq 1$. For the third term we have the lower bound $\ln(Z(M)) \geq \ln(1-R) \geq -C R$. 
	
	To derive a bound for the second term, we first note that it equals
	\begin{align}
		&-\beta \Tr[ \chi(\mathcal{M} < M) ( \mathcal{H}_{\mathrm{G}} - \mu_0 \mathcal{N}_{\mathrm{G}} ) \widetilde{\Gamma}_{\mathrm{G}} ]/Z(M) \nonumber \\
		&\hspace{1cm} - \Tr[ \chi(\mathcal{M} < M) \widetilde{\Gamma}_{\mathrm{G}} ] \ln( \Tr[ \exp(-\beta( \mathcal{H}_{\mathrm{G}} - \mu_0 \mathcal{N}_{\mathrm{G}} )) ]) /Z(M). \label{eq:probabilisticEntropy3}
	\end{align}
	The term in the first line is bounded from above by
	\begin{equation}
		\beta \sqrt{ \Tr[ \chi(\mathcal{M} < M) \widetilde{\Gamma}_{\mathrm{G}} ] } \sqrt{ \Tr[ ( \mathcal{H}_{\mathrm{G}} - \mu_0 \mathcal{N}_{\mathrm{G}} )^2 \widetilde{\Gamma}_{\mathrm{G}} ] } /Z(M) %
		\label{eq:probabilisticEntropy4}
	\end{equation}
	where the term inside the second square root can be estimated as follows
	\begin{align}
		&\Tr[ ( \mathcal{H}_{\mathrm{G}} - \mu_0 \mathcal{N}_{\mathrm{G}} )^2 \widetilde{\Gamma}_{\mathrm{G}} ] = \sum_{p,q \in P_{\mathrm{G}}} (p^2-\mu_0) (q^2-\mu_0) \Tr[a_p^* a_p a_q^* a_q \widetilde{\Gamma}_{\mathrm{G}}] \nonumber \\
		&\hspace{2cm}= \sum_{p,q \in P_{\mathrm{G}} \atop p \neq q} (p^2-\mu)(q^2-\mu) \mathbf{E}(N_p) \mathbf{E}(N_q) + \sum_{p \in P_{\mathrm{G}}} (p^2-\mu)^2 \mathbf{E}(N_p^2) \nonumber \\
		&\hspace{2cm} \leq \left( \sum_{p \in P_{\mathrm{G}}} \frac{p^2-\mu_0}{\exp(\beta(p^2-\mu_0))-1} \right)^2 + \frac{1}{2} \sum_{p \in P_{\mathrm{G}}} \frac{(p^2-\mu_0)^2}{\sinh^2\left( \frac{\beta(p^2-\mu)}{2} \right) }.
		\label{eq:probabilisticEntropy4b}
	\end{align}
	To come to the second line we used the independence of the family $\{ N_p \}_{p \in P_{\mathrm{G}}}$ of random variables in \eqref{eq:probabilisticLemma2}, and the bound in the third line follows from 
	\begin{equation}
		\mathbf{E}(N_p) = \frac{1}{\exp(\beta(p^2-\mu_0)) -1 } \quad \text{ and } \quad \mathbf{E}(N_p^2) = \frac{1+\exp(-\beta(p^2-\mu_0))}{4 \sinh^2(\beta(p^2-\mu_0)/2)}.
		\label{eq:probabilisticEntropy4c}
	\end{equation}
	An application of Lemma~\ref{lem:approximateMomentumSumsByIntegrals} shows that the right-hand side of \eqref{eq:probabilisticEntropy4b} is bounded from above by a constant times $\beta^{-5}$. When we additionally use Lemma~\ref{lem:probabilisticLemma} to bound the term inside the first square root on the right-hand side of \eqref{eq:probabilisticEntropy4}, we find
	\begin{equation}
		\beta \sqrt{ \Tr[ \chi(\mathcal{M} < M) \widetilde{\Gamma}_{\mathrm{G}} ] } \sqrt{ \Tr[ ( \mathcal{H}_{\mathrm{G}} - \mu_0 \mathcal{N}_{\A} )^2 \widetilde{\Gamma}_{\mathrm{G}} ] } /Z(M) \leq C \sqrt{R} \beta^{-3/2}.
		\label{eq:probabilisticEntropy4d}
	\end{equation}
	
	Finally, another application of Lemmas~\ref{lem:approximateMomentumSumsByIntegrals} and \ref{lem:probabilisticLemma} shows that the term in the second line of \eqref{eq:probabilisticEntropy3} is bounded from above by a constant times
	\begin{equation}
		- R \sum_{p \in P_{\mathrm{G}}} \ln( 1 - \exp(-\beta(p^2 - \mu_0)) ) \leq C R \beta^{-3/2}. 
		\label{eq:probabilisticEntropy5}
	\end{equation}
	In combination, these considerations and \eqref{eq:conditionOnM} prove the claim.
\end{proof}

We are now prepared to give the proof of Lemma~\ref{lm:Gamma0}.

\begin{proof}[Proof of Lemma~\ref{lm:Gamma0}]

In this proof we use notation from the proof of Lemma~\ref{lem:probabilisticLemma} without additional reference and denote the right-hand side of \eqref{eq:probabilisticLemma} by $R$. We start with the proof of \eqref{lem:N0UncorrelatedTrialState}. The parameter $N_0$ is chosen such that
\begin{equation}
	N = \Tr[\mathcal{N} \Gamma_0] = N_0 + \Tr[\mathcal{N} \Gamma_{\mathrm{G}}]
	\label{eq:N0TrialState1}
\end{equation}
holds  (compare also with \eqref{eq:Nplus}), where
\begin{equation}
	\Tr[\mathcal{N} \Gamma_{\mathrm{G}}] = \frac{\sum_{\mathbf{n}} |\mathbf{n}| \exp(-\beta \sum_{p \in P_{\mathrm{G}}} (p^2-\mu_0) n_p ) \chi(Y(n) \geq M)  }{\sum_{\mathbf{n}}  \exp(-\beta \sum_{p \in P_{\mathrm{G}}} (p^2-\mu_0) n_p ) \chi(Y(n) \geq M) }
	\label{eq:N0TrialState2}
\end{equation}
with $|n| = \sum_{p \in P_{\mathrm{G}}} n_p$. Our goal is to get rid of the characteristic function $\chi(Y(n) \geq M)$. Assume that $\mathbf{n}$ is an occupation number configuration with $|\mathbf{n}| \geq 1$. If we add this configuration to both sums on the right-hand side of \eqref{eq:N0TrialState2} the term increases. If we additionally add $1$ to the numerator to compensate for the weight of the vector $\mathbf{n}=(0,0,0,...)$ that we need to add in the denominator we find the inequality
\begin{align}
	&\frac{\sum_{\mathbf{n}} |\mathbf{n}| \exp(-\beta \sum_{p \in P_{\mathrm{G}}} (p^2-\mu_0) n_p ) \chi(Y(n) \geq M)  }{\sum_{\mathbf{n}}  \exp(-\beta \sum_{p \in P_{\mathrm{G}}} (p^2-\mu_0) n_p ) \chi(Y(n) \geq M) } \nonumber \\
	&\hspace{3cm} \leq \frac{1 + \sum_{\mathbf{n}} |\mathbf{n}| \exp(-\beta \sum_{p \in P_{\mathrm{G}}} (p^2-\mu_0) n_p )   }{\sum_{\mathbf{n}}  \exp(-\beta \sum_{p \in P_{\mathrm{G}}} (p^2-\mu_0) n_p ) }.
	\label{eq:N0TrialState3}
\end{align}
In combination with \eqref{eq:N0TrialState2} and the fact that the denominator is bounded from below by one, \eqref{eq:N0TrialState3} proves 
\begin{align}
	\Tr[\mathcal{N} \Gamma_{\mathrm{G}}] \leq 1 + \Tr[\mathcal{N} \widetilde{\Gamma}_{\mathrm{G}}] &= 1 + \sum_{p \in P_{\mathrm{G}}} \frac{1}{\exp(\beta(p^2-\mu_0(\beta,N)))-1} \nonumber \\
	&\leq 1 + \sum_{p \in \Lambda^* \backslash \{ 0 \}} \frac{1}{\exp(\beta(p^2-\mu_0(\beta,N)))-1}.
	\label{eq:N0TrialState4}
\end{align} 
Eq.~\eqref{lem:N0UncorrelatedTrialState} is a direct consequence of \eqref{eq:N0TrialState1}, \eqref{eq:N0TrialState4}, the definition of $N_0^{\mathrm{id}}$ in \eqref{eq:condensateNumberIdealGasTorus} and \eqref{eq:chemicalPotentialIdealGasTorus}. It remains to prove \eqref{lem:energyUncorrelatedTrialState}. 
 
We need to derive an upper bound for the expectation of the energy in the state $\Gamma_0$. In the following we use notation that has been introduced above \eqref{eq:probabilisticEntropy2}. We also denote $\gamma_0(p) = (\exp(\beta(p^2-\mu_0))-1)^{-1}$. With $\mathcal{H}_N \geq 0$ and Lemma~\ref{lem:probabilisticLemma}, we check that
\begin{equation}
	\Tr[\mathcal{H}_N \Gamma_0] \leq \Tr[\mathcal{H}_N \widetilde{\Gamma}_0](1+ C R).
	\label{eq:EnergyTrialState0}
\end{equation}
Here $\widetilde{\Gamma}_0$ equals the state $\Gamma_0$ with $\Gamma_{\mathrm{G}}$ replaced by $\widetilde{\Gamma}_{\mathrm{G}}$.

A straightforward computation that uses the Wick rule shows
\begin{align}
	\Tr[ \mathcal{H}_N \widetilde{\Gamma}_0 ] =& \sum_{p \in P_{\mathrm{G}}} p^2 \gamma_0(p) + \frac{N_0}{N^{1-\kappa}} \sum_{p \in P_{\mathrm{G}}}  \widehat{V}( p/N^{1-\kappa} )\gamma_0(p) + \frac{\widehat{V}(0)}{2 N^{1-\kappa}} N_0 (N_0 -1)   \nonumber \\
	&+ \frac{\widehat{V}(0)}{2 N^{1-\kappa}} \left[ 2 N_0 \sum_{u \in P_{\mathrm{G}}} \gamma_0(u) + \left( \sum_{u \in P_{\mathrm{G}}} \gamma_0(u) \right)^2 + \sum_{u \in P_{\mathrm{G}}} \gamma_0^2(u) \right] \nonumber \\
	&+ \frac{1}{2 N^{1-\kappa}} \sum_{p \in \Lambda^* \backslash \{ 0 \}; u,u+p \in P_{\mathrm{G}} } \widehat{V}( p/N^{1-\kappa} ) \gamma_0(u+p) \gamma_0(u).
	\label{eq:EnergyTrialState1}
\end{align}
Let us denote the inverse Fourier transform of $\gamma_0(p) \chi(p \in P_{\mathrm{G}})$ by $g(x)$. The second term on the right-hand side of \eqref{eq:EnergyTrialState1} can be written as
\begin{align}
	N_0 \int_{[-1/2,1/2]^3} N^{2(1-\kappa)} V(N^{1-\kappa} x) g(x) \mathrm{d}x &\leq \frac{N_0}{N^{1-\kappa}} \int_{\mathbb{R}^3} V(x) \mathrm{d} x \sum_{p \in P_{\mathrm{G}}} \gamma_0(p) \nonumber \\
	&\leq  \frac{\widehat{V}(0)}{N^{1-\kappa}} N_0 (N-N_0+ C e^{-c \beta^{-3/2}}  ).
	\label{eq:EnergyTrialState2}
\end{align}
To obtain the second bound, we used
\begin{align}
	\sum_{p \in P_{\mathrm{G}}} \gamma_0(p) &\leq \Tr[ \mathcal{N}_+ \Gamma_{\mathrm{G}} ] + \sqrt{\Tr[ \mathcal{N}_+^2 \Gamma_{\mathrm{G}} ] \Tr[ \chi(\mathcal{M} < M) \Gamma_{\mathrm{G}} ]} \nonumber \\
	&\leq N-N_0 + C e^{-c \beta^{-3/2}}.
	\label{eq:EnergyTrialState2b}
\end{align}
The first bound in \eqref{eq:EnergyTrialState2b} follows from the Cauchy--Schwarz inequality and the second from the bound we used in the last step in \eqref{eq:propParticleNumberPowers8b}.

Similarly, we check that the term in the last line of \eqref{eq:EnergyTrialState1} satisfies
\begin{equation}
	\frac{1}{2} \int_{[-1/2,1/2]^3} N^{2(1-\kappa)} V(N^{1-\kappa} x)  g^2(x) \mathrm{d}x \leq \frac{\widehat{V}(0)(N-N_0+C e^{-c \beta^{-3/2}})^2}{N^{1-\kappa}}.
	\label{eq:EnergyTrialState3}
\end{equation}
Finally, using $\gamma_0(p) \leq 1/(\beta p^2)$, we check that the last term in the second line of \eqref{eq:EnergyTrialState1} is bounded from above by
\begin{equation}
	\frac{\widehat{V}(0)}{2 \beta^2 N^{1-\kappa}} \sum_{u \in 2 \pi \mathbb{Z}^3 \backslash \{ 0 \}} \frac{1}{|p|^4}. 
	\label{eq:EnergyTrialState4}
\end{equation}
In combination, the above considerations and \eqref{lem:N0UncorrelatedTrialState} show
\begin{equation}
	\Tr[\mathcal{H}_N \Gamma_0] \leq \sum_{p \in P_{\mathrm{G}} } p^2 \gamma_0(p) + \frac{\widehat{V}(0)}{2 N^{1-\kappa}} ( 2 N^2 - (N^{\mathrm{id}}_0)^2 ) + C N^{\kappa} ( 1 + \beta^{-2} N^{-1} + e^{-c \beta^{-3/2}} ).
	\label{eq:EnergyTrialState5}
\end{equation}
An application of \eqref{lem:removeCutoffEntropy} allows us to bound the entropy of $\Gamma_0$ in terms of that of $\widetilde{\Gamma}_0$, which results in an additional error term that is bounded by a constant times $\exp(-c \beta^{-3/2})$.

In the last step we put the kinetic energy and the entropy of $\widetilde{\Gamma}_0$ together and remove the condition $p \in P_{\mathrm{G}}$ in the relevant momentum sums. Since $\widetilde{\Gamma}_0 = |N_0 \rangle \langle N_0 | \otimes \widetilde{\Gamma}_{\mathrm{G}}$ and $S(|N_0 \rangle \langle N_0 | \otimes \widetilde{\Gamma}_{\mathrm{G}}) = S(\widetilde{\Gamma}_{\mathrm{G}})$ we have
\begin{align}
	\sum_{p \in P_{\mathrm{G}} } p^2 \Tr[ a_p^* a_p \widetilde{\Gamma}_0 ] - \frac{1}{\beta} S(\widetilde{\Gamma}_0) = \frac{1}{\beta} \sum_{p \in P_{\mathrm{G}}} \ln\left( 1 - \exp(\beta(p^2 - \mu_0)) \right) + \mu_0 \sum_{p \in P_{\mathrm{G}}} \gamma_0(p).
	\label{eq:EnergyTrialState6}
\end{align}
Both terms decrease if we add more terms to the momentum sum. That is, we have to estimate the error for this replacement. Using the exponential decay of the above function, it is not difficult to check that there exists a constant $c>0$ such that the sum of all terms with momenta  $|p| > \beta^{-1/2(1 + \delta_2)}$ is bounded by a constant times $\exp(-c \beta^{-\delta_2})$. Moreover, the term with $p = 0$ is bounded by a constant times $\ln(N)/\beta$.

Putting together these bounds, \eqref{eq:EnergyTrialState5}, and \eqref{eq:EnergyTrialState6}, 
we find
\begin{align}
	\Tr (\cH_N \G_0) - \b^{-1} S(\G_0) \leq&  F_0(\b, N)  +\frac 12 N^{\k-1} \widehat V(0) \big(2 N^2 - [N_0^{\rm{id}}(\beta,N)]^2 \big) \nonumber \\
	&\hspace{-0.8cm}+ C \bigg\{  N^{\kappa} ( 1 + \beta^{-2} N^{-1} + e^{-c \beta^{-3/2}} ) + \frac{e^{-c \beta^{-\delta_2}} + \ln(N)}{\beta} \bigg\}.  
	\label{lem:energyUncorrelatedTrialState-v0}
\end{align} 
When we use $\beta \leq N^{\sigma}$, $\delta_2 > 0$ and $\kappa > 1/3$, this bound implies \eqref{lem:energyUncorrelatedTrialState}. 

\end{proof}

\section{Properties of the solution to the scattering equation}
\label{app:scatteringEquation}
In this appendix we prove statements related to the functions $f_{\ell}$ in \eqref{eq:scatl} and $\eta$ in \eqref{eq:etap}.
\subsection{Proof of Lemma~\ref{lm:scatt-eq}}
The proof of items i)--iii) in Lemma~\ref{lm:scatt-eq} can be found in \cite[Appendix A]{BBCS3} (replacing $N$ by $N^{1-\k}$). To show iv) we observe that
\begin{align}
	&\sum_{r\in \L^*_+} N|(\widehat V_N\ast \widehat f_{N,\ell})(r-p)||r|^{-2} \nonumber \\
	&\hspace{3cm}\leq C N^\k \hskip -0.2cm \sum_{\substack{r\in \L^*_+: \\|r|\leq N^{1-\k}}}\hskip -0.2cm |r|^{-2} +  N\int_{|\tl r|\geq 1}\hskip -0.1cm |(\widehat{Vf_{\ell }})(\tl r-\tl p)||\tl r|^{-2} d \tl r
	\label{eq:appScatteringEquation1}
\end{align}
where we defined $\tl r=r/N^{1-\k}, \tl p=p/N^{1-\k}$ in the second term. We also
used that the sum over  $\tl r\in \L^*/N^{1-\k}$ can be interpreted as a Riemann sum that approximates the integral in \eqref{eq:appScatteringEquation1}. On the other hand,  we have
\begin{align}
\int_{|r|\geq 1} |(\widehat{Vf_{\ell }})(r-\tl p)||r|^{-2} d r 
\leq C \|\widehat {Vf_{\ell }}\|_q\, \Big(\int_{|r|\geq 1} |r|^{-2q'} dr \Big)^{1/q'} \leq C \|\widehat {V f_{\ell }}\|_q
\end{align}
for any $q<3$, where $q'=q/(q-1)$. With the assumption $V \in L^{q'}(\bR^3)$ for some $q'> 3/2$, then \eqref{BCS-bounds} follows by the Hausdorff-Young inequality using $|f_{\ell }|\leq 1$.  

The bound involving $|\widehat V_N(r-p)|$ follows from a similar argument.
\subsection{Proof of Lemma~\ref{lm:eta}}
The bounds for $\| \eta_H \|_2$, $\| \check{\eta}_H \|_{H^1}$ and $\| \eta_H \|_\io$ were shown in \cite[Lemma 2.4]{BaCS2021}. It therefore only remains to show the bound for $\|\eta_H\|_1$. With \eqref{eq:scatl-etap}, \eqref{BCS-bounds} and $N^{3-3\k}\l_\ell \leq C$ (which follows from Lemma~\ref{lm:scatt-eq}, i)) we obtain
		\[ \label{eq:etaH-L1}
		\sum_{r \in P_{\mathrm{H}}} |\eta_r| \leq  C N + C \sum_{r\in P_{\mathrm{H}}} {  \big(\widehat \chi_\ell \ast \widehat f_{N,\ell}\big)(r)} |r|^{-2} \leq C N\,.
		\]
		To bound the second term we used that $\|\widehat{\chi}_\ell \ast \widehat{f}_{N,\ell} \|_2 = \| \chi_\ell f_{N,\ell}\|_2 \leq C $.
	


\vspace{0.5cm} 

\noindent (Giulia Basti), Department of Mathematics, Sapienza University of Rome \\
Piazzale Aldo Moro 5, 00185 Roma, Italy \\
E-mail address: \texttt{giulia.basti@uniroma1.it} \\

\vspace{-0.1cm} 

\noindent (Chiara Boccato), Department of Mathematics, University of Pisa \\
Via Buonarroti 1/c, 56127 Pisa, Italy \\
E-mail address: \texttt{chiara.boccato@unipi.it} \\

\vspace{-0.1cm} 

\noindent (Serena Cenatiempo), Mathematics Division, Gran Sasso Science Institute \\
Viale Rendina 24, 67100 L'Aquila, Italy \\
E-mail address: \texttt{serena.cenatiempo@gssi.it} \\

\vspace{-0.1cm} 

\noindent (Andreas Deuchert) Department of Mathematics, Virginia Tech \\ 
225 Stanger Street, Blacksburg, VA 24060-1026, USA \\ 
E-mail address: \texttt{andreas.deuchert@vt.edu} \\


\begin{thebibliography}{55}

\def\bibskip{\\[-0.66cm]}

\bibitem{Basti2022} G. Basti, \textit{A second order upper bound on the ground state energy of a Bose gas beyond the Gross--Pitaevskii regime}, J. Math. Phys. \textbf{63}, 071902 (2022) \bibskip

\bibitem{BCOPS2023} G. Basti, S. Cenatiempo, A. Olgiati, G. Pasqualetti, B. Schlein, \textit{A Second Order Upper Bound for the Ground State Energy of a Hard-Sphere Gas in the Gross–Pitaevskii Regime}, Commun. Math. Phys. \textbf{399}, 1–55 (2023) \bibskip

\bibitem{BaCS2021} G. Basti, S. Cenatiempo, B. Schlein, {\it A new second order upper bound for the ground state energy of dilute Bose gases}, {Forum Math. Sigma} {\bf 9}, e74 (2021) \bibskip

\bibitem{BGOPS24} G. Basti, S. Cenatiempo,  A. Giuliani, A. Olgiati, G. Pasqualetti, B. Schlein,  {\it Upper bound for the ground state energy of a dilute Bose gas of hard spheres},
{Arch. Ration. Mech. Anal} {\bf 248}, 100 (2024) \bibskip

\bibitem{Bijl}
 A. Bijl, {\it The lowest wave function of the symmetrical many particles system}, Physica {\bf 7} no. 9, 869--886, (1940)  \bibskip

\bibitem{BocBreCeSchl2019} C. Boccato, C. Brennecke, S. Cenatiempo, B. Schlein, \textit{Bogoliubov Theory in the Gross--Pitaevskii Limit}, Acta Math. \textbf{222}, 219--335 (2019)  \bibskip

\bibitem{BocDeuSto2024} C. Boccato, A. Deuchert, D. Stocker, \textit{Upper bound for the grand canonical free energy of the Bose gas in the Gross--Pitaevskii limit}, SIAM J. Math. Anal. \textbf{56}, No. 2, 2611--2660 (2024)  \bibskip

\bibitem{BBCS3}
C. Boccato, C. Brennecke, S. Cenatiempo, B. Schlein, {\it Optimal rate for Bose-Einstein condensation in the Gross--Pitaevskii regime.} Commun. Math. Phys {\bf 376}, 1311–1395 (2020)  \bibskip

\bibitem{BOSAS}
M. Brooks, J. Oldenburg, D. Saint Aubin, B. Schlein, {\it Third Order Upper Bound for the Ground State Energy of the Dilute Bose Gas}, arXiv:2506.04153 (2025)  \bibskip

\bibitem{CapDeu2024} M. Caporaletti, A. Deuchert, \textit{Upper bound for the grand canonical free energy of the Bose gas in the Gross--Pitaevskii limit for general interaction potentials}, Ann. Henri Poincaré (2024)  \bibskip

\bibitem{CarOlgSA2025} C. Caraci, A. Olgiati, D. Saint Aubin, B. Schlein, \textit{Third Order Corrections to the Ground State Energy of a Bose Gas in the Gross–Pitaevskii Regime}, Commun. Math. Phys. \textbf{406}, 153 (2025) \bibskip

\bibitem{Dingle}
R. Dingle. {\it The zero-point energy of a system of particles}, The London, Edinburgh, and
Dublin Philosophical Magazine and Journal of Science {\bf 40}, no. 304, 573–578 (1949) \bibskip

\bibitem{DeuSei-20} A. Deuchert, R. Seiringer, \textit{Gross--Pitaevskii limit of a homogeneous Bose gas at positive temperature}, Arch. Rational Mech. Anal. \textbf{236}, 1217–1271 (2020)   \bibskip

\bibitem{DeuSeiYng2019} A. Deuchert, R. Seiringer, J. Yngvason, \textit{Bose-Einstein Condensation for a Dilute Trapped Gas at Positive Temperature}, Commun. Math. Phys. \textbf{368}, 723 (2019) \bibskip

\bibitem{Dyson1957} F. J. Dyson, \textit{Ground-State Energy of a Hard-Sphere Gas}, Physical Review \textbf{106.1}, 20 (1957) \bibskip

\bibitem{ErdSchlYau2008} L. Erd\"os, B. Schlein, H.-T. Yau, \textit{Ground-state energy of a low-density Bose gas: A second-order upper bound}, Phys. Rev. A \textbf{78}, 053627 (2008) \bibskip

\bibitem{Foll1989} G. Folland, \textit{Harmonic Analysis in Phase Space}, Princeton University Press, Princeton, New Jersey (1989) \bibskip

\bibitem{FS}
S. Fournais, J. P. Solovej, \textit{The energy of dilute Bose gases}, {\it Ann. Math.} {\bf 192}(3), 893--976 (2020) \bibskip

\bibitem{FS2} S. Fournais, J. P. Solovej, \textit{The energy of dilute Bose gases II: The general case}, Invent. Math. {\bf 232}(2), 863--994 (2023) \bibskip

\bibitem{FGJMO2024} S. Fournais, T. Girardot, L. Junge, L. Morin, M. Olivieri, \textit{The Ground State Energy of a Two-Dimensional Bose Gas}, Commun. Math. Phys. \textbf{405}, 59 (2024) \bibskip

\bibitem{HHNST} F. Haberberger, C. Hainzl, P. T. Nam, R. Seiringer, A. Triay, \textit{The free energy of dilute Bose gases at low temperatures}, arXiv:2304.02405 (2024)  \bibskip

\bibitem{HHST} F. Haberberger, C. Hainzl, B. Schlein, A. Triay, \textit{Upper Bound for the Free Energy of Dilute Bose Gases at Low Temperature}, arXiv: 2405.03378  (2024) \bibskip

\bibitem{HuPi1959} N. M. Hugenholtz, D. Pines, \textit{Ground-State Energy and Excitation Syectrum of a System of Interacting Bosons}, Phys. Rev. \textbf{116}, no. 3, 489–506 (1959) \bibskip

\bibitem{Jastrow1955}
R. Jastrow, \textit{Many-body problem with strong forces}, Phys. Rev. {\bf 98}, no. 5, 1479--1484 (1955) \bibskip

\bibitem{LHY1957} T. D. Lee, K. Huang, C. N. Yang, \textit{Eigenvalues and Eigenfunctions of a Bose System of Hard Spheres and Its Low-Temperature Properties}, Phys. Rev. \textbf{106}, 6,
1135--1145 (1957) \bibskip

\bibitem{LiSeiYng2000} E. H. Lieb, R. Seiringer, J. Yngvason, \textit{Bosons in a trap: A rigorous derivation of the Gross-Pitaevskii energy functional}, Phys. Rev. A \textbf{61}, 043602 (2000)  \bibskip


\bibitem{LiYng98}
E. H. Lieb, J. Yngvason, {\it Ground State Energy of the low density Bose Gas.} Phys.
Rev. Lett. {\bf 80}, 2504--2507 (1998) \bibskip

\bibitem{MaySei2020} S. Mayer, R. Seiringer, \textit{The free energy of the two-dimensional dilute Bose gas. II. Upper bound}, J. Math. Phys. \textbf{61}, 061901 (2020) \bibskip

\bibitem{NapReuSol2017} M. Napiórkowski, R. Reuvers, J. P. Solovej, \textit{The Bogoliubov free energy functional II: the dilute limit}, Commun. Math. Phys. \textbf{360}, 347 (2017) \bibskip

\bibitem{Robinson1971} D. W. Robinson, \textit{The Thermodynamic Pressure in Quantum Statistical Mechanics}, Springer Lecture Notes in Physics, Vol. \textbf{9} (1971)  \bibskip

\bibitem{Ruelle1999} D. Ruelle, \textit{Statistical Mechanics. Rigorous Results}, World Scientific (1999) \bibskip

\bibitem{Sa1959} K. Sawada, \textit{Ground-State Energy of Bose-Einstein Gas with Repulsive Interaction}, Phys. Rev. \textbf{116}, no. 6, 1344–1358 (1959) \bibskip

\bibitem{Seiringer2006} R. Seiringer, \textit{The thermodynamic pressure of a dilute Fermi gas}, Commun. Math. Phys. \textbf{261}, 729 (2006) \bibskip

\bibitem{Sei-T}
R. Seiringer, \textit{Free Energy of a Dilute Bose Gas: Lower Bound}, Commun. Math. Phys. {\bf 279}, 595--636 (2008) \bibskip

\bibitem{Wu1959} T. T. Wu, \textit{Ground State of a Bose System of Hard Spheres}, Phys. Rev. \textbf{115}, no. 6, 1390--1404 (1959) \bibskip

\bibitem{YauYin2009} H.-T. Yau, J. Yin, \textit{The Second Order Upper Bound for the Ground Energy of a Bose Gas}, J. Stat. Phys. \textbf{136}, 453 (2009) \bibskip

\bibitem{Yin2010} J. Yin, \textit{Free Energies of Dilute Bose Gases: Upper Bound}, J. Stat. Phys. \textbf{141}, 683–726 (2010) 


\end{thebibliography}
\end{document}